\documentclass[10pt]{article}
\usepackage{multicol}
\usepackage{graphicx}
\usepackage{amsmath}
\usepackage[a4paper]{geometry}
\usepackage{hyperref}
\usepackage{rotating}
\usepackage{xcolor}
\usepackage{caption}
\usepackage{csquotes}

\begin{document}

\begin{center}


{\Large \bf Comparing EPOS-4, EPOS-LHC, and SMASH for identified-hadron observables in the NICA energy range}

\vskip1.0cm
Murad Badshah$^{1,}${\footnote{Corresponding author: murad\_phy@awkum.edu.pk; muradbadshah25295@gmail.com}},
Haifa I. Alrebdi$^{2,}$\footnote{hialrebdi@pnu.edu.sa},
Sana Raza Khan$^{3}$,
Muhammad Ajaz$^{1,}${\footnote{Corresponding author: ajaz@awkum.edu.pk}}\\
{\small\it $^1$Department of Physics, Abdul Wali Khan University Mardan, Mardan 23200, Pakistan \\
$^2$Department of Physics, College of Science, Princess Nourah bint Abdulrahman University, P.O. Box 84428, Riyadh 11671, Saudi Arabia\\
$^3$Government Post Graduate College for Women Mardan, Mardan 23200, Pakistan \\
}
\end{center}

{\bf Abstract:}
 We report on a systematic simulation study of identified-hadron production in minimum bias $Au+Au$ collisions at $\sqrt{s_{NN}} = 6,~7$ and $8~GeV$. Event samples were produced using three state-of-the-art theoretical frameworks that represent different microscopic assumptions: the hybrid/string-based event generators EPOS-LHC and EPOS-4, and the purely hadronic transport model SMASH. We study a set of complementary observables probing longitudinal baryon transport, transverse collective motion, hadronization dynamics, and strangeness production: rapidity $dN/dy$, transverse-momentum spectra $dN/dp_T$, two-dimensional $p_T-y$ density maps, $v_2/n_q$ vs. $p_T/n_q$ and a set of particle-yield ratios ($\pi^-/\pi^+,~K^-/K^+,~\bar{p}/p,~K^+/\pi^+,~K^-/\pi^-,~p/\pi^+~\Lambda/\pi^+)$. Charged and neutral pion yields and shapes ($dN/dy$ and $dN/dp_T$) are similar in all three models at these energies (resonance feed-down and isospin constraints wash out model-dependent early-time dynamics); strange mesons and baryons are highly model sensitive (EPOS-4 yields the highest midrapidity $K^+$ and $\Lambda$ yields and hardest kaon and strange-baryon $p_T$ spectra, EPOS-LHC intermediate, SMASH suppressed); two-dimensional $p_T-y$ maps show that EPOS populate higher $p_T$ over a wider rapidity interval. The NCQ-scaled elliptic flow shows the best approximate collapse in EPOS-LHC, partial collapse in SMASH and EPOS-4, indicating that EPOS-LHC transports a more coherent quark-level anisotropy to the hadron stage at these energies. Mostly the difference between the models increases with collision energy over the $6-8~GeV$ region, and the difference between hadronic and hybrid/partonic scenarios identifies several observables, intermediate-$p_T$ baryon/meson ratios, $\Lambda/\pi^+$, $y$ dependence of $p_T$, $K^-/K^+$ ($p_T$ dependence), and NCQ-scaled $v_2$, that are particular discriminators in the NICA energy domain.\\


\vskip1.0cm

\begin{multicols}{2}

{\section{Introduction}}
Relativistic collisions of heavy ions are a unique way to investigate the properties of strongly interacting matter in extreme temperature and density conditions. The theory of strong interactions, Quantum Chromodynamics (QCD), predicts a rich phase structure of such matter. Lattice QCD computations have found that at low baryon chemical potential ($\mu_B\approx 0$), hadronic matter passes through a smooth crossover to a deconfined quark-gluon plasma (QGP) at a critical temperature $T_c\approx155~MeV$ \cite{aoki2006order, borsanyi2010there, bazavov2012chiral}. This has been experimentally verified in the Relativistic Heavy Ion Collider (RHIC) and the Large Hadron Collider (LHC) where matter undergoes collective hydrodynamic flow, partonic energy loss (jet quenching), strangeness enhancement, and long-range correlations, which is a reflection of the creation of a strongly coupled QGP \cite{adams2005experimental, adcox2005formation, stephanov2004qcd, fodor2004critical, rajagopal2001condensed}.

However, at large baryon density, which corresponds to smaller collision energies, there is a divergence in theoretical predictions. The crossover transition itself can be indicated to shift to a first-order phase transition, with a critical endpoint \cite{magnetbeam, yang2017star, afanasiev2002energy} at the end. Model studies and QCD-inspired methods indicate that this transition can be made. One of the most important outstanding issues in the field is to establish the location of this transition line and the QCD critical point \cite{hades2009high}. With little data available and dense baryonic matter, experimental advances in this regime are difficult. 

In the past 20 years, there have been tremendous strides in the programs in the current facilities. The CERN SPS investigated $\sqrt{s_{NN}} = 6-17~GeV$ and found evidence of anomalies in strangeness production, fluctuations, and event-by-event observables \cite{blaschke2013searching}. HADES at GSI has also concentrated on lower energies ($\sqrt{s_{NN}} = 2-3~GeV$), including detailed measurements of hadronic spectra, flow, and dilepton production, which are needed as input to hadronic transport models \cite{hades2018centrality}. At RHIC, the systematic study of $\sqrt{s_{NN}} = 7.7-62~GeV$ has been done with the Beam Energy Scan (BES-I and BES-II) \cite{adamczyk2017bulk}. Results of STAR and PHENIX show that observables like $v_2$, kurtosis of net-proton distributions, and ratios of particles show non-monotonic behavior, which is expected to be caused by critical behavior and the alteration of baryon transport \cite{adamczyk2017bulk, luo2015energy, adare2016measurement, mitchell2013rhic}.

Although this has been achieved, the region $\sqrt{s_{NN}}= 4-11~GeV$ is relatively untouched by contemporary experiments. This energy window is the maximum net-baryon densities that can be reached in heavy-ion collisions, before reaching the lower $\mu_B$ accessible at even greater energies. Herein, the onset of deconfinement and the first-order phase transition are most likely to show up. This gap is envisioned by the upcoming NICA (Nuclotron-based Ion Collider fAcility) at JINR Dubna \cite{kekelidze2017nica, kekelidze2017feasibility}. NICA will accelerate heavy ions (Au, Bi) at $\sqrt{s_{NN}}= 4-11~GeV$ with high luminosity and a modern detector package, which will allow measuring hadronic spectra, collective flow coefficients, strangeness yields, and fluctuations of conserved charges with unprecedented precision.

Theoretical modelling is still important even though already $Au+Au$ data in the same energy window have been obtained by RHIC in the BES program \cite{adamczyk2017bulk}. These measurements, e.g., of net-proton fluctuations and flow observables in the collision energy range of $7.7 - 62.4~GeV$, provide the first constraints for the QCD phase structure in the region of high baryon density and provide motivation for an extended program at NICA. We proceed to use our calculations in this work as a model-to-model comparison in the NICA energy range and not as a definitive measure against existing RHIC data. We want to find observables on which the three descriptions have the greatest differences, and thereby future measurements at NICA and additional dedicated comparisons to RHIC BES data are able to distinguish efficiently between hadronic and hybrid descriptions. Simulations may be used to give a baseline expectation, determine which observables are discriminating, and inform the analysis strategy to be developed to analyze NICA results. Importantly, various theoretical approaches have different assumptions regarding the degrees of freedom which are relevant: purely hadronic transport models assume that the medium is an interacting hadron gas, whereas hybrid models include partonic string dynamics and hydrodynamic collective flow. Comparing predictions in such frameworks, it is possible to identify what characteristics of the results are strong and what are model assumptions.

We perform a comparative study of three state-of-the-art theoretical models, EPOS-LHC, EPOS-4, and SMASH, in this work, to systematically compare hadronic observables at $\sqrt{s_{NN}} = 6, 7$, and $8~GeV$. They both are unique modeling philosophies. EPOS-LHC is a hybrid event generator that is initially adapted to LHC data, using a core-corona representation in which dense regions are effectively hydrodynamically evolved, and dilute regions break down into strings. Its modern successor is EPOS-4, which includes better viscous hydrodynamics, event-by-event variations, and better hadronization schemes. SMASH (Simulating Many Accelerated Strongly-interacting Hadrons) is a microscopic hadronic transport model, which gives a purely hadronic start point with no explicit partonic or hydrodynamic step.

Both EPOS-LHC and EPOS-4 are here made as downward extrapolations of their standard tunes. In our $Au+Au$ simulations at $6, 7$ and $8~GeV$ collision energies, we vary the beam energy, but not model parameters. It is noteworthy that in these low energies, the predictions are to be treated as exploratory ones rather than as descriptions of existing data. Moreover, SMASH is used not as a hydrodynamic or partonic stage but as a pure hadronic transport model, and thus the three frameworks represent conceptually different models.

Our observables of interest include: rapidity ($y$) distributions, transverse momentum ($p_T$) spectra, two-dimensional $p_T-y$ correlations, elliptic flow scaling with the number of constituent quarks (NCQ scaling), and ratios of particle yields ($\pi^-/\pi^+, K^-/K^+, K^+/p, \bar{p}/p$ and $\Lambda/p$). These observables probe different parts of the collision dynamics. Baryon transport and baryon stopping are experimentally studied by the rapidity distributions. Transverse momentum spectra measure radial flow and hadronization. $p_T-y$ maps are a combination of both effects, and they can be used to identify local structures in phase space. NCQ scaling tests the collectivity of the degrees of freedom at the quark level. The ratios of the particles are sensitive to strangeness enhancement, baryon-meson competition, as well as the chemical freeze-out conditions.

This work provides a detailed theoretical background to the NICA program by a methodical study of the observables of these three models and the identification of discriminators that are capable of differentiating between the hadronic and hybrid dynamics. The findings identify the most promising observables that can serve as probes of the baryon-rich QCD medium, and they also establish the basis of future experimental validation.

{\section{Models Discussion}} 
\subsection{General remarks}
Theoretical descriptions of heavy-ion collisions generally pass through several phases: initial interactions between nucleons, initial pre-equilibrium dynamics, bulk medium dynamics, hadronization, and late-time hadronic interactions. The physics content of the predictions depends on the way of modeling these stages.

Hybrid models like EPOS are a combination of collective and microscopic string-based fragmentation. They model dense regions as fluid-like media that are hydrodynamically expanding, and dilute regions are modeled by string fragmentation. They contain partonic as well as hadronic phases, which makes them appropriate to study the interaction between collective flow and hadronization. By contrast, hadronic transport models like SMASH execute the Boltzmann equation in a test-particle Monte Carlo method with purely hadronic degrees of freedom in the current work, and do not have an explicit hydrodynamic or partonic phase in the setup we make, as EPOS-LHC/EPOS-4 does.

Comparing EPOS-LHC, EPOS-4, and SMASH, it becomes possible to evaluate the behaviour of collective effects, strangeness production and baryon stopping in the presence of a medium, when the latter is considered to be either hybrid or hadronic. This is especially noteworthy in the NICA energy window, where either or both of the hadronic and deconfined mechanisms can be present.

\subsection{EPOS-LHC}
EPOS-LHC \cite{pierog2015epos} is an event generator, which is a Gribov-Regge multiple scattering model with parton-based string construction. This method introduces a parton ladder into every parton-parton interaction and views the parton ladder as a color flux tube or string. These strings possess energy and momentum and then break up into hadrons. Various scattering processes are always done with energy-momentum conservation.

One of the key characteristics of EPOS is the core-corona approach \cite{werner2007core}. The overlapping strings merge to create a dense core in high-density regions. Fragments of strings are produced on their own in dilute regions (corona). This two-column treatment enables EPOS to extrapolate between heavy-ion collision hydrodynamic-like behavior and simple string fragmentation behavior in $pp$ or $pA$ collisions.

In the present work, we use the public EPOS-LHC tune \cite{pierog2015epos} implemented in the CRMC package. In the default tune, as already stated, EPOS-LHC is based on the standard EPOS Gribov-Regge multiple-scattering scheme. Each elementary scattering is represented by a parton ladder realized as a flux tube (string). String segments in high-density regions are assigned to a \enquote{core} which is hadronized statistically using parameterized collective flow, while low-density \enquote{corona} segments hadronize using standard string fragmentation. This produces collective-like spectra and strangeness enhancement without the need for a full (3+1)D hydrodynamic evolution with a follow-up hadronic transport afterburner. In our simulations, we generate EPOS-LHC events via CRMC in this default configuration by specifying only the colliding system, the centre-of-mass energy, and the number of events; we do not attach any external hadronic afterburner such as UrQMD.

EPOS-LHC was designed to be used at very high energies, but can be used at lower energies by extrapolation. By extrapolation, we mean that we vary the beam energy, but not model parameters. At $\sqrt{s_{NN}} = 6-8~GeV$, the core fraction is lower than at the LHC conditions, although moderate collective flow and strangeness enhancement are still included in the model. This has rendered it a handy hybrid baseline model of the NICA regime.

\subsection{EPOS-4}
The newest version of the EPOS framework is EPOS-4 \cite{jahan2022beam}, which has significant improvements to the hydrodynamic and hadronization steps. It has been created to provide a single description of the small and large systems across a wide spectrum of collision energies, ranging from several $GeV$ to several $TeV$.
Some of the important enhancements of EPOS-4 over EPOS-LHC are:

\noindent \textbf{Viscous hydrodynamics:} EPOS-4 has a complete viscous hydrodynamic step with realistic shear and bulk viscosities, which are needed to give accurate flow observables.

\noindent \textbf{Refined core-corona separation:} Core formation criteria and string fusion dynamics are better described, and offer a more realistic picture of baryon stopping.
 
\noindent \textbf{Enhanced hadronization scheme:} EPOS-4 uses a more flexible hadronization process; recombination and fragmentation are both allowed.
 
\noindent \textbf{Event-by-event fluctuations:} With better treatment of fluctuations, EPOS-4 is able to recreate higher-order flow harmonics and correlations. Here, the better treatment of fluctuations in EPOS-4 means that it generates fluctuating initial conditions event-by-event due to multiple parallel parton-ladder (Pomeron) scatterings, which leads to a fluctuating flux-tube/string configuration. After the core corona separation, it generates fluctuating hydrodynamic initial conditions for the core. This cannot be taken as a pure Monte-Carlo Glauber initialization of the energy density; a Glauber-like geometry may be involved in the sampling of nucleon configurations in nuclear collisions, but the fluctuations that are important to the further evolution are dominated by the multiple-scattering / flux-tube dynamics of microscopic size, and the core-corona process \cite{werner2024parallel}.
 
\noindent \textbf{Coupling to hadronic afterburners:} EPOS-4 easily connects to modern hadronic afterburners. In our simulations, EPOS-4 is executed with its standard hadronic cascade, i.e., with UrQMD afterburner ON.

These properties enhance the EPOS-4 to be more predictive in the intermediate energy range. EPOS-4 generally gives harder $p_T$ spectra, stronger baryon-to-meson enhancement, and stronger strangeness yields than EPOS-LHC. At $6-8~GeV$, EPOS-4 predictions will exhibit obvious departures from hadronic models, and this is an early collectivity test.

Past more comprehensive comparisons with STAR $Au+Au$ data indicate that the default EPOS-4 tune gives a qualitatively reasonable account of identified-hadron spectra to intermediate BES energies, but at $\sqrt{s_{NN}}=7.7~GeV$ it over predicts low-$p_T$ yields and gives spectra that are too soft to describe several species, and would then require a dedicated retuning to give a quantitatively reliable description at this energy \cite{werner2025heavy}. Moreover, a separate analysis comparing EPOS-4, EPOS-LHC, and Pythia 8.3 with STAR charged pions, kaons, and $p$(anti-$p$) spectra at $\sqrt{s_{NN}}$=7.7 and 11.5 GeV was already done by some of the current authors (Ref. \cite{alrebdi2026insights,waqas2026}) using the STAR centrality definition of BES as it is defined in Rivet. That paper has concluded that EPOS-4 provides the most consistent overall agreement among those three generators, especially in central events, but in the most peripheral events, EPOS-4 has noticeable differences for the studied species of particles at these energies \cite{alrebdi2026insights,waqas2026}.

\subsection{SMASH}

The Simulating Many Accelerated Strongly-interacting Hadrons (SMASH) model \cite{weil2016particle} is a microscopic model of hadronic transport intended to model heavy-ion collisions at both low and intermediate energies. SMASH is a computer code that provides a microscopic, numerical implementation of the relativistic Boltzmann equation of the hadrons and resonances \cite{tindall2017equilibration}.
\begin{equation}
    \frac{df_i}{dt}=C_i[f],
\end{equation}
with $f_i$ the distribution of the hadron species $i$, and $C_i$ the collision integral of all the allowed scattering and decay processes.

SMASH contains the entire range of hadrons and resonances to about $2.35~GeV$. It evolves test particles by binary scatterings, resonance formation and decay, and mean-field effects (if enabled), using tabulated cross sections and decay channels. Its cross sections are restricted by experimental information, where it is possible. Besides this, SMASH is capable of including mean-field nuclear potentials, which are significant at SIS energies in the context of collective flow.

\noindent \textbf{Key features of SMASH:}
\noindent i) Resonance-based interactions: The majority of the inelastic reactions are through resonance formation and decay.
 ii) Detailed hadronic rescattering: Final-state hadrons are rescattered many times, re-allocating energy and forming spectra.
 iii) None of the partonic stages: When used as stand alone, SMASH lacks deconfined degrees of freedom and hydrodynamics. Any collectivity is only the result of rescattering.
iv) Versatility: In current work, we use SMASH as a pure hadronic transport model; however, it has also found application as a hadronic afterburner in hybrid schemes, including the SMASH-vHLLE hybrid scheme \cite{schafer2022particle} and related low-energy schemes \cite{goes2025pushing}, which are directly applicable to NICA/BES energies. Being an all-hadronic minimum, SMASH serves as a necessary point of reference. It is predicted to have strong baryon stopping, wide rapidity distributions of protons, and steeply falling $p_T$ distributions in the $6-8 ~GeV$ regime. The comparison of SMASH and EPOS models shows the importance of partonic and collective dynamics.

\subsection{Modeling Philosophy of intermediate energies}
The three models are not only technically different, but have different philosophical assumptions regarding the active degrees of freedom at intermediate energies ($\sqrt{s_{NN}} = 4-11~GeV$). EPOS-LHC and EPOS-4 presuppose that collective dynamics and hydrodynamic expansion, and fusion of strings are already important even at these relatively low energies, albeit to a smaller degree than at RHIC or LHC. They are thus early pressure build-up, partonic collectivity, and enhanced strangeness production mechanisms. In SMASH, in contrast, the medium is assumed to be a strongly interacting hadronic gas at all times. Any collective effects are only due to rescattering amongst hadrons. This is a fundamental difference. In the case of future NICA measurements of hard $p_T$ spectra, large baryon-to-meson ratios, and nearly NCQ scaling of elliptic flow, this would support the hybrid paradigm. On the other hand, when spectra are steep, strangeness is suppressed, and flow scaling is weak, the hadronic transport picture is supported. In this way, the use of these models in the NICA window makes it possible to test the rival theoretical philosophies in a real way.

\section{Results and discussion}
In this work, $50,000$ $Au+Au$ events were analyzed per model and per collision energy. The generator outputs were stored in HepMC format and processed with C++ analysis codes. Only final-state particles were used in all analyses, requiring status equal to $ 1$. Identified particle species were selected by PDG codes. Transverse-momentum spectra were filled over $p_T$ from $0$ to $5~GeV/c$ (only $p_T \leq 3~GeV/c$ is shown in Fig. 4—6; bins at higher $p_T$ are generated in the analysis but are omitted from the plots due to limited or almost no statistics), and rapidity distributions over $y$ from $-3.5$ to $+3.5$, where rapidity was computed from the particle four-momentum as $y = 0.5~\ln[(E + p_z)/(E - p_z)]$, with safeguards for numerically ill-defined cases.

One-dimensional distributions were normalized per event and per bin width, producing per-event differential yields, while the two-dimensional $y-p_T$ maps were normalized per event only. Particle ratios were computed as bin-by-bin divisions of the corresponding normalized $p_T$ spectra. The minimum-bias character of the samples is defined at the generator level: EPOS-4 was run with centrality set to $0$, and SMASH sampled impact parameter over $b = 0-15~fm$ with areal (“quadratic”) sampling. For EPOS-LHC generated through CRMC, the collision is by default set to minimum bias.

\subsection{Rapidity distribution}
\begin{figure*}
\centering
\includegraphics[width=0.322\textwidth]{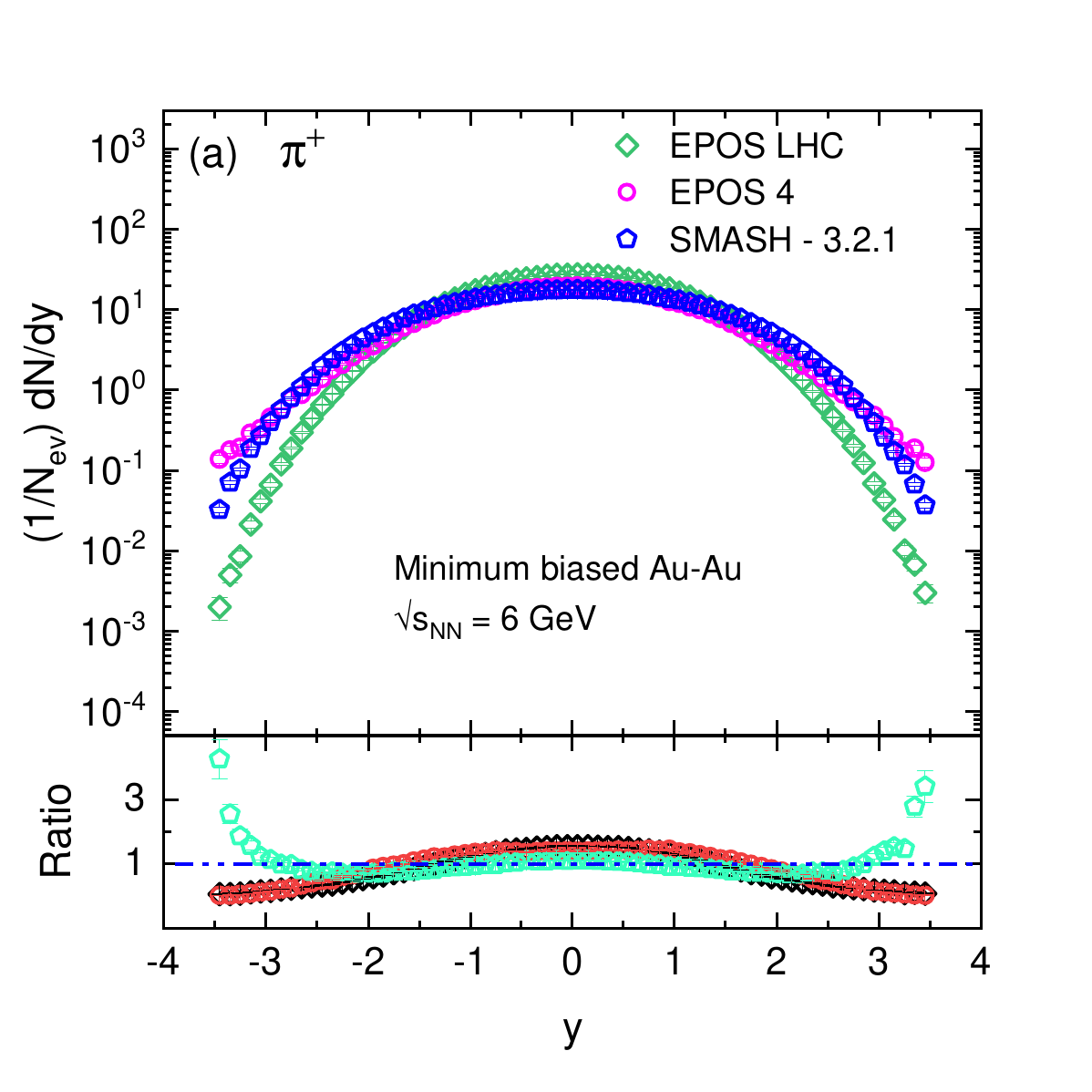}
\includegraphics[width=0.322\textwidth]{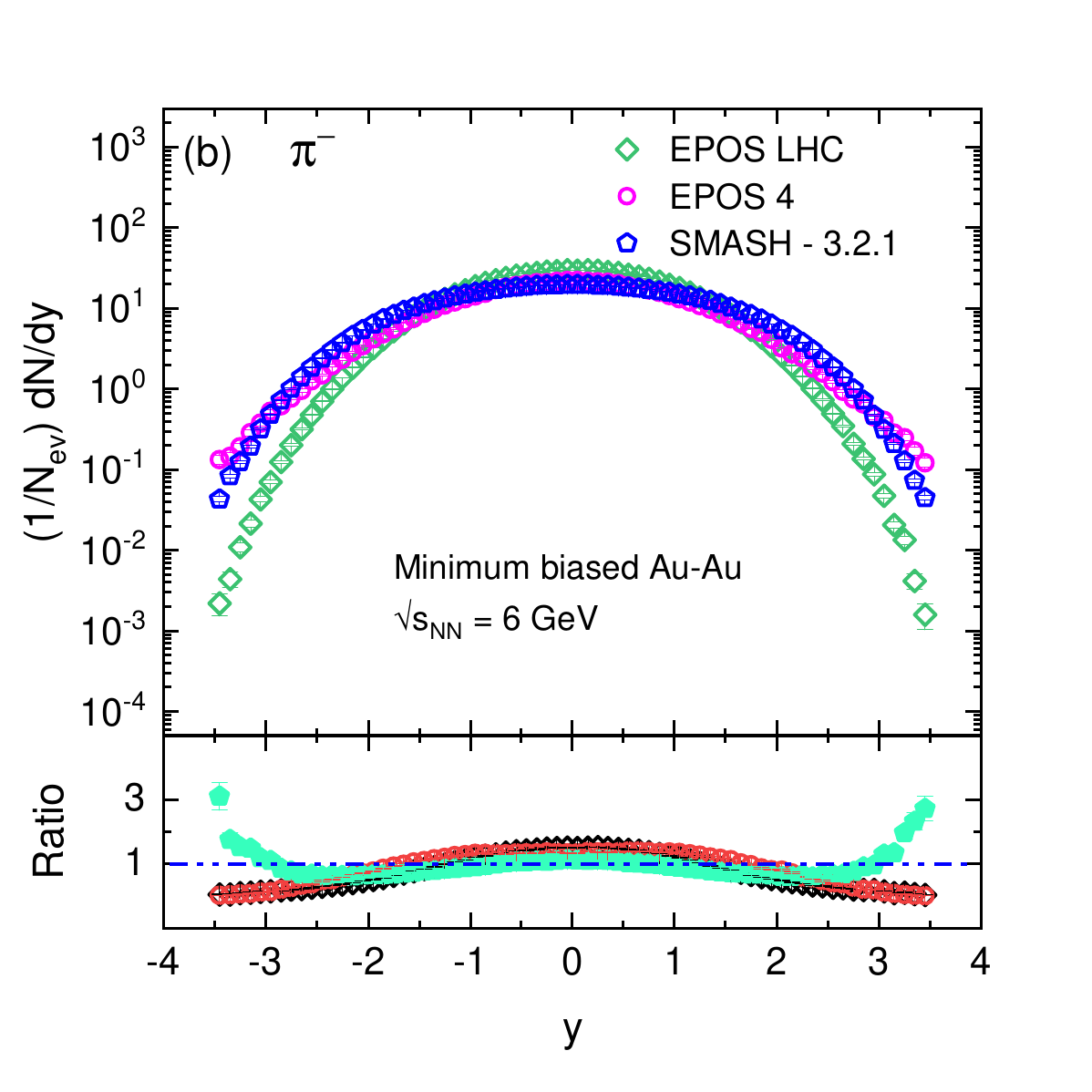}\vspace{-0.35cm} 
\includegraphics[width=0.322\textwidth]{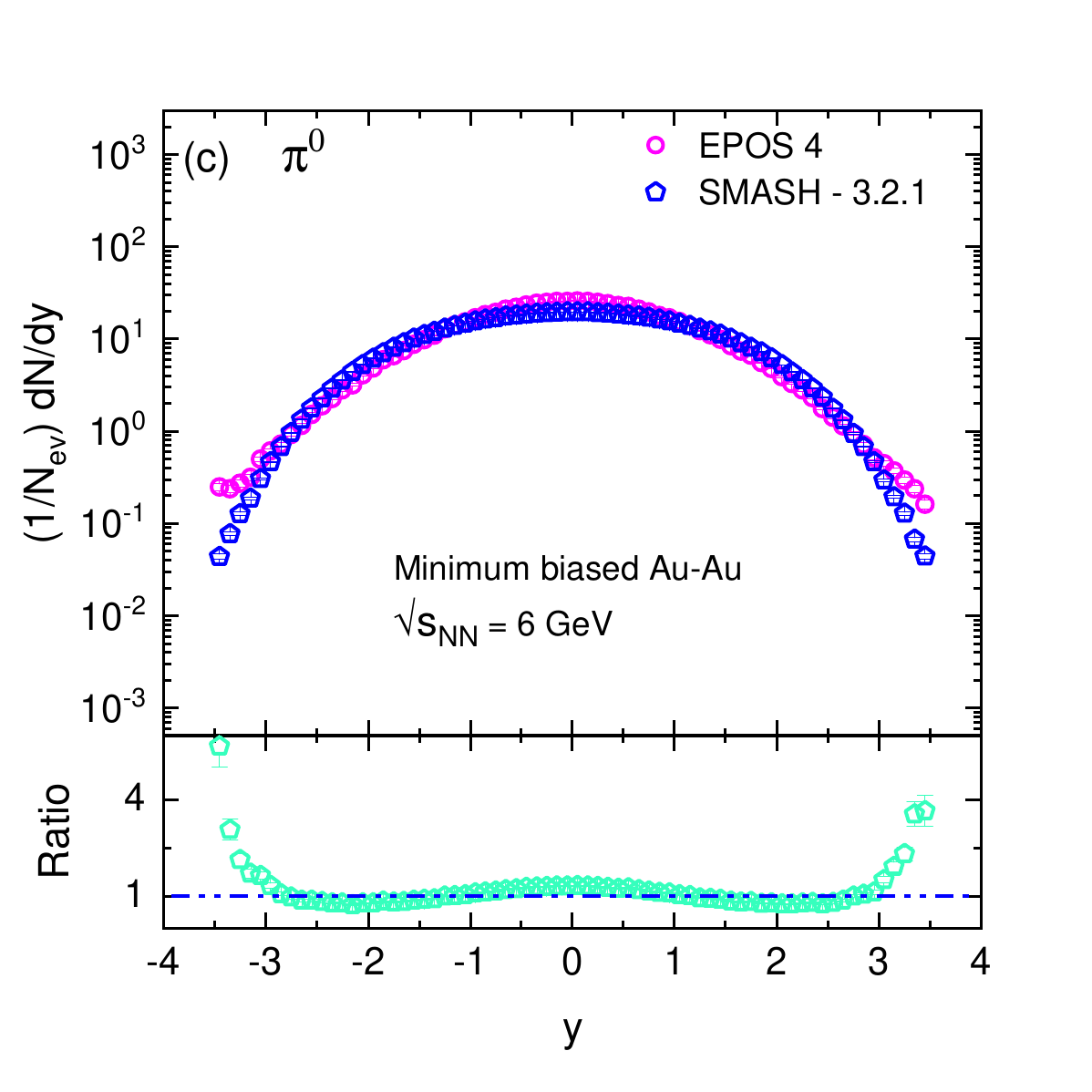}
\includegraphics[width=0.322\textwidth]{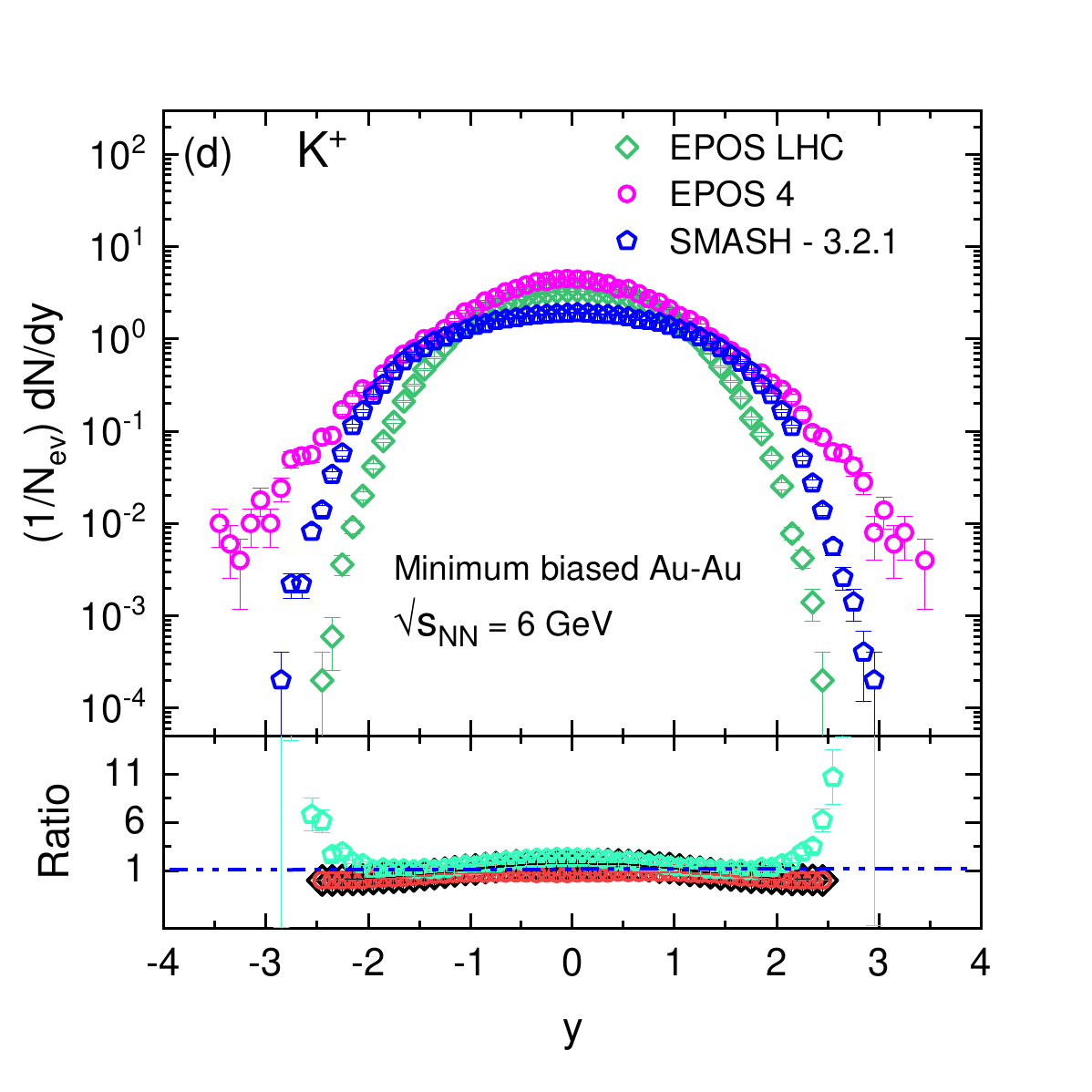}\vspace{-0.35cm}
\includegraphics[width=0.322\textwidth]{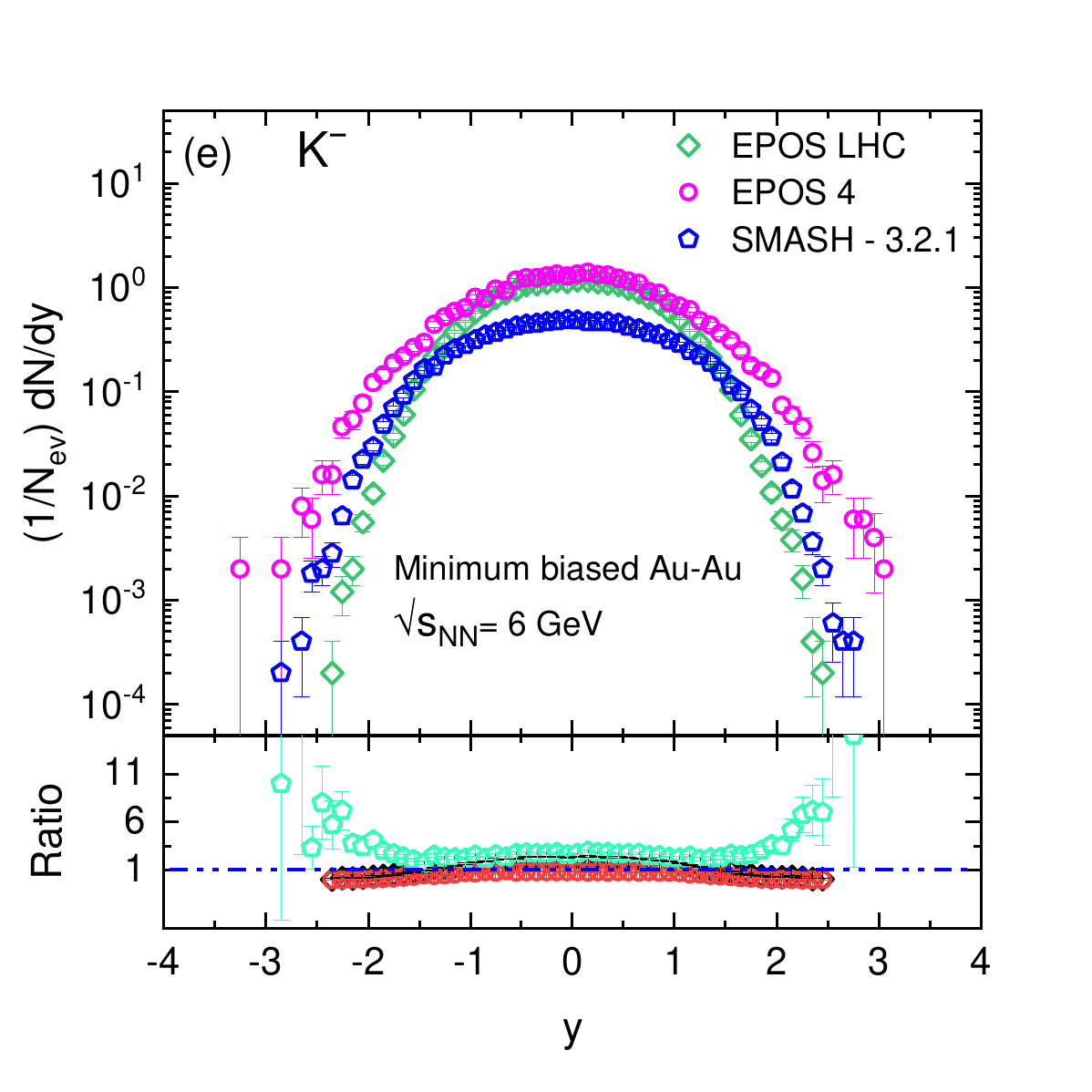}
\includegraphics[width=0.322\textwidth]{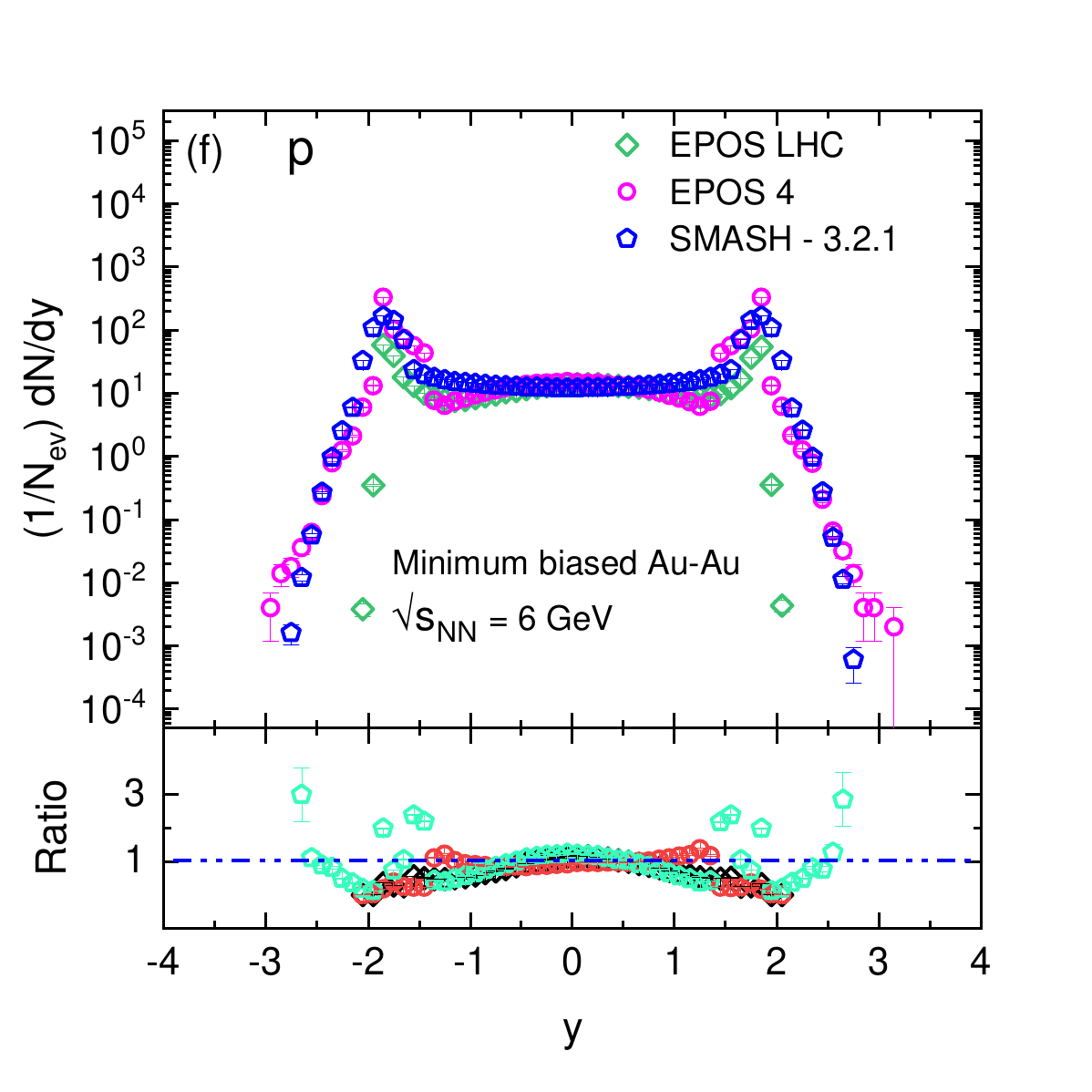}\vspace{-0.35cm}
\includegraphics[width=0.322\textwidth]{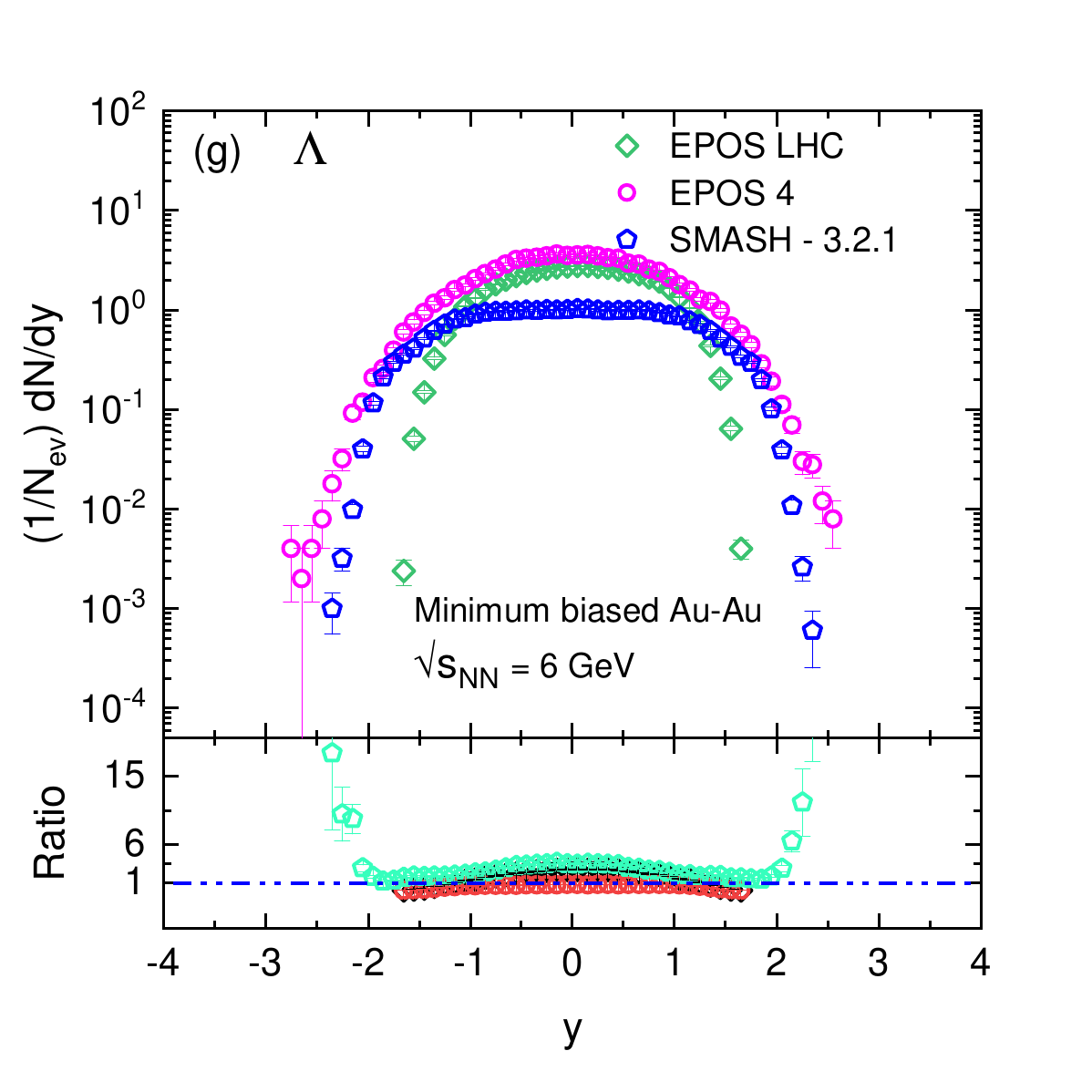}
\includegraphics[width=0.322\textwidth]{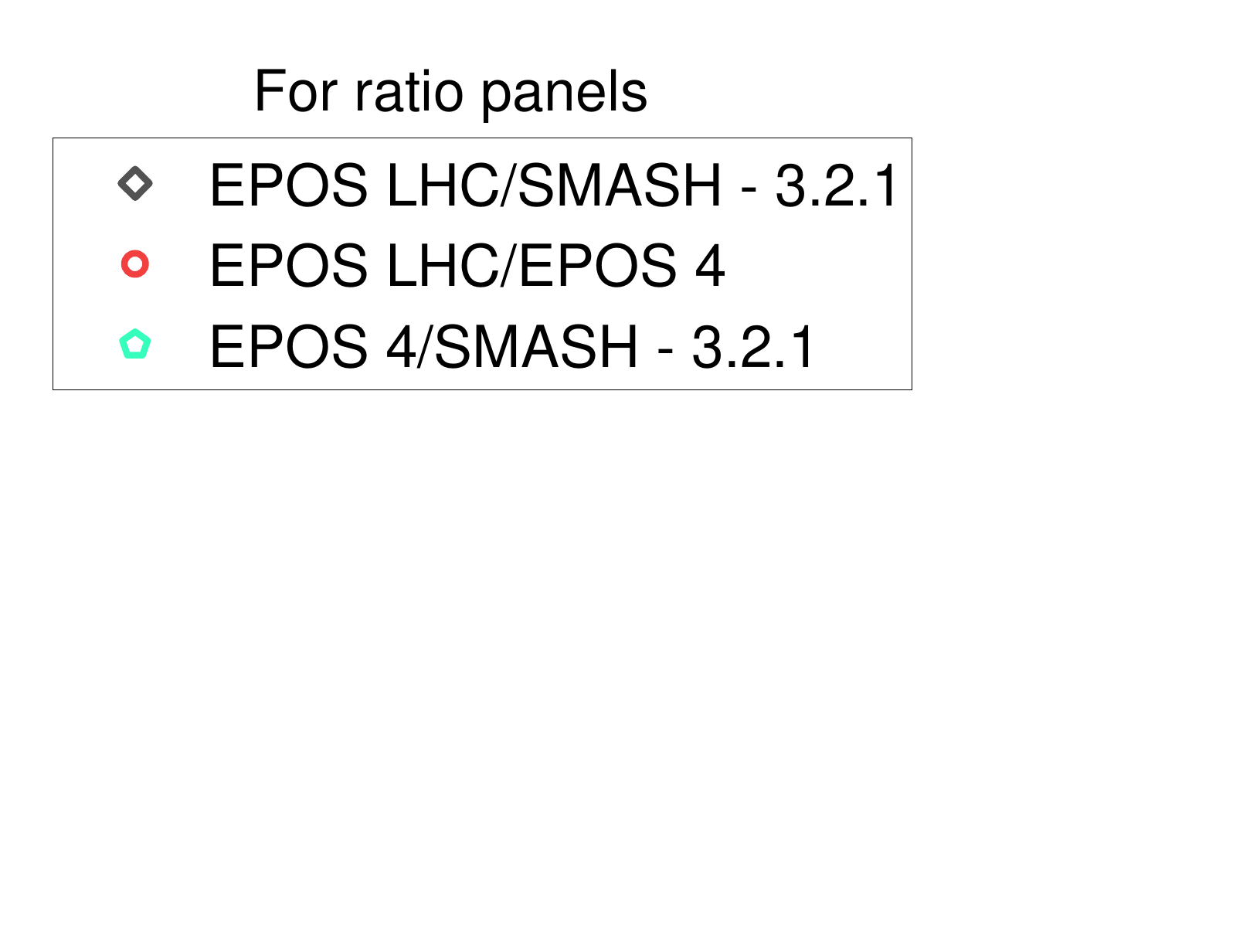}\vspace{-0.35cm}
\caption {Rapidity distributions (event-normalized $dN/dy$) of identified hadrons at minimum-bias in $Au+Au$ collisions at $\sqrt{s_{NN}}=6~GeV$ forecasted by the three models. The results of EPOS-LHC, EPOS-4, and SMASH for various species of hadrons are presented in panels (a)-(g) to highlight the differences in total yield, width in rapidity, and mid-rapidity behaviour. Model legends are in the panels. Each plot has a ratio panel at its bottom, simplifying the model to model comparison.}
\end{figure*}

\begin{figure*}
\centering
\includegraphics[width=0.322\textwidth]{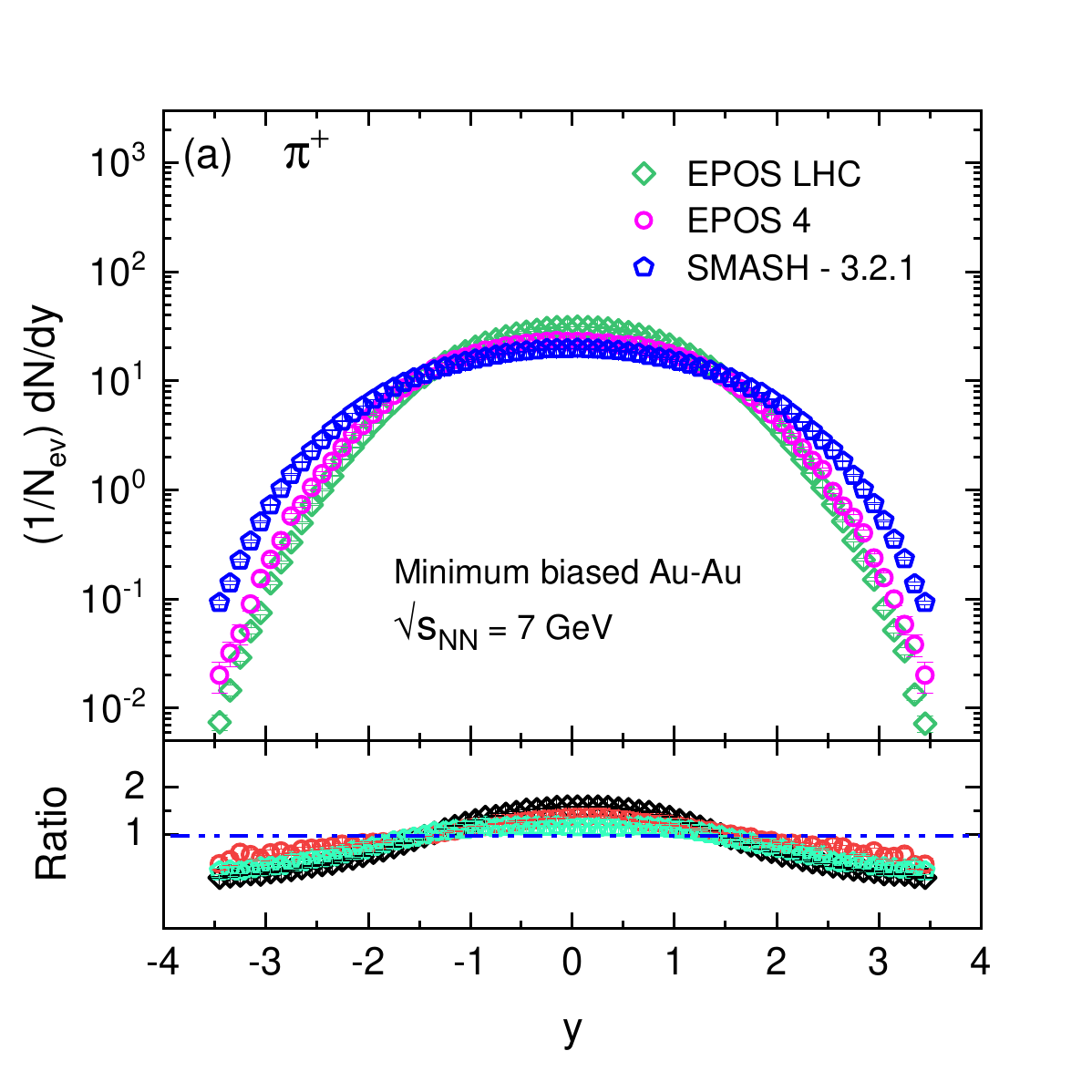}
\includegraphics[width=0.322\textwidth]{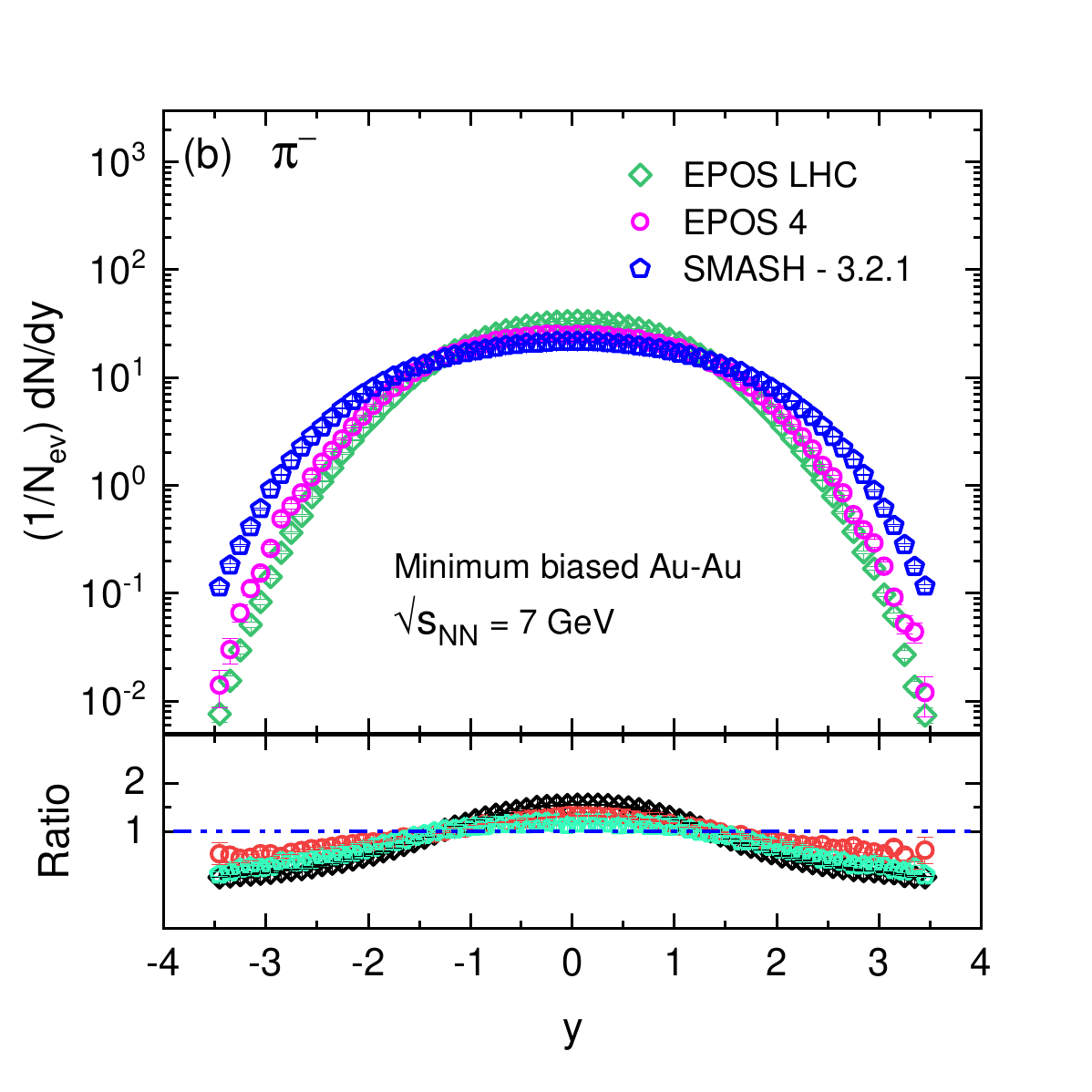}\vspace{-0.35cm} 
\includegraphics[width=0.322\textwidth]{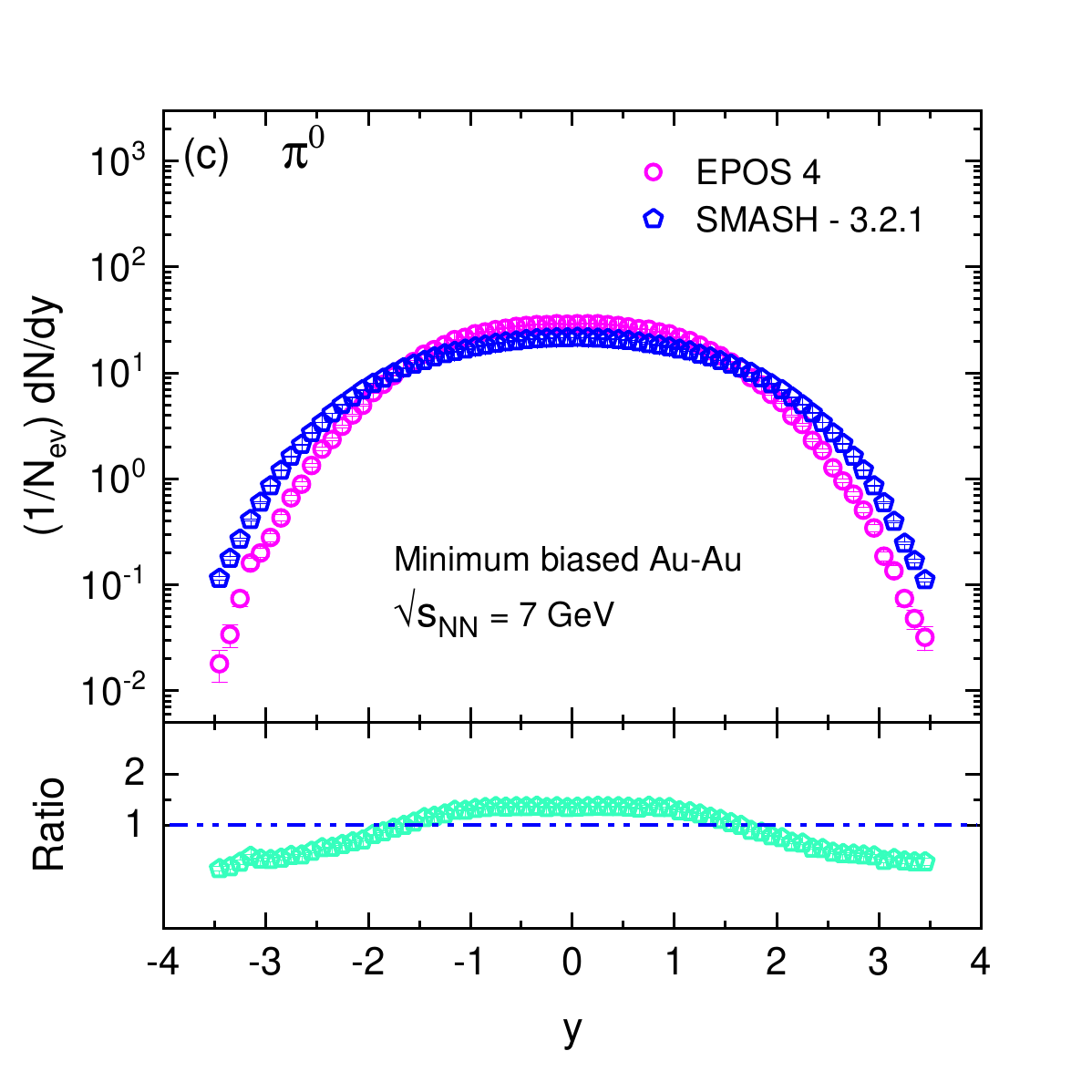}
\includegraphics[width=0.322\textwidth]{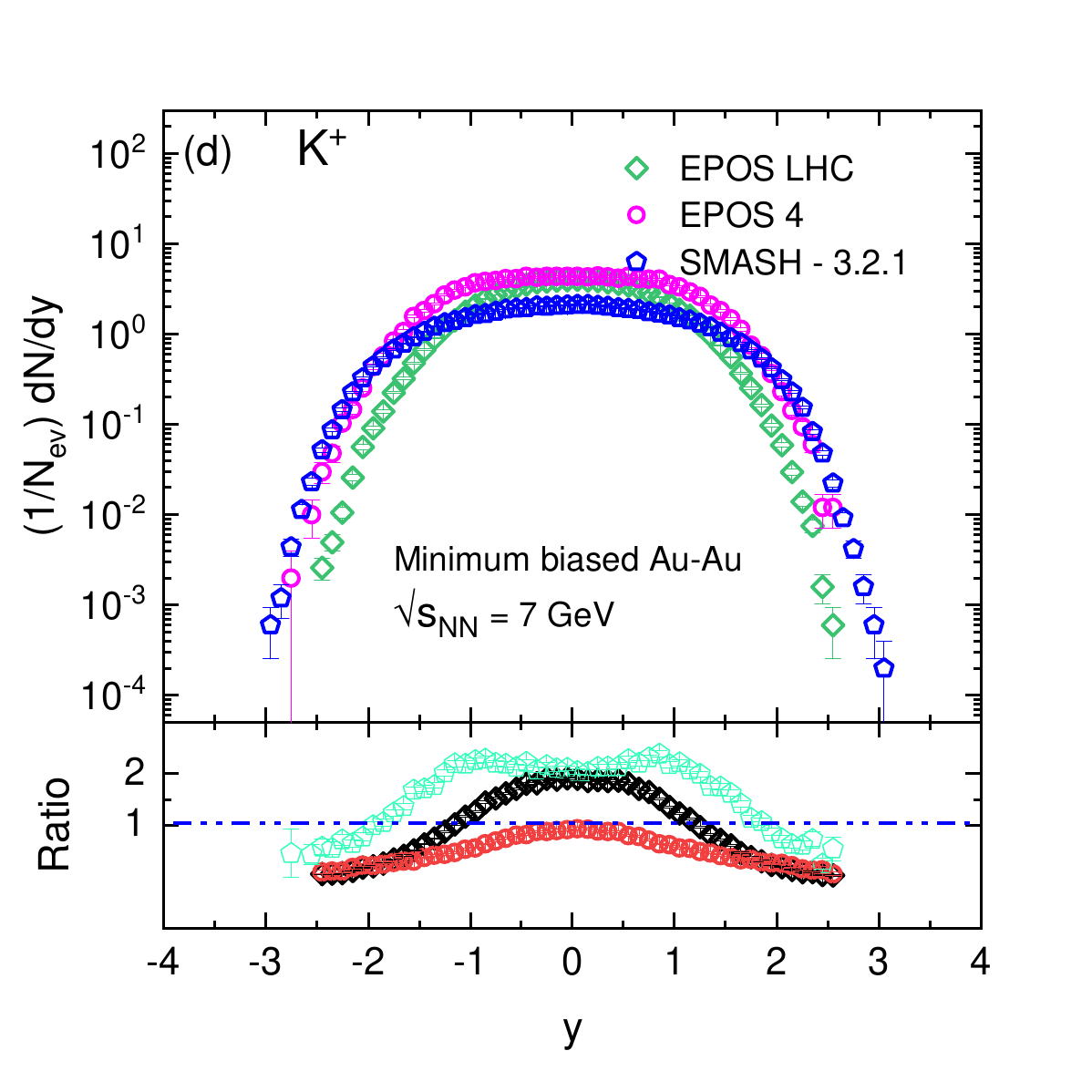}\vspace{-0.35cm}
\includegraphics[width=0.322\textwidth]{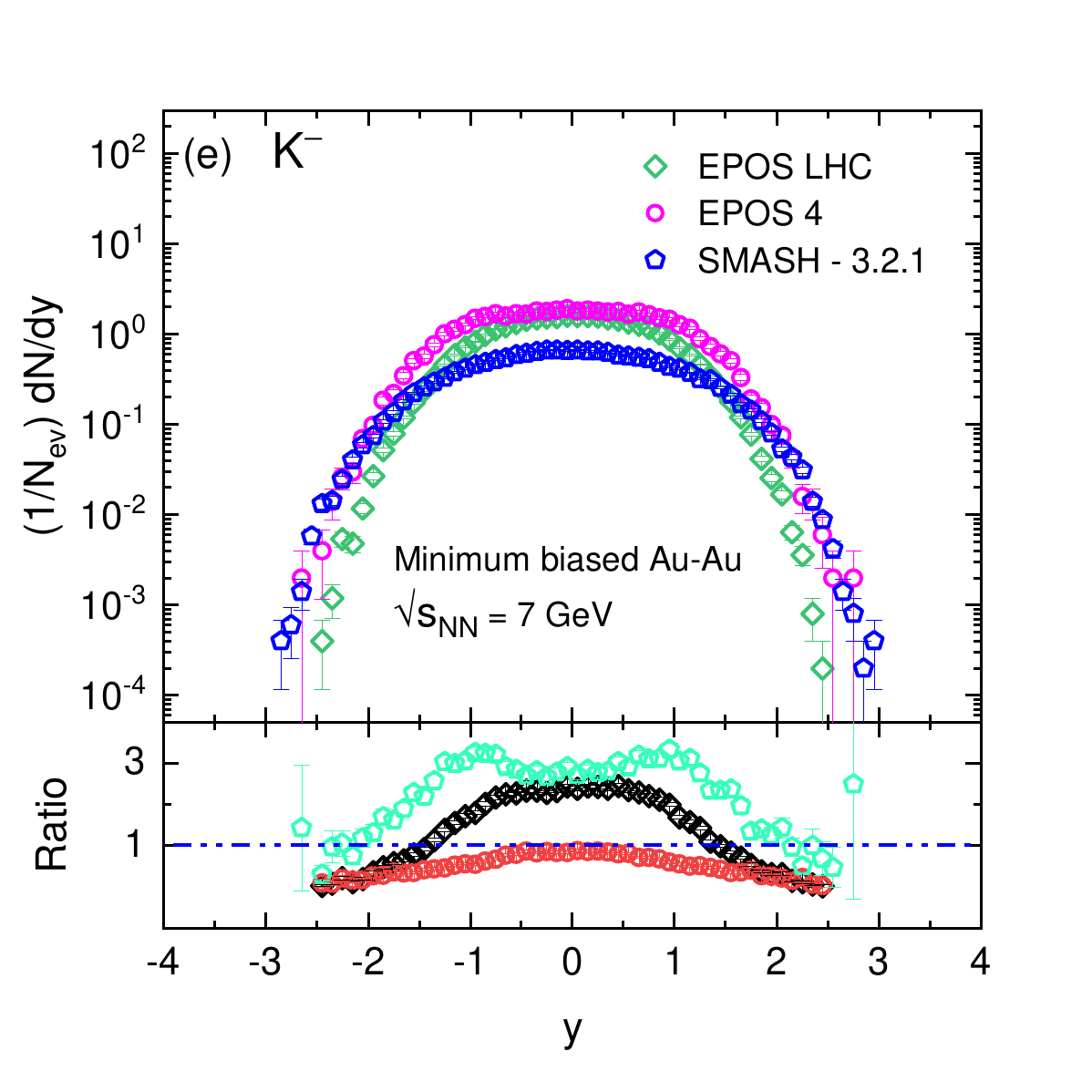}
\includegraphics[width=0.322\textwidth]{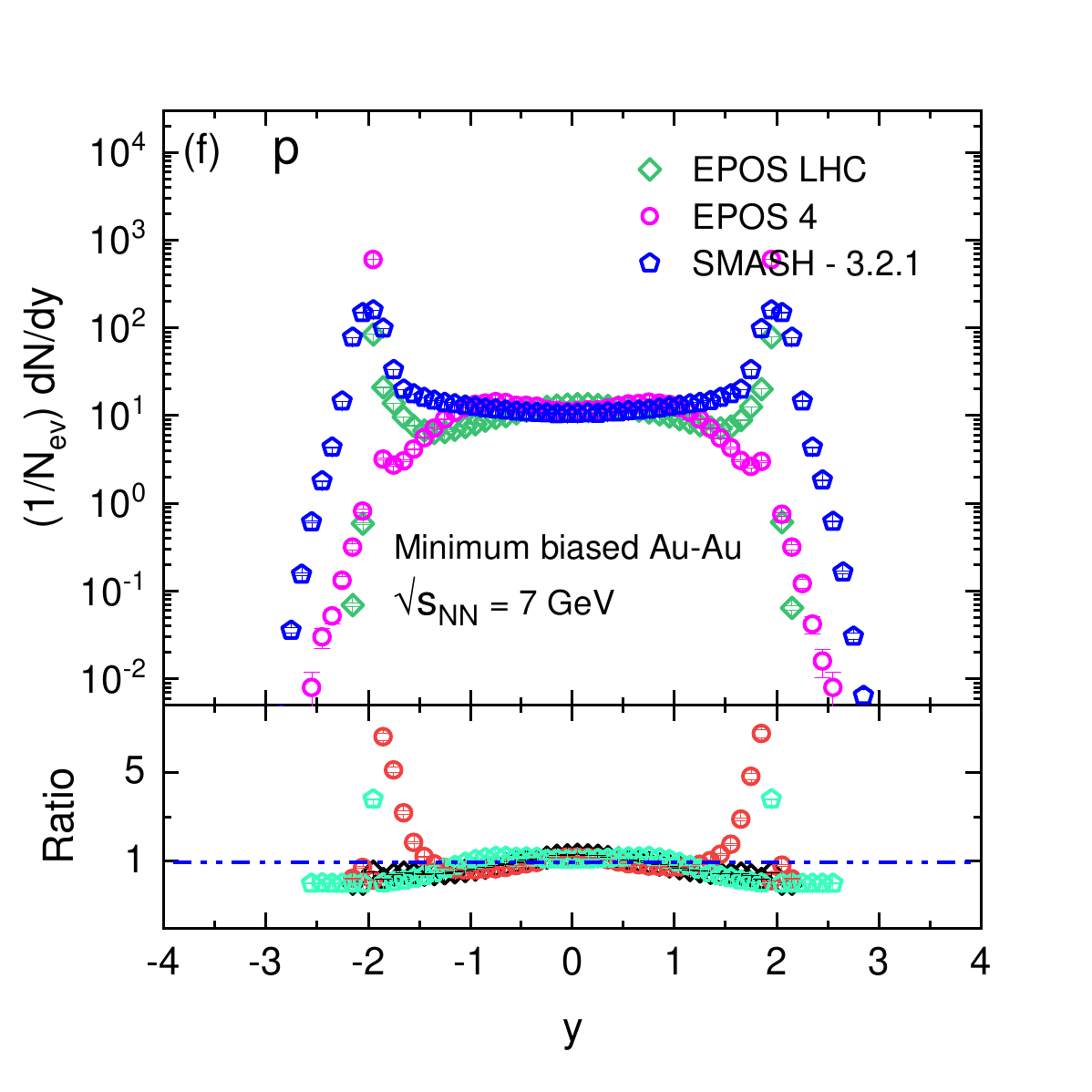}\vspace{-0.35cm}
\includegraphics[width=0.322\textwidth]{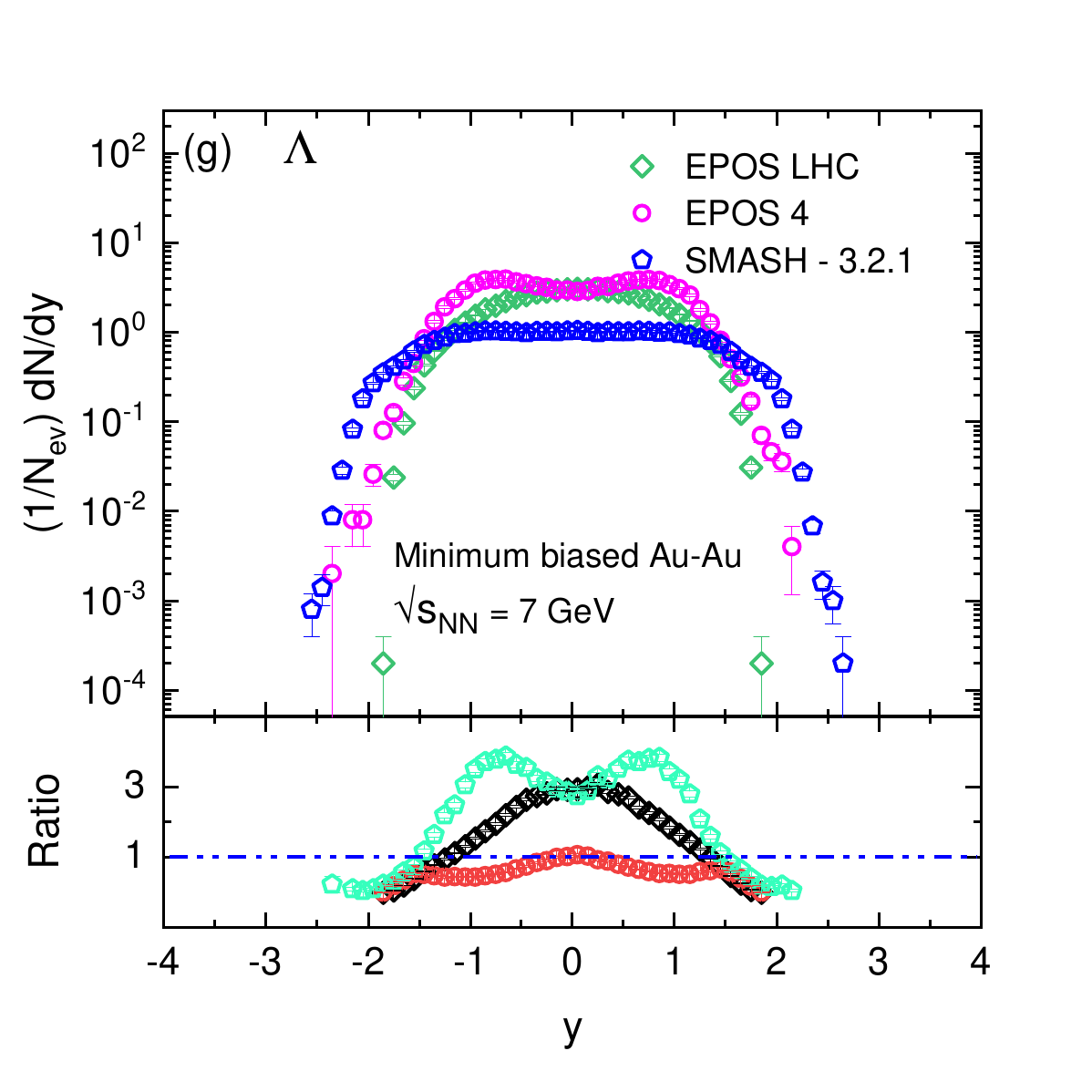}
\includegraphics[width=0.322\textwidth]{legends.pdf}\vspace{-0.35cm}
\caption {Rapidity distributions (event-scaled $dN/dy$) of hadrons (identified) in minimum-bias $Au+Au$ collisions, on a scale of $\sqrt{s_{NN}}=7~GeV$. The three model predictions (EPOS-LHC, EPOS-4, SMASH) are illustrated to highlight the energy evolution between 6 and 7 GeV and model sensitivity in strange/higher-mass hadrons.}
\end{figure*}
\begin{figure*}
\centering
\includegraphics[width=0.3222\textwidth]{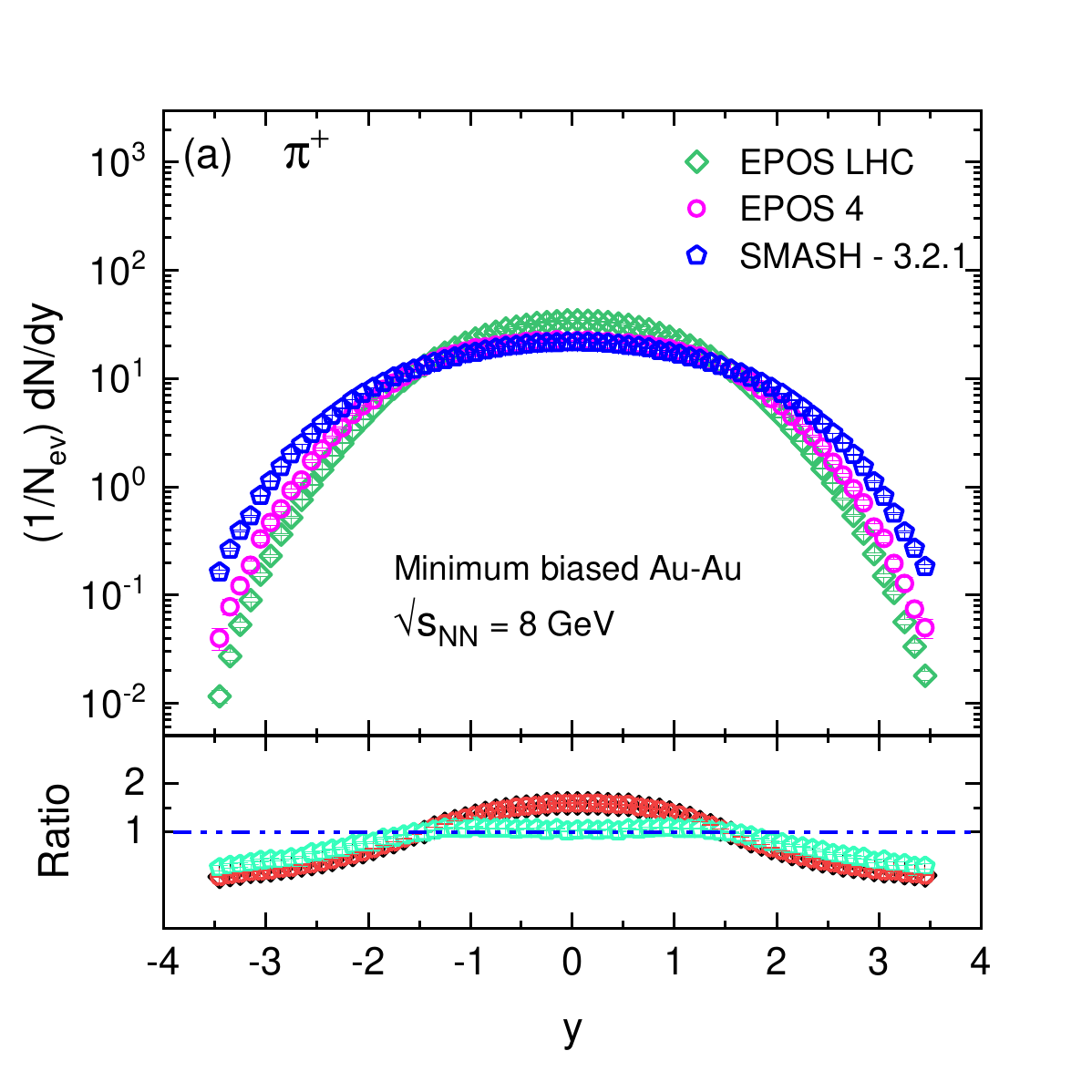}
\includegraphics[width=0.3222\textwidth]{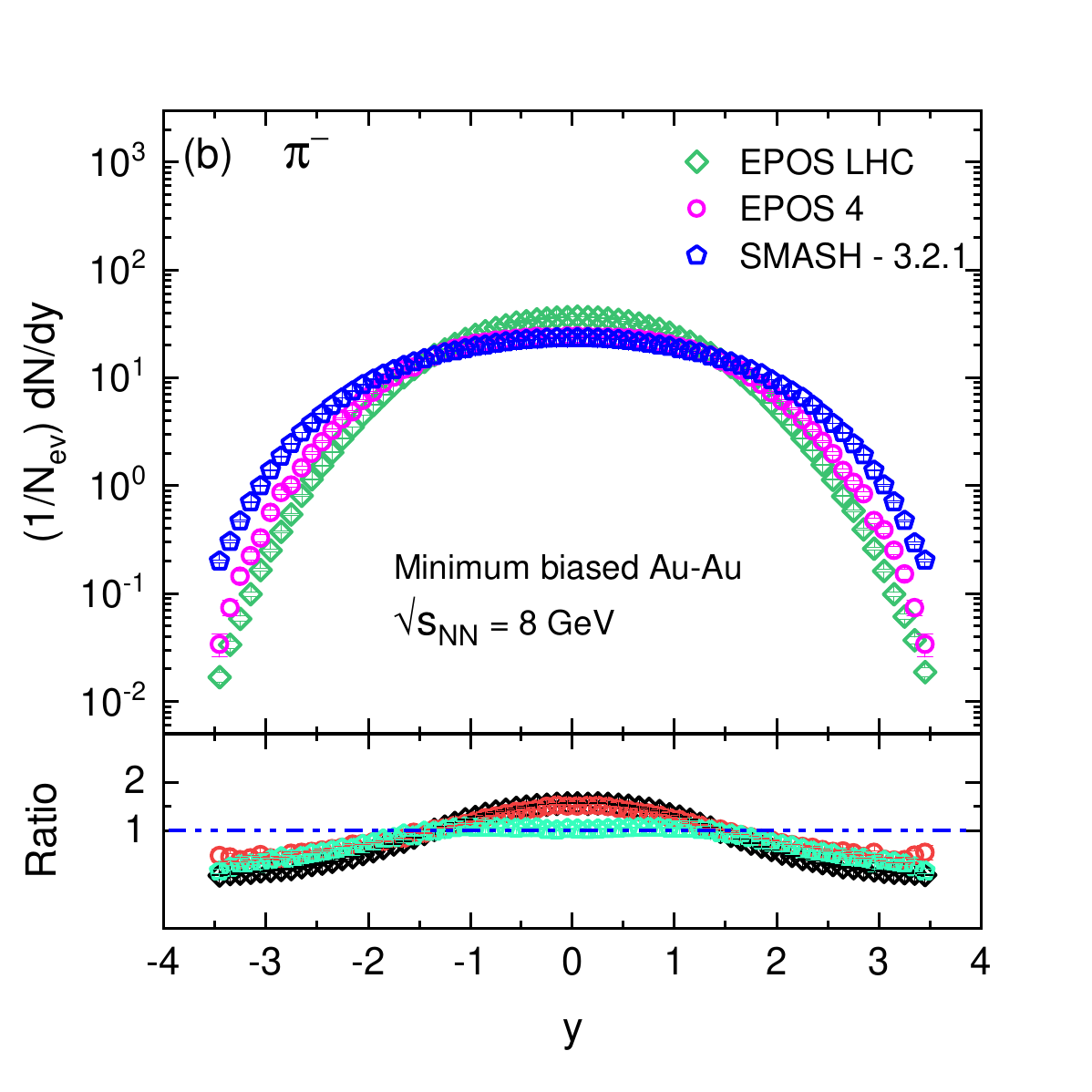}\vspace{-0.35cm}     
\includegraphics[width=0.3222\textwidth]{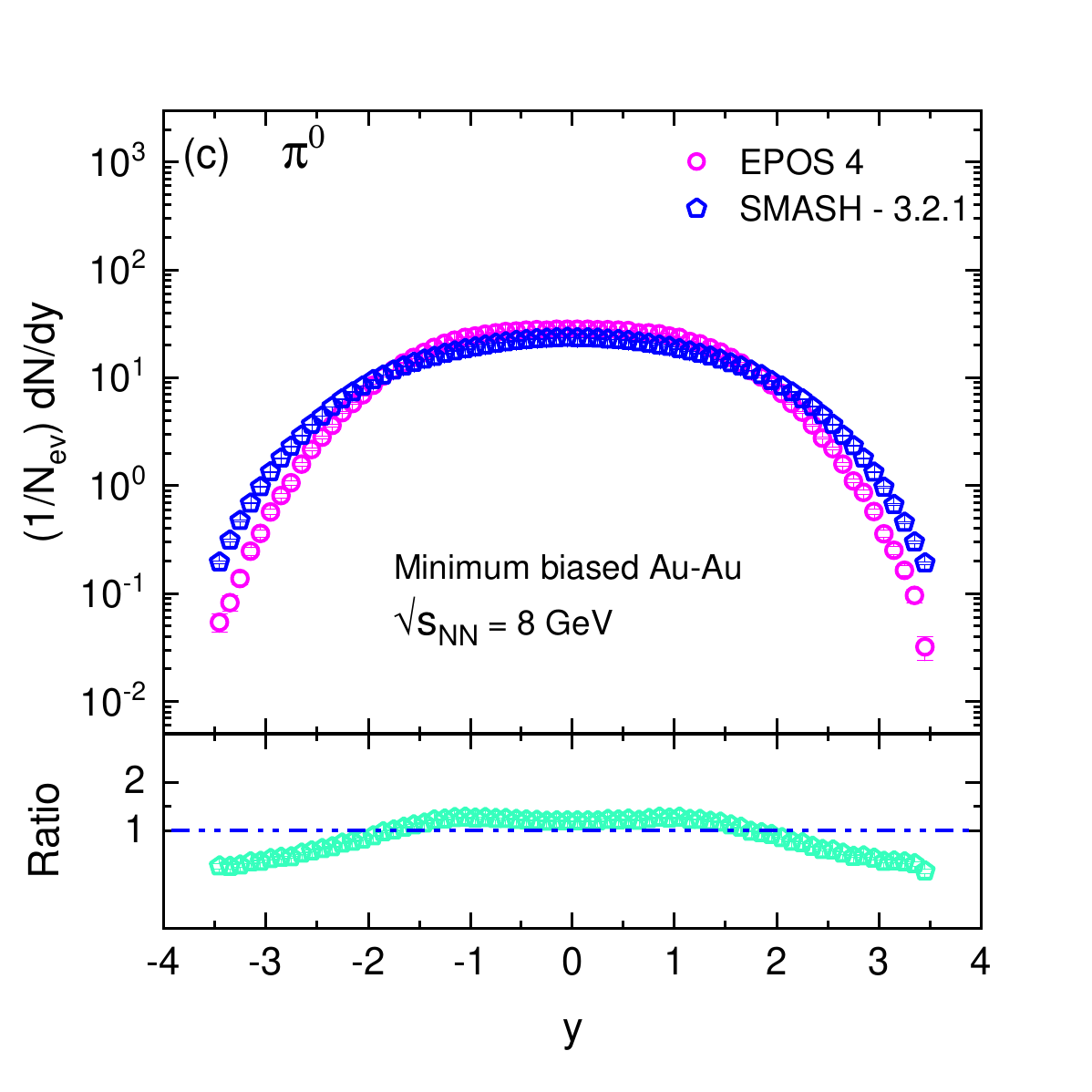}
\includegraphics[width=0.3222\textwidth]{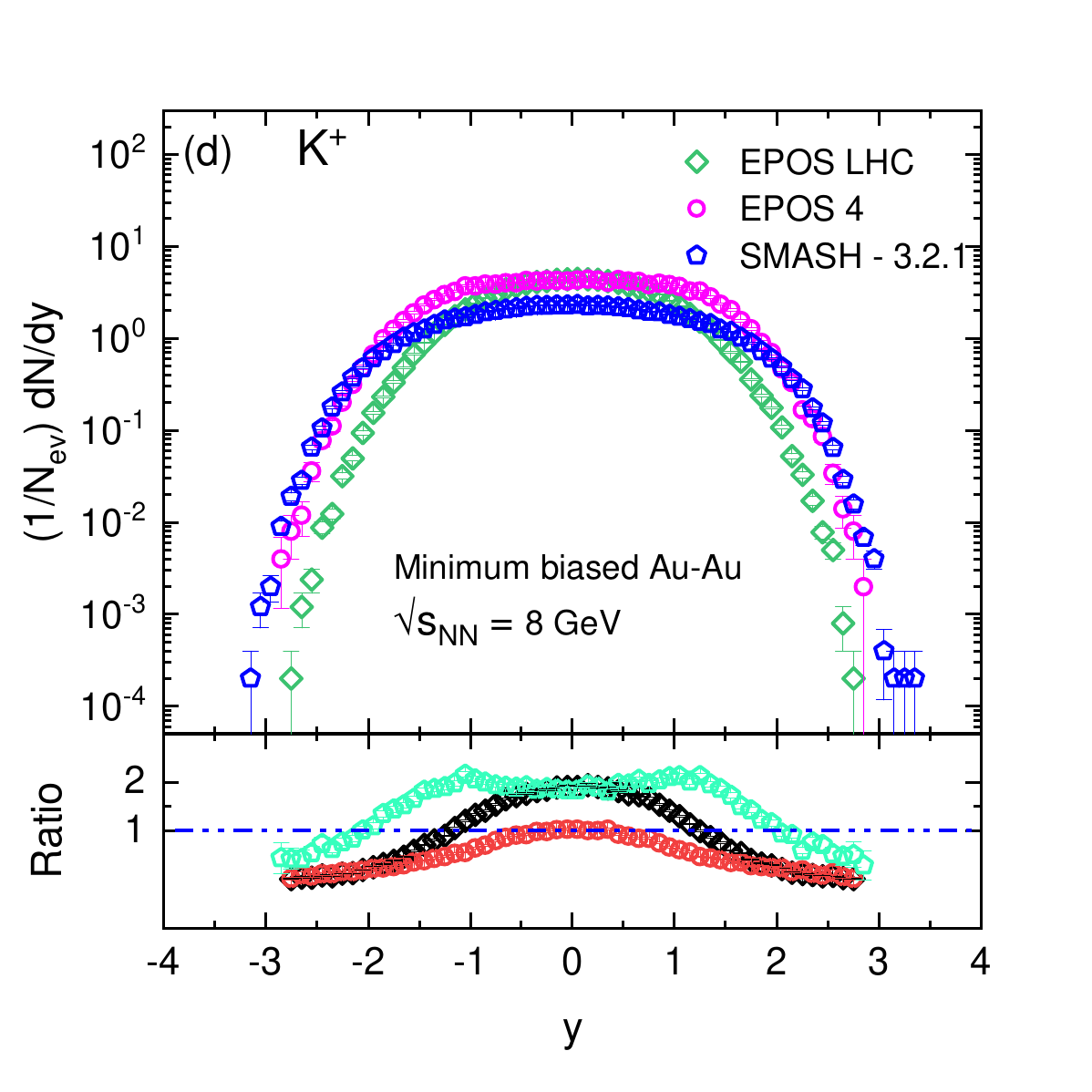}\vspace{-0.35cm}
\includegraphics[width=0.3222\textwidth]{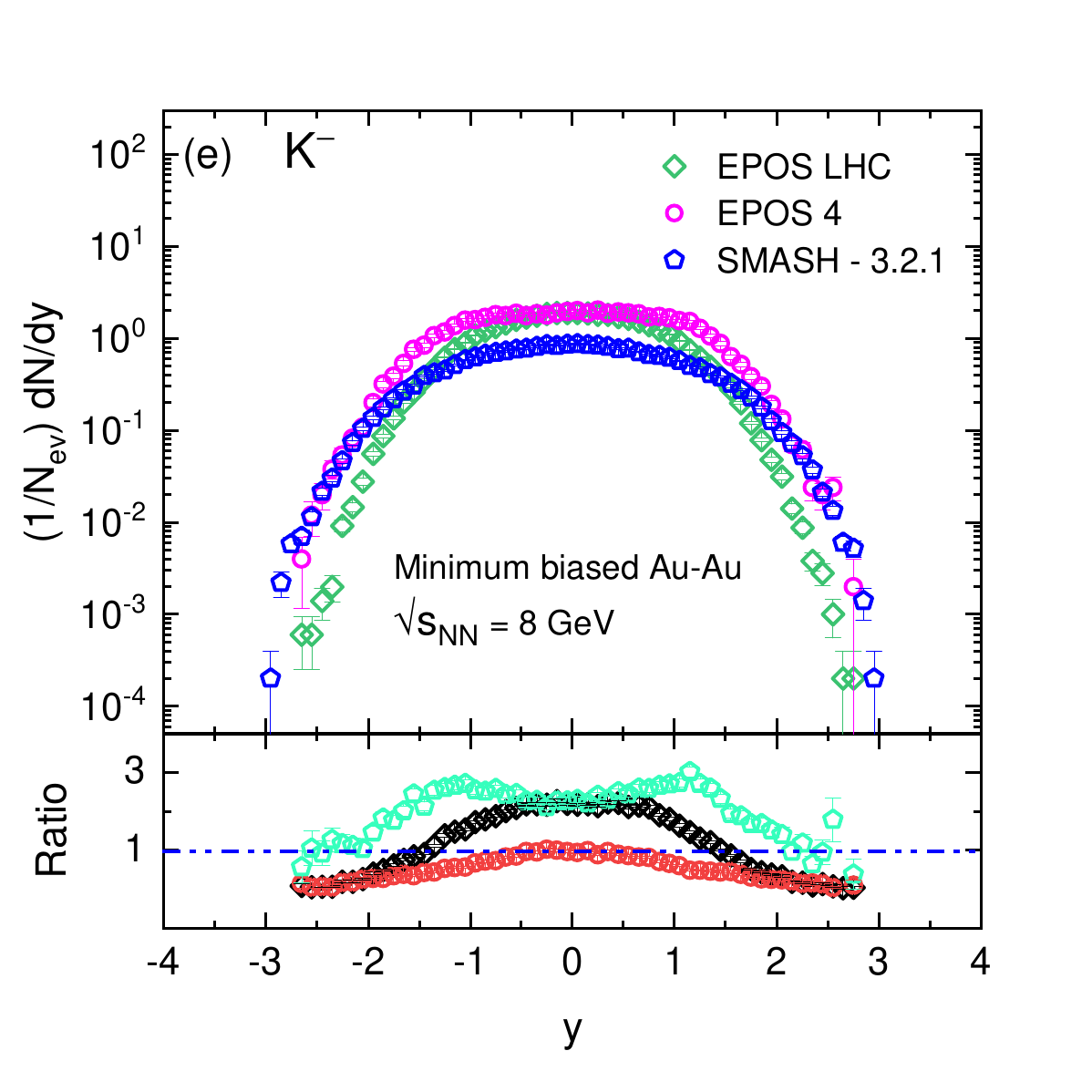}
\includegraphics[width=0.3222\textwidth]{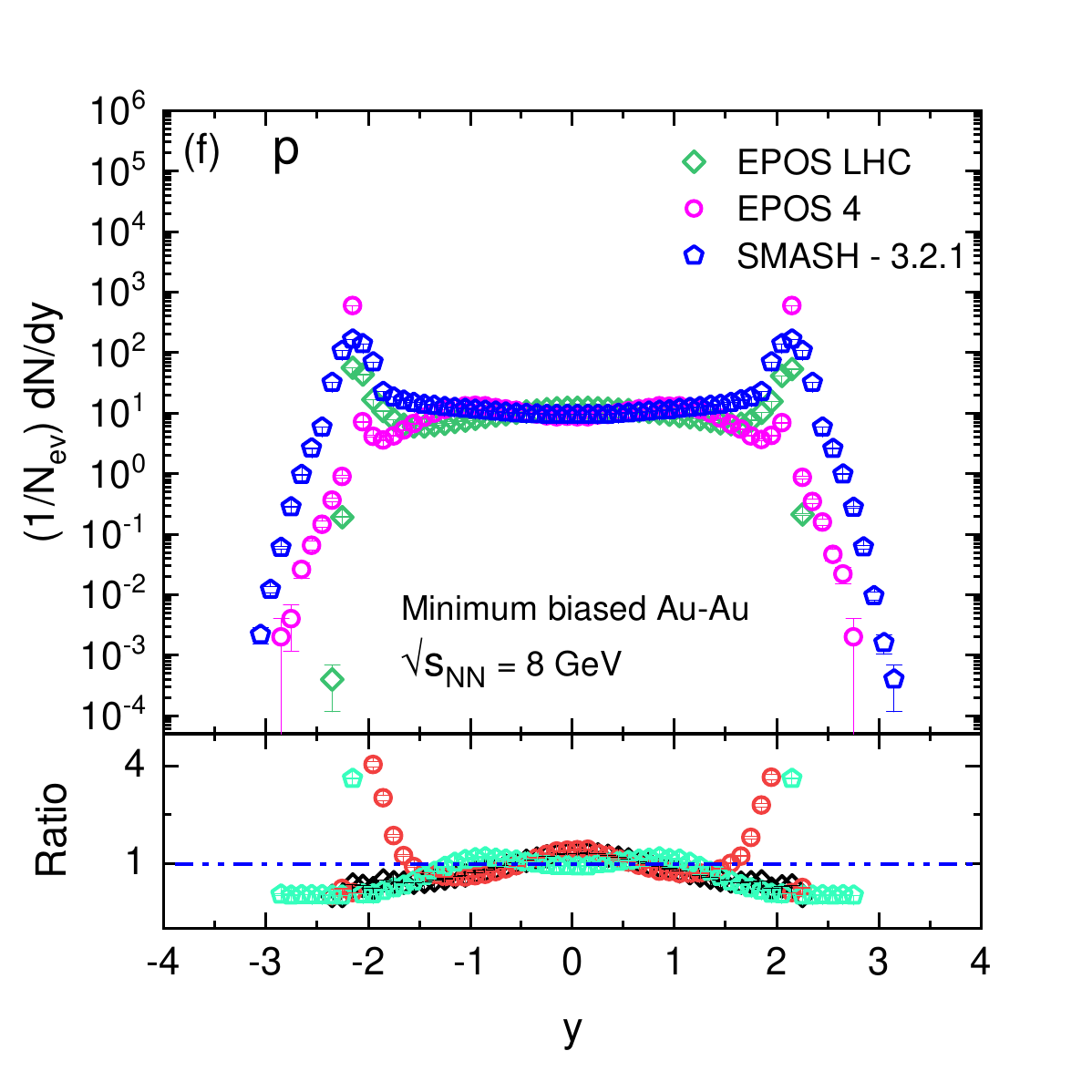}\vspace{-0.35cm}
\includegraphics[width=0.3222\textwidth]{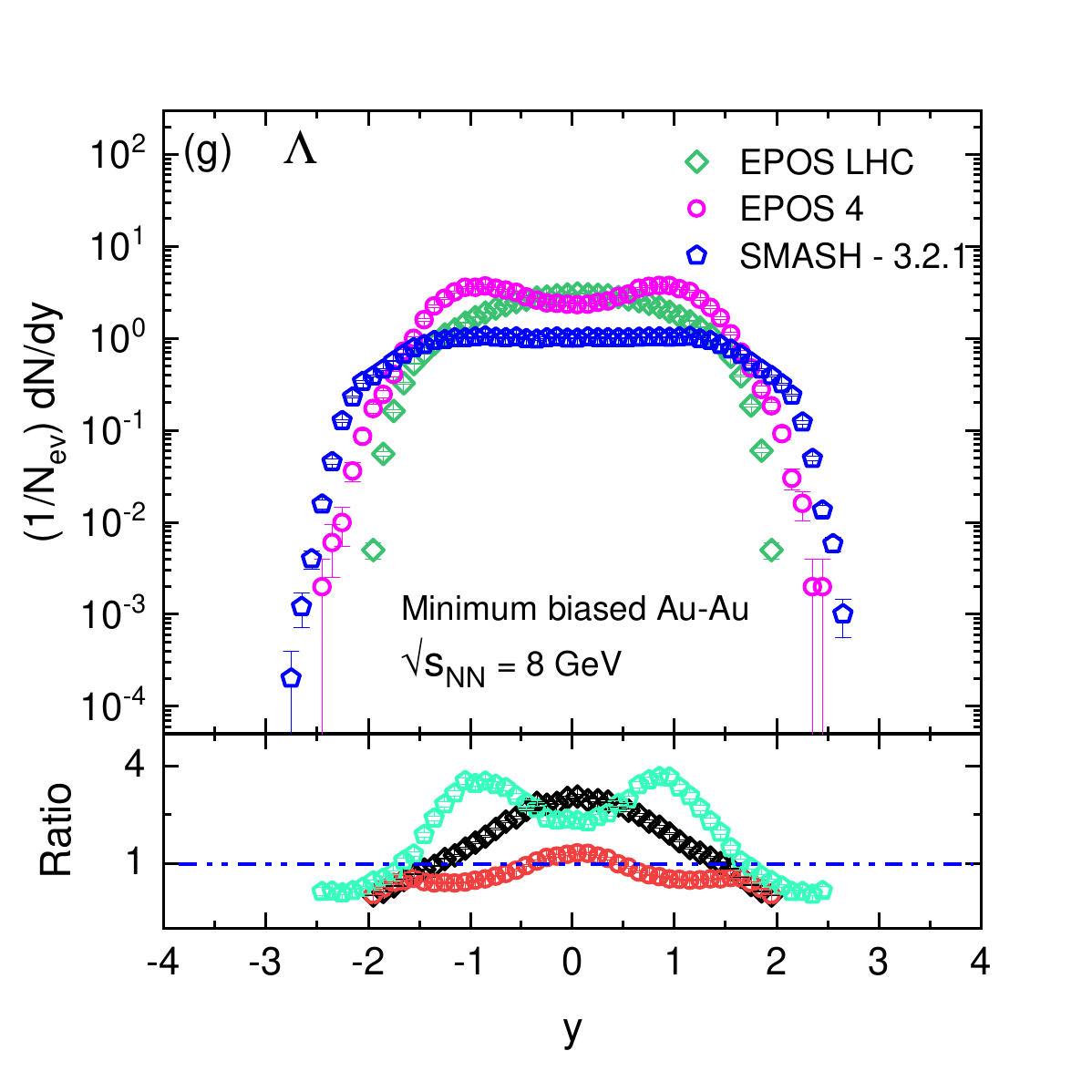}
\includegraphics[width=0.322\textwidth]{legends.pdf}\vspace{-0.35cm}
\caption {Rapidity distributions (event-normalized $dN/dy$) of identified hadrons at minimum-bias in $Au+Au$ collisions at $\sqrt{s_{NN}}=8~GeV$ forecasted by the three models. The results of EPOS-LHC, EPOS-4, and SMASH for various species of hadrons are presented in panels (a)-(g) to highlight the differences in total yield, width in rapidity, and mid-rapidity behaviour. Model legends are in the panels.}
\end{figure*}
Fig. 1(a) represents the rapidity distribution of $\pi^+$ at $\sqrt{s_{NN}} = 6~GeV$ predicted by EPOS-LHC, EPOS-4, and SMASH. The $\pi^+$ rapidity distribution at 
$\sqrt{s_{NN}} = 6~GeV$ is reproduced nearly identically in SMASH, EPOS-LHC, and EPOS-4 with 
Gaussian-shaped \cite{wong2008landau} distributions with similar magnitudes centred at midrapidity. This good 
agreement is not a coincidence, at these energies the overwhelming share of $\pi^+$ is made by 
resonance and higher baryonic resonance decays, e.g., $\Delta^{++}\xrightarrow{}p\pi^+$ and $\Delta^+\xrightarrow{}n\pi^+$. Since these three models are 
similar in their treatment of resonance excitation and decay kinematics, the final-state $\pi^+$ 
distributions are similar by default. The other reason is that the multiplicities of pions are incredibly 
large relative to those of other hadron species, therefore, statistical fluctuations or processes unique 
to a model are washed away. The physical meaning behind this is that pion rapidity profiles are 
mainly sensitive to the overall energy deposition physics as well as resonance decay physics, which 
is strong in modeling methods. It implies that although $\pi^+$ distributions ensure that both models 
respect energy and charge, they do not discriminate much between hadronic-only transport 
(SMASH) and hybrid/string-based models (EPOS). Their introduction of $\pi^+$ is rather novel as a 
reference, rather than as a probe of new physics, because their behavior determines the scale on 
which kaons, protons, and hyperons will be measured.

The $\pi^-$ distribution, displayed by Fig. 1(b), is similar to the $\pi^+$ distribution, and has almost the same shapes and magnitudes across models. The 
$\pi^-/\pi^+$ ratio is nearly unity with rapidity, as it should be in Au+Au collisions because of isospin 
symmetry \cite{bearden2004nuclear}. At rapidities forward/backward, small deviations may occur due to neutron excess in 
the nuclei of gold, though in the case of the rapidity studied, these deviations are small. This is because 
the $\pi^-$ production also occurs through resonance decays $\Delta^-,~N^*\xrightarrow{}\pi^-+N$, where $N^*$ is an excited nucleon state, and these are also 
included in all the models in the same way. The consistency of the descriptions of both $\pi^+$ and 
$\pi^-$ means that both charge conservation and chains of resonance decay are well-implemented. 
Physically, this accord indicates that the yields of charged pions at $\sqrt{s_{NN}} = 6~GeV$ are such that 
at that energy scale, resonance feed-down is of the trivial nature and not the ones that are sensitive 
to the initial stage of the collision. From the novelty perspective, this once again points out the 
drawback of pious as probes: although rich and experimentally accessible to quantify, they fail to 
discriminate between transport and hybrid methods. They are primarily useful in creating a 
standard level of comparison with rarer species.

Please note that the EPOS-LHC model does not provide precise predictions for neutral pions in our calculations. This does not happen because the physics of the model suppresses the production of $\pi^0$, but it is merely a technical drawback of the public EPOS-LHC implementation. Because EPOS-LHC was originally optimized for high-energy LHC collisions, its output tables do not explicitly give the yield of neutral pions, but instead reconstruct neutral pions indirectly through their electromagnetic decay channel ($\pi^0\xrightarrow{}\gamma\gamma$). However, both EPOS-4 and SMASH do use $\pi^0$ as an explicit particle species in their event records: EPOS-4 as one of the extensive hadronic spectrums that they tune to the FAIR/NICA energy scale, and SMASH via its detailed hadronic resonance and decay chains. This is why the comparison of the $\pi^0$ rapidity and transverse momentum distributions in this work is limited to EPOS-4 and SMASH. The rapidity distributions of neutral pions at $\sqrt{s_{NN}} = 6, 7$ and $8~GeV$, given in panels (c) of Figs. 1-3 are generally consistent with those of charged pions, and also provide a valuable test of isospin symmetry. 

The model has a clear dependence on the $K^+$ distribution as can be seen in Fig. 1(d). EPOS-4 forecasts visibly greater yields at midrapidity than SMASH and 
EPOS-LHC, which are closer to each other. The physics explanation is the fact that the channel of 
production of $K^+$ is dominated by associated production ($NN\xrightarrow{}N\Lambda K^+$). The constraint of $K^+$ 
yields in SMASH are the presence of baryon-baryon on-threshold scatterings and the following 
hadronic interactions. In comparison, EPOS-4 uses a string fragmentation and core-corona 
mechanism, which enables more efficient production of strangeness, hence increasing the yields 
of $K^+$. EPOS-LHC is tuned at higher energy, thus indicating an intermediate behavior. This is a 
significant difference: $K^+$ is the first strange hadron species on which the models differ, that is, it 
is a sensitive probe of the extent to which the system is enhanced in strangeness. Novelty-wise, 
the fact that the $K^+$ in EPOS-4 is already getting enhanced at $\sqrt{s_{NN}} = 6~GeV$ is an indication that 
the strangeness enhancement could be coming in earlier in energy than we might think, and that 
can be conclusively determined at NICA.

In Fig. 1(e), $K^-$ production is extremely suppressed 
relative to $K^+$, and the inhibition is greatest in SMASH. This is simply because $K^-$ cannot be 
formed by associated production, but they need to be formed by the use of $s\bar{s}$ pair 
production in reactions like $\pi N\xrightarrow{}NK\bar{K}$. These reactions are energetically expensive in baryon-rich environments, and $\sqrt{s_{NN}}=6~GeV$ is quite small there. EPOS-4 projects somewhat larger midrapidity 
$K^-$ production, as the production of string breakings is an extraneous source of $s\bar{s}$ creation. 
EPOS-LHC once again is intermediate, with rather modest pair production. The model separation 
of $K^-$ is because these mesons are heavy and hence would have a direct 
bearing on the production mechanism on a microscopic level. Physically, it is significant since 
$K^-$ produces reflectance, probabilities of getting pairs, which very sensitively depend on the 
presence or absence of string-like dynamics. The new thing is that the ratios of $K^-/K^+$ at such 
energies are one of the most sensitive tests of the production environment, and data in the future 
NICA can directly differentiate between hadronic and string-based pictures.

The proton $dN/dy$ at $6~GeV$ is shown in Fig. 1(f) and exhibits evident peaks in the fragmentation regions ($|y|\geq 1.5-2$), as well as a lower yield at midrapidity. The total mid-rapidity yields of the three models are similar in magnitude; however, the fragmentation peaks differ distinctly across models. SMASH exhibits one of the widest rapidity distributions, with fragmentation distributions that are relatively large and tails that extend more towards the beam rapidity. That is, SMASH puts a larger fraction of protons in the forward/backward rapidity regions than in midrapidity regions than some versions of EPOS. EPOS-4 also generates quite large fragmentation peaks and is close to SMASH in the fragmentation region, but its yield at midrapidity is similar to that of SMASH. EPOS-LHC has the fewest fragmentation peaks compared to the other two and a slightly less general width; the forward/backward maxima of EPOS-LHC are smaller than those of SMASH and EPOS-4. Even at low energy ($6~GeV$), the baryon stopping remains important, though the models vary in the extent to which the baryon number remains in the fragmentation regions. The larger distribution in SMASH comes about since, in a microscopic hadronic cascade, baryons are more likely to be subject to multiple scattering and resonance-mediated transport without the same processes that concentrate baryons at a narrow region in the midrapidity plane. EPOS-LHC (tuned to high energies) exhibits rather weaker fragmentation peaks at this point, whereas EPOS-4, which is intended to treat intermediate energies more effectively, follows SMASH through the fragmentation region more closely.

The $\Lambda$ rapidity distribution, shown in Fig. 1(g), 
also displays the differences related to the baryon stopping and strangeness production. EPOS-4 
predicts a greater midrapidity $\Lambda$ density than SMASH or EPOS-LHC. The physical explanation is 
that $\Lambda$ production is strictly interconnected with $K^+$ production through parallel processes; with 
the increase of $K^+$, $\Lambda$ is also increased. The $\Lambda$ yields in SMASH are restricted to hadronic 
scattering thresholds, thereby producing less. In EPOS-4, there are string fragmentation and 
partonic-like dynamics, which produce an increased number of $\Lambda$s. The fact that $\Lambda$ is specifically 
fascinating: whereas in protons, which are the particles that are just sensitive to baryon transport, 
at the same time, $\Lambda$ is sensitive to both baryon stopping and strangeness physics. The new here is 
the high degree of model dependence of $\Lambda$ yields at such low energies, not so evident in the 
previous SPS data. This demonstrates that $\Lambda$ measurements at NICA will act as a twofold 
measurement: of the redistribution of baryon charge and of the creation of strangeness.

In Fig. 2(a), the production of $\pi^+$ has a slight increase 
in overall magnitude at $\sqrt{s_{NN}}=7~GeV$, but the shape of the rapidity distribution and the model
comparison are the same as in $6~GeV$. The three models lead to almost the same Gaussian-shaped curves with the peaks at midrapidity, which shows that pion yields are still dominated by 
the decay of resonances and bulk hadronic scattering. The physical explanation is the same: the 
increase in $\sqrt{s_{NN}}$ of $1 ~GeV$ does not significantly change the pion production process; the system 
still spends the largest part of the lifetime in the baryon-rich hadronic phase, in which the $D$ and $N^*$ 
decays are the most coincident ones. Accordingly, $\pi^+$ are not discriminants of microscopic 
physics, but bulk carriers of entropy. The similarity between the models is significant since it 
demonstrates that every model manages the conservation of soft bulk particles in a similar way. 
But the insensitiveness is also used to highlight the reason why pions are not adequate to 
differentiate competing pictures of the early stage. What is new in this is showing explicitly that 
adding $1~GeV$ of $\sqrt{s_{NN}}$ to this window does not alter the position of pions as baseline probes.
 
Fig. 2(b) represents $\pi^-$ rapidity distribution, which once again resembles $\pi^+$, and the overall yield is slightly enhanced, 
and the shapes are almost model independent. The $\pi^-/\pi^+$ ratio remains near unity as a function 
of rapidity, indicating that isospin symmetry is maintained. This is because $\pi^-$ also comes out 
strongly due to resonance decays like $\Delta^-\xrightarrow{}n\pi^-$, and these modes are not sensitive to the small 
variation in beam energy. The physical meaning is that ratios of charged pions at these energies 
measure global isospin equilibrium and not finer details of collision dynamics. This feature is 
reproduced by all the models, which once again confirms the reliability of their isospin 
conservation and resonance feed-down models. Similar to the $\pi^+$ case, the novelty in this case is 
minimal, only to establish that the baseline is robust enough to interpret the results of strangeness 
and baryons.

The $K^+$ distribution, given in Fig. 2(d), becomes larger and broader at $7~GeV$ than at $6~GeV$. EPOS-4 yields much higher midrapidity, 
compared to SMASH, with EPOS-LHC intermediate. The physical explanation is that with 
increased $\sqrt{s_{NN}}$, associated production ($NN\xrightarrow{}N\Lambda K^+$) can be done more often, and the string-based 
processes in EPOS-4 enhance this channel compared to what can be done in the purely hadronic 
SMASH \cite{sturm2001evidence, fuchs2006kaon}. This is directly connected with the physics of strangeness enhancement, where it is found 
that partonic or string-based processes enhance the strangeness yields more effectively than 
hadronic scattering. Literature-wise, SPS experiments indicated a steep increase in the $K^+/\pi^+$ ratio 
with energy, the so-called horn, which has been attributed to the beginning of deconfinement \cite{alt2008pion}. 
What is new about our results, however, is that at $6-8~GeV$ the difference in model results in the 
production of $K^+$ is already dramatic, and NICA can use this as a sensitive probe of these 
processes.

Fig. 2(e) shows the $K^-$ rapidity distribution at $7~GeV$, which is increasing as compared to 6 GeV, 
but yields are heavily suppressed in comparison with $K^+$. EPOS-4 is found to be predicting 
stronger $K^-$ midrapidity yields than SMASH, with EPOS-LHC taking an intermediate position. 
This is because increasing $\sqrt{s_{NN}}$ can allow more channels of $s\bar{s}$ pairs to be created, and $K^-$ 
production is less suppressed, but SMASH is limited to hadronic thresholds, whereas EPOS-4 can 
do more pair creation by fragmenting strings. This difference is crucial physically: since none of 
the associated production can be made to produce $K^-$, their production is a clean probe into the 
efficiency of the formation of $s\bar{s}$ pairs \cite{sturm2001evidence, fuchs2006kaon, forster2007production}. The meaning is that the sensitivity of the $K^-$ gives 
more clearly the underlying microscopic degrees of freedom than $K^+$. This is novel in the sense 
that even with a $1~GeV$ increase in $\sqrt{s_{NN}}$, the difference in $K^-$predictions among models is 
measurable, and therefore this observable is of top priority in future NICA analyses.

The double-hump structure is still present in the proton rapidity spectra at $\sqrt{s_{NN}}=7~GeV$, given in Fig. 2(f), however, the fragmentation peaks tend to be higher than at $6~GeV$ (the plots show an increase in forward/backward maxima with energy) and the midrapidity valley is slightly more filled in in some models \cite{appelshauser1999baryon}. SMASH once again exhibits a relatively broad proton rapidity distribution that features prominent fragmentation peaks and long tails. The width in comparison to the EPOS curves is greater (i.e., yields at greater $|y|$ are greater in SMASH). EPOS-4 remains between SMASH and EPOS-LHC in the fragmentation region; in particular, the $y$ bin at which EPOS-4 shows fragmentation peaks is similar to SMASH (or a bit different in some cases, depending on the specific $y$).
EPOS-LHC has the smallest distribution of the three, with smaller fragmentation peaks and a midrapidity shape that is more flattened but less extended. As $\sqrt{s_{NN}}$ increases, the phase space of the longitudinal evolution becomes larger, and therefore models are more visibly different in the distribution of baryon number. The hadronic transport in SMASH causes further protons to spread in the fragmentation areas. That behavior is reproduced in part by EPOS-4, which uses more realistic string/core dynamics at intermediate energies, and at which EPOS-LHC is the most midrapidity-concentrated (i.e., least baryon flux into the fragmentation wings).

The $\Lambda$ rapidity distribution, shown in Fig. 2(g), is largest at $7~GeV$ with 
EPOS-4 gives the largest yields and SMASH the smallest. The physical explanation is that $\Lambda$s, as 
well as $K^+$, are highly affected by production channels associated with them, which are more 
active as $\sqrt{s_{NN}}$ increases. This effect is enhanced through string dynamics in EPOS-4, but not that much in SMASH. The importance is that $\Lambda$s are a complementary probe to protons: via protons, it is 
possible to measure baryon stopping, but with $\Lambda$s, both stopping and strangeness production are 
encoded. In the literature view, the observation of improved $\Lambda$ yields was a classical SPS 
observation associated with strangeness enhancement \cite{andersen1999strangeness, antinori2006enhancement, rafelski1982strangeness}, and our findings indicate that this behavior 
can be observed in the $6-8~GeV$ region. The novelty is that model variations in $\Lambda$ production are 
already strong at $7~GeV$, and this species is an obvious experimental goal of NICA.

The rise in $\pi^+$ yields at $8~GeV$, shown in Fig. 3(a), further increases. 
However, the shape and inter-model consistency are similar. The overlapping of the Gaussian-like 
profiles with peaked midrapidity occur in all the models. The cause is the same, viz. that resonance 
decays, notwithstanding the decrease of $\sqrt{s_{NN}}$ in this low-energy regime, predominate in the 
production of pions. This stability, physically, emphasizes that the pion distributions do not 
indicate the dynamics of interest on the microscopic level but indicate the generation of entropy. 
What is new is the confirmation that in three successive energies, the pions are nondiscriminating 
probes and hence experiments should now concentrate on rarer hadrons to test models.

The $\pi^-$ distribution, shown in Fig. 3(b), is the same as $\pi^+$, only with a slightly higher yield than at $7~GeV$, 
and almost the same shapes according to models. The ratio of isospin symmetry is still seen in the 
$\pi^-/\pi^+$ ratio and no significant disparities arise. The physical meaning is once again that pions are 
a consistency check and not a discriminator. The only difference here is that pion distributions are 
robust throughout the $6-8~GeV$ window, and it proves their position as a baseline observable.

Fig. 3(d) displays the $K^+$ rapidity distribution at $\sqrt{s_{NN}}=8~GeV$. The model disagreement is most apparent here: EPOS-4 is much higher in its midrapidity $K^+$ yields than SMASH and EPOS-LHC. This is because with $\sqrt{s_{NN}}$, the production channels 
related to it increase, and the string mechanisms of EPOS enhance this effect, but SMASH has the 
restriction of hadronic scattering. Physically, this implies that the enhancement of strangeness is 
stronger in EPOS-4, which can be attributed to partonic-like dynamics. The importance is that even 
within this narrow band of energy, the $K^+/\pi^+$ ratio, which is a typical SPS observation, is 
strongly model-dependent \cite{alt2008pion}. The novelty is that one of the most significant observables to 
distinguish between hadronic and string-based mechanisms at NICA is the production of $K^+$, which has been predicted by our simulations.

$K^-$ yields further increase at $8~GeV$, evident from Fig. 3(e), but are still 
suppressed compared to $K^+$. EPOS-4 has the highest $K^-$ yields, with SMASH the lowest. The 
physical explanation is that increasing $\sqrt{s_{NN}}$ increases the number of channels of creating pairs by 
creating $s\bar{s}$ pairs, but only string dynamics (in EPOS) can increase this production 
significantly. This emphasizes that $K^-$ is a special probe of pair creation in that it cannot be 
created through other related channels. The meaning is that there are not only quantitative 
differences in the production of $K^-$, but also qualitative ones, as they are images of different 
microscopic images. What is new here is that already in the low-energy regime, $K^-$ distributions 
are very model-dependent and, thus, high-priority observables to NICA.

The shape of the hump of the proton rapidity spectra is the most prominent at $8~GeV$ as shown in Fig. 3(f).  All three yield very similar magnitudes in midrapidity, although the fragmentation asymptotes are different amongst models. SMASH generates the most widespread distribution, with the highest relative yields in the fragmentation regions, though smaller than EPOS-4 at only two points, and the longest tails. SMASH has the largest width enhancement at $6- 8~GeV$. EPOS-4 also exhibits serious fragmentation peaks, and is less distant to SMASH than EPOS-LHC in the wings, with few discrepancies in details (peak heights and absolute $y$ positions are model dependent). EPOS-LHC is the thinnest and least peaked in the fragmentation regions; midrapidity yields are the same as other models, but EPOS-LHC gives relatively less emphasis to the forward/backward maxima. Here, the separation between SMASH and EPOS-LHC is the highest. The broadening of SMASH with energy suggests that baryons in a hadronic cascade are more likely to be left towards their original longitudinal momentum and populate fragmentation regions more heavily, where EPOS variants, especially EPOS-LHC, provide mechanisms (core formation, string junction dynamics and stronger effective stopping in the model parametrization) that move more baryon density to midrapidity or at least weaken the weight of the fragmentation peaks.

The growth in $\Lambda$ yields is even stronger at $8~GeV$, given in Fig. 3(g), and EPOS-4 generates the largest increase 
compared to SMASH. The physical explanation is that the more $\Lambda$s one has, the more frequent 
the related production and fragmentation of strings is, preferring the more $\Lambda$ production. 
SMASH is still limited, resulting in fewer $\Lambda$s in total. Physically, this proves that $\Lambda$ production is 
an integrated indicator of baryon transportation and strangeness enhancement, and it is among the 
most informative species. From the literature point of view, $\Lambda$ enhancement played a central role in 
SPS and RHIC, when it comes to the diagnosis of strangeness physics. Our results are novel in that 
they demonstrate that $\Lambda$ distributions highly deviate among models even within the $6-8~GeV$ 
window, and this is a sensitive probe of NICA.
\begin{figure*}
\centering
\includegraphics[width=0.32\textwidth]{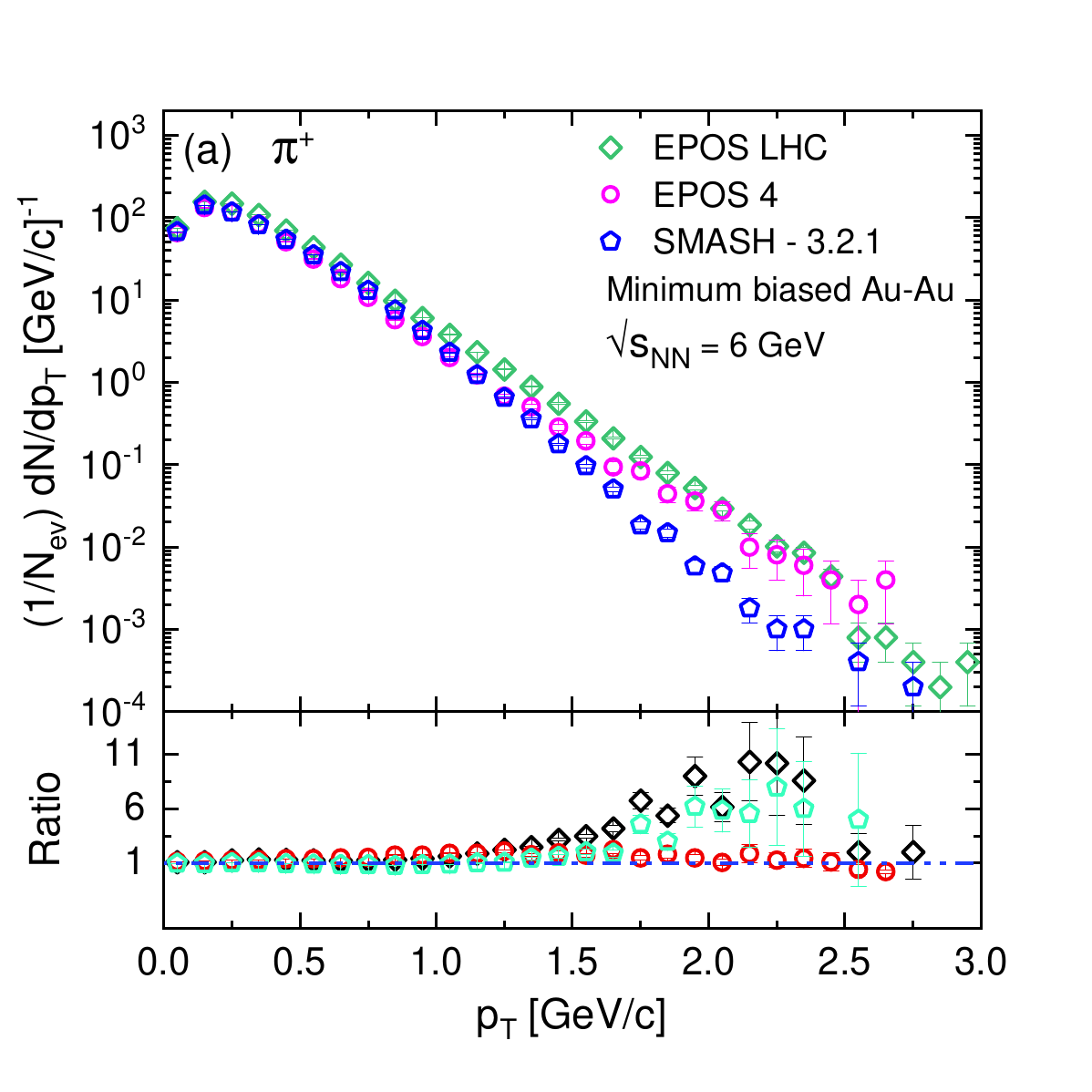}
\includegraphics[width=0.32\textwidth]{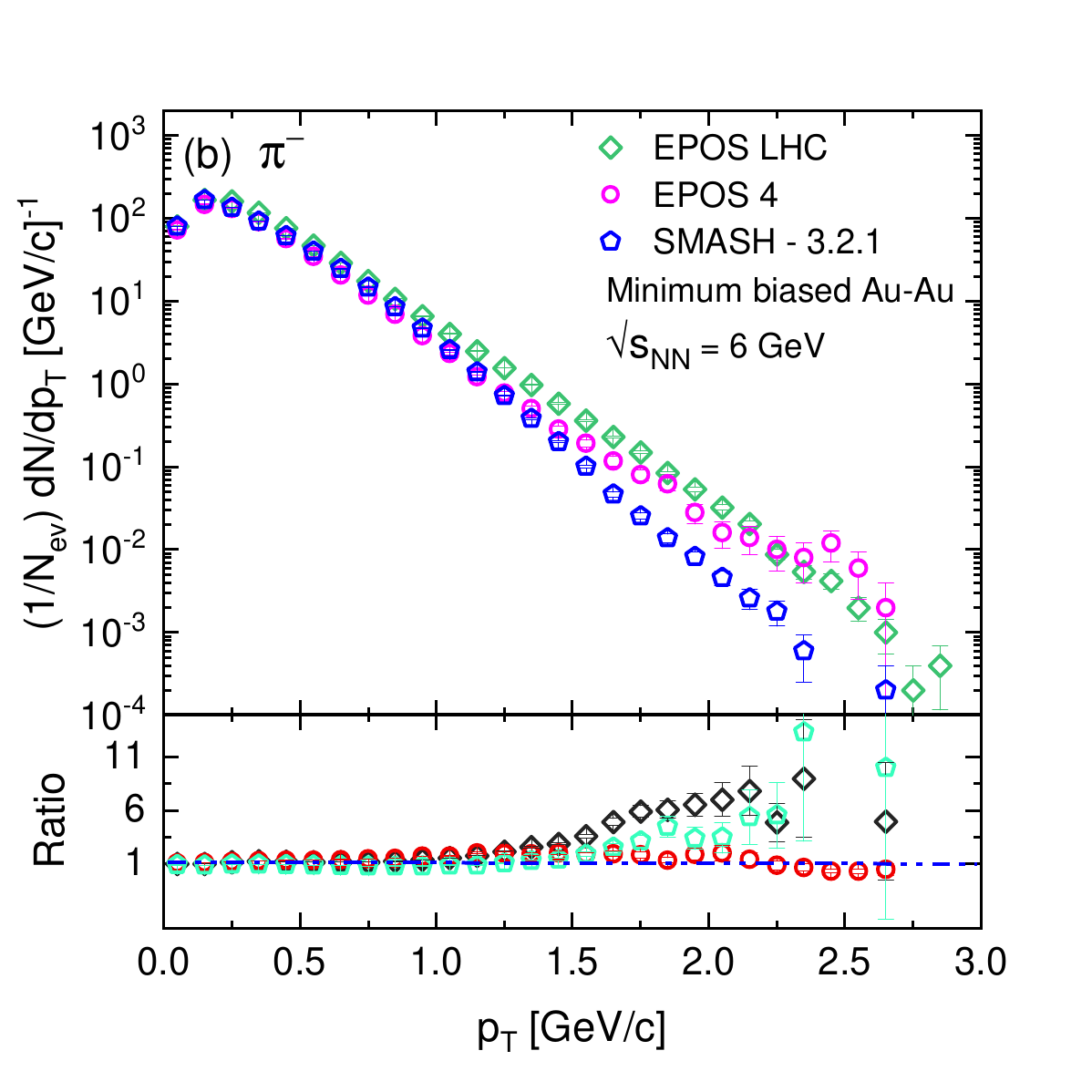}\vspace{-0.35cm} 
\includegraphics[width=0.32\textwidth]{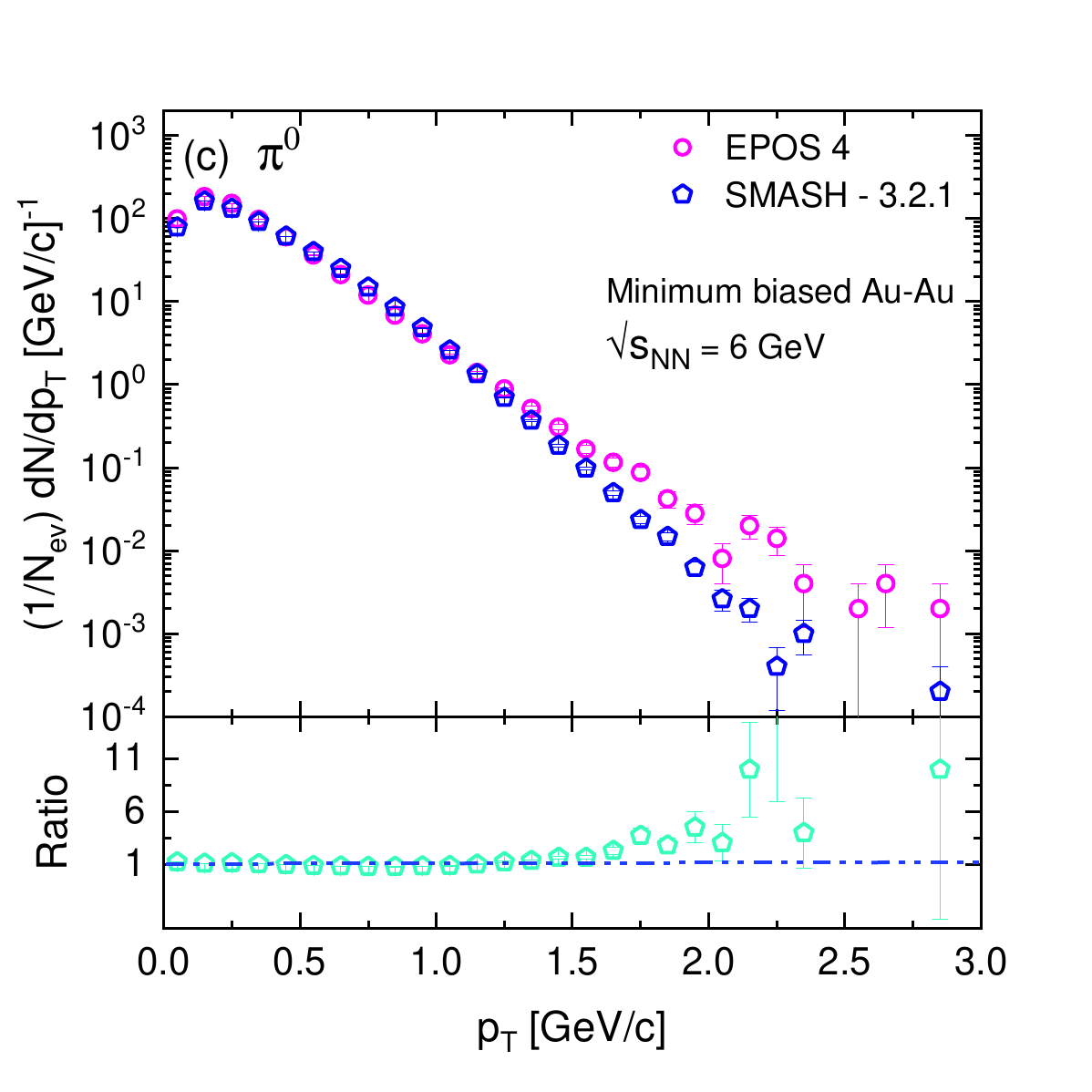}
\includegraphics[width=0.32\textwidth]{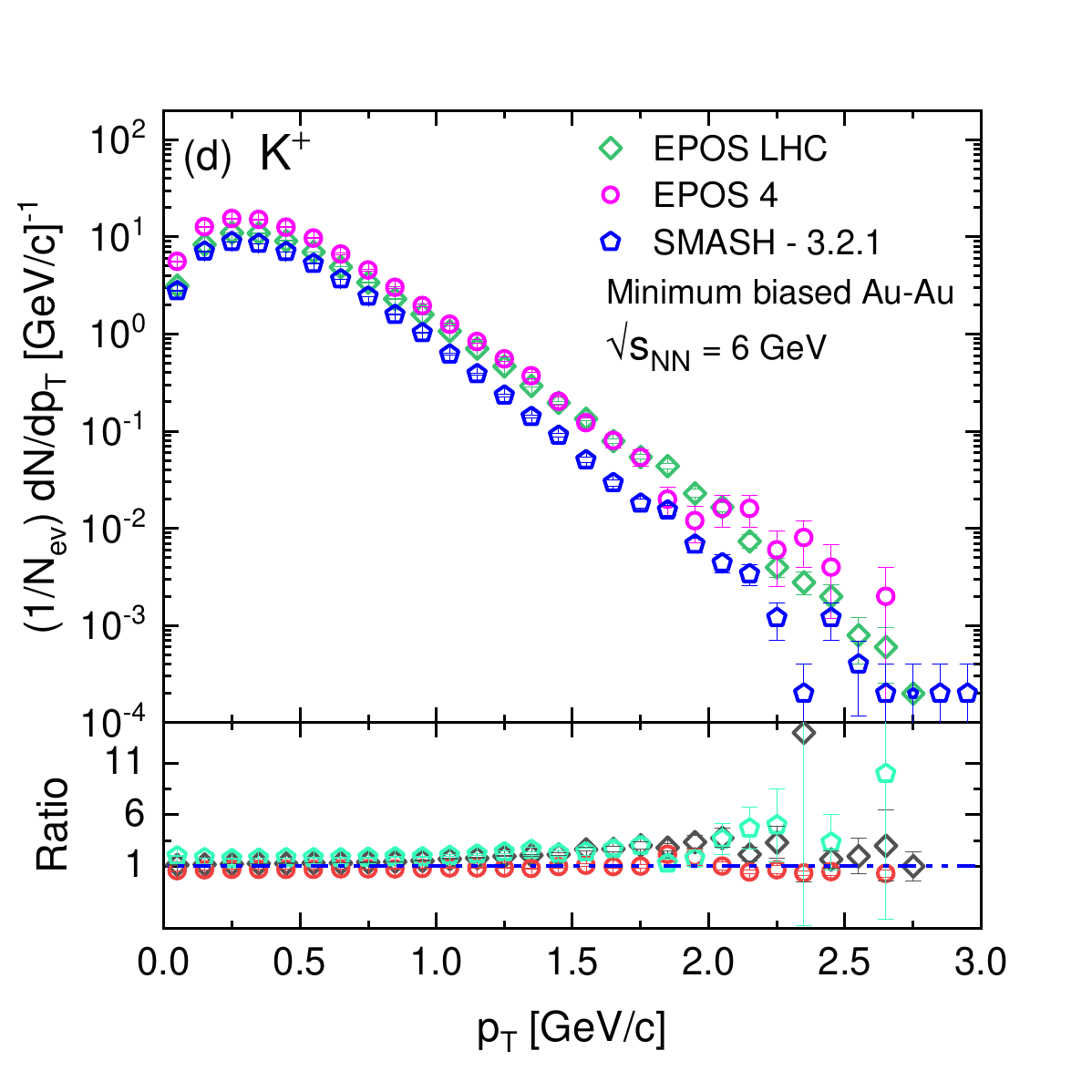}\vspace{-0.35cm}
\includegraphics[width=0.32\textwidth]{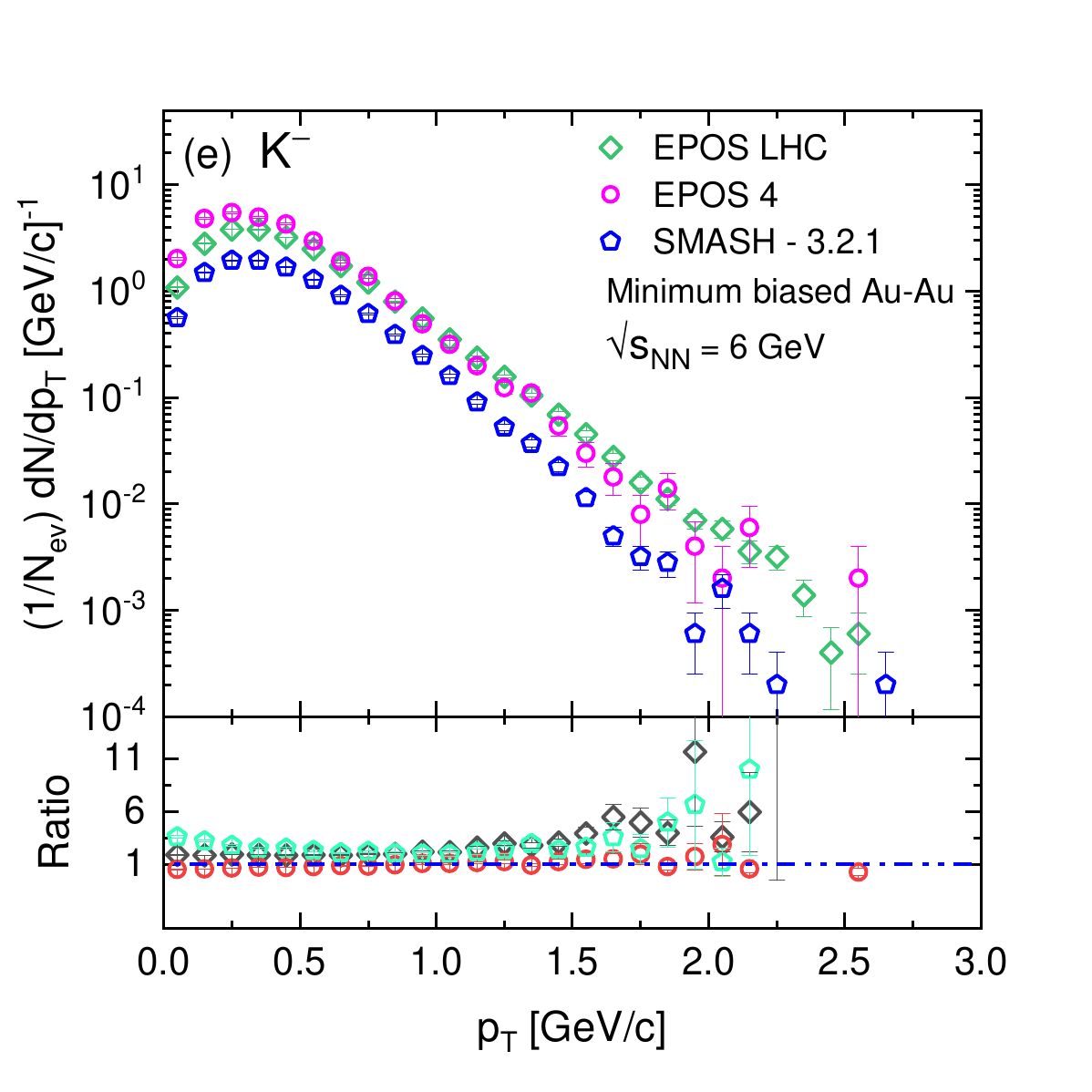}
\includegraphics[width=0.32\textwidth]{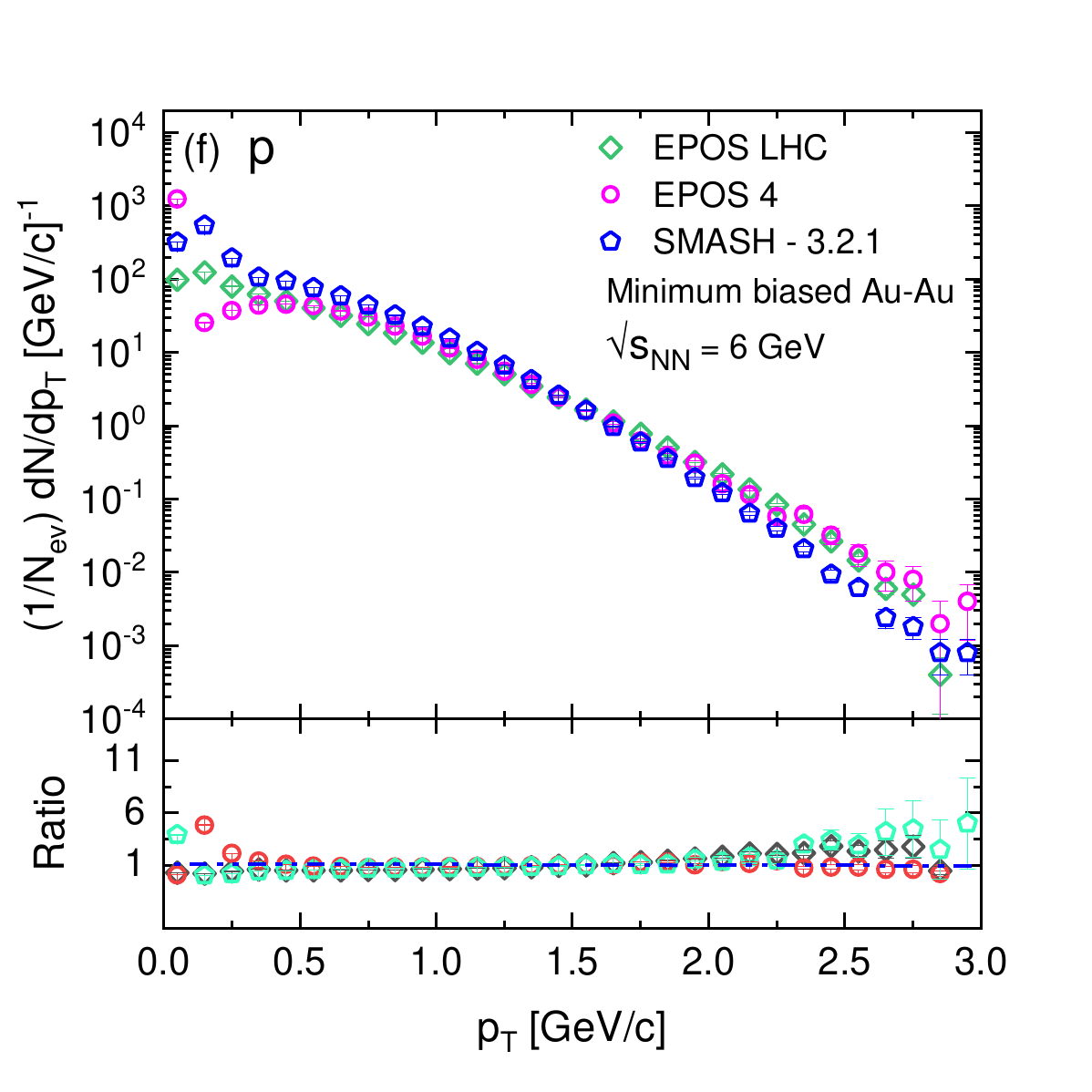}\vspace{-0.35cm}
\includegraphics[width=0.32\textwidth]{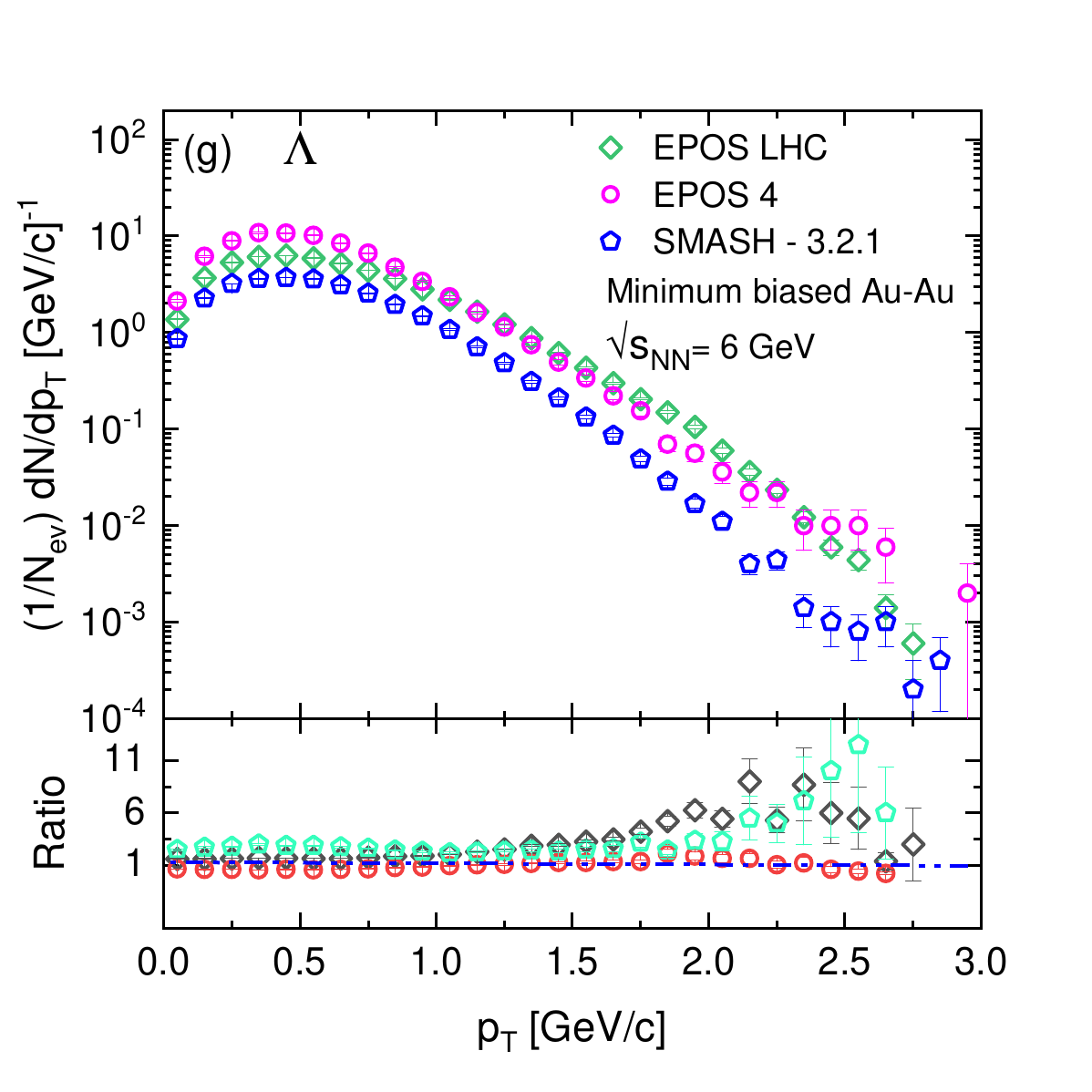}
\includegraphics[width=0.32\textwidth]{legends.pdf}\vspace{-0.35cm}
\caption {Event-normalized $dN/dp_T$ transverse-momentum spectra of identified hadrons in minimum-bias $Au+Au$ collisions at $\sqrt{s_{NN}}=6~GeV$. The EPOS-LHC, EPOS-4, and SMASH model predictions are compared to demonstrate variations in the shape of the spectrum, slope (soft vs. hard components), and relative yield at low and intermediate $p_T$. Each plot has a ratio panel at its bottom, simplifying the model to model comparison.}
\end{figure*}
\begin{figure*}
\centering
\includegraphics[width=0.32\textwidth]{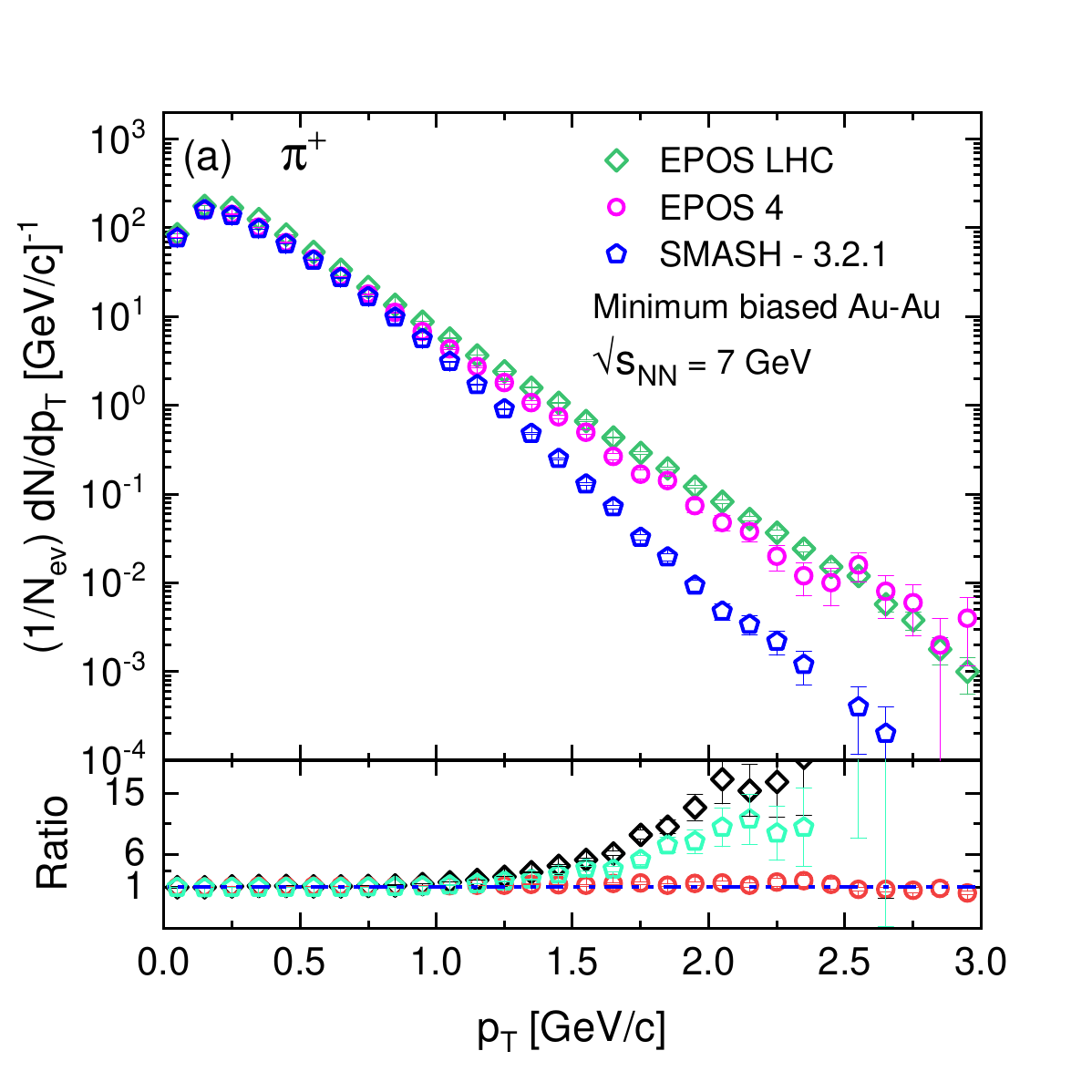}
\includegraphics[width=0.32\textwidth]{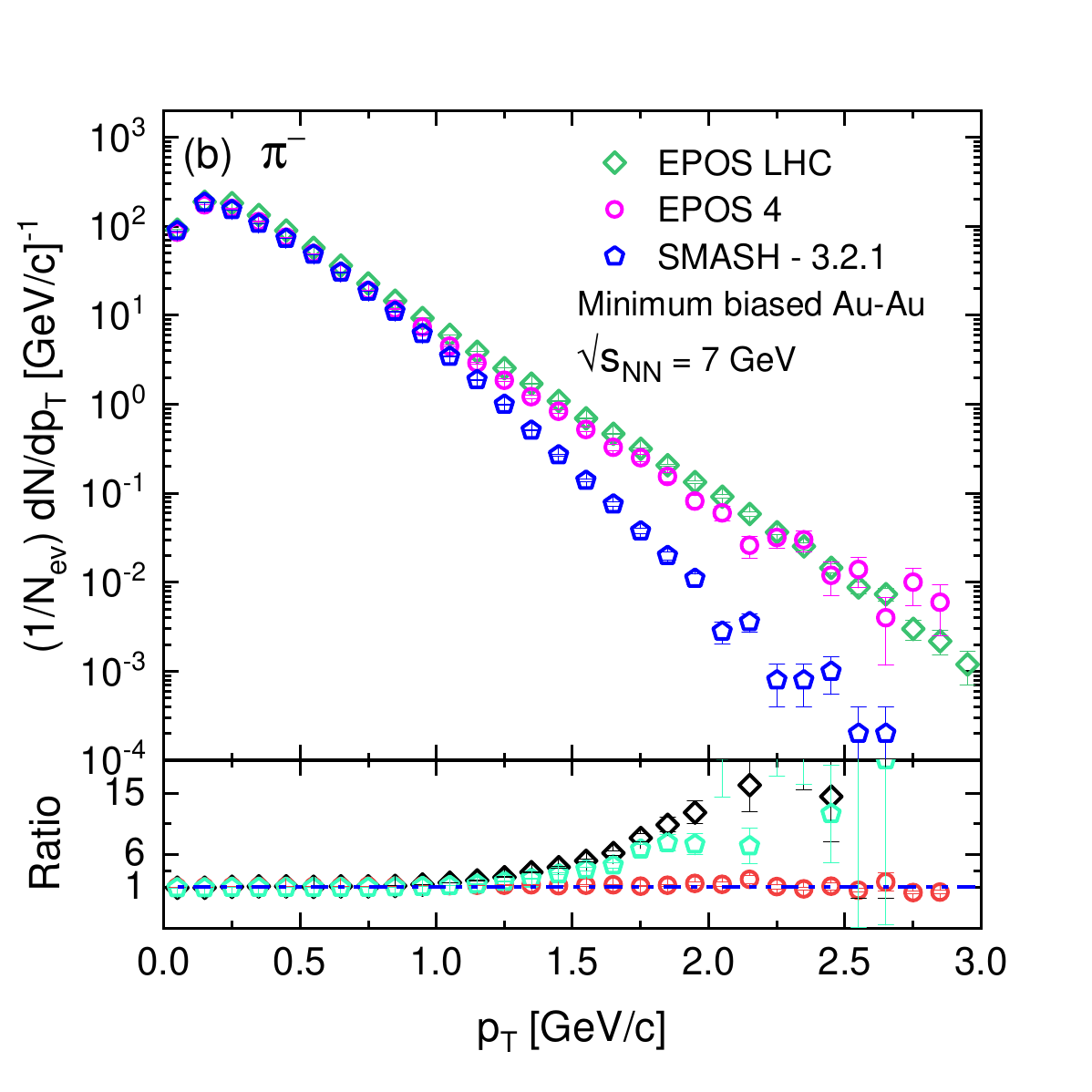}\vspace{-0.35cm} 
\includegraphics[width=0.32\textwidth]{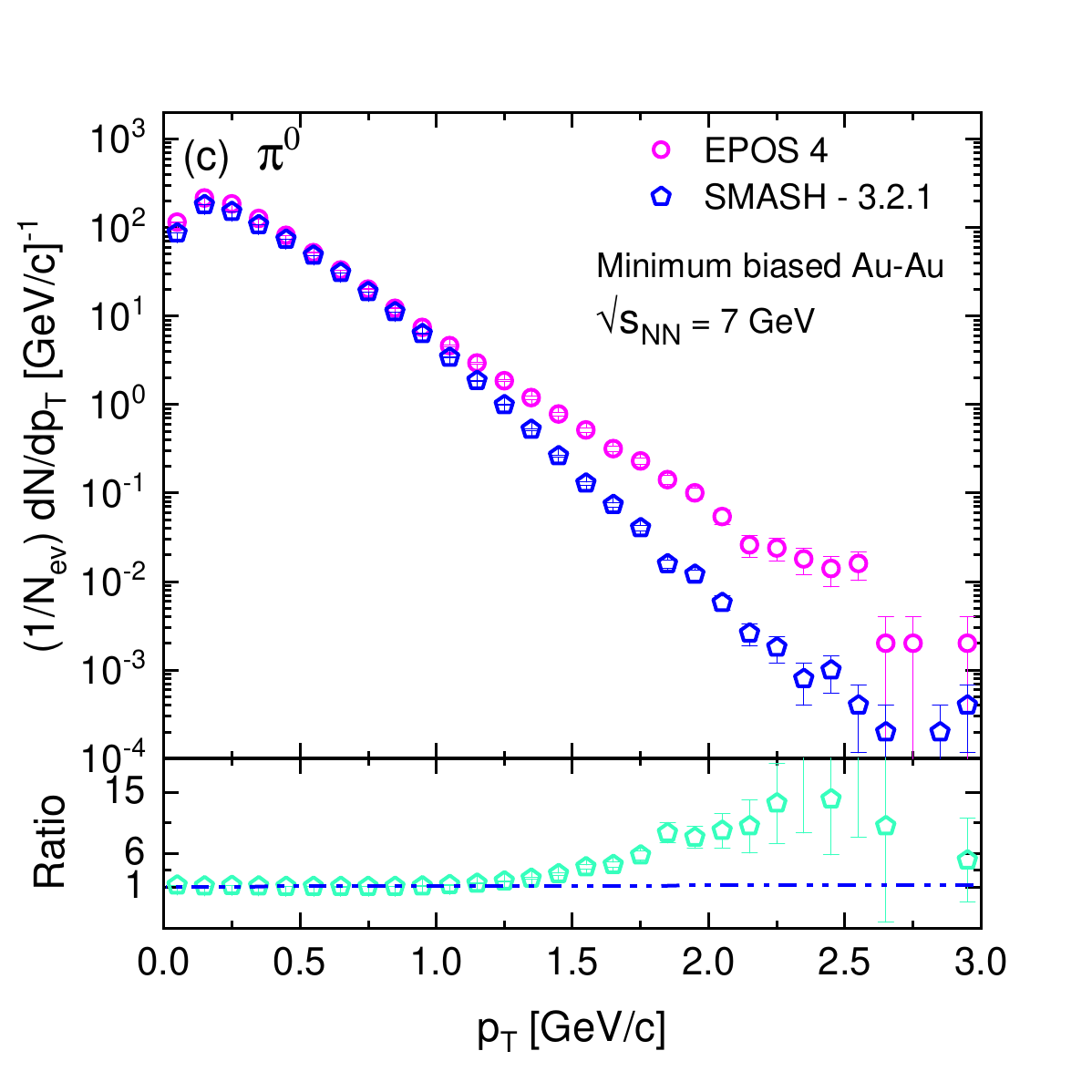}
\includegraphics[width=0.32\textwidth]{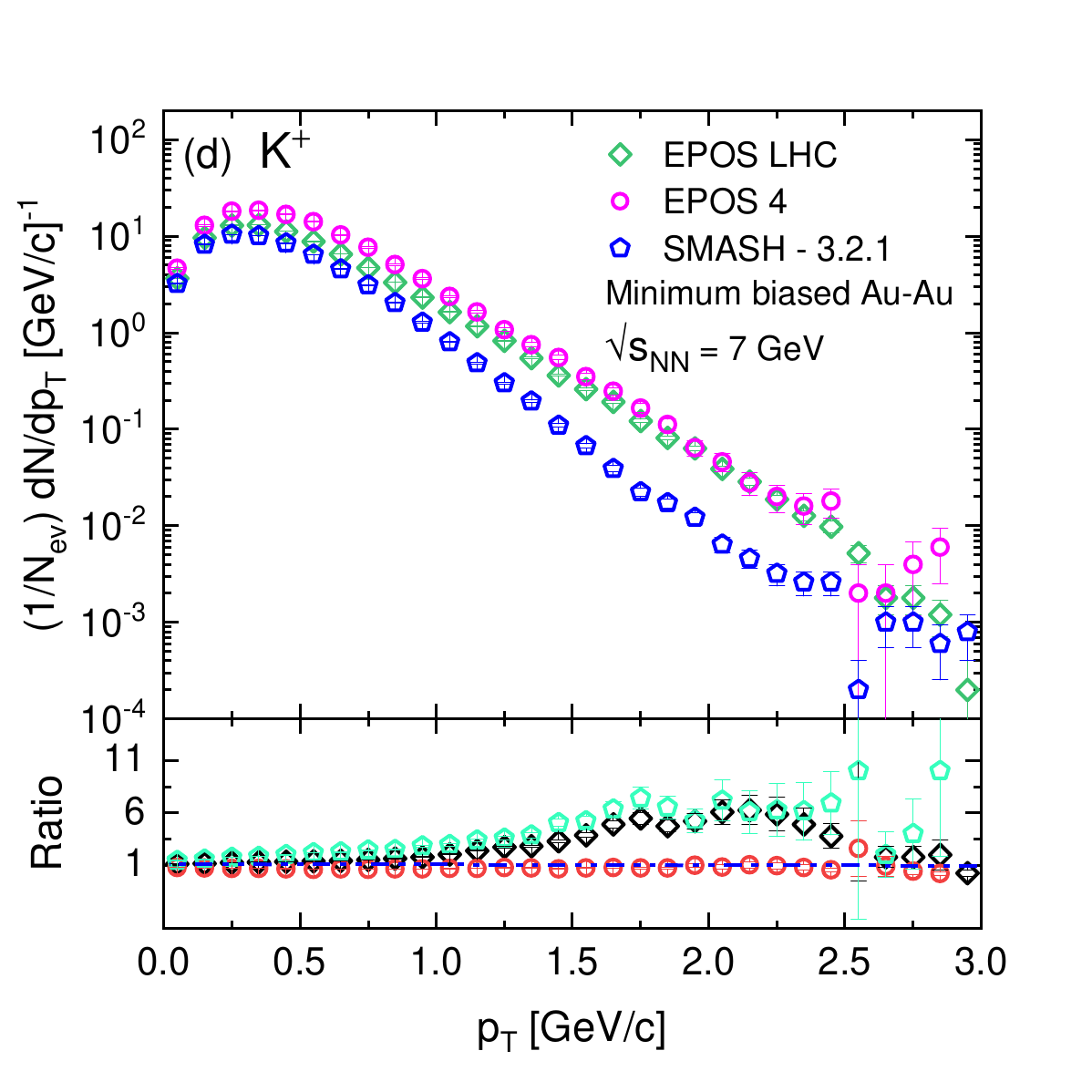}\vspace{-0.35cm}
\includegraphics[width=0.32\textwidth]{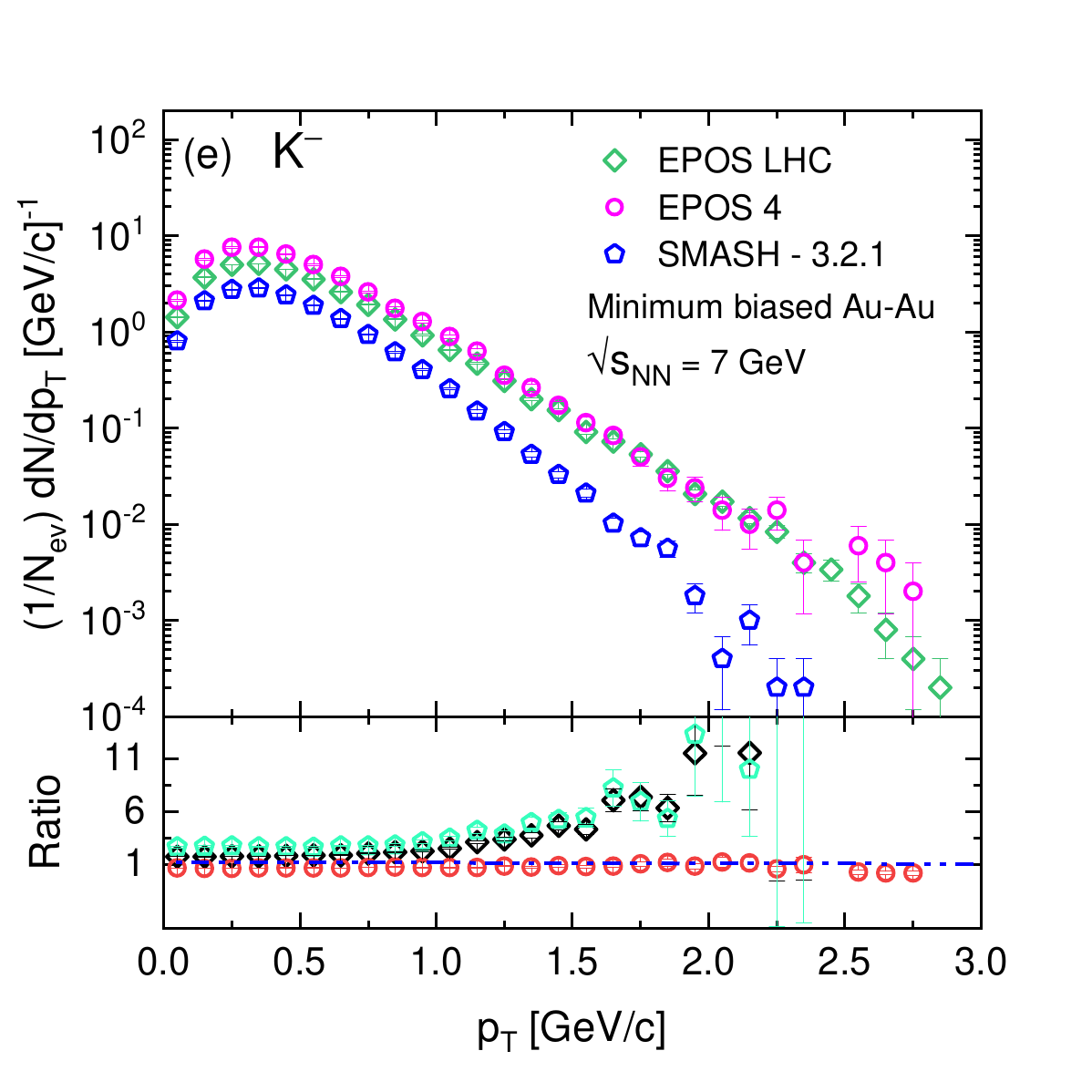}
\includegraphics[width=0.32\textwidth]{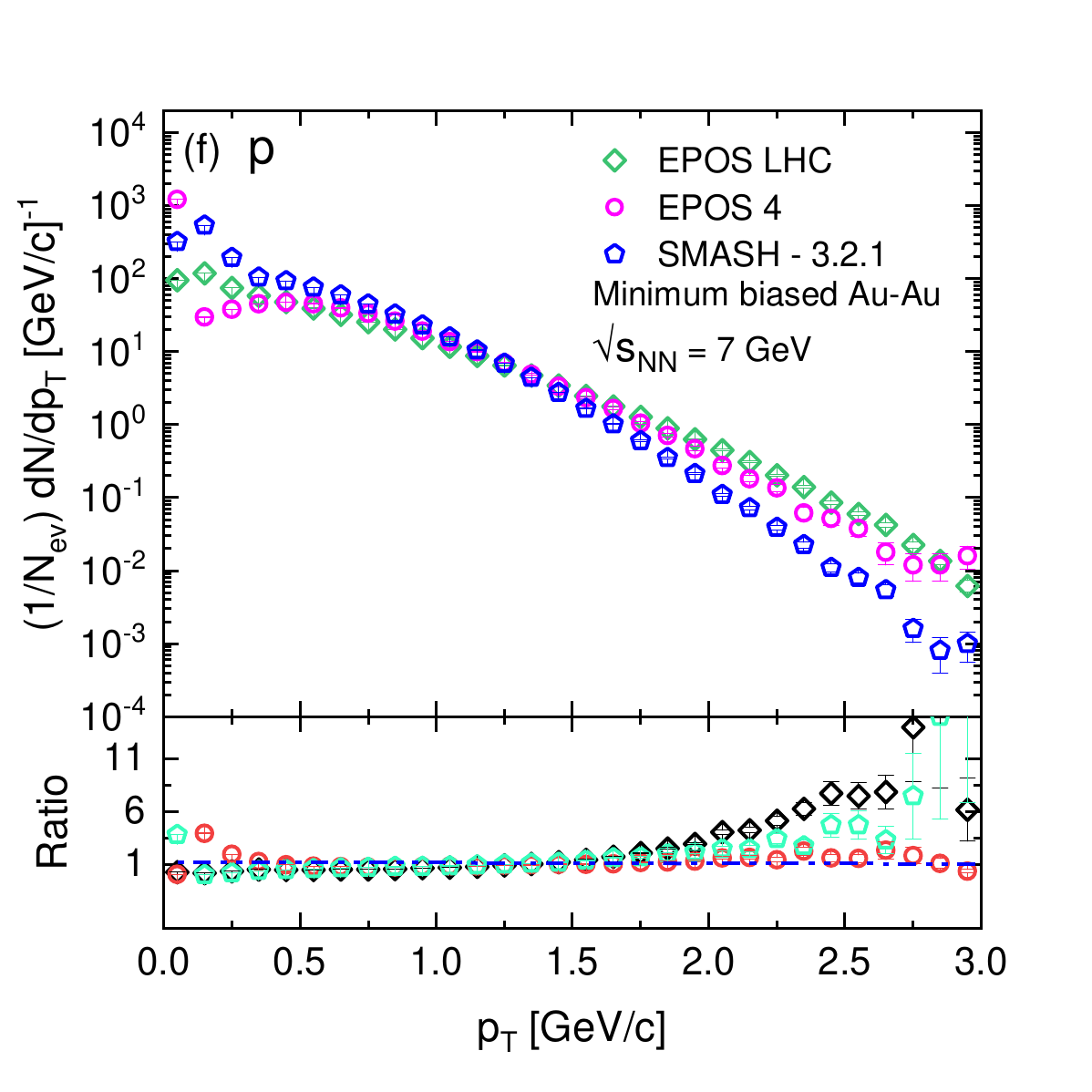}\vspace{-0.35cm}
\includegraphics[width=0.32\textwidth]{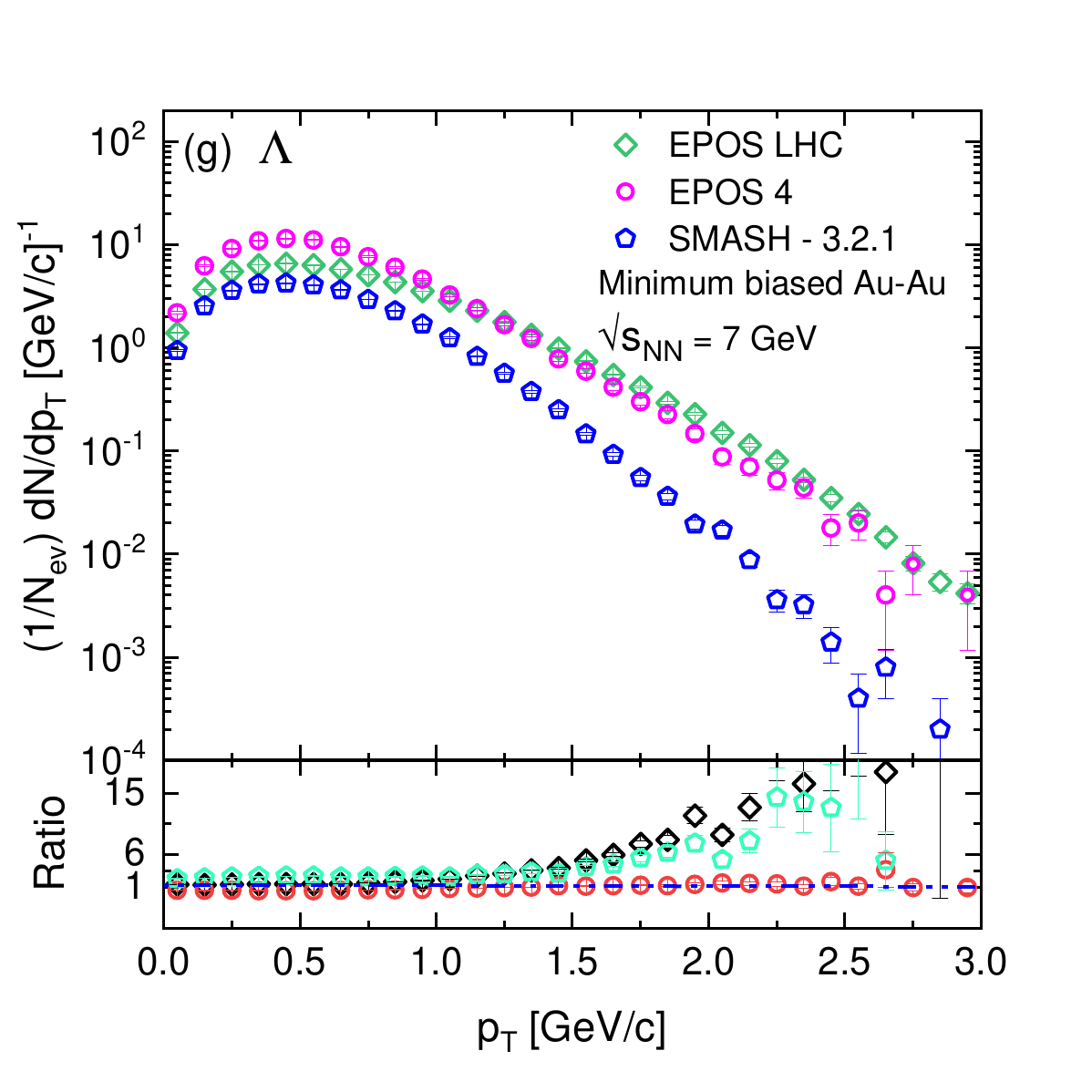}
\includegraphics[width=0.32\textwidth]{legends.pdf}\vspace{-0.35cm}
\caption {Event-normalized $dN/dp_T$ transverse-momentum spectra of identified hadrons in minimum-bias $Au+Au$ collisions at $\sqrt{s_{NN}}=7~GeV$. The EPOS-LHC, EPOS-4, and SMASH model predictions are compared to demonstrate variations in the shape of the spectrum, slope (soft vs. hard components), and relative yield at low and intermediate $p_T$.}
\end{figure*}
\begin{figure*}
\centering
\includegraphics[width=0.32\textwidth]{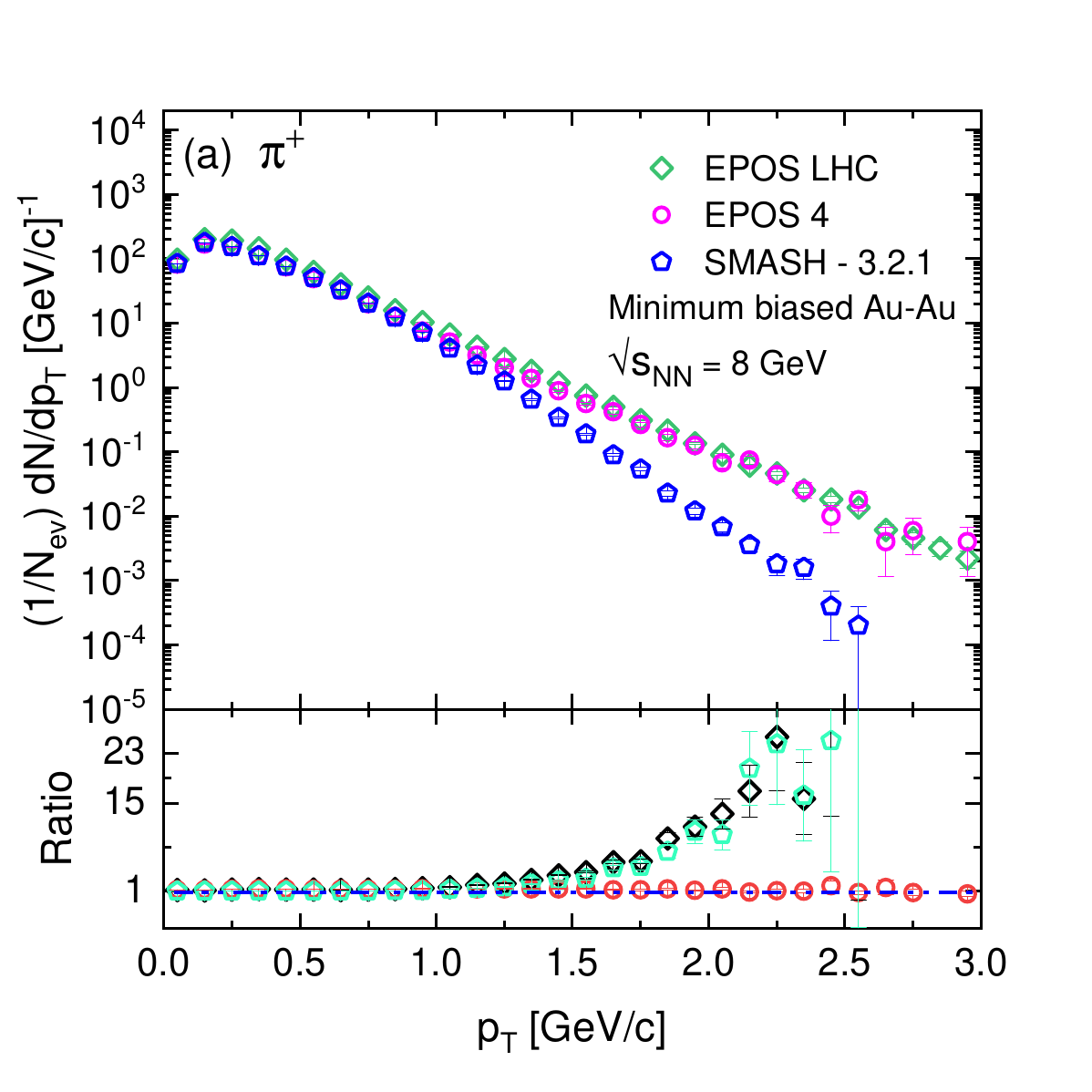}
\includegraphics[width=0.32\textwidth]{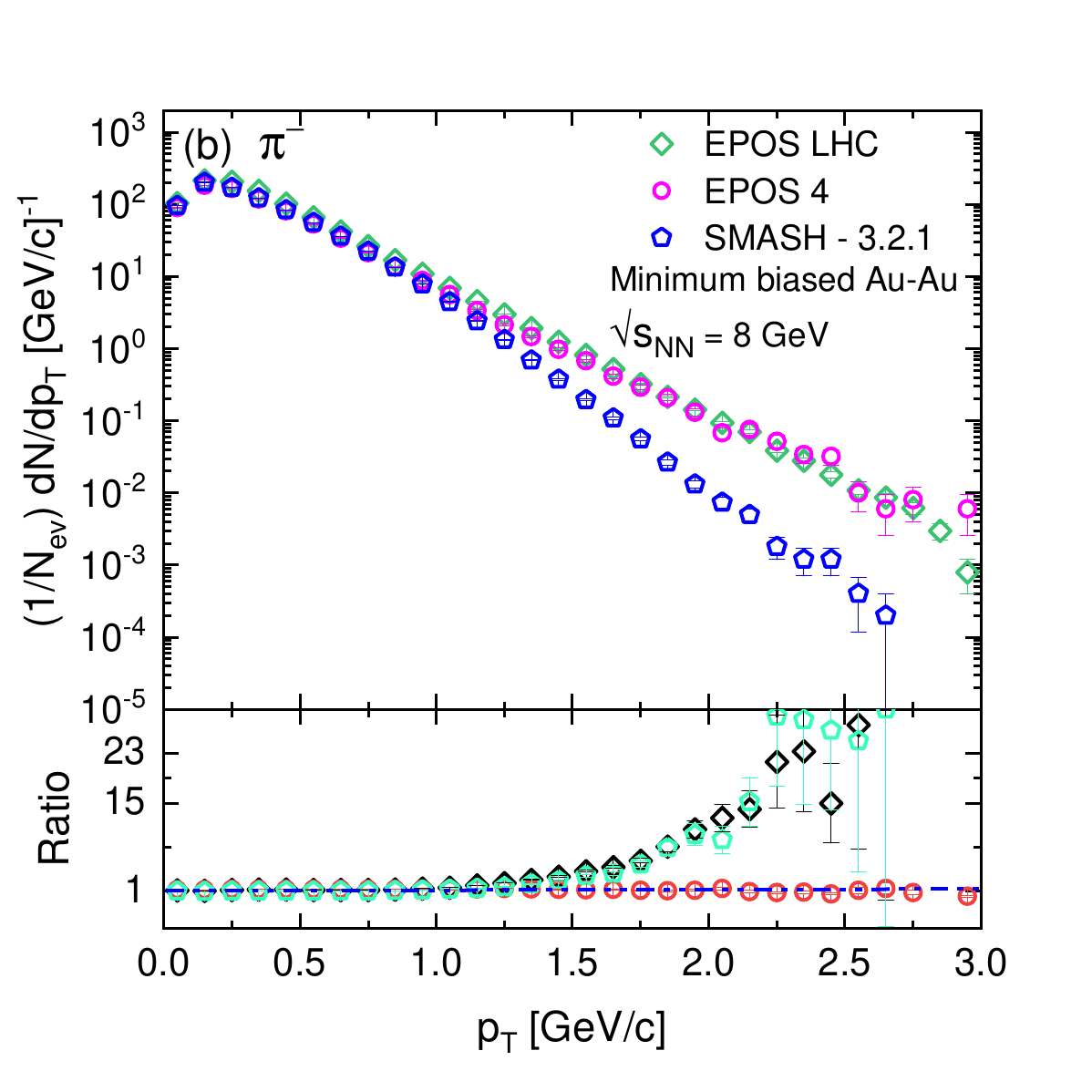}\vspace{-0.35cm} 
\includegraphics[width=0.32\textwidth]{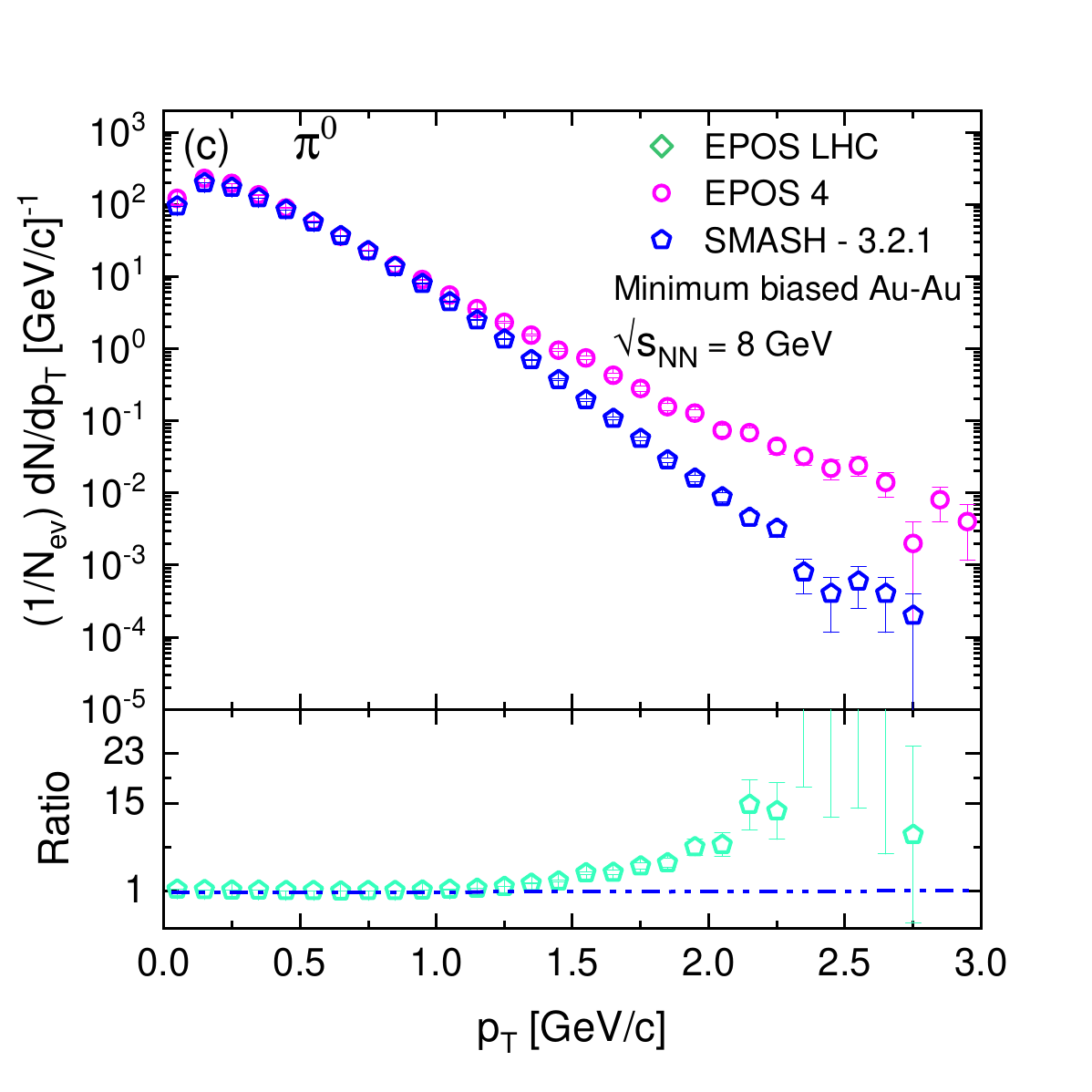}
\includegraphics[width=0.32\textwidth]{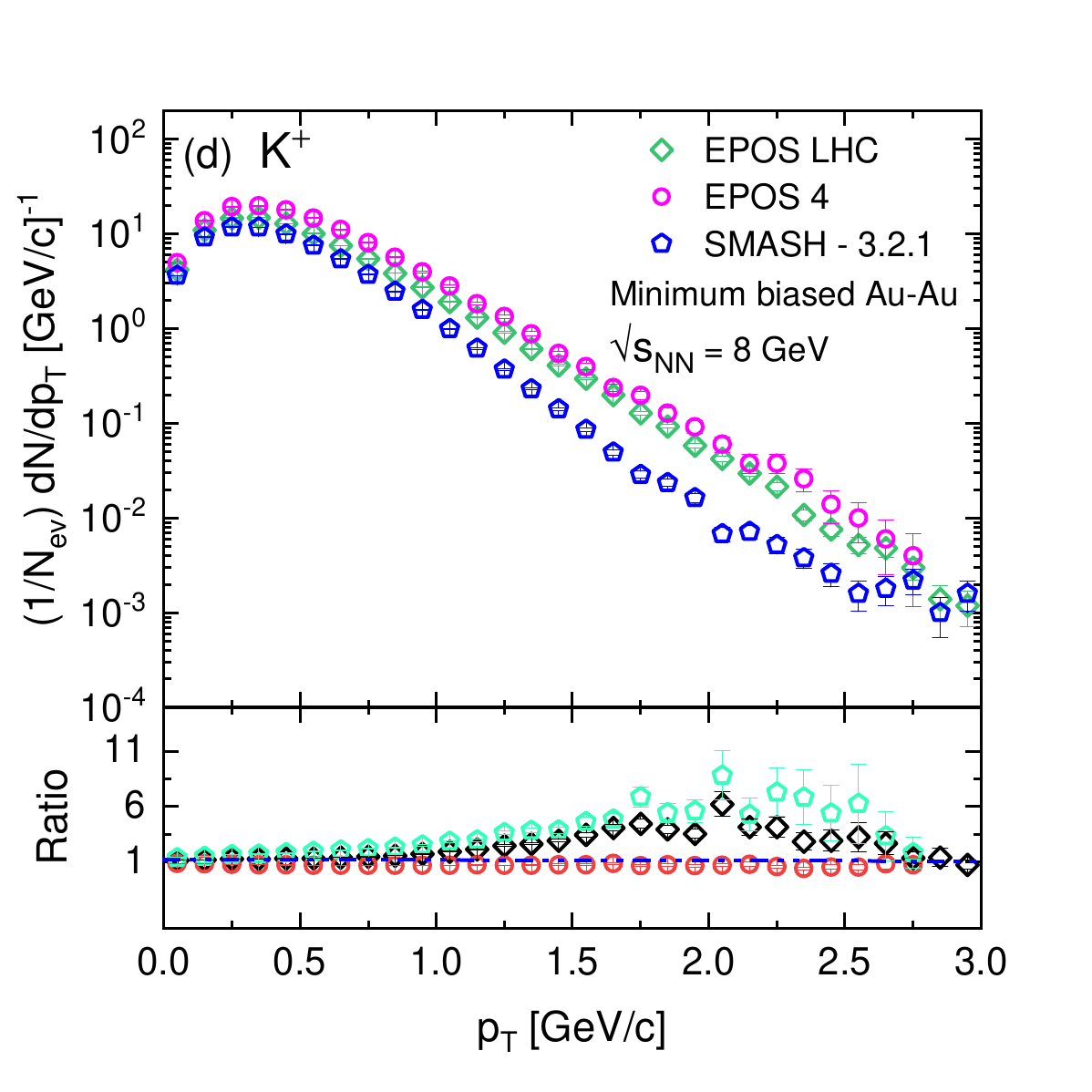}\vspace{-0.35cm}
\includegraphics[width=0.32\textwidth]{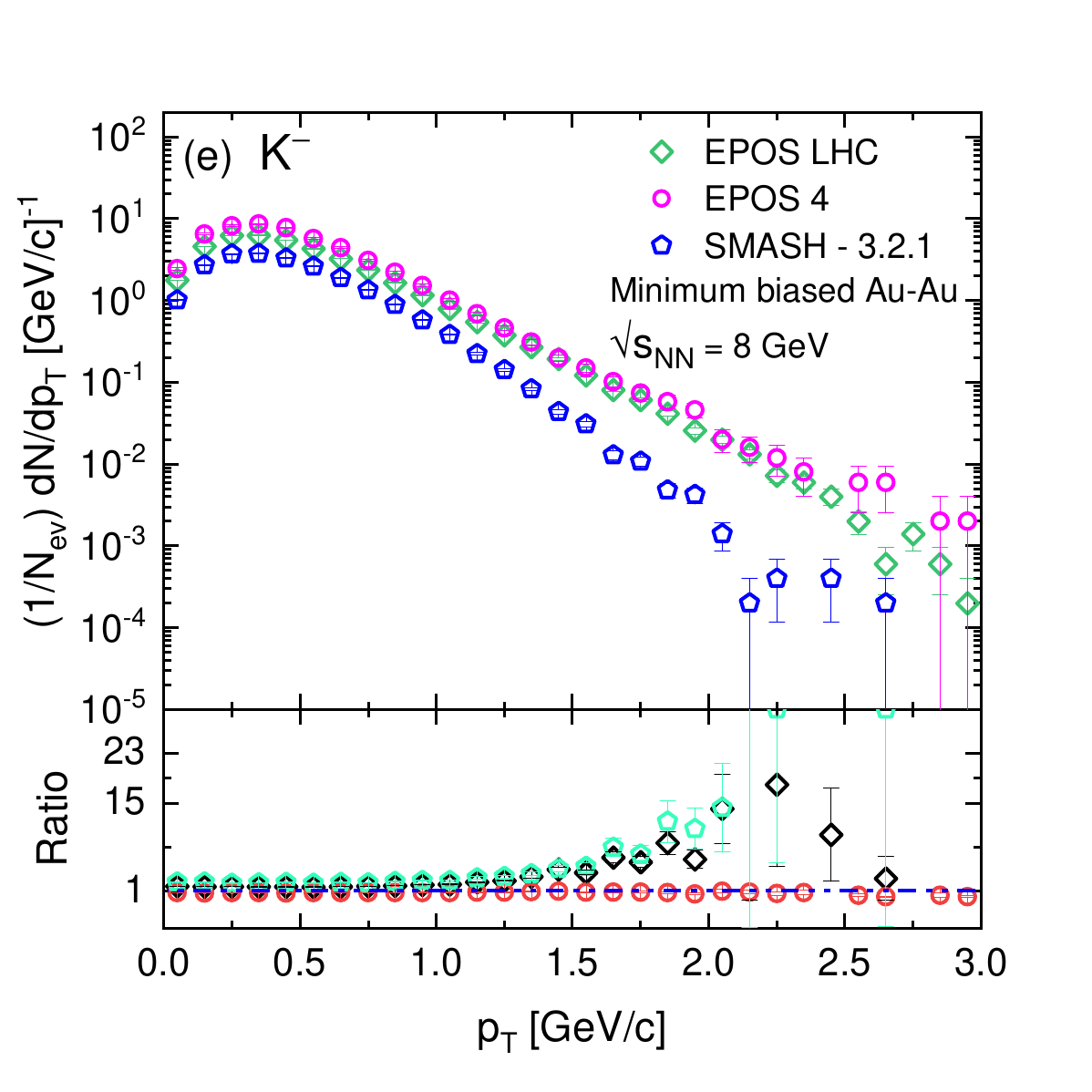}
\includegraphics[width=0.32\textwidth]{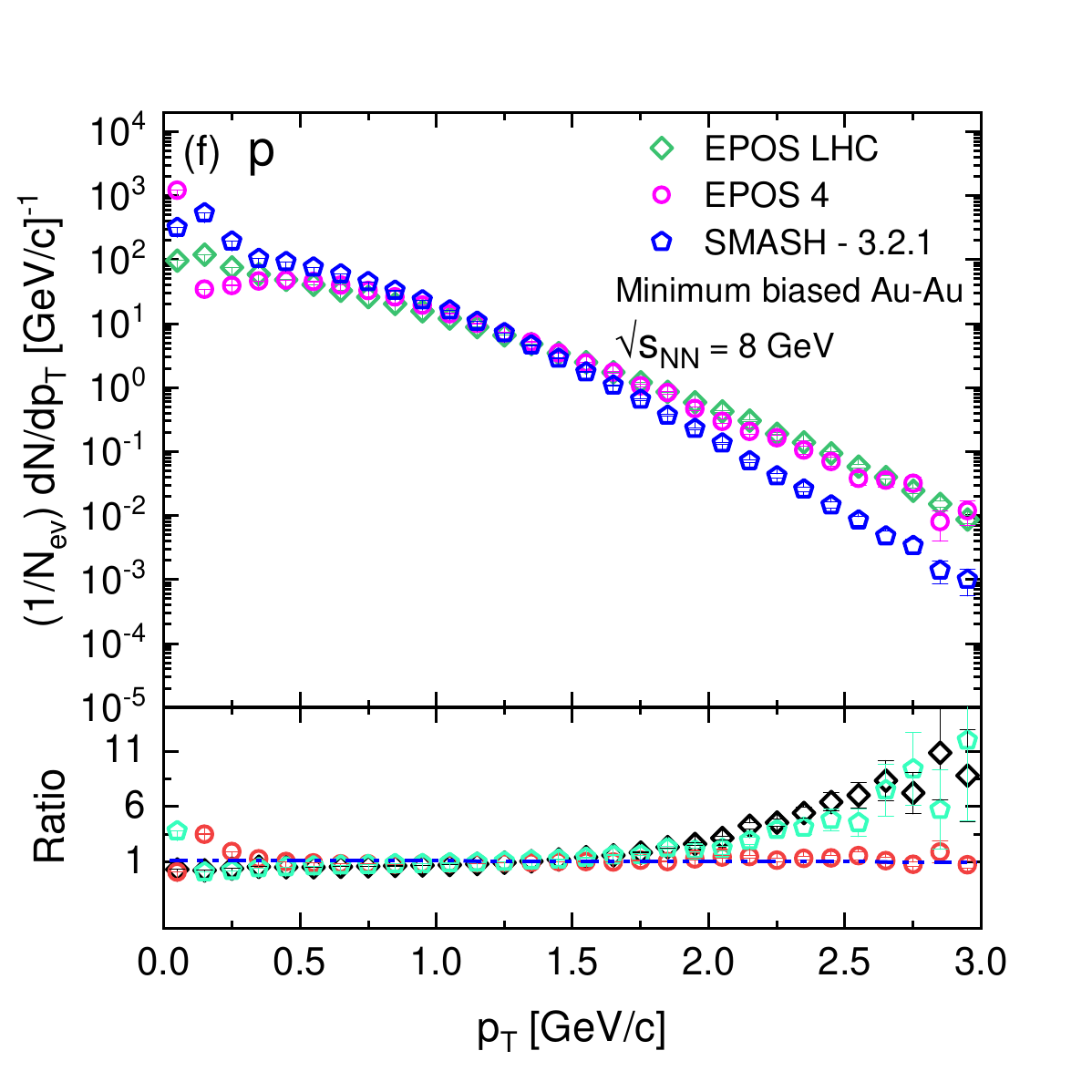}\vspace{-0.35cm}
\includegraphics[width=0.32\textwidth]{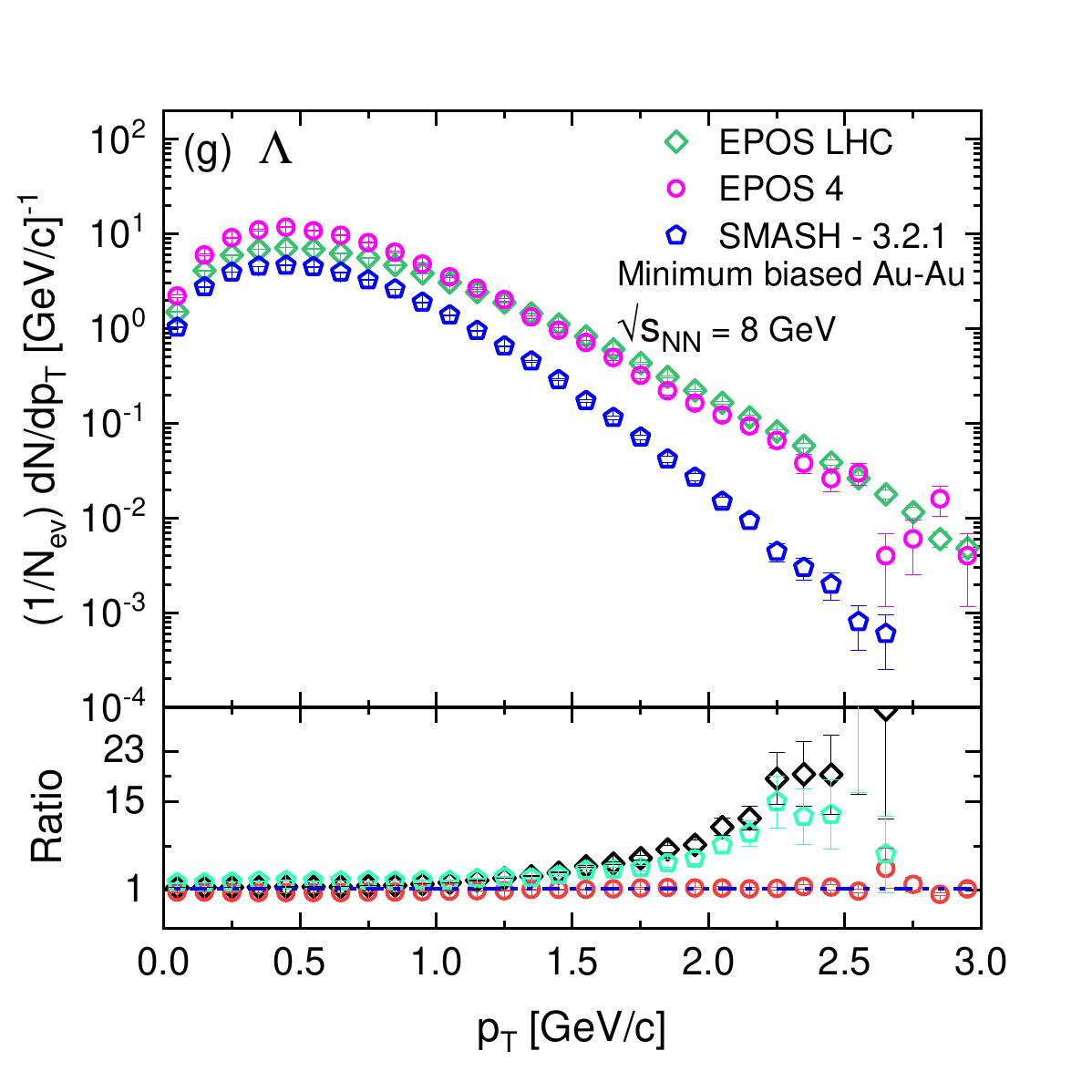}
\includegraphics[width=0.32\textwidth]{legends.pdf}\vspace{-0.35cm}
\caption {Transverse-momentum spectra (event-normalized $dN/dp_T$) of identified hadrons in minimum-bias $Au+Au$ collisions at $\sqrt{s_{NN}}=8~GeV$.}
\end{figure*}
\subsection{Transverse momentum distributions}
The $\pi^+$ production is shown in panel (a) of Fig. 4 as a function of $p_T$ at $6~GeV$ (event-normalized, on a log scale). There seems to be a distinct hierarchy, with SMASH giving the steepest spectrum (fastest fall-off with $p_T$), and EPOS-LHC and EPOS-4 giving spectra that are more flattened at intermediate $p_T$ (In the current energy scale, the range between $0.8~ GeV/c$ and $2.5~GeV/c$ is the intermediate-$p_T$. The reason this was chosen is that below $0.8~GeV/c$ resonance feed-down and bulk freeze-out are a significant contribution to the spectra, and above $0.8~GeV/c$ collective and string dynamics start to make a greater contribution. At such energies, there exists no actual high-$p_T$ domain in the RHIC/LHC sense, hence the range to a scale of about $3~GeV/c$ can be regarded as intermediate), EPOS-LHC being generally the flattest of the three. The physical reason is that SMASH is a microscopic hadronic cascade, which generates pions mostly through resonance decays and limited rescattering, hence it does not have a strong radial push, and it generates fewer pions at intermediate $p_T$. EPOS versions also feature string fragmentation and a working core area that adds extra transverse momentum, which changes part of the yield to higher $p_T$, creating a flatter shape. This distinction is significant since the intermediate-$p_T$ area is exposed to collective-like transverse processes: a steeper slope indicates more transfer of momentum perpendicular to the beam either through increased radial expansion or a more complex piece of string. This finding is consistent with slope analyses at energies \cite{alt2008energy, schnedermann1993thermal}, and the intermediate-$p_T$ window is identified as the most discriminating region to pion spectra at NICA energies. A slope measurement at this region, in an experiment, will distinguish hadronic and hybrid/string-inspired dynamics.

The same model ordering is observed in the $\pi^-$ $p_T$ spectrum at $6~GeV$, given in Fig. 4(b): SMASH, EPOS-4, and EPOS-LHC are steep at higher $p_T$, and any numerical low-order variation is negligible. This can be explained using the same physical argument: resonance feed-down and weak rescattering in SMASH put pions at low $p_T$, and EPOS models are an extension of this to pions with a finite amount of additional transverse momentum by a subgroup of pions through core dynamics or string fragmentation. This panel supports the statement that the use of intermediate-$p_T$ charged pions is a sensitive probe of transverse dynamics in cases where the species are rich. It also notes that mere agreement at very low $p_T$ does not mean that the physics are the same; it is the intermediate band that is the discriminating power.

The $\pi^0$ transverse momentum distributions are displayed in panels (c) of Figs. 4--6, at $\sqrt{s_{NN}} = 6~GeV,~7~GeV$ and $8~GeV$, respectively. At $6~GeV$, $p_T$ spectra are steeply falling in SMASH, with the bulk of the distributions concentrated in soft pions, $p_T < 1 ~GeV/c$, whereas EPOS-4 provides much flatter distributions extending into the intermediate region ($0.8-3 ~GeV/c$). This hardening is a manifestation of the incorporation of collective transverse expansion and string fragmentation in EPOS, not present in the hadronic-only SMASH. This trend continues at $\sqrt{s_{NN}} = 7~GeV$ and $8~GeV$, where the model differences increase at the high tail of $p_T$, i.e., $p_T>1.3~GeV/c$ with increasing the collision energies.

Much more explicit model dependence can be observed in the $K^+$ transverse momentum
spectrum given in Fig. 4(d). EPOS-4 yields better results in terms of overall predicting higher 
yields and a harder (flatter) spectral slope than SMASH, and EPOS-LHC is between the two. The physical 
cause is that kaons are not as resonance-dominated and are more influenced by the production 
process. In SMASH, threshold-limited associated production generates kaons, which lead to steep 
spectra. Other channels of production in EPOS-4 include strings, partonic-like dynamics, and so 
the kaons have increased multiplicities, as well as increased radial push, and thus flatter spectra. 
The importance lies in the fact that kaons are a pure probe of both the production of strangeness 
and collective acceleration, and therefore the spectra of the kaons directly probe the assumptions 
that the models make concerning physics. The kaon spectra at the energies of SPS and BES are sensitive probes of collectivity in baryon-rich matter, as far as the literature is 
concerned. The new aspect of this is that already at $\sqrt{s_{NN}}=6~GeV$, the kaons will exhibit strong 
model separation, and so their significance as discriminators at NICA.

The $K^-$ $p_T$ spectra in minimum-bias $Au+Au$ collision at $\sqrt{s_{NN}}=6~GeV$ is shown in Fig. 4(e). The overall 
strong suppression of $K^-$ spectra varies across models, although the shape and slope appear to 
vary significantly. EPOS-4 is predictive of high yields and a steep slope, whereas SMASH is 
predictive of steep, damped spectra. This can be attributed to their microscopic production 
mechanisms: $K^-$ need the creation of $s\bar{s}$ pairs, which is inefficient in SMASH but not in 
EPOS because of the creation of string fragments. The steeper slope in EPOS-4 is because partonic-like processes can give more transverse momentum to the produced pairs. Physically, this renders $K^-$ especially sensitive to the issue of whether the system has early 
partonic-like dynamics or is limited to hadronic processes. This is important because $K^-$ spectra 
can be used as a direct measure of the level of strangeness equilibration in the system. What is new 
is to demonstrate that, at low $\sqrt{s_{NN}}$ when $K^-$ yields are small, the spectral shape has a strong 
message regarding underlying dynamics, and this is critical to NICA.

The proton $p_T$-spectrum at $6~GeV$, displayed in Fig. 4(f), demonstrates some model 
differences. SMASH creates steeper spectra, and EPOS-4 generates flatter distributions at 
intermediate $p_T$. The physical explanation is that in SMASH, the protons still maintain much 
of the original momentum in hadronic scattering, with little collective acceleration, whereas in 
EPOS-4, the core-corona picture, instills radial flow in the core, and it becomes flatter. EPOS-LHC 
is in the middle as it is tuned as a hybrid. This difference plays an important role since both baryon 
transport and radial collectivity are encoded in proton spectra. The wider context results of RHIC 
BES demonstrated that baryon spectra hardened with energy, which can be interpreted as a 
manifestation of collectivity growth. Such hardening is already anticipated to be observable in the 
$6~GeV$ region in EPOS-4, but not in SMASH, and proton spectra are expected to be a decisive 
observable.

We note that the differences in the low and intermediate $p_T$ proton spectra between EPOS-LHC, EPOS-4, and SMASH are expected to be affected not only by their initial state and possible hydrodynamic dynamics, but also by their different treatment of the late hadronic stage. EPOS-4 features an explicit hadronic rescattering phase (UrQMD) following hydrodynamics. In contrast, SMASH incorporates hadronic transport throughout the entire evolution, while EPOS-LHC employs a parametrized approach for flow and resonance decays, without a separate transport afterburner. The baryon $p_T$ spectra comparison must hence be taken as a combined probe of the dynamics in early times as well as a probe of the model-dependent hadronic phase.

Fig. 4(g) displays the $\Lambda$ $p_T$ distribution in minimum-bias $Au+Au$ collisions at $6~GeV$. The distinct difference between the three models arises: SMASH has the steepest slope, EPOS-4 and EPOS-LHC have flatter, harder spectra, EPOS-LHC being a little harder than EPOS-4. It means that SMASH focuses on $\Lambda$ yield at low $p_T$, whereas EPOS versions still have a large population in the medium-$p_T$ range ($0.8- 3~GeV/c$). Physically, this is a manifestation of the production processes: SMASH produces $\Lambda$ in the associated hadronic processes (such as $NN\xrightarrow{}N\Lambda K$), as well as in the second rescattering, which is scarcely able to deposit large $p_T$. Conversely, the EPOS models include a hydrodynamic-like core and string fragmentation dynamics, which increases some of the created $\Lambda$ and moves the yield to higher $p_T$. Another contribution in this case is resonance feed down of the heavier strange baryons (e.g. $\Sigma^*\xrightarrow{}\Lambda+\pi,~\Xi\xrightarrow{}\Lambda+K$) which slightly broadens the spectrum. This late-stage hadronic freeze-out component is increased by SMASH, but EPOS hydrodynamics freeze $\Lambda$ earlier, and are found to maintain the higher-$p_T$ push. The predictions in EPOS at $6~GeV$ already indicate the beginning of nontrivial collective or string-like effects, compared to AGS and SPS systematics, with most of the $\Lambda$ spectra being soft and exponential. NICA $\Lambda$ $p_T$ measurements will therefore present a useful initial test of whether or not strangeness in a baryon-rich regime may be found to be collectively hardened.

The $p_T$ spectrum of $\pi^+$ at $7~GeV$ is shown in Fig. 5(a). The same pattern is observed at $7~GeV$, except that all spectra have been slightly hardened (flattered): EPOS-LHC and EPOS-4 are flatter than SMASH in the intermediate-$p_T$ window, and absolute yields are higher. The physical explanation is that small changes in available energy allow more and more secondary scatterings and a little stronger collective effects in EPOS models, slightly increasing the yields up to higher $p_T$. In SMASH, hadronic rescattering is the strongest mechanism, and without an explicit early dense medium, it cannot succeed in increasing the intermediate-$p_T$ tail, which occurs in EPOS. This panel thus demonstrates that the model separation is strong to minor variations in energy. In practice, the energy scan of NICA must be capable of tracing this evolution and determining whether hardening with $\sqrt{s_{NN}}$ is consistent with EPOS-like processes or has to be consistent with hadronic transport.

The $p_T$ spectrum of $\pi^-$ at $7~GeV$ is shown in Fig. 5(b). Observations of $\pi^+$ are verified by the $\pi^-$ distribution at $7~GeV$: the models of EPOS predict a more significant intermediate-$p_T$ component compared to SMASH. Physically, this indicates that processes causing the transfer of transverse momentum, such as hydrodynamic pressure gradients in a core, more energetic string fragmentation, or multiple low-angle scatterings, are more effective in EPOS. Minor variations between EPOS-LHC and EPOS-4 are related to implementation and tuning options (core fraction, string tension, viscous damping), but both are evidently out of SMASH predictions. This panel further intensifies the suggestion that the intermediate-$p_T$ area should be a target in the initial NICA analyses.

In the $K^+$ spectrum at $7~GeV$, shown in Fig. 5(d), there is an 
improvement in yield compared to $6~GeV$ because the frequency of associated production increases 
at the higher energy. EPOS-4 is still predicting higher normalization and stronger slopes than 
SMASH, with EPOS-LHC intermediate once again. This difference is explained by the fact that 
kaons are far less affected by resonance decays than pions; the slopes of kaons are highly 
dependent on the dynamics of their production. In EPOS-4, these two effects of increased 
production channels and radial acceleration lead to flatter and more extended distributions. Kaons 
are confined to hadronic production at near threshold in SMASH, which is less steep. This 
difference, physically, makes kaons clean strangeness and flow probes. From the literature 
perspective, the kaon slopes were already known to be significant in collectivity diagnosis at SPS \cite{alt2008pion, afanasiev2002energy}. What is new with this is that even at the $7~GeV$ regime, kaons are highly model dependent, which 
implies that they will be determinative in the future measurements by NICA.

Fig. 5(e) represents $K^-$ $p_T$ distributions at $7~GeV$. The 
increase of $K^-$ spectra at $7~GeV$ is compared with the decrease at $6~GeV$, but the overall increases 
are suppressed, with SMASH projecting the steepest slope and lowest yields. EPOS-4, on the other 
hand, has better yields and less skewed distributions, with more efficient $s\bar{s}$ pair production. The physical explanation is that the hadronic thresholds of SMASH do not allow the production of $K^-$ easily, while the string dynamics of 
EPOS-4 allows its production. The increased slope in EPOS-4 indicates that such pairs are produced in more 
energetic processes with greater momentum transfer. The importance of this observable is that the 
production of $K^-$ directly measures the possibility of pair creation being confined to hadronic 
scatterings or not. Based on the literature, $K^-$ spectra played a significant role in SPS in detecting 
non-hadronic behavior \cite{alt2008pion}. The novelty of this is that $K^-$ at $7~GeV$ already develops such deviations in model predictions, and thus is a particularly sensitive probe to NICA.

In SMASH and EPOS-4, protons at $7~GeV$ show an increase in normalization, shown in Fig. 5(f), but still retain the steep 
slope and flatter slope pattern, respectively. The physical explanation is that in SMASH, protons 
freeze out mostly via hadronic scatterings with little radial flow. In contrast, in EPOS-4, the 
hydrodynamic-like expansion increases protons more effectively, such that they have more complex  
spectra. EPOS-LHC is once again in the middle, as it has hybrid physics. Its importance 
is that protons can provide direct information on baryon transport and collectivity, and spectral 
slopes of protons can give a clear difference between hadronic and hybrid models. In a more general view, radial flow sensitivity was also found to be sensitive to baryon slopes at RHIC BES \cite{adamczyk2017bulk}, and the findings are now extended to the 7 GeV scale. The novelty here is that proton spectra, even at these relatively low energies, can give intense discrimination between rival dynamics.

The $\Lambda$ spectrum at $7~GeV$ is shown in Fig. 5(g). The hierarchy remains the same; i.e., SMASH provides the steepest distribution, while EPOS-LHC and EPOS-4 are flatter, with EPOS-LHC being slightly harder. The total yield is greater than at $6~GeV$, as a result of a greater phase space for the production of strange baryons. The energy dependence is subtle but systematic, and as the intermediate-$p_T$ yield in EPOS increases with respect to SMASH, it indicates the growing role of collective-like dynamics in hybrid models. Physically, Lambda particles are made both directly and by the feed-down of higher mass states. Feed-down even at $7~GeV$ includes $\Sigma$ and $\Xi$, and the transverse impulse in EPOS is stronger, so even the secondary $\Lambda$ receives more momentum than SMASH. More so, the time of kinetic freeze-out is essential: in SMASH, $\Lambda$ decouple later, when they have lost some of their momentum due to rescattering, whereas in EPOS, they decouple earlier, when they remain attached to the expanding core and have a higher $p_T$. These measurements are related to those of the SPS era \cite{antinori2004study}, in which $\Lambda$ hardening was used as an indicator of collective expansion; the simulations show that these characteristics could also be observed in the energy scale of NICA. Fig. 5(g) predicts that differences in $\Lambda$ slopes between intermediate $p_T$ are predictable and could distinguish between hadronic and core-corona dynamics.

Fig. 6(a) shows the $p_T$ distribution of $\pi^+$ at $8~GeV$. Once again, at $8~GeV$, the intermediate-$p_T$ hardening of EPOS-LHC and EPOS-4 compared to SMASH is a bit stronger than at lower energies. EPOS-LHC is slightly flatter than EPOS-4, but both have higher yields in the intermediate $p_T$ range than SMASH. The physical origin is the same: the effective core and string dynamics of EPOS result in a fraction of pions with larger transverse momentum, whereas the hadronic cascade of SMASH does not. The increase in energy contributes to the increased visibility of this behavior, as with increased energy per nucleon, there is an increased transverse energy conversion. Experimentally, this panel shows that any hardening found with energy in the $\pi^+$ spectrum would be a critical discriminant: a tendency of the sort of EPOS would suggest increasing collective or semi-hard contributions as energy increases.

The same conclusions are drawn from the $\pi^-$ $p_T$ spectrum at $8~GeV$, as shown in Fig. 6(b): EPOS models continue to exhibit flatter intermediate-$p_T$ slopes compared to SMASH, and the relative differences remain unchanged or change only slightly with energy. Since the charged-pion spectra are by far the most accurate data that experiments can obtain, these panels provide a practical avenue for limiting the existence and strength of collective-like transverse dynamics at NICA energies. Using it in conjunction with baryon and strangeness measures, the distribution over these $\pi^+$ and $\pi^-$ panels will provide a comprehensive account of the initial and final phases of the collision.

The best separation of models is observed at $8~GeV$ 
with $K^+$ spectra as shown in Fig. 6(d). EPOS-4 yields higher and has far more difficult slopes, whereas SMASH 
is far steeper. The physical explanation is that associated production channels are more common 
now, and the string dynamics of EPOS-4 increase the yield and flow. Kaons in SMASH are 
confined to hadronic scattering, and therefore, they are limited in spectra. What is important is that 
kaons give a direct measurement of strangeness enhancement and collective dynamics, and their 
slopes at $8~GeV$ are dramatically model dependent. The new thing is that this race offers possibly 
the cleanest test of rival models at NICA.

The increase in $K^-$ yields goes further 
at $8~GeV$ as displayed in Fig. 6(e), but still suppression relative to $K^+$ is present. The EPOS-4 generates greater yields and flatter slopes, and the SMASH yields are steep and suppressed. The physical cause is that pair 
creation is more effective at large collision energies, yet in the models that allow string dynamics, it is the process 
that takes place at the required rate. The importance is that $K^-$ spectra can be a pure identifier of 
hadronic-only and hybrid approaches. In a larger sense, the argument of nonhadronic dynamics 
was based on the slope of $K^-$ at SPS, and we have done the same at $8~GeV$. The difference is that 
$K^-$ spectra at NICA energies can unambiguously determine whether or not partonic-like 
processes are already present.

Proton $p_T$ spectra at $8~GeV$, given in Fig. 6(f), have once again steeper 
slopes in SMASH than in EPOS-4 and EPOS-LHC. The physical cause is that 
collective radial expansion is more powerful at higher collision energies in the EPOS structure, which increases 
protons to higher $p_T$. Protons in SMASH are confined to hadronic scatterings, which have 
softer slopes. The importance is that protons act as carriers of baryons and directly probe the 
stopping and radial flow, and models clearly distinguish their slopes. In terms of the 
literature, the findings of baryon slopes at RHIC BES were considered to be the key to the 
diagnosis of the collectivity appearance \cite{adamczyk2017bulk}, and our results show this sensitivity in the NICA range. 
The new thing is that proton slopes at $8~GeV$ can be among the most obvious distinguishing 
between hadronic and partonic processes.
 
The spectrum of the $\Lambda$ is shown in Fig. 6(g) at $\sqrt{s_{NN}} = 8~GeV$. The results are most clearly differentiated into SMASH, EPOS-4, and EPOS-LHC. SMASH is the only one to generate a steep slope, while EPOS-4 and, more so, EPOS-LHC result in very flat spectra, retaining yield in the $0.8-3~GeV/c$ region to a substantial degree \cite{hagedorn1965statistical}. The cumulative $\Lambda$ yield is once more increased, which is in line with increased energy supply to produce strange baryons. Significantly, the EPOS curves gain a conspicuous shoulder in the intermediate-$p_T$ range that is not present in SMASH, which is an accumulated impact of stronger string excitations and collective transverse expansion at this energy. Physically, this pattern is persistent at $6-8~GeV$ and, therefore, suggests that the dynamics of transverse expansion and freeze-out are sensitized by $\Lambda$ spectra, as well as the production of strangeness. The contributions of feed-down still exist, albeit with the hard momentum distribution of the parent states inherited in EPOS by the feed-down Lambda, and in SMASH by the hadronic feed-down at a late stage, being mostly soft. The results of the EPOS at $8~GeV$ indicate that NICA can potentially easily identify the same collectivity-like behavior at lower energies than we had previously suggested, due to the systematic hardening of $\Lambda$ with centrality and energy. Experimentally establishing the presence of such hardening would provide strong evidence that the system is establishing pressure gradients on a large enough scale to accelerate strange baryons, which is difficult to believe within a purely hadronic scenario.
\begin{figure*}
\centering
\includegraphics[width=0.32\textwidth]{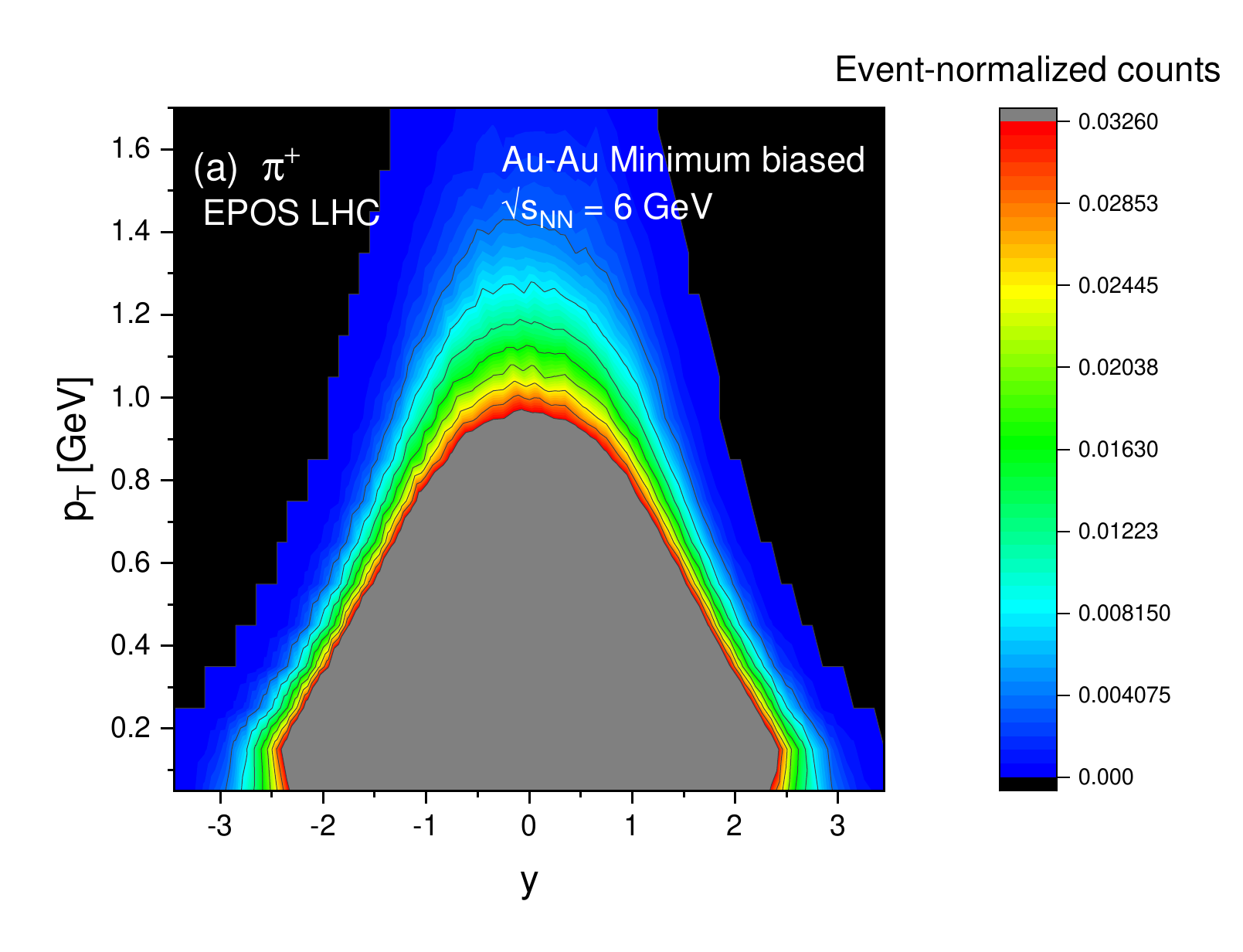}
\includegraphics[width=0.32\textwidth]{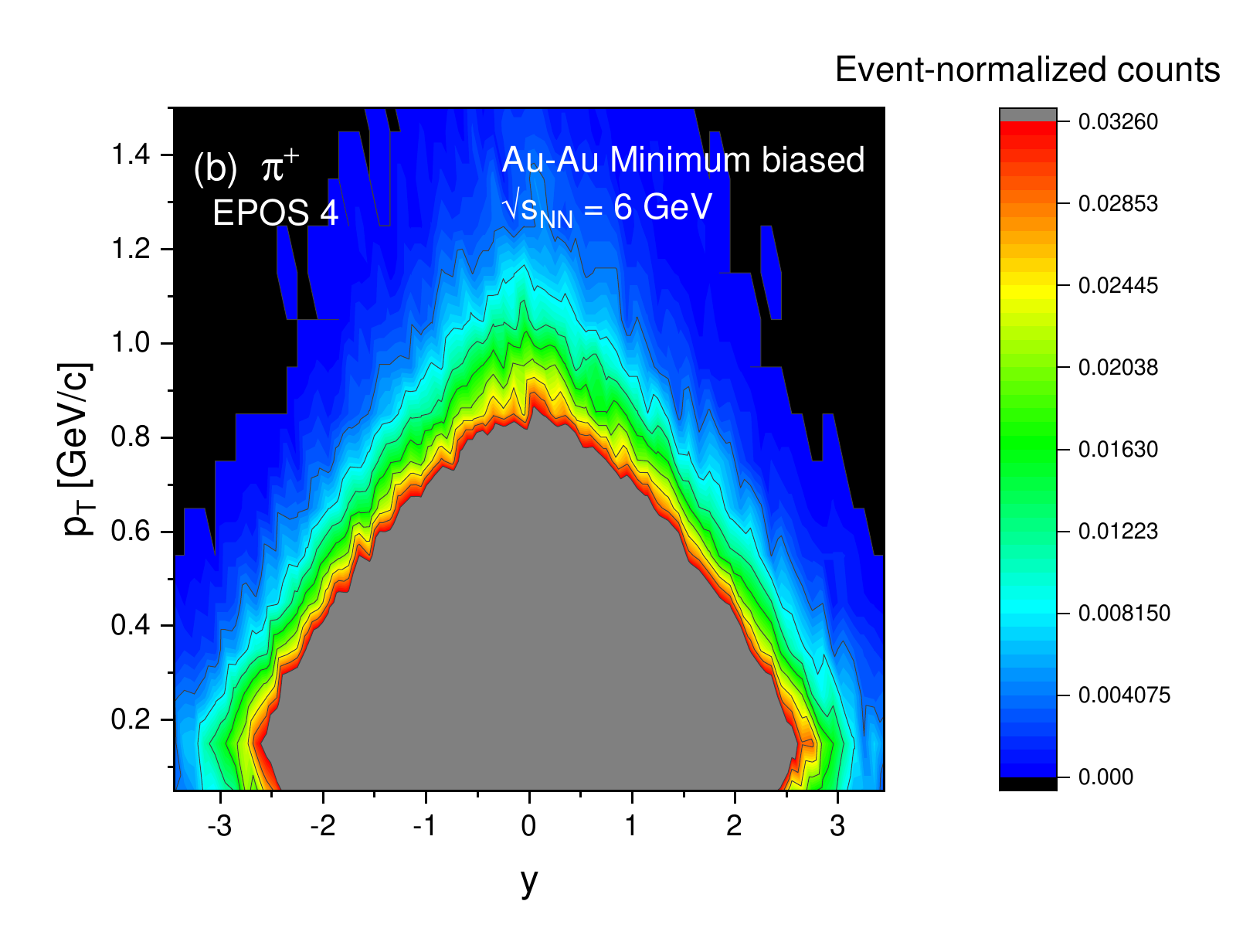} 
\includegraphics[width=0.32\textwidth]{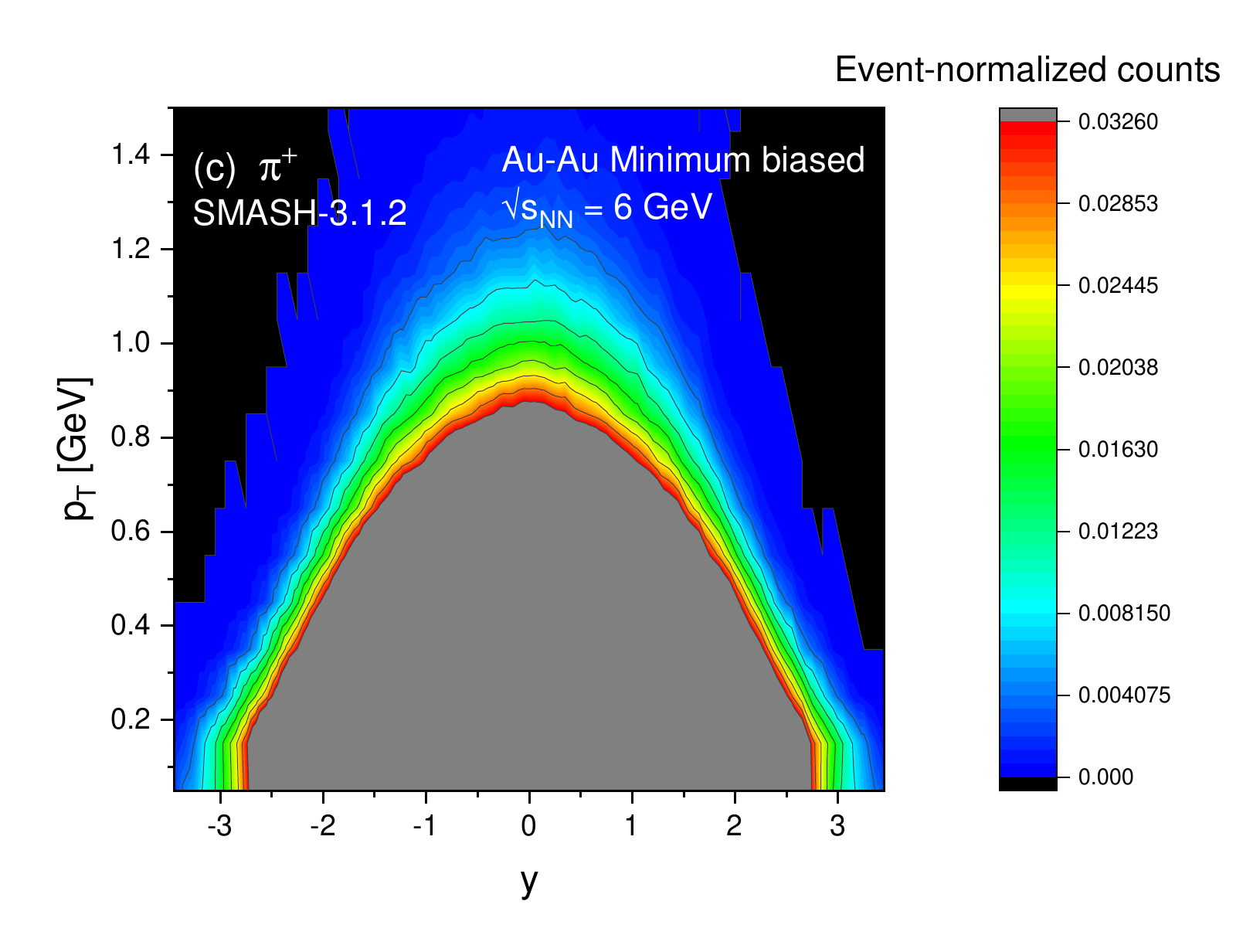}\vspace{-0.35cm}
\includegraphics[width=0.32\textwidth]{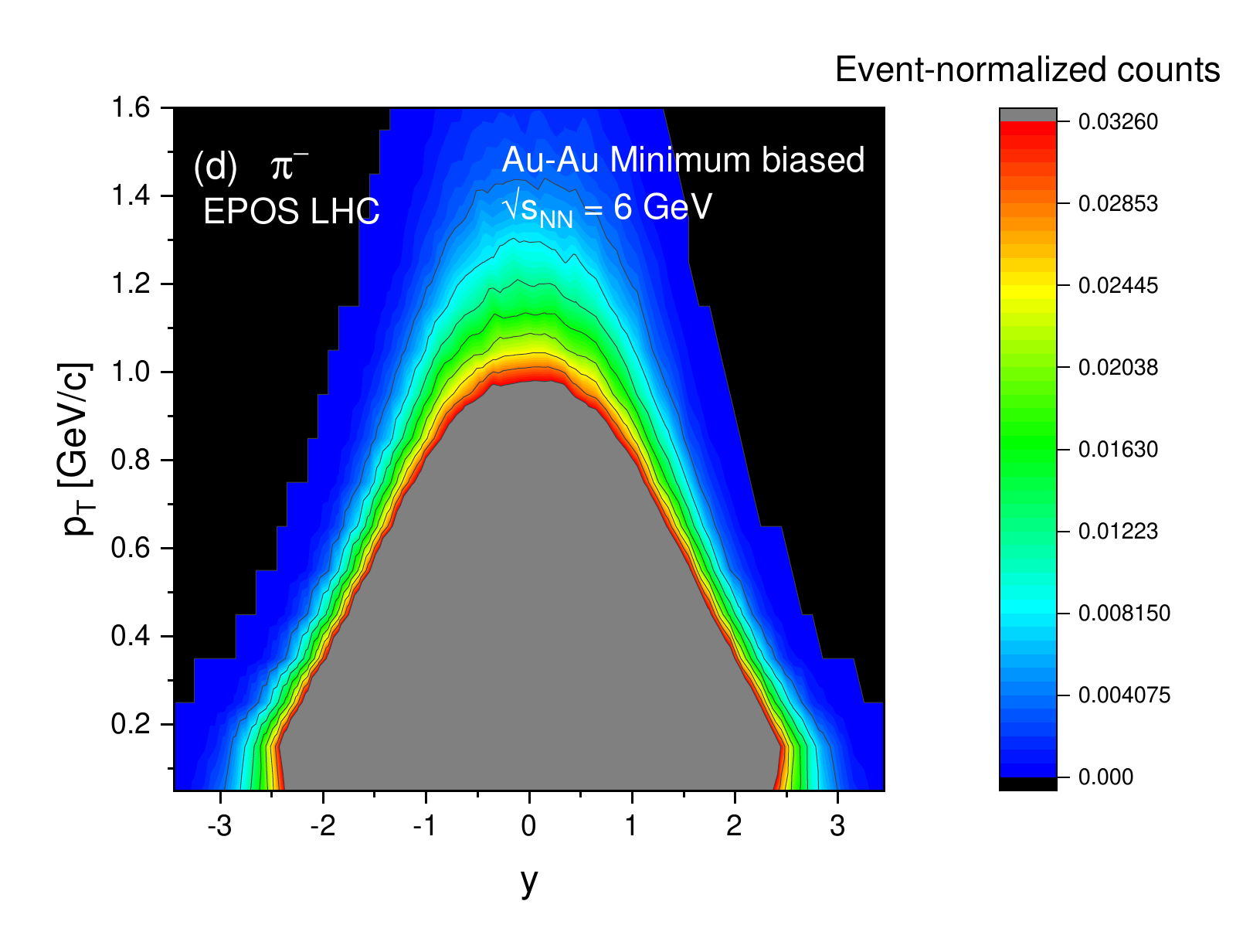}
\includegraphics[width=0.32\textwidth]{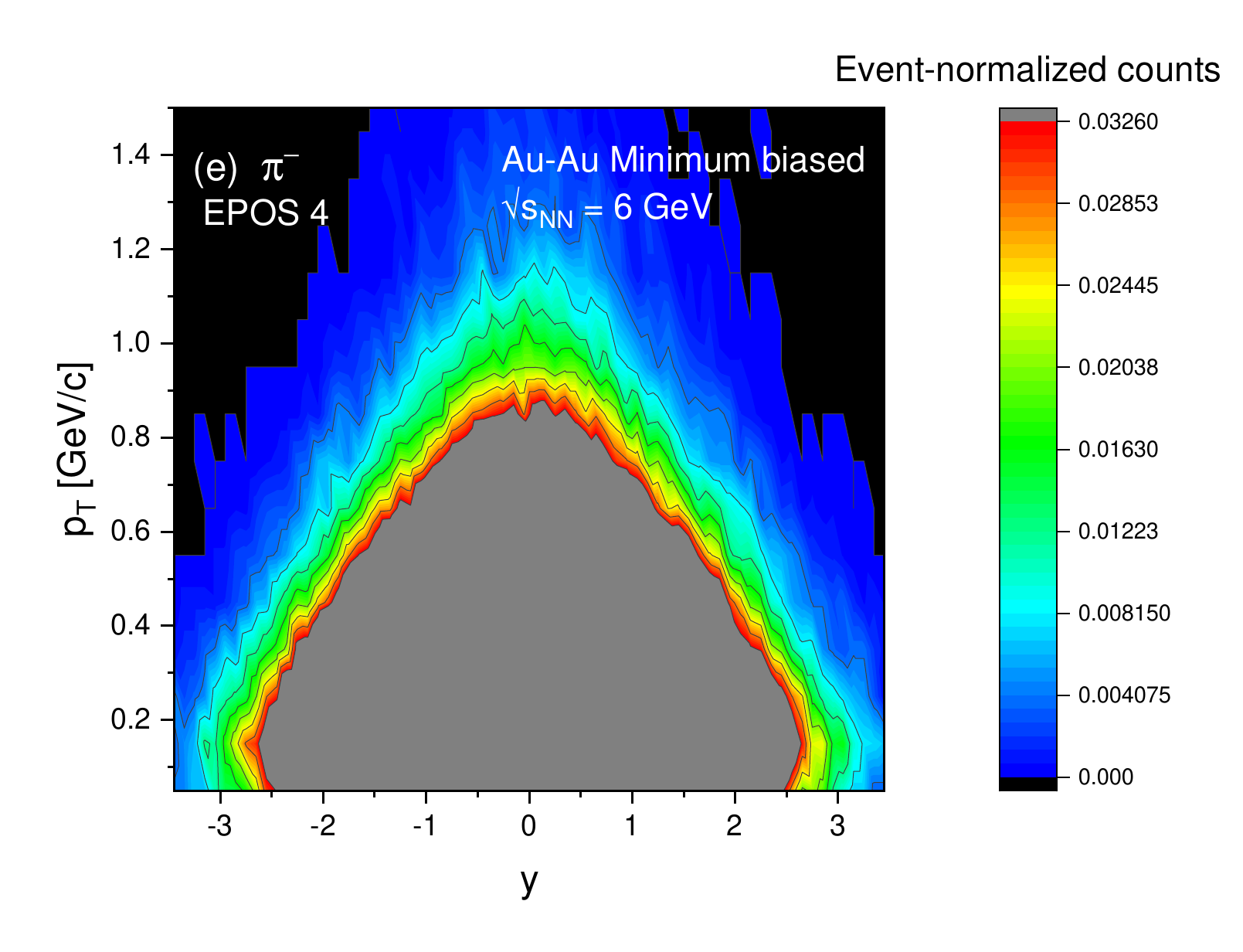} 
\includegraphics[width=0.32\textwidth]{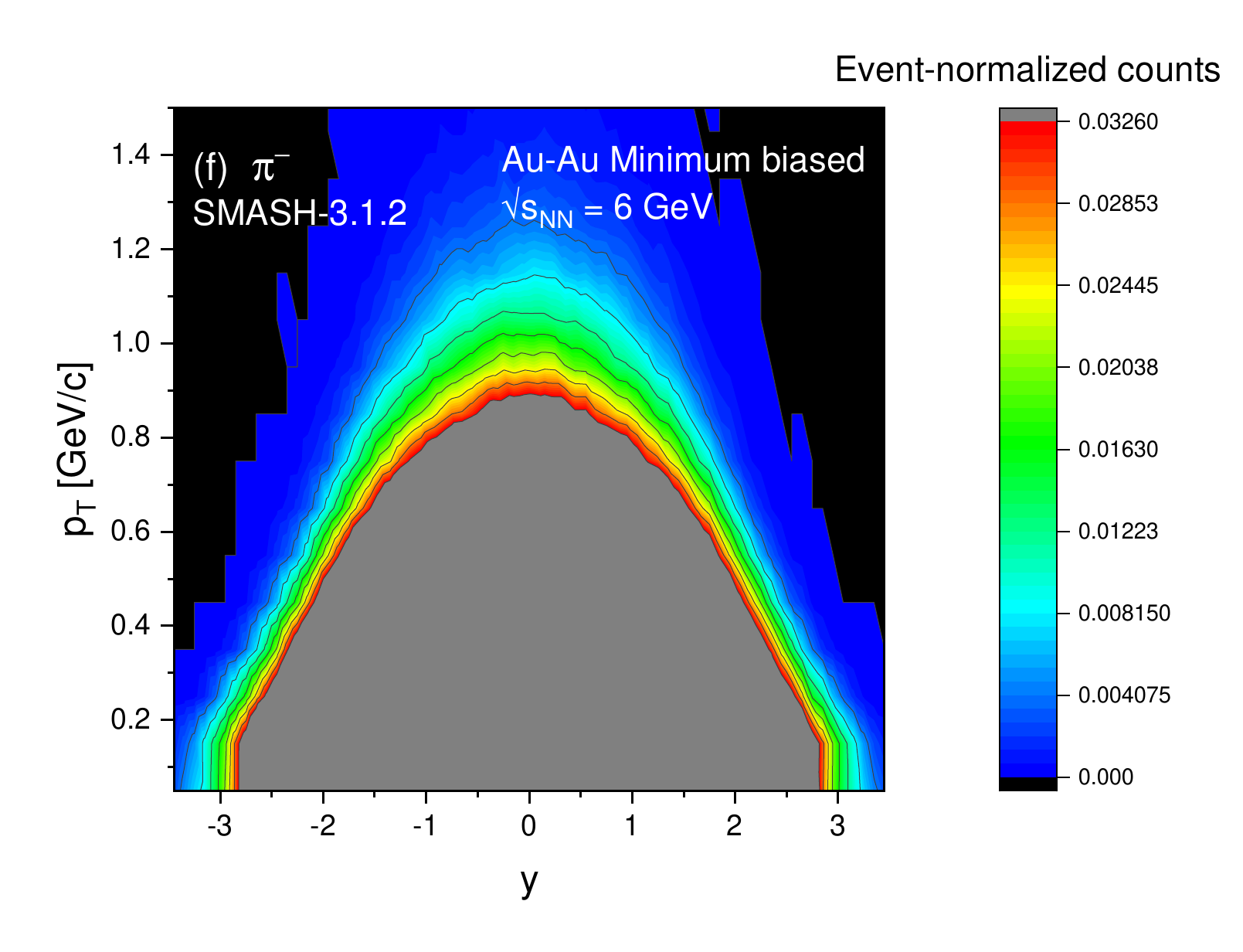}\vspace{-0.35cm}
\includegraphics[width=0.32\textwidth]{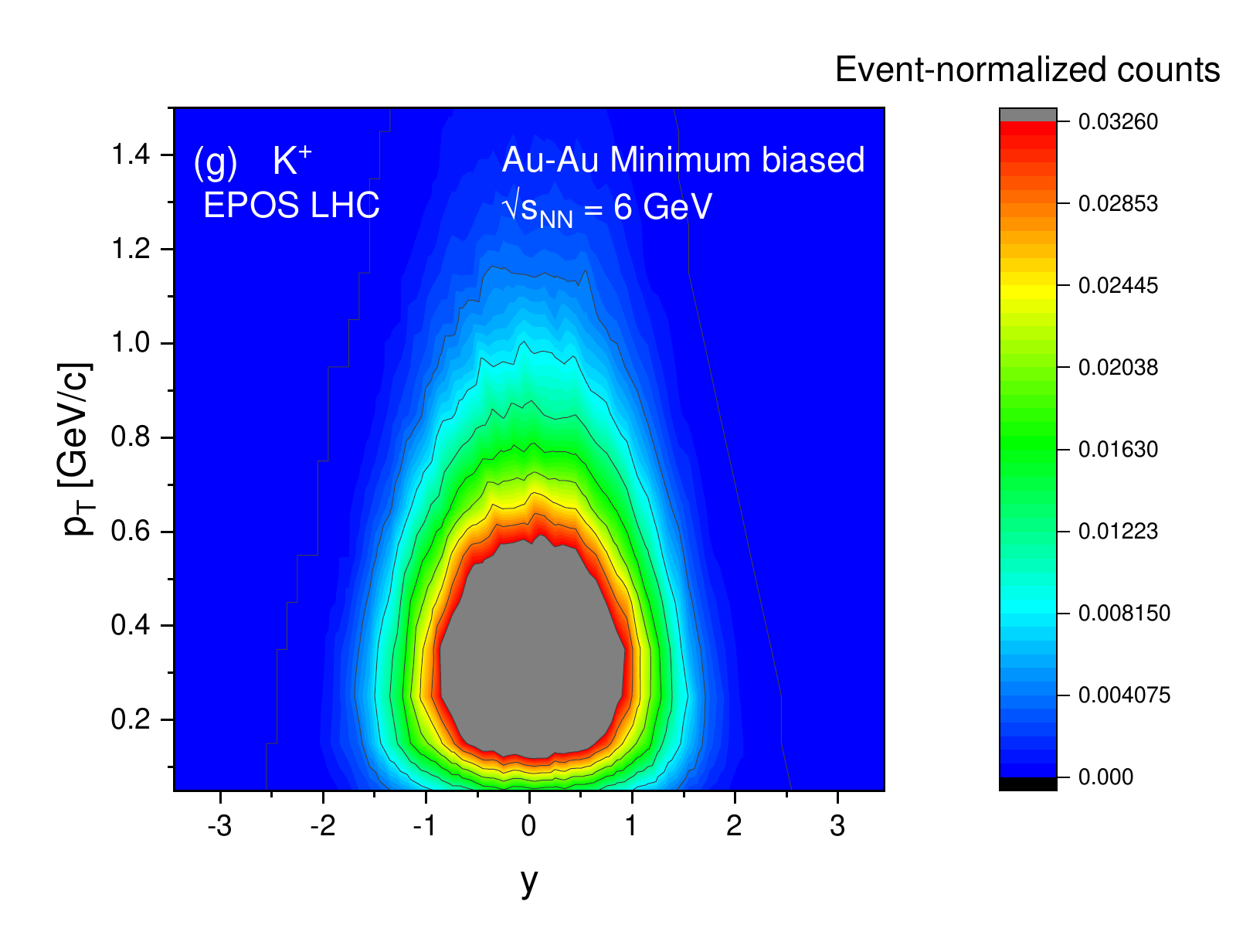}
\includegraphics[width=0.32\textwidth]{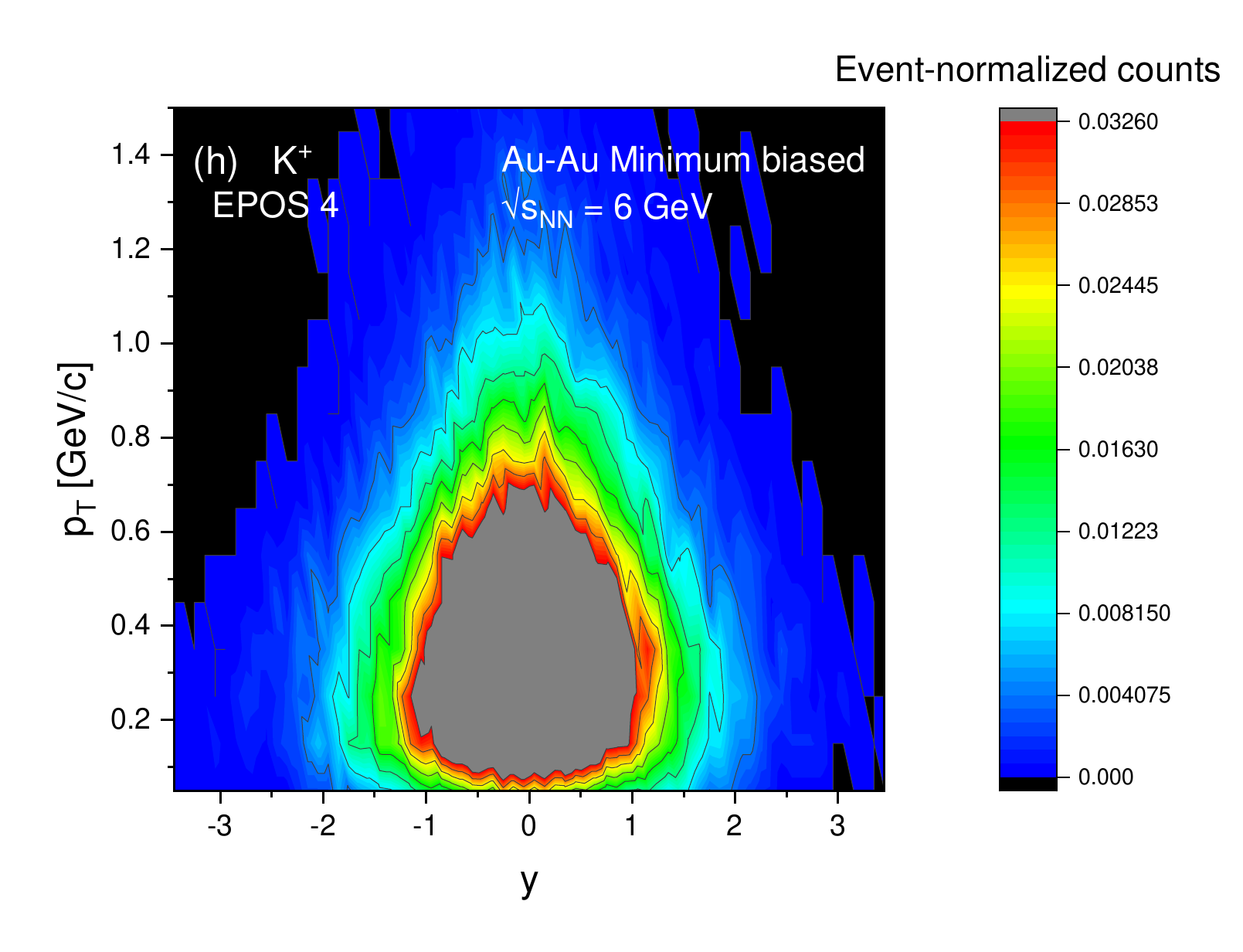}
\includegraphics[width=0.32\textwidth]{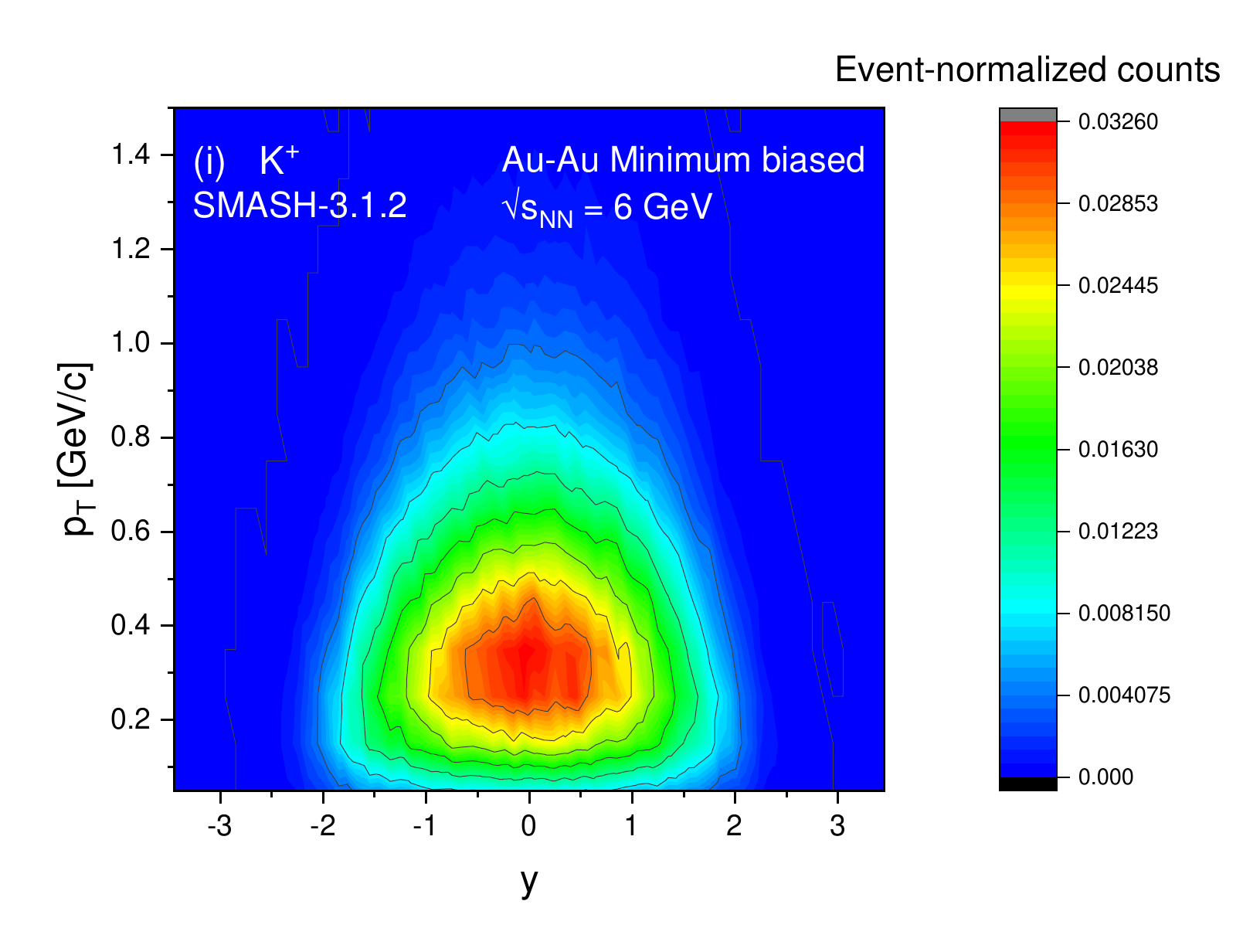}\vspace{-0.35cm}
\includegraphics[width=0.32\textwidth]{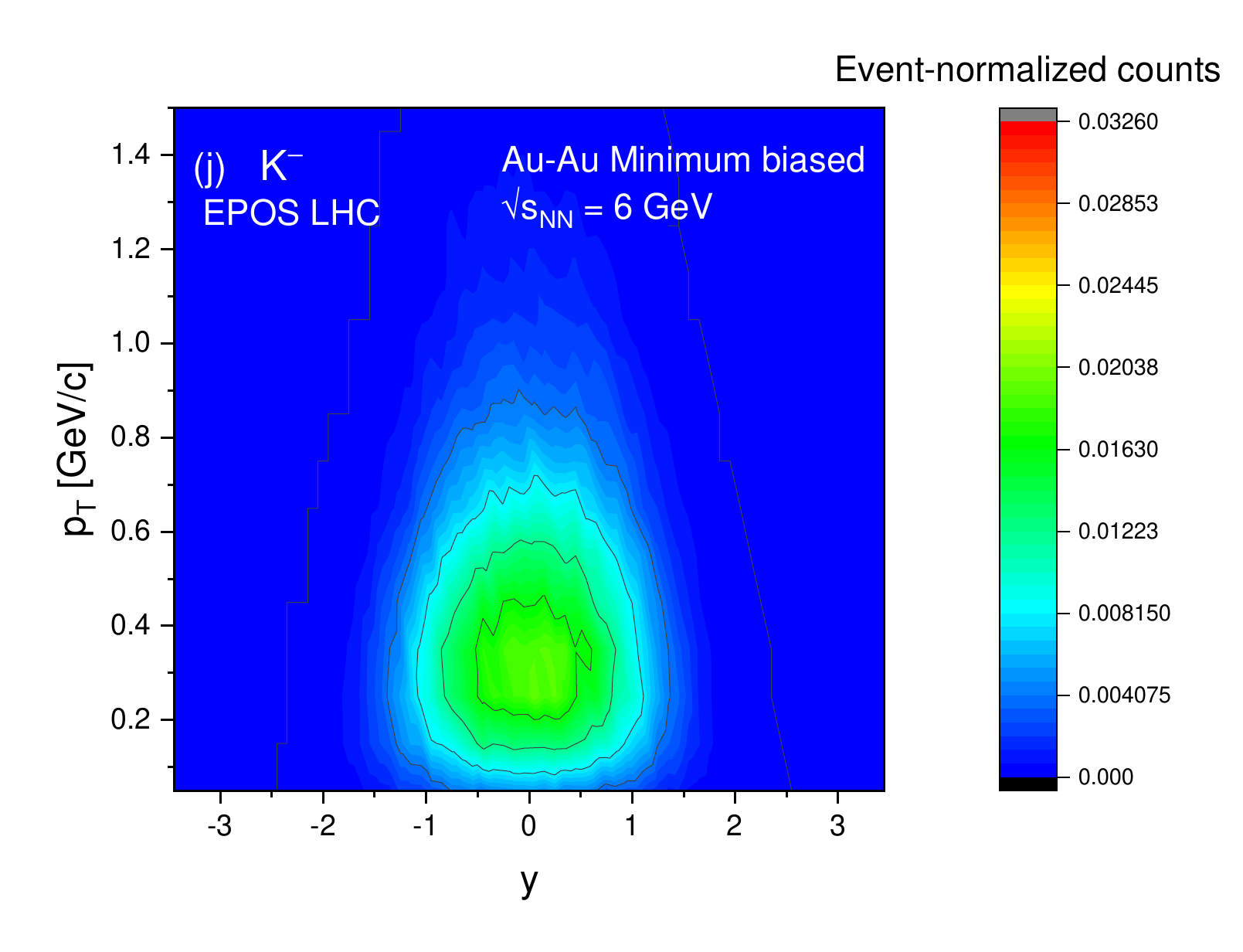}
\includegraphics[width=0.32\textwidth]{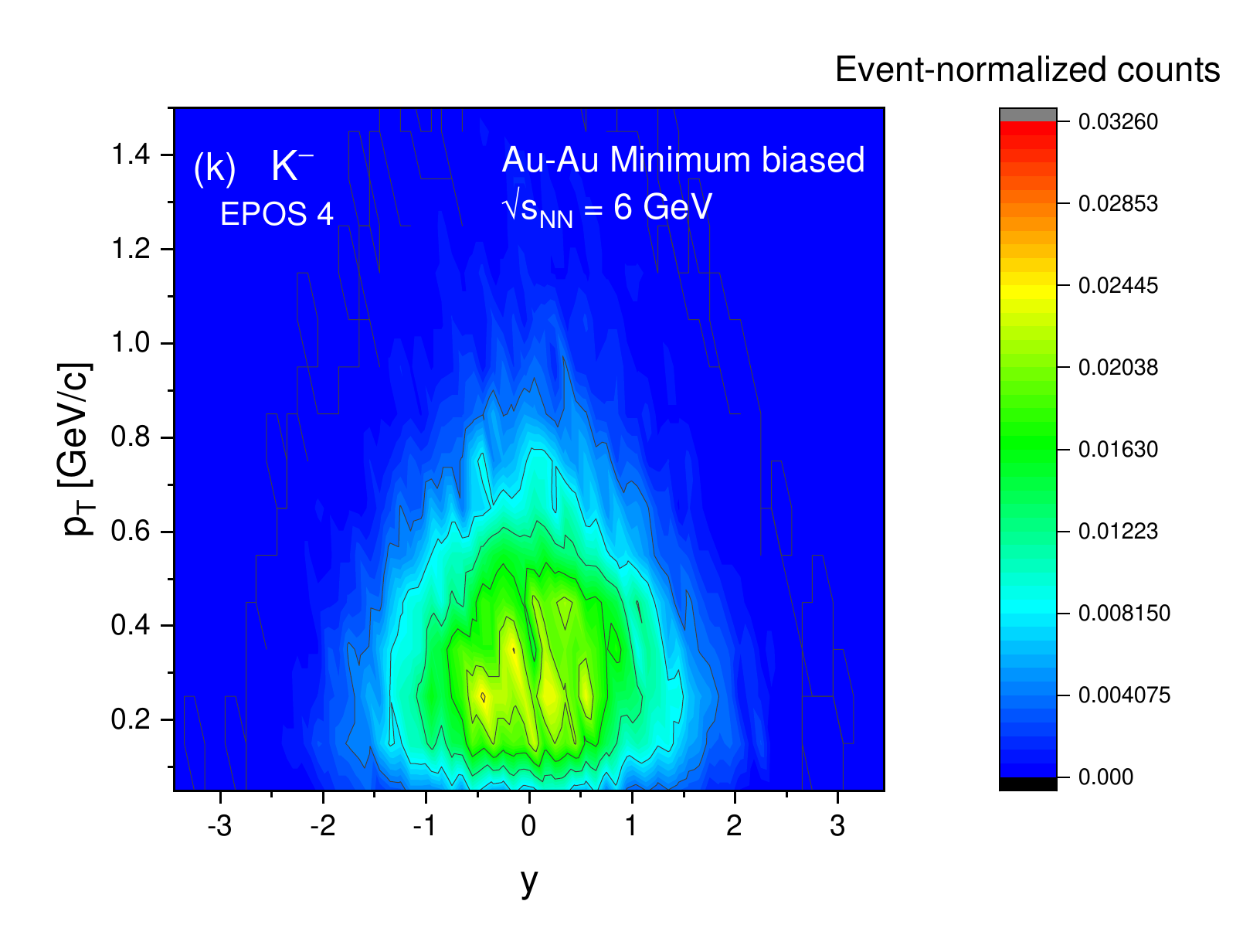}
\includegraphics[width=0.32\textwidth]{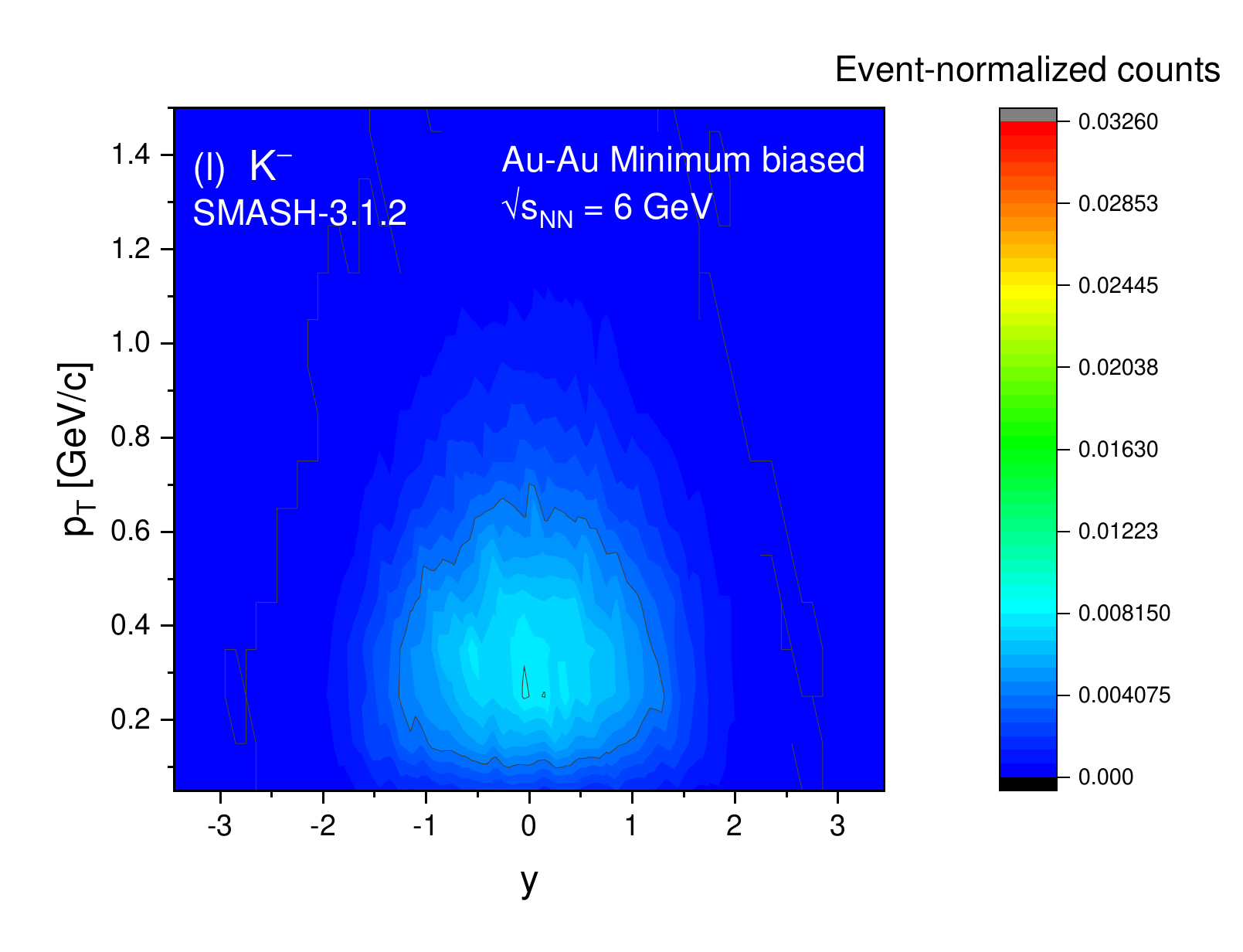}\vspace{-0.35cm}
\includegraphics[width=0.32\textwidth]{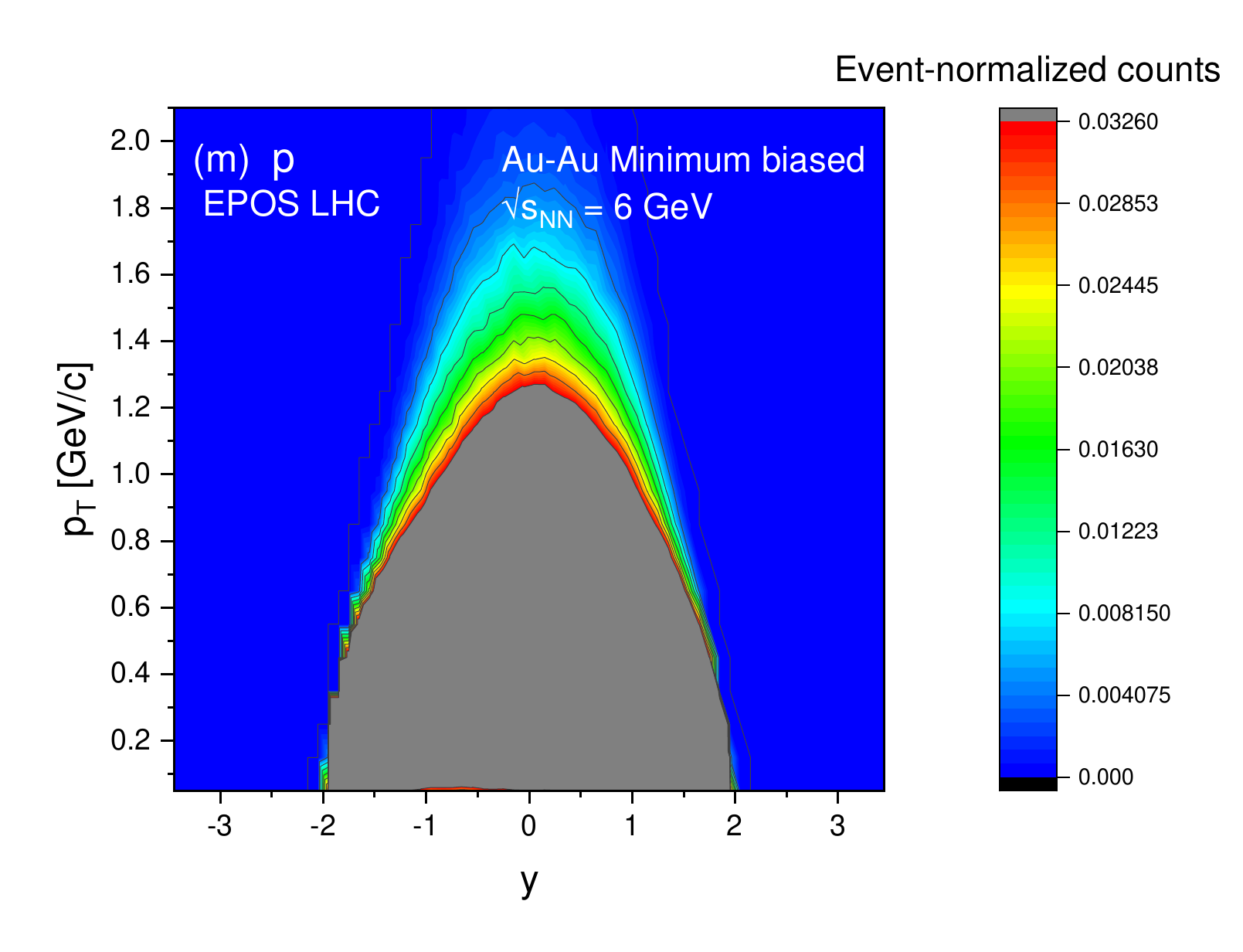}
\includegraphics[width=0.32\textwidth]{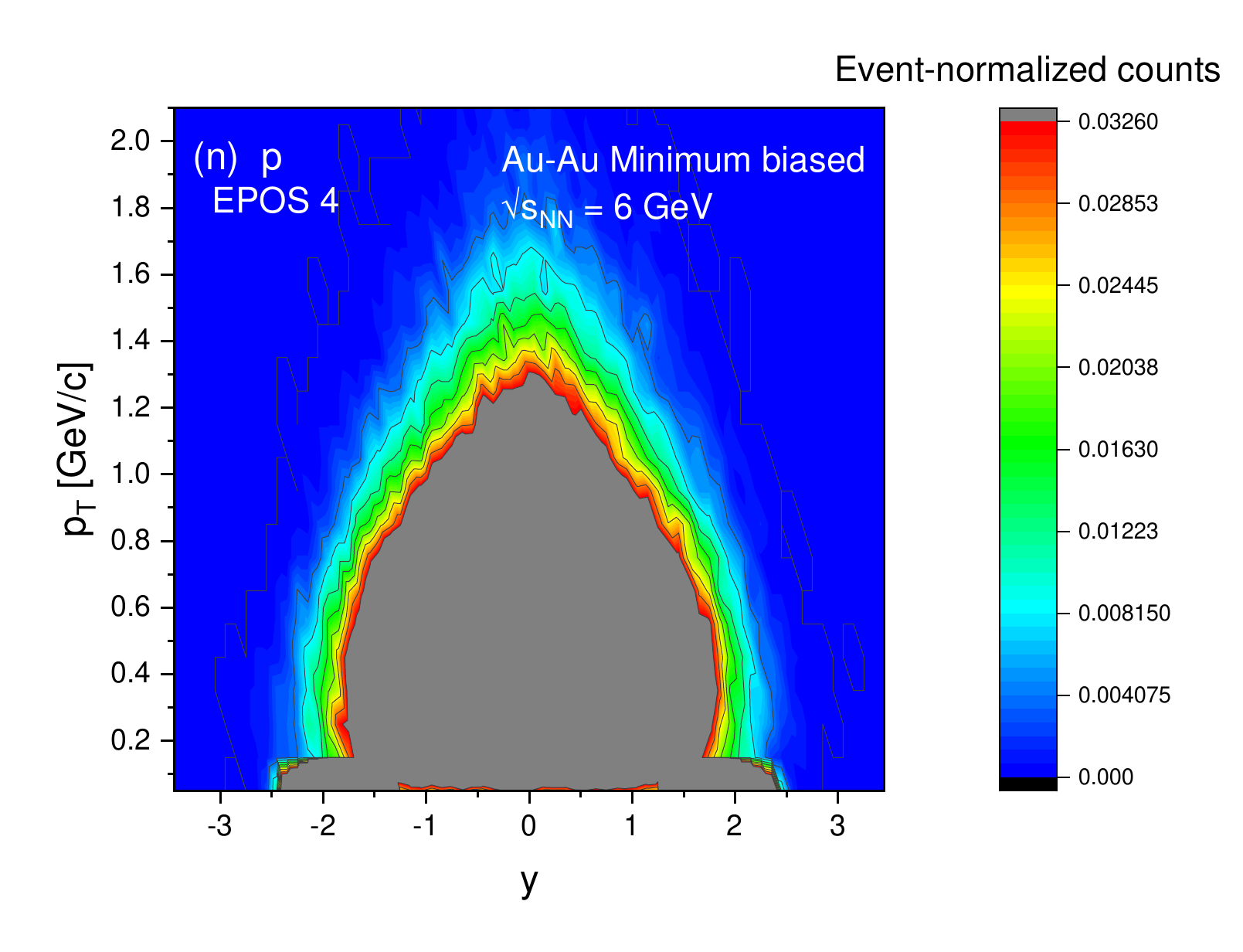}
\includegraphics[width=0.32\textwidth]{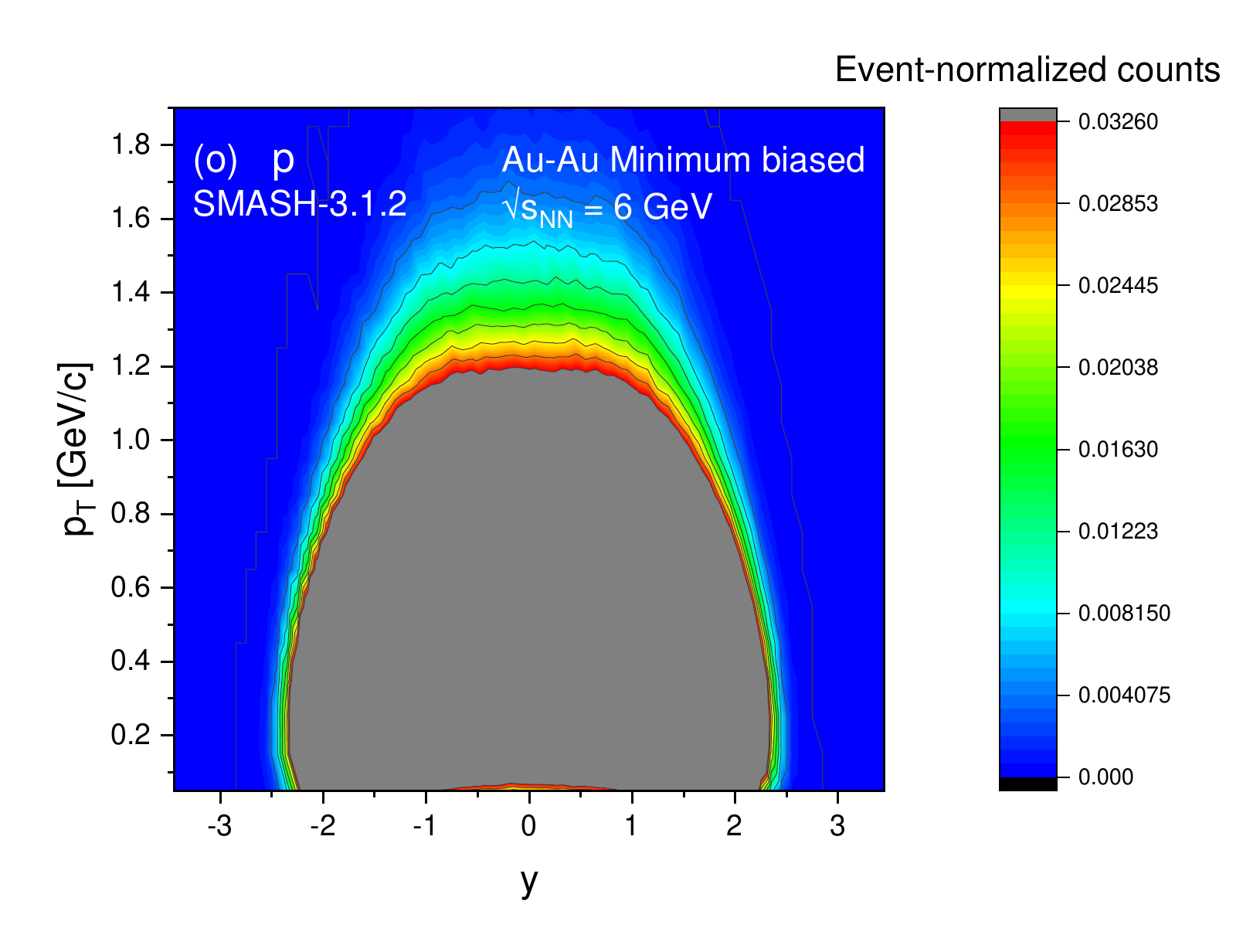}\vspace{-0.35cm}
\includegraphics[width=0.32\textwidth]{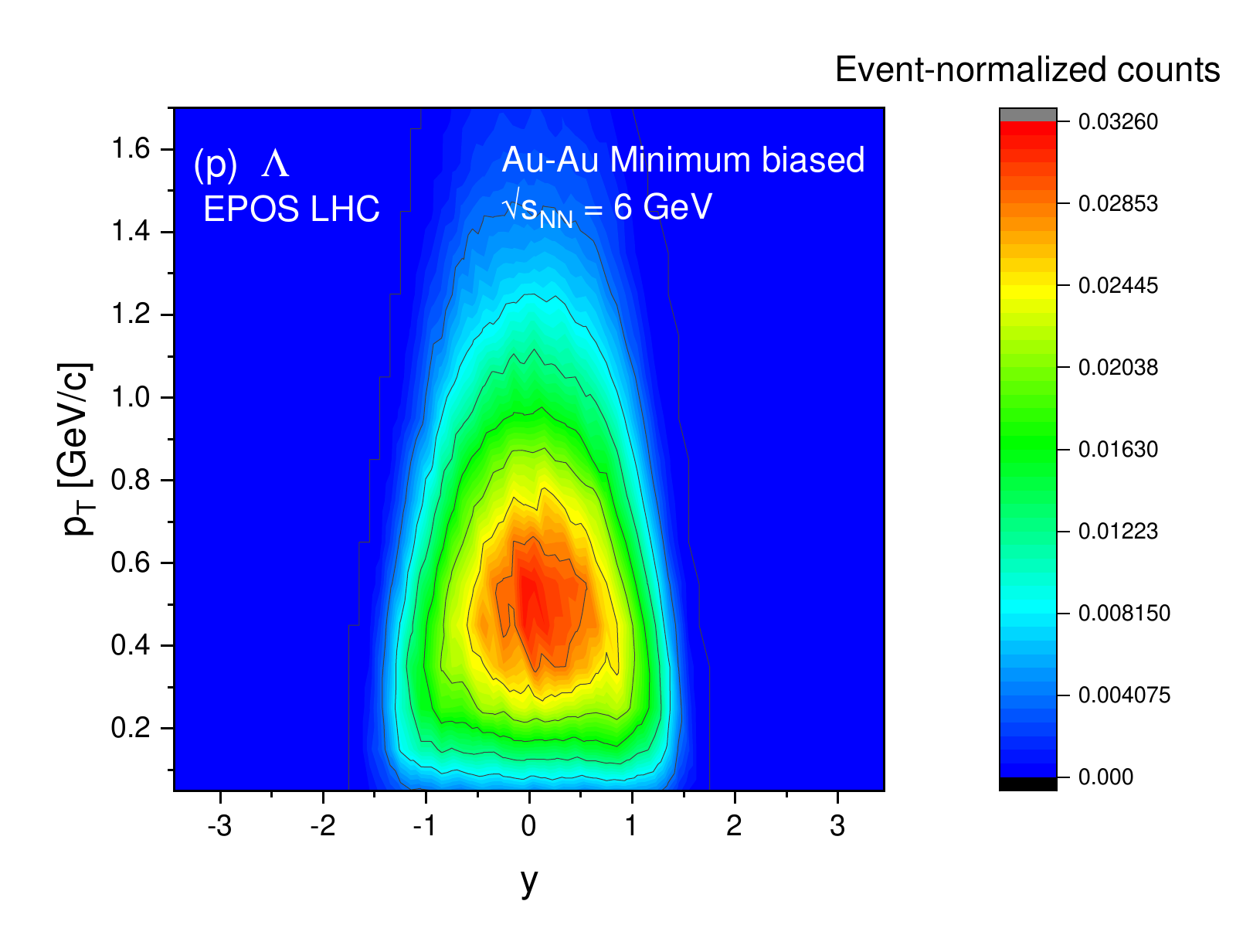}
\includegraphics[width=0.32\textwidth]{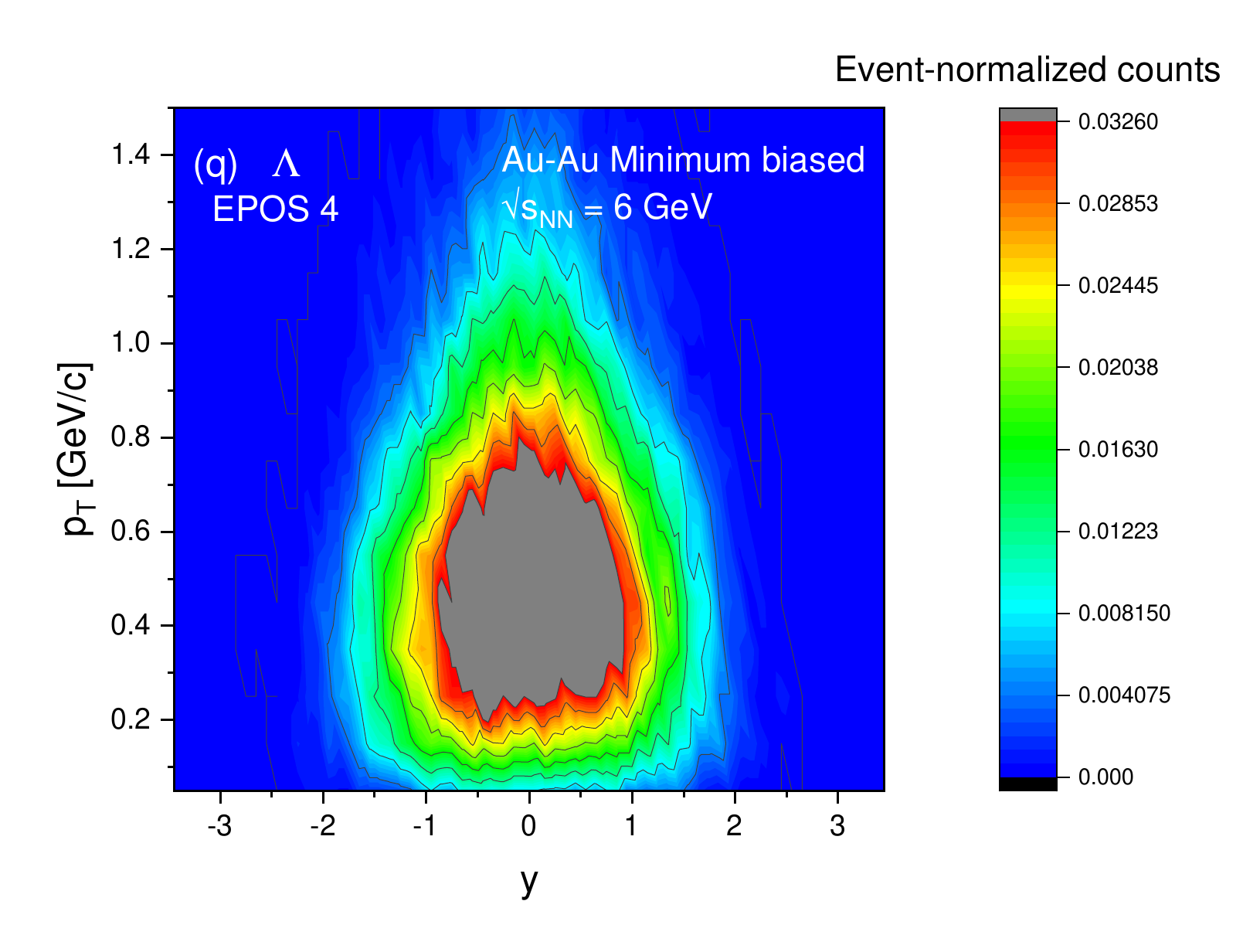}
\includegraphics[width=0.32\textwidth]{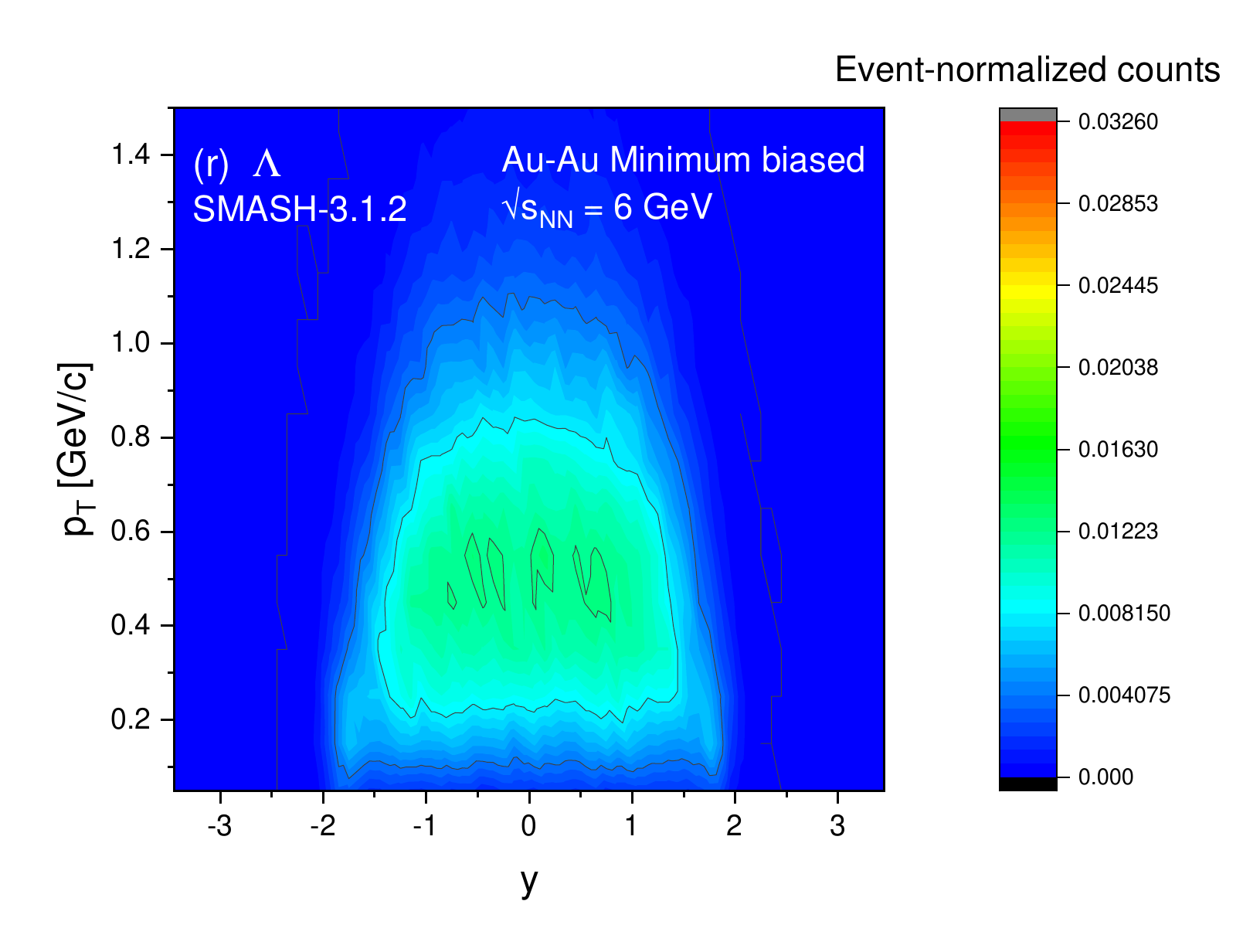}\vspace{-0.35cm}
\caption {Two-dimensional $p_T$ vs rapidity density maps (color scale = event-normalized counts) of various identified hadrons in minimum-bias collision of $Au+Au$ in $\sqrt{s_{NN}}=6~GeV$. The arrangement of the panels is three columns (left-right: EPOS-LHC, EPOS-4, SMASH) and numerous rows (by species). Particularly: (a)-(c) $\pi^+$ (EPOS-LHC / EPOS-4 / SMASH), (d)-(f) $\pi^-$, (g)-(i) $K^+$, (j)-(l) $K^-$, (m)-(o) $p$ and (p)-(r) $\Lambda$. The color scale indicates relative event-normalized population in the $p_T-y$ plane and points out variations in the manner each model populates intermediate $p_T$ and non-central rapidities.}
\end{figure*}

\begin{figure*}
\centering
\includegraphics[width=0.32\textwidth]{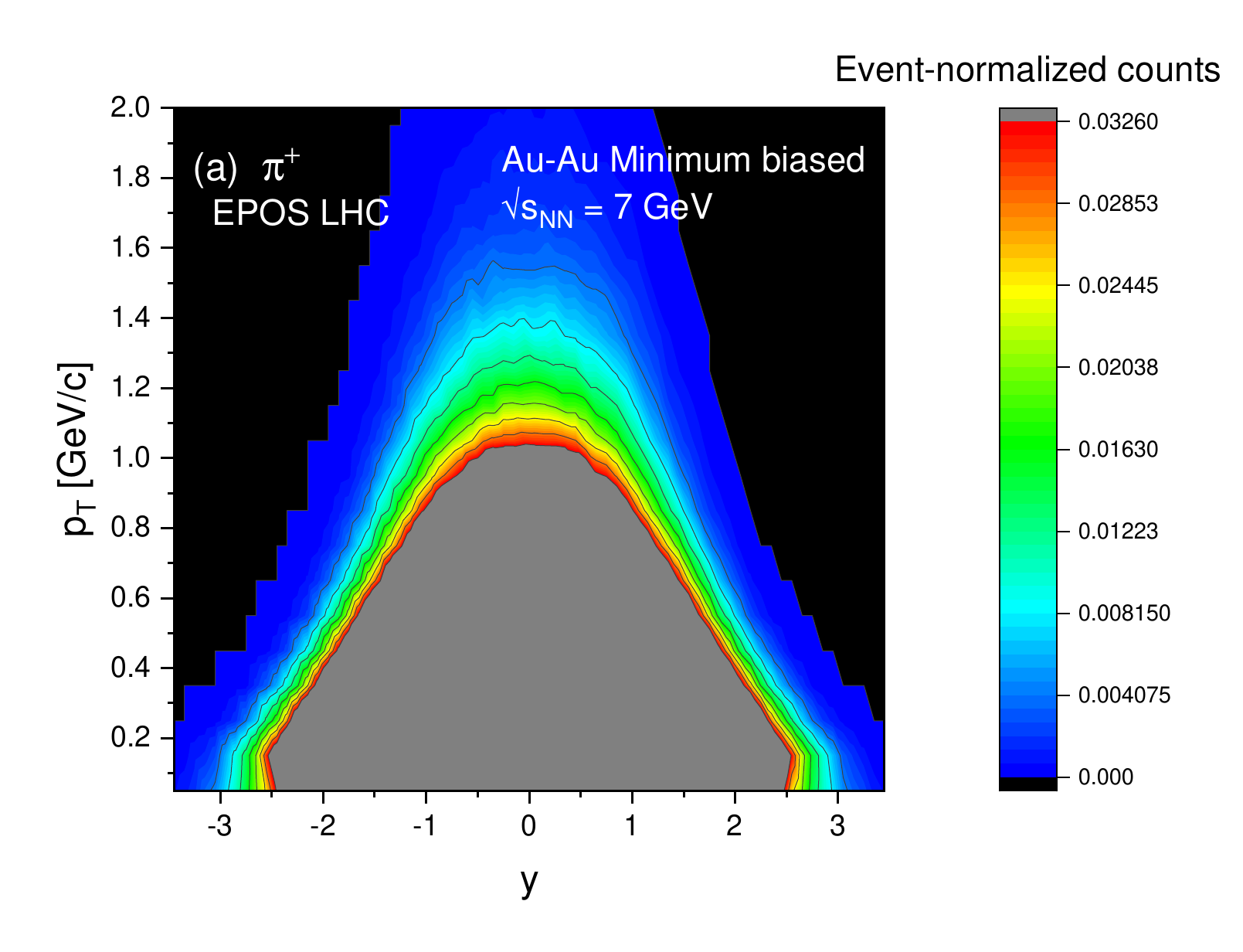}
\includegraphics[width=0.32\textwidth]{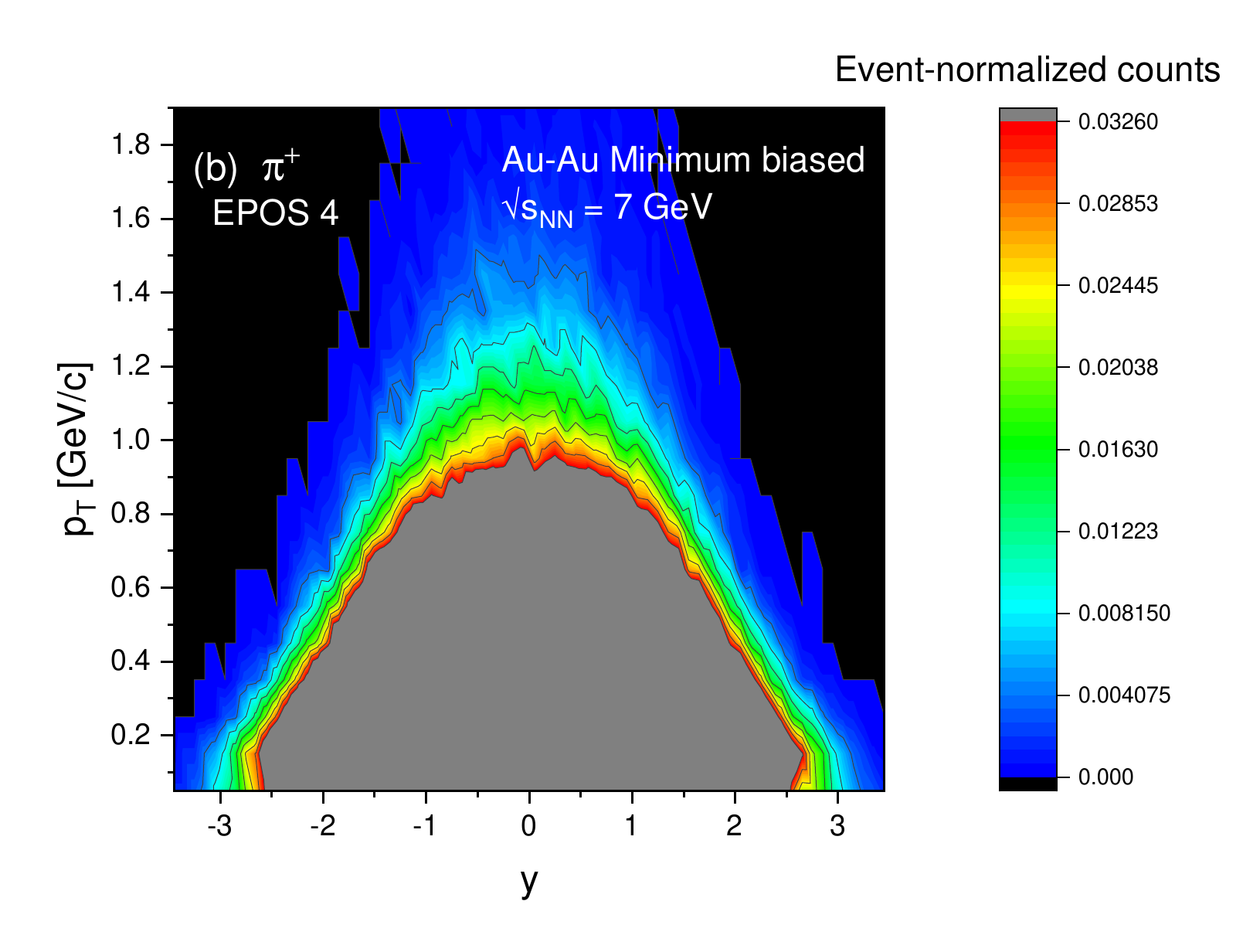} 
\includegraphics[width=0.32\textwidth]{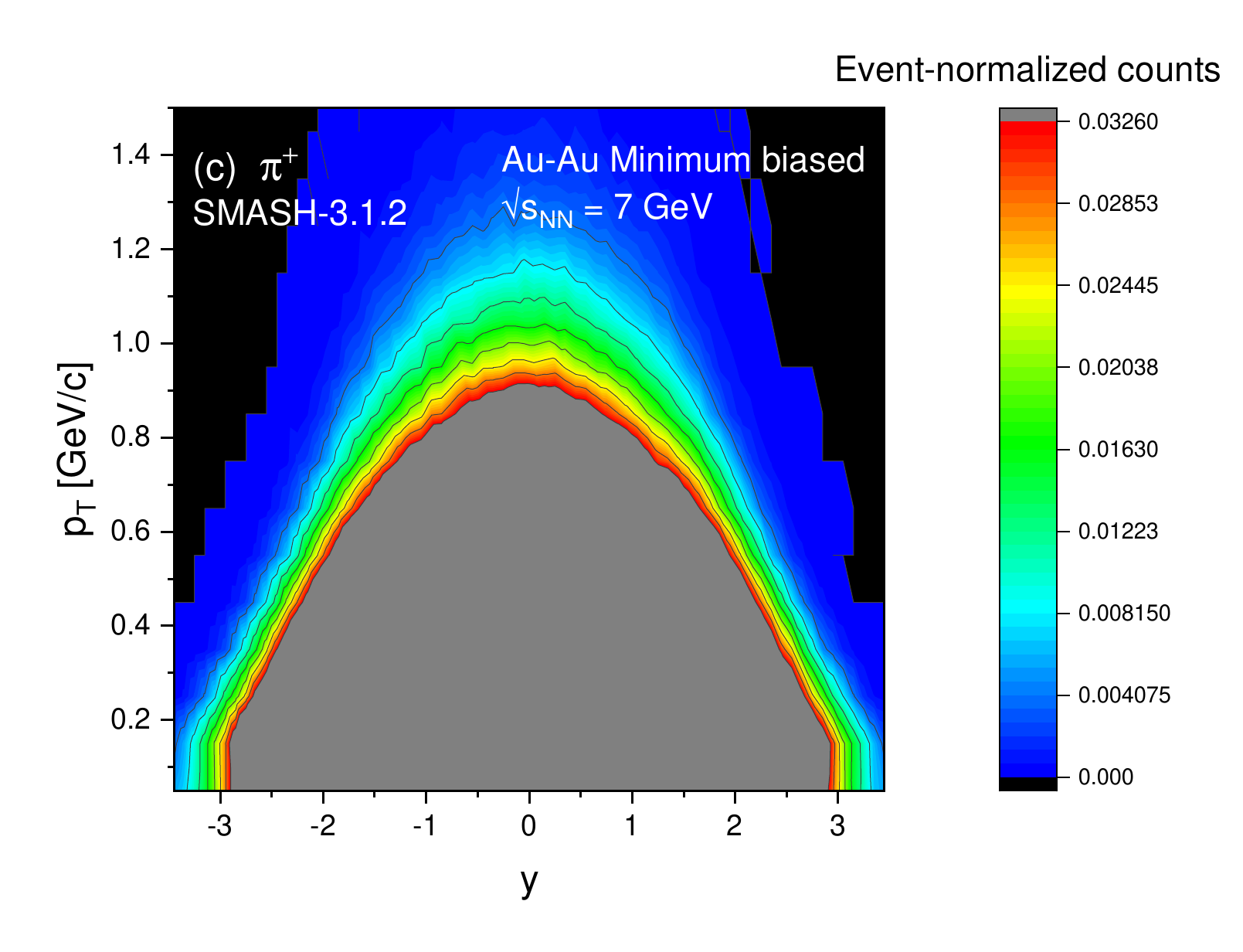}\vspace{-0.35cm}
\includegraphics[width=0.32\textwidth]{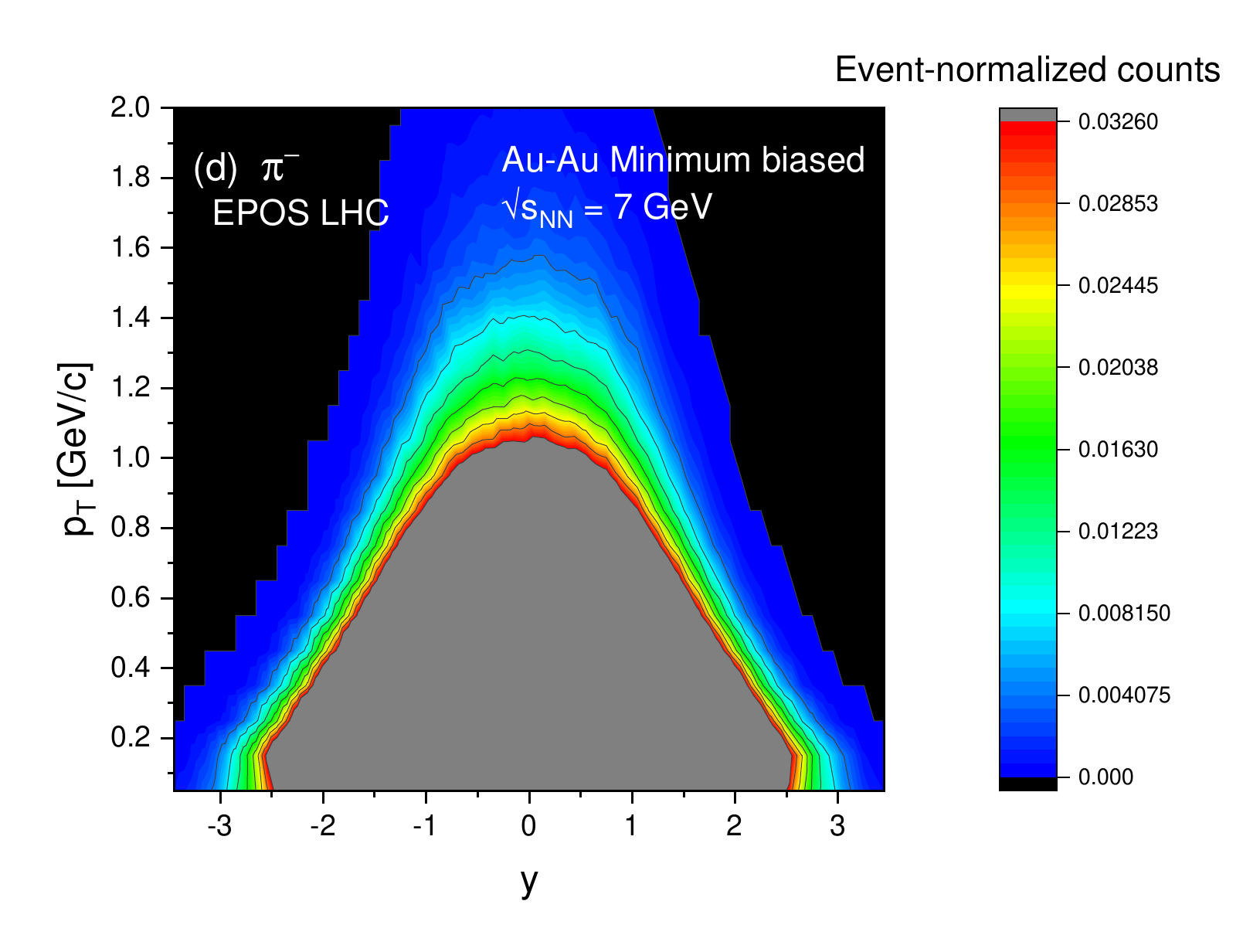}
\includegraphics[width=0.32\textwidth]{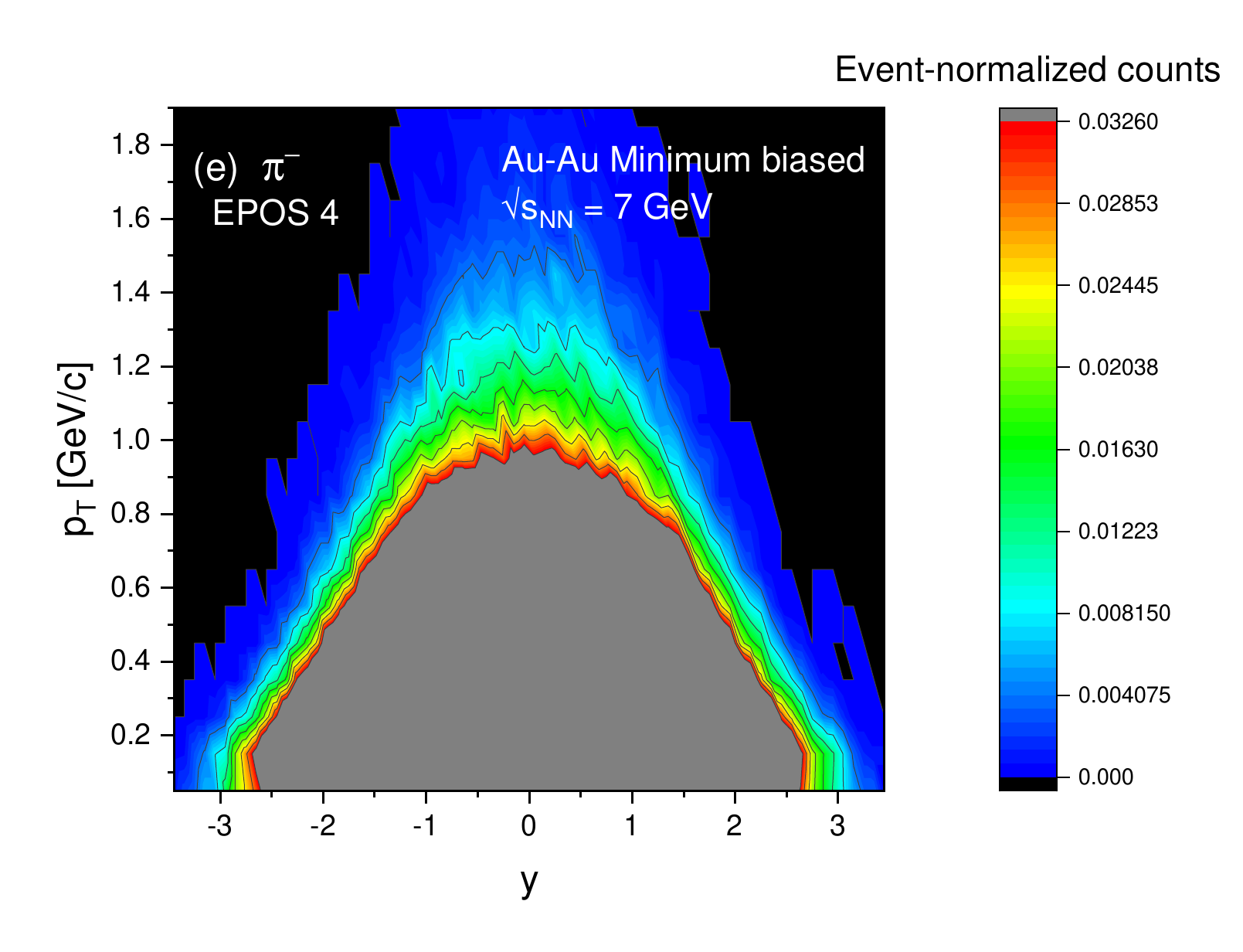} 
\includegraphics[width=0.32\textwidth]{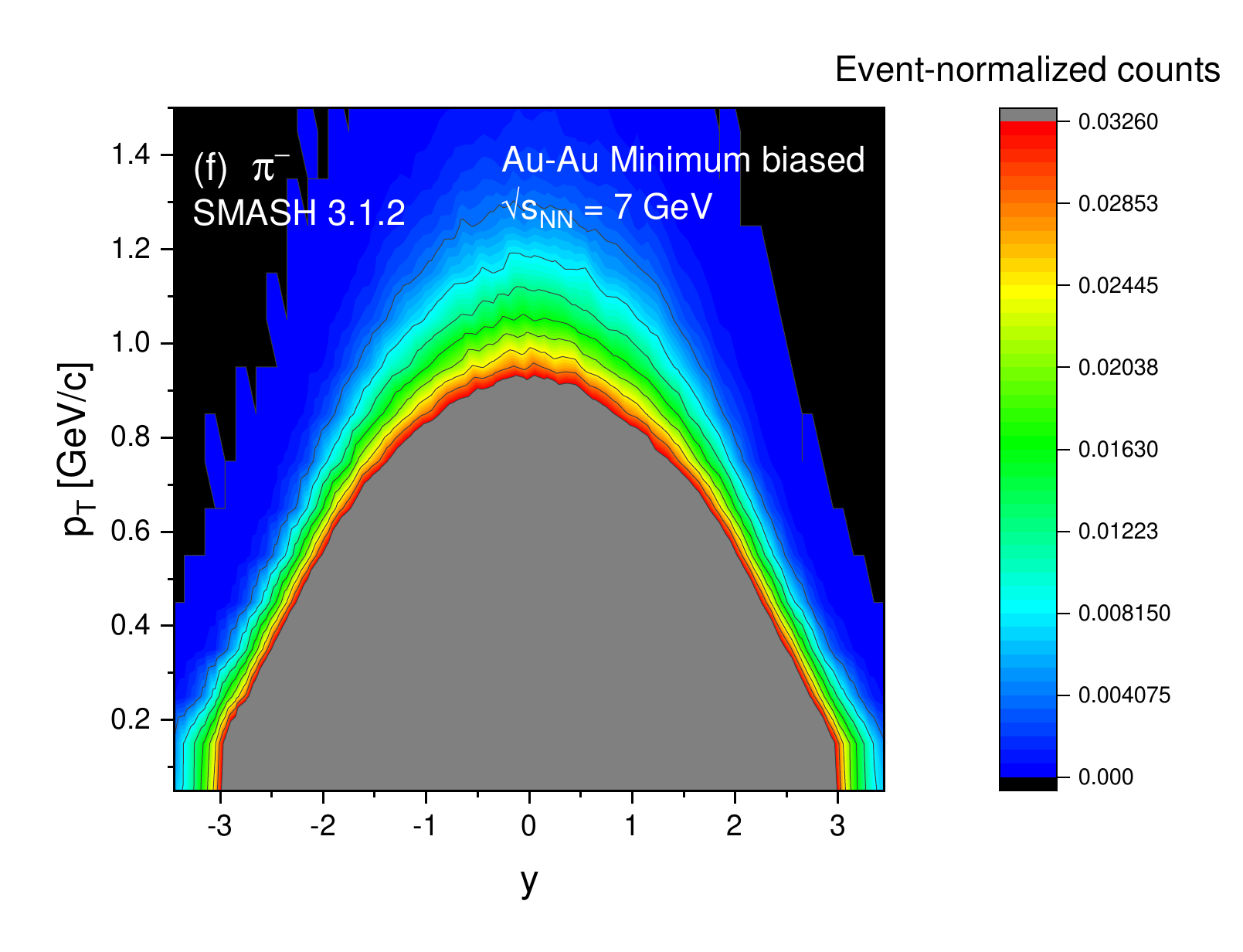}\vspace{-0.35cm}
\includegraphics[width=0.32\textwidth]{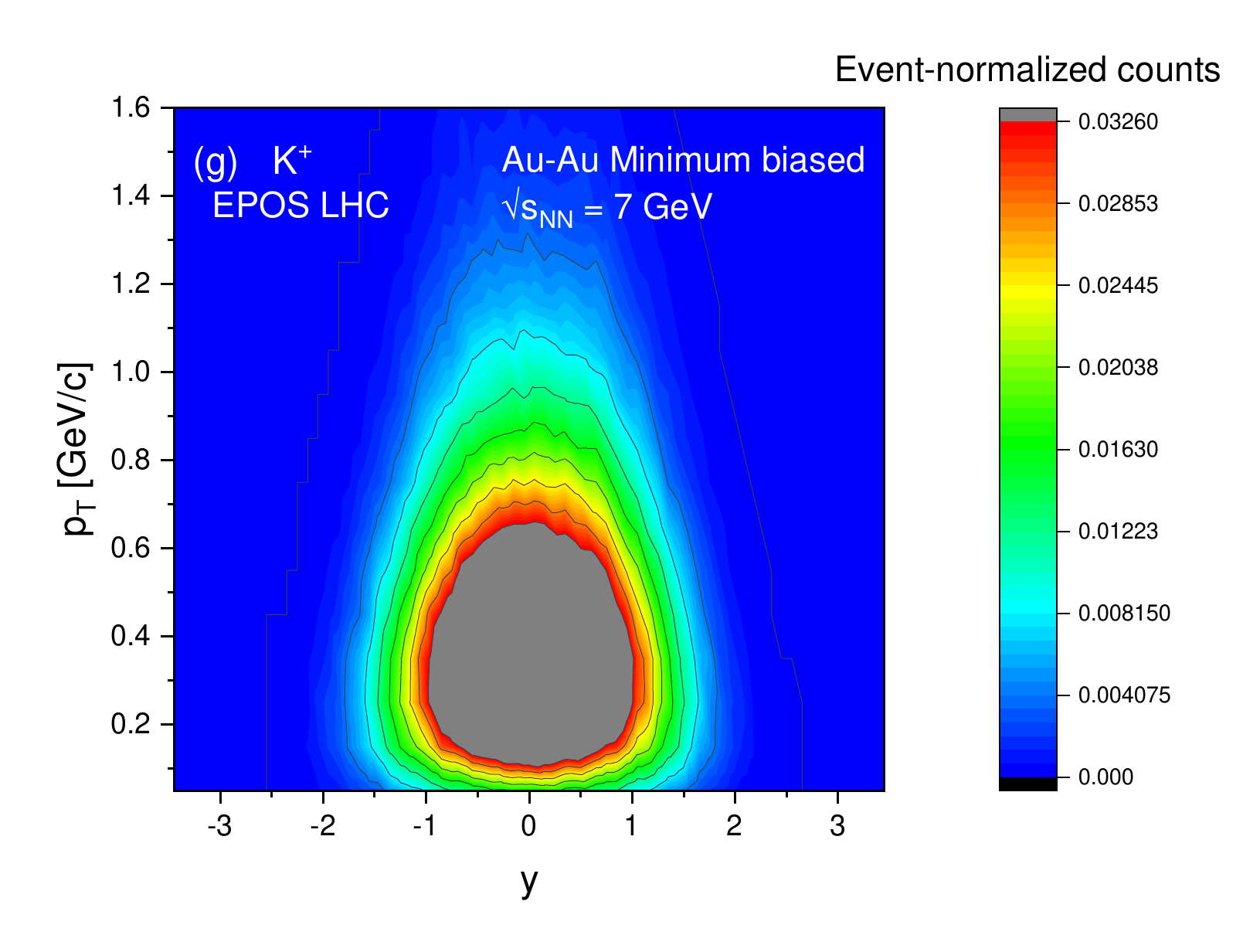}
\includegraphics[width=0.32\textwidth]{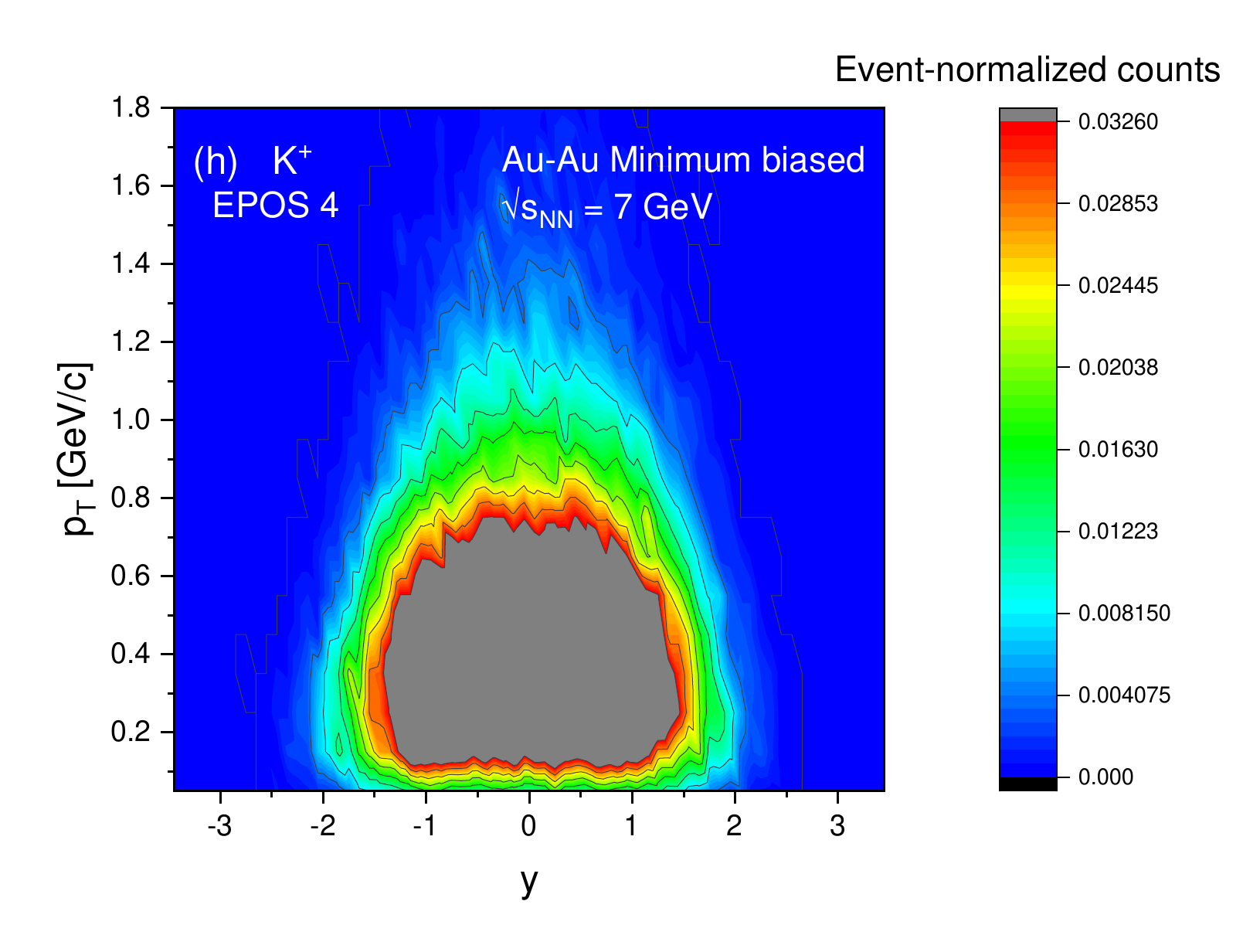} 
\includegraphics[width=0.32\textwidth]{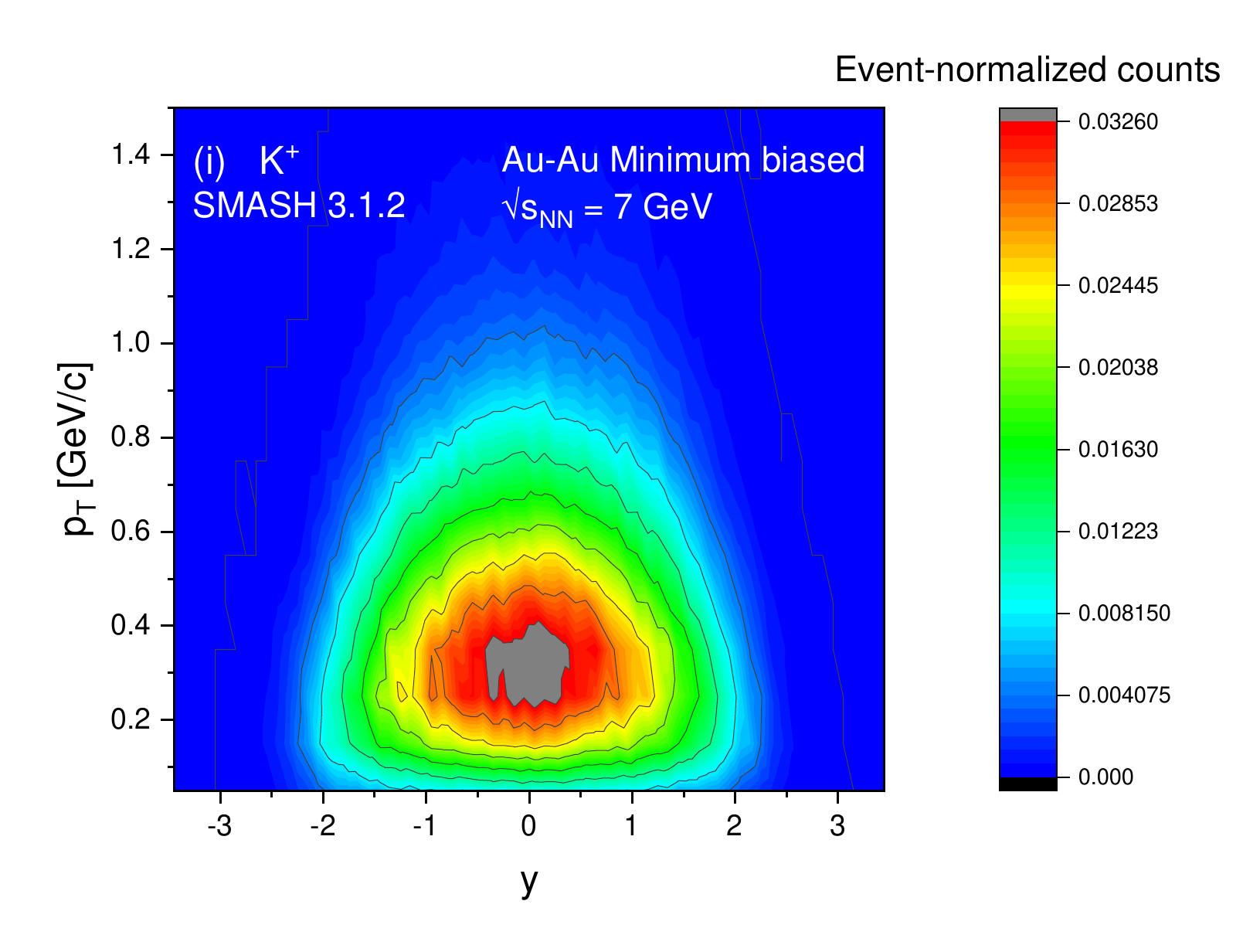}\vspace{-0.35cm}
\includegraphics[width=0.32\textwidth]{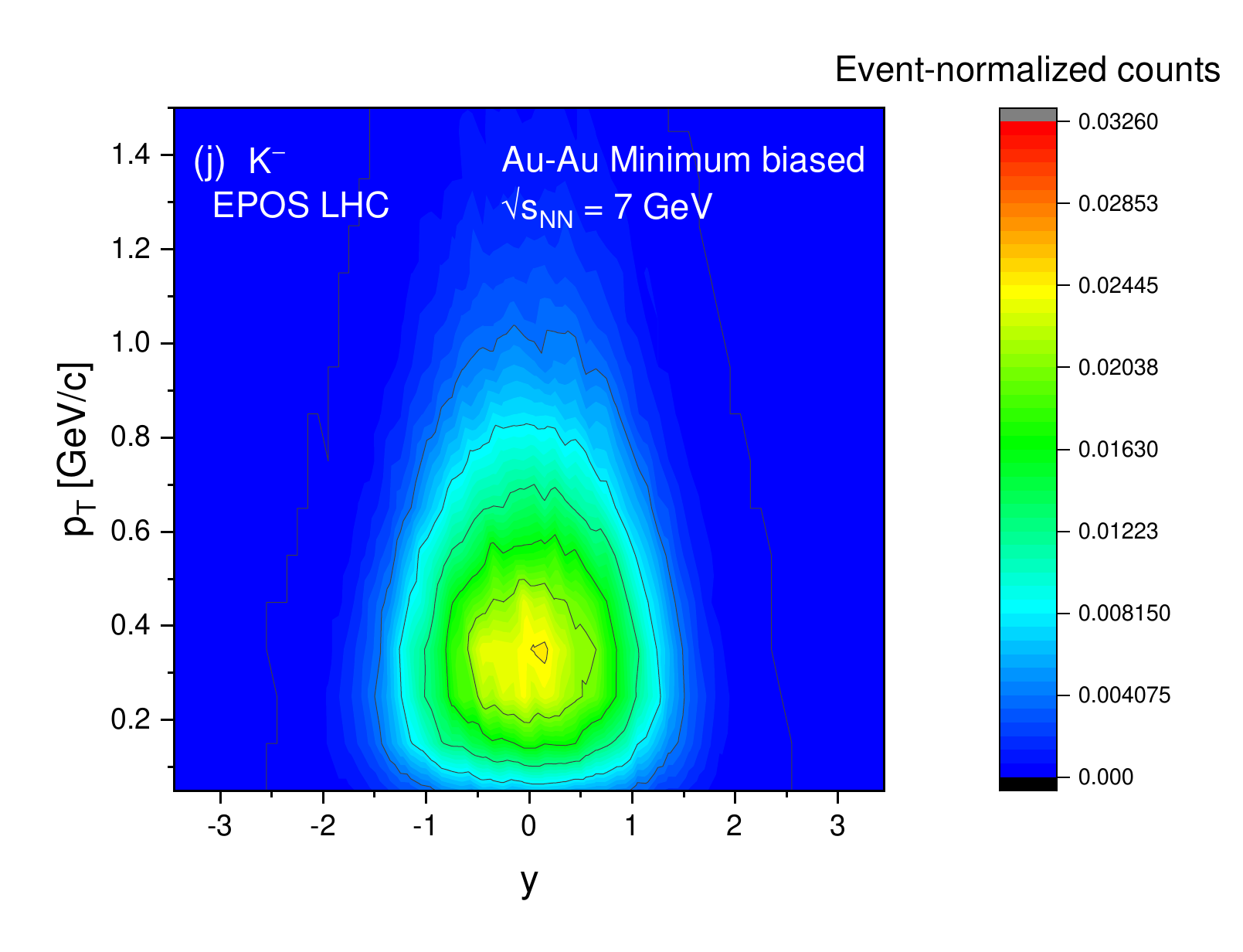}
\includegraphics[width=0.32\textwidth]{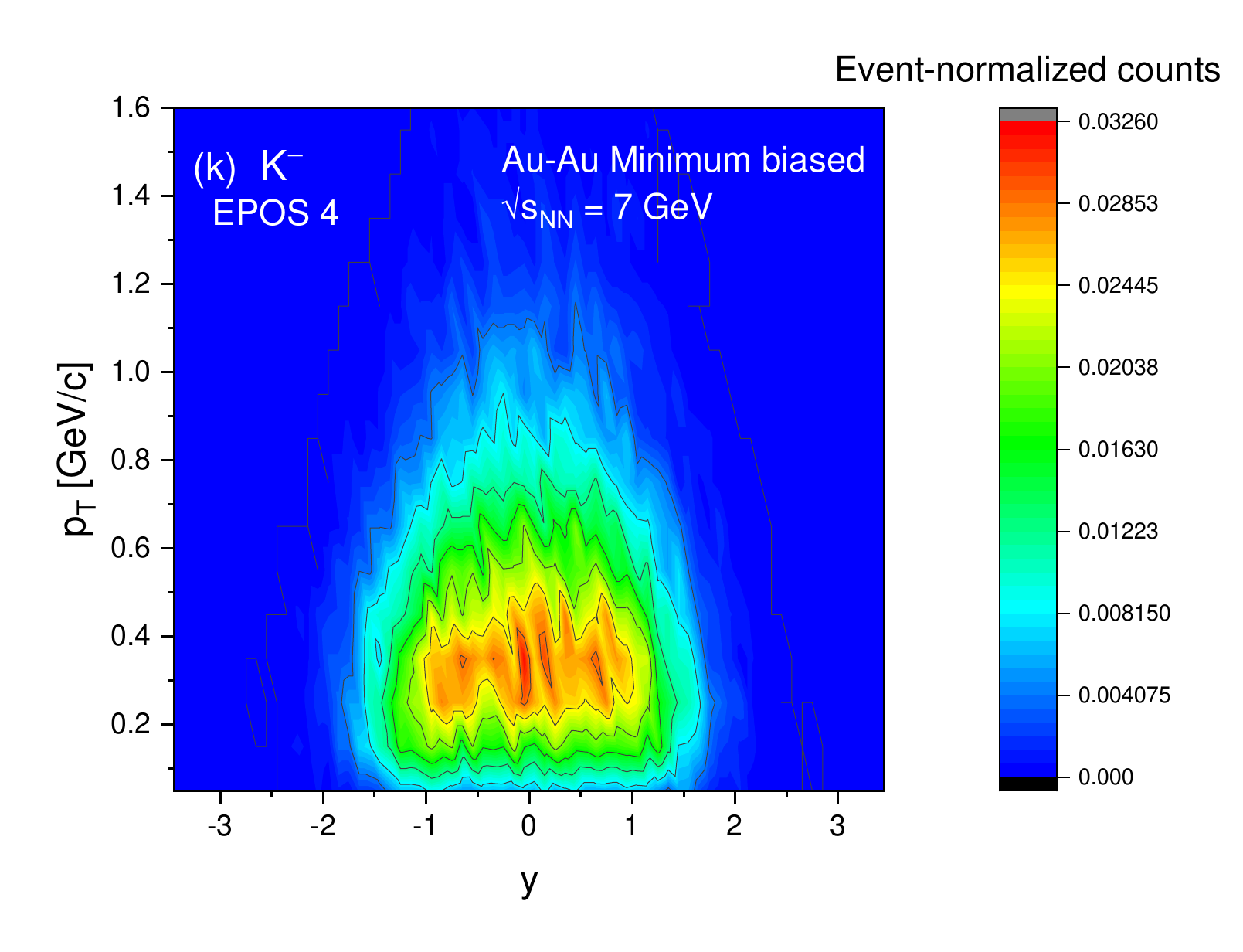} 
\includegraphics[width=0.32\textwidth]{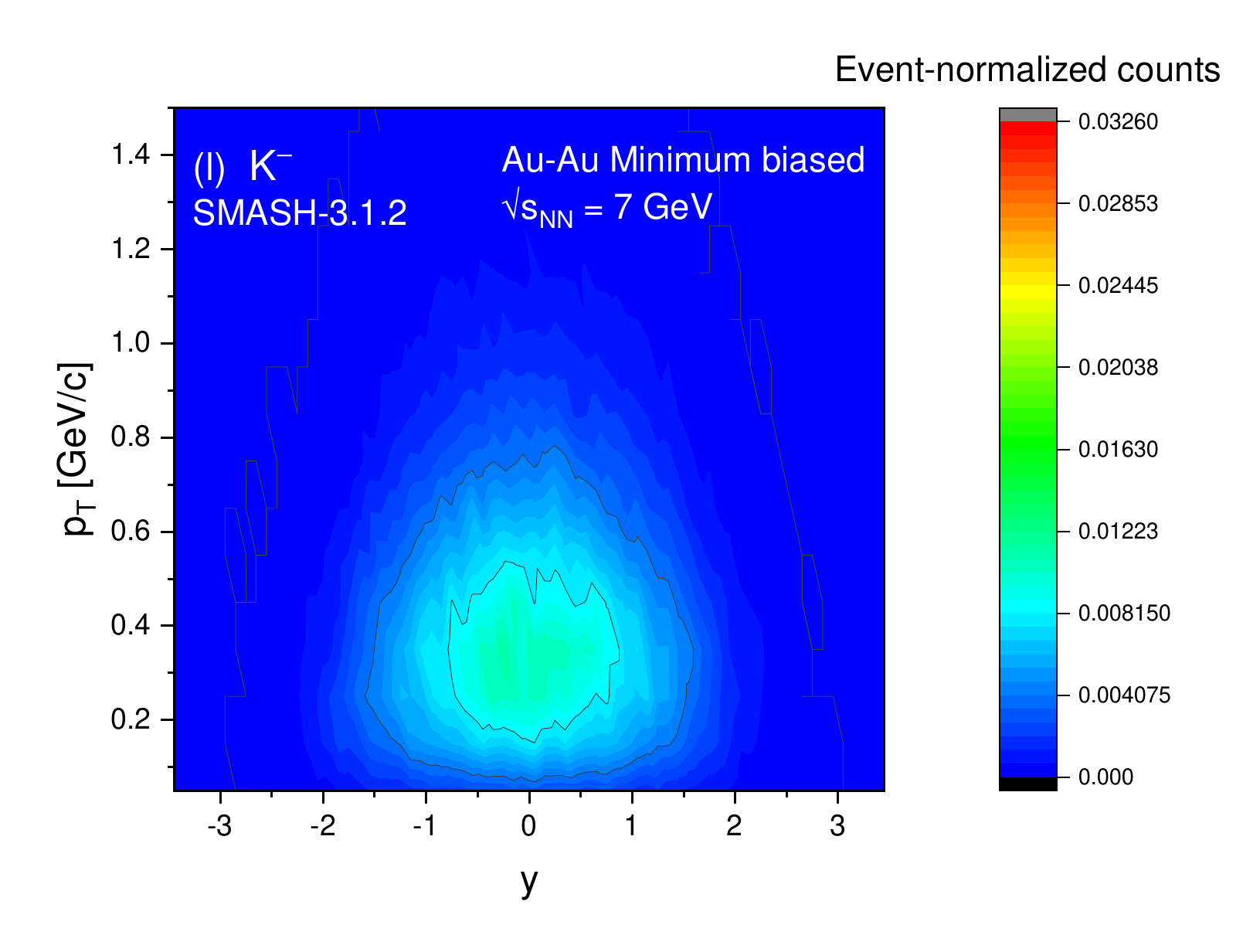}\vspace{-0.35cm}
\includegraphics[width=0.32\textwidth]{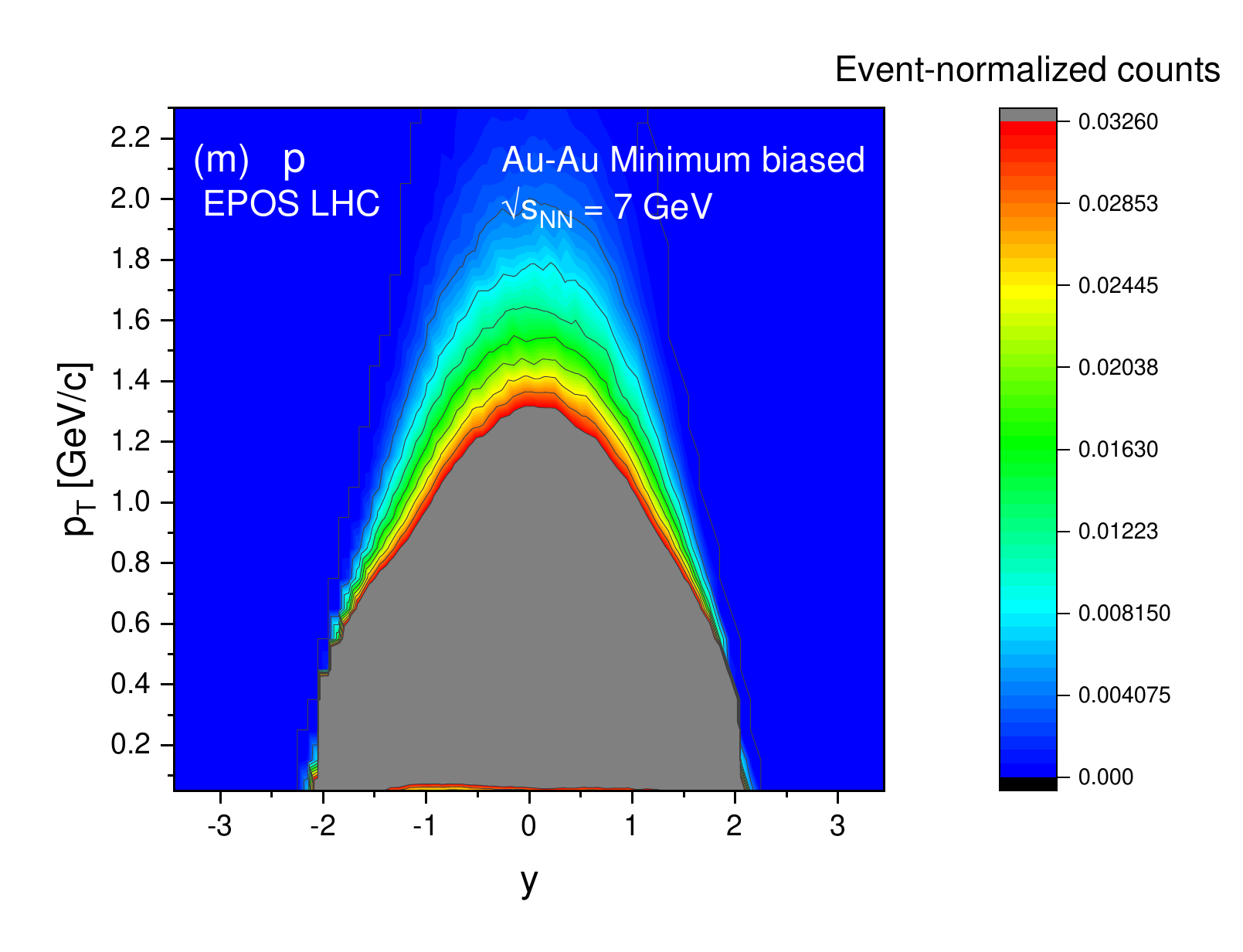}
\includegraphics[width=0.32\textwidth]{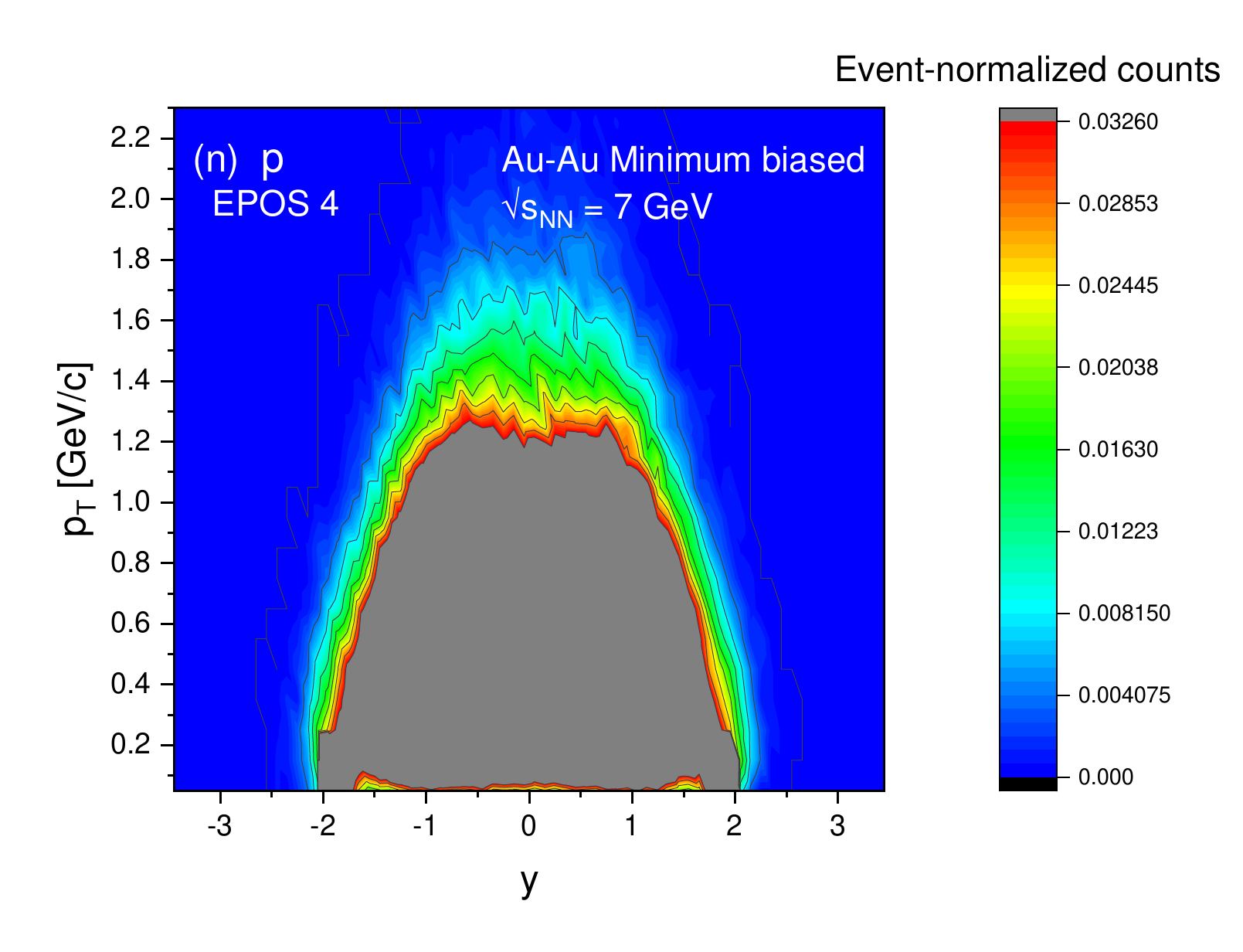} 
\includegraphics[width=0.32\textwidth]{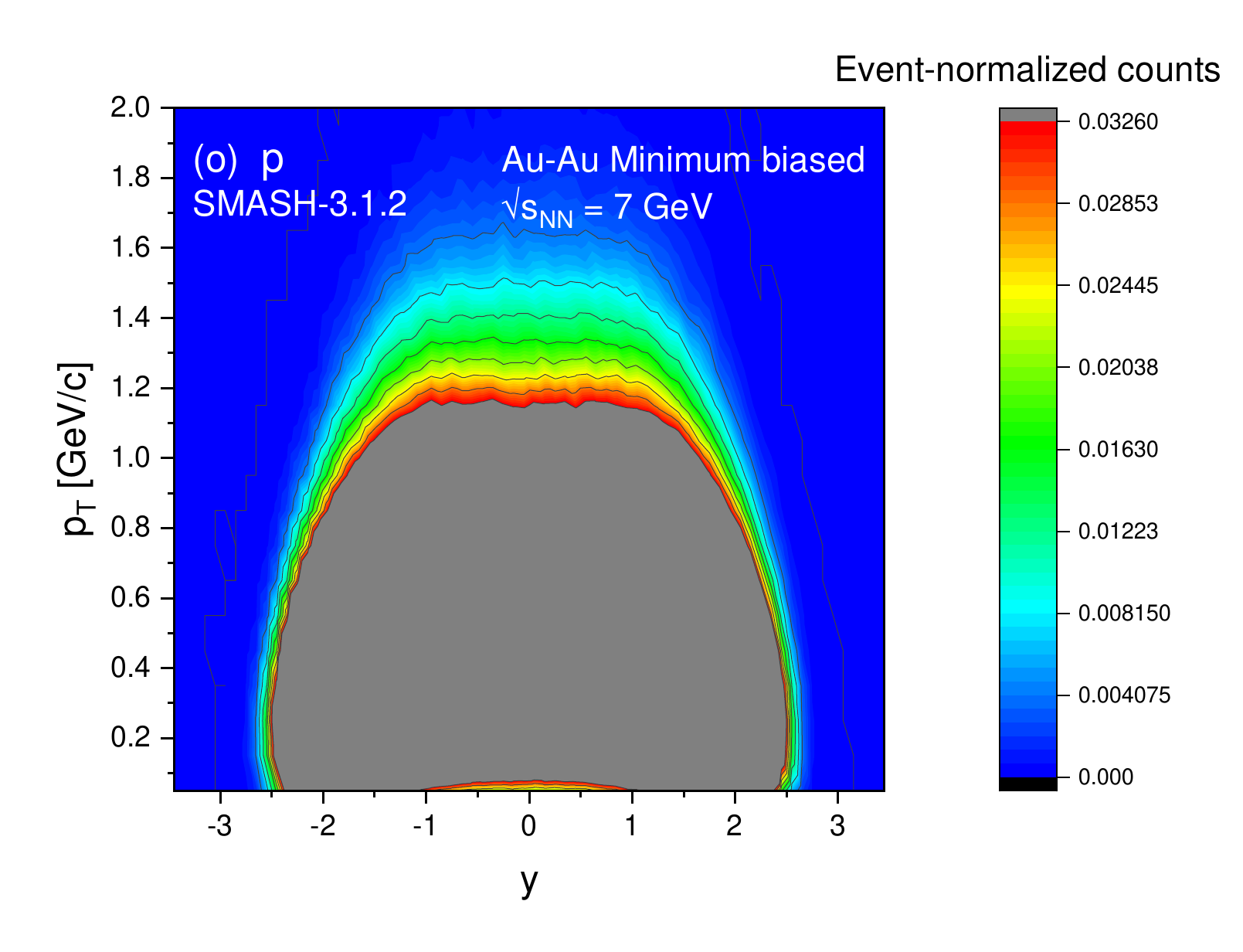}\vspace{-0.35cm}
\includegraphics[width=0.32\textwidth]{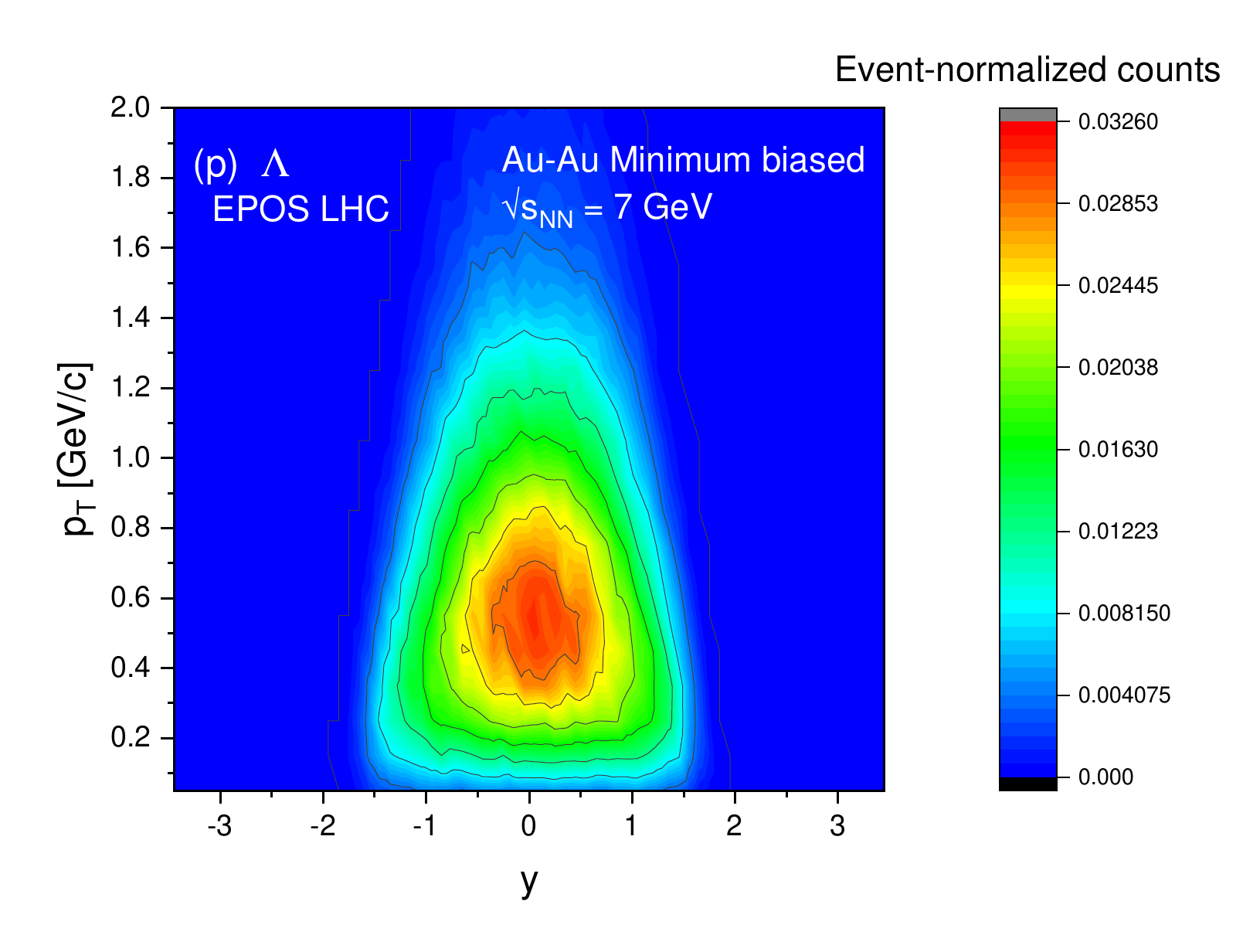}
\includegraphics[width=0.32\textwidth]{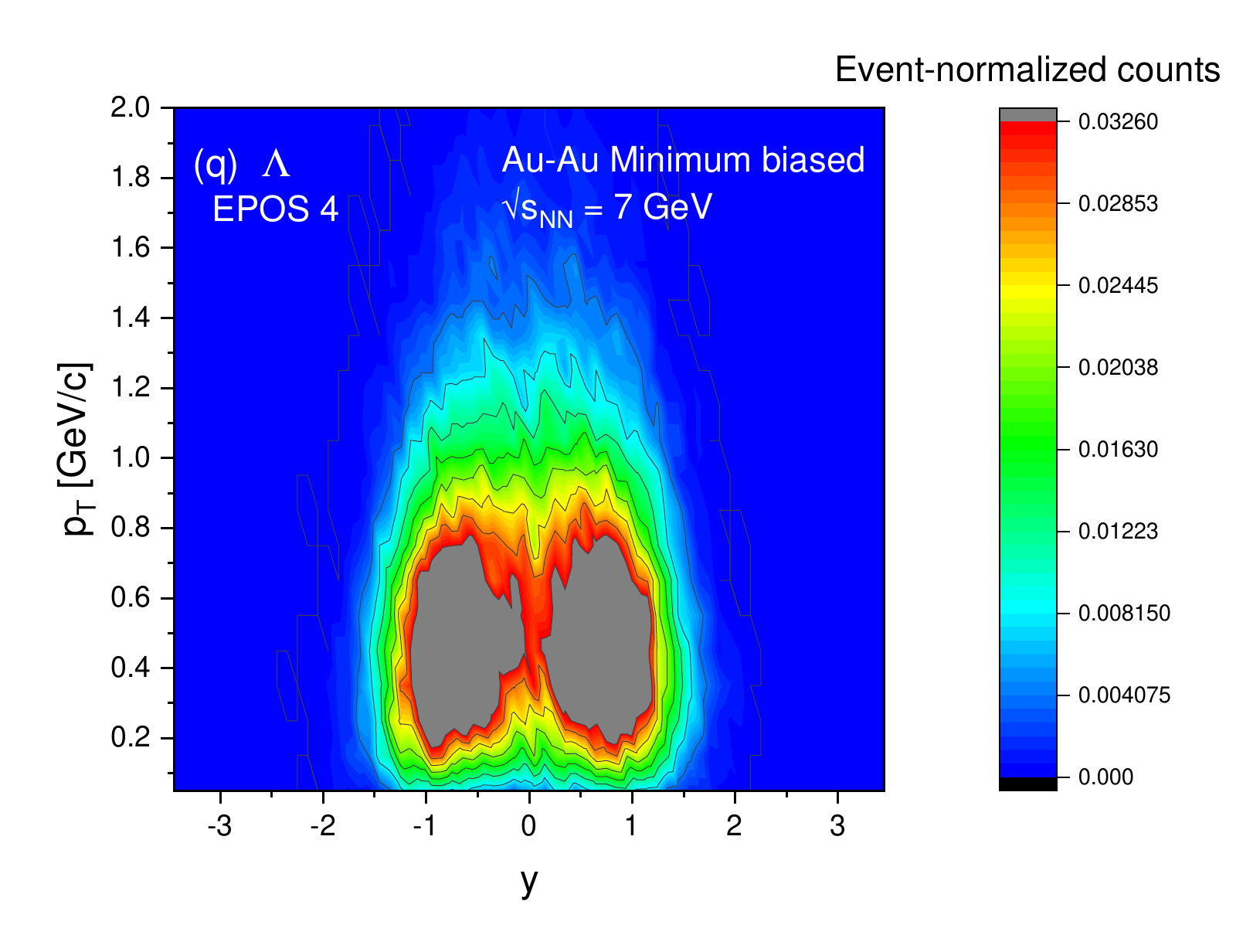} 
\includegraphics[width=0.32\textwidth]{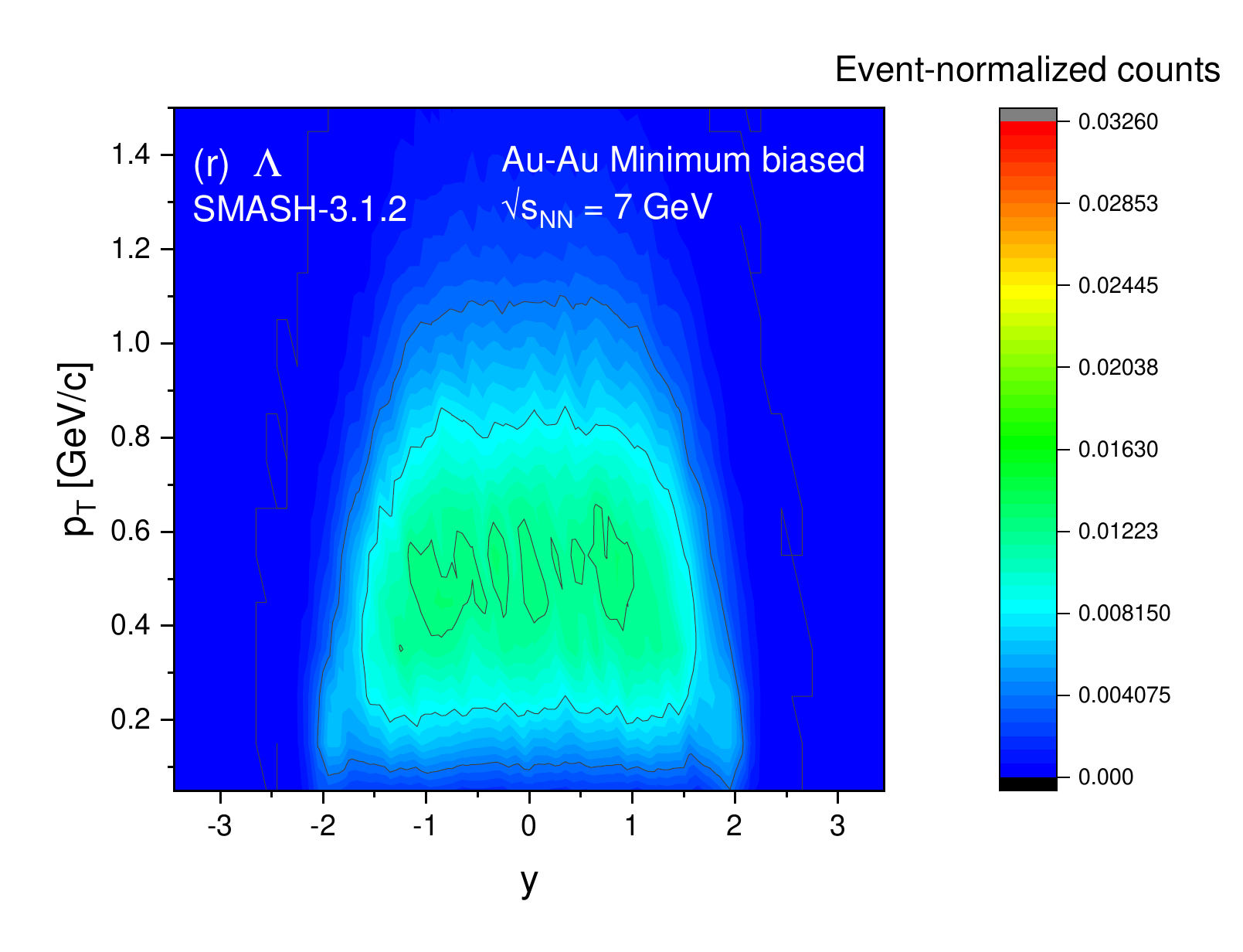}\vspace{-0.35cm}
\caption {Two-dimensional $p_T$ vs rapidity density maps (color scale = event-normalized counts) of various identified hadrons in minimum-bias collision of $Au+Au$ in $\sqrt{s_{NN}}=7~GeV$. The arrangement of the panels is similar to Fig. 7.}
\end{figure*}
\begin{figure*}
\centering
\includegraphics[width=0.32\textwidth]{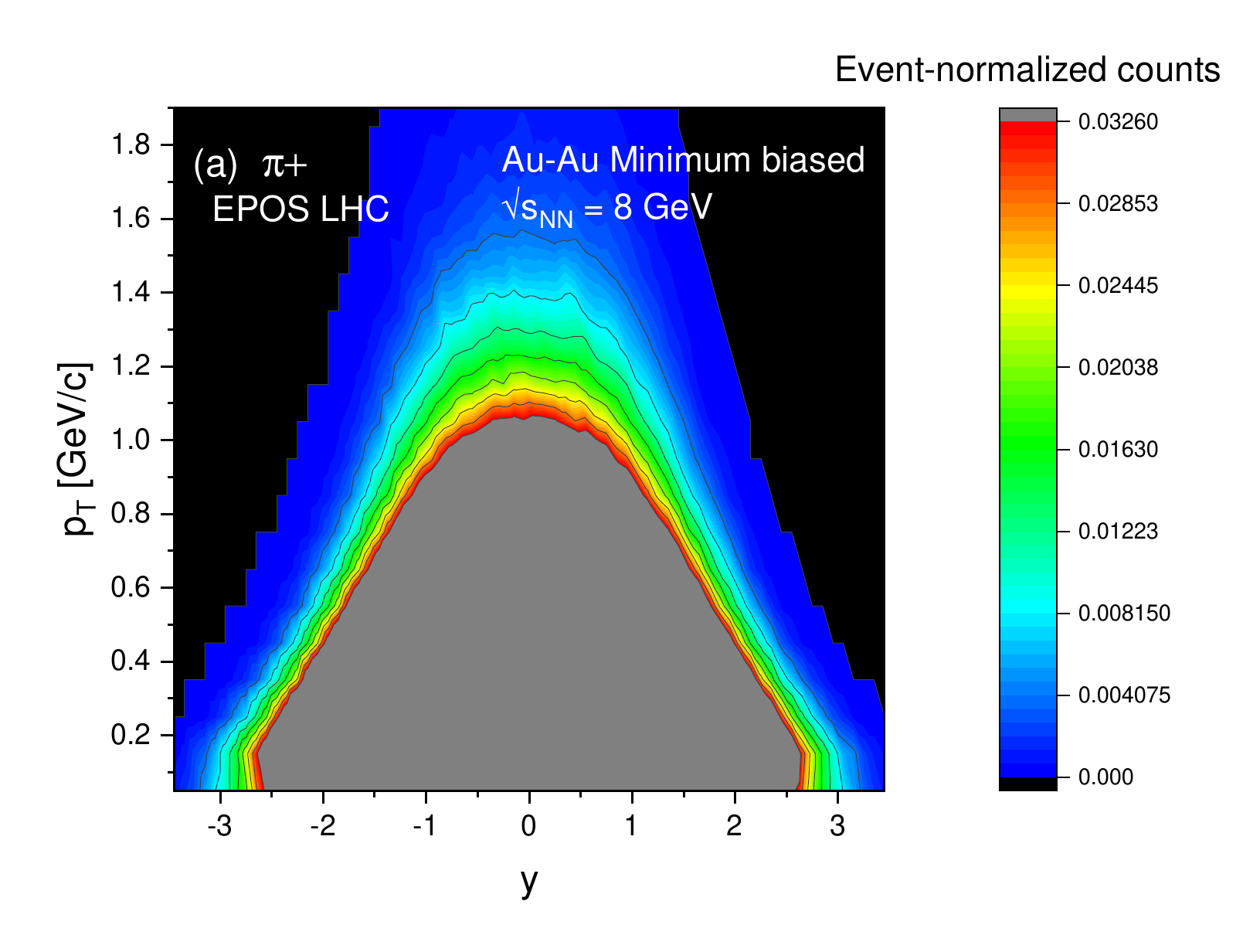}
\includegraphics[width=0.32\textwidth]{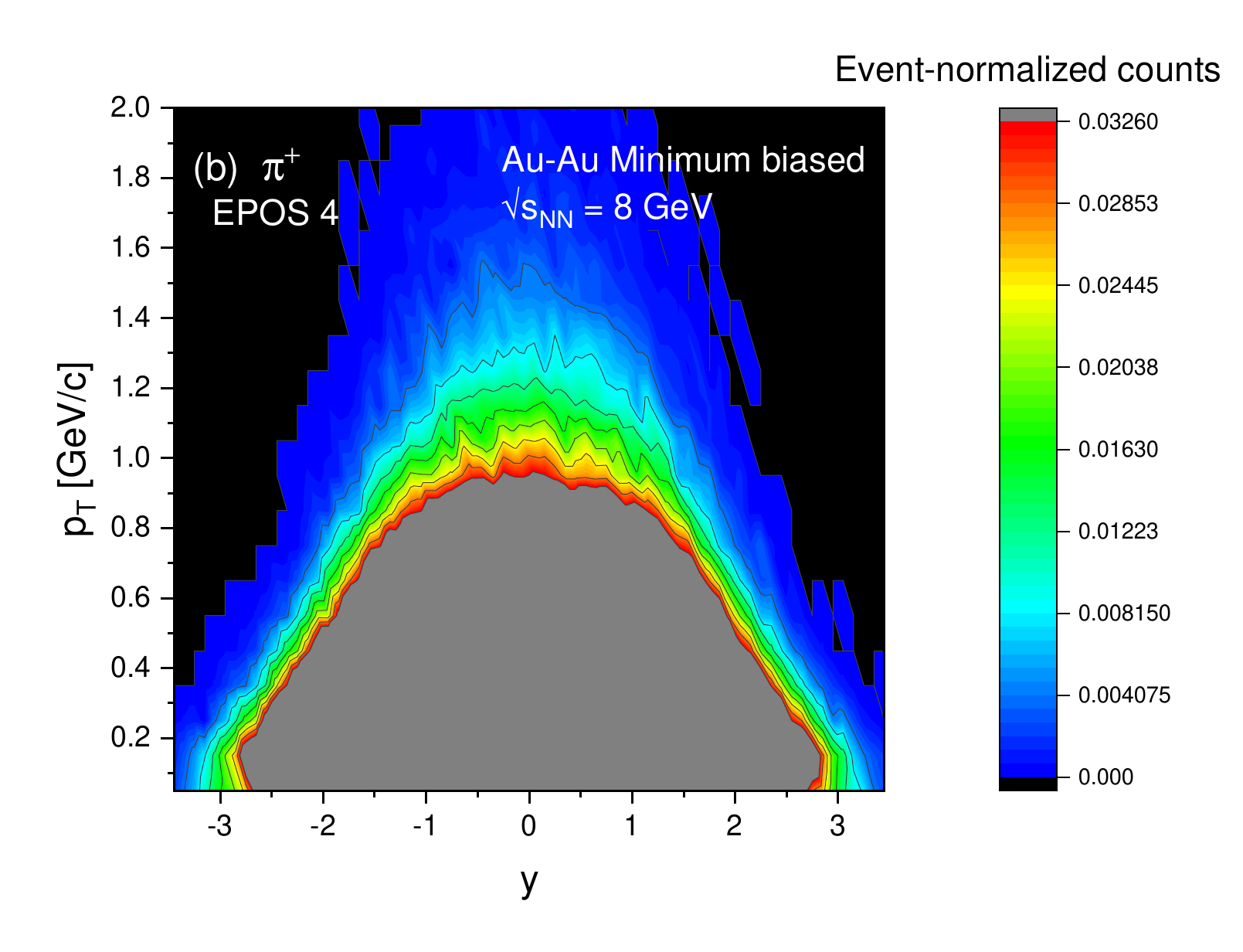} 
\includegraphics[width=0.32\textwidth]{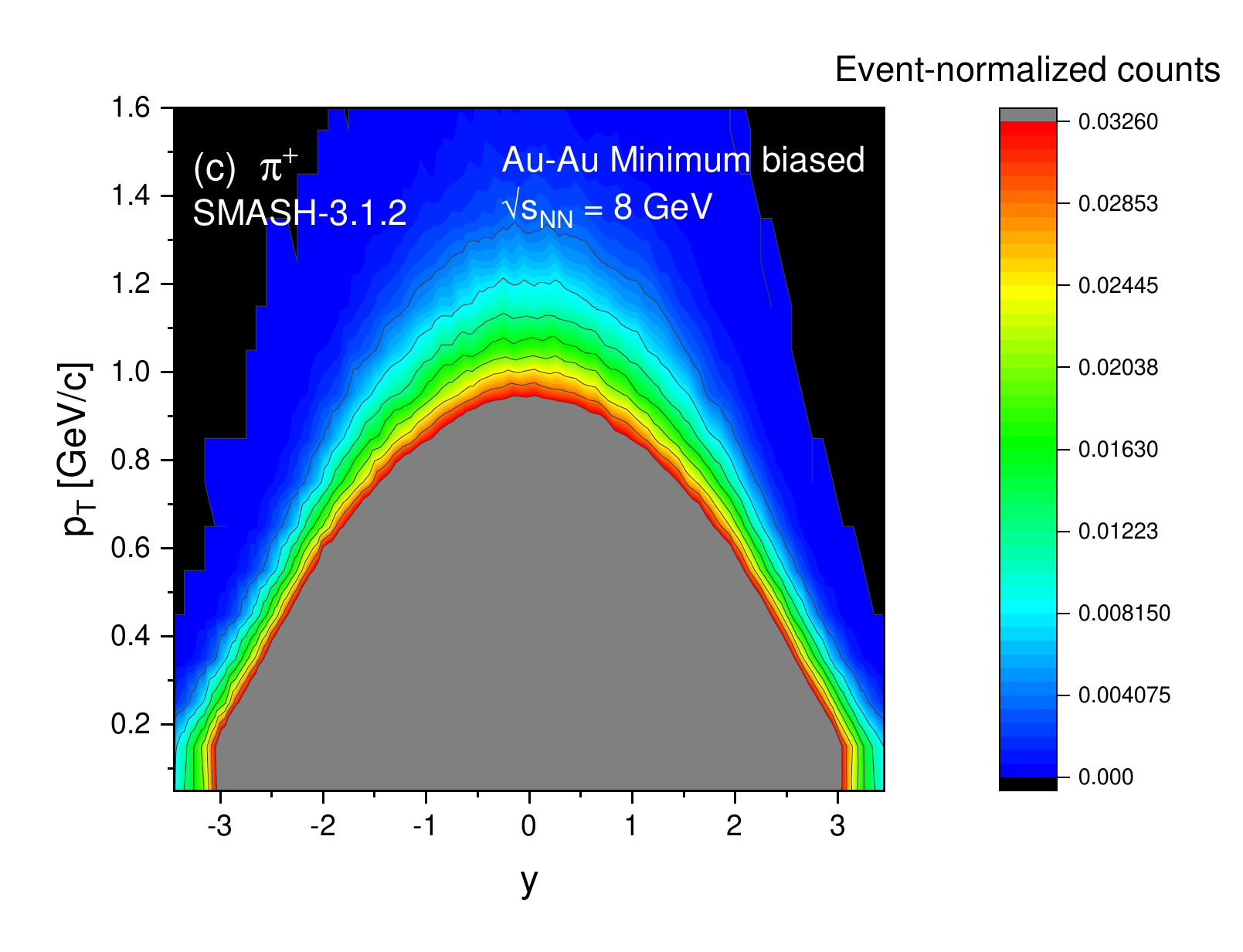}\vspace{-0.35cm}
\includegraphics[width=0.32\textwidth]{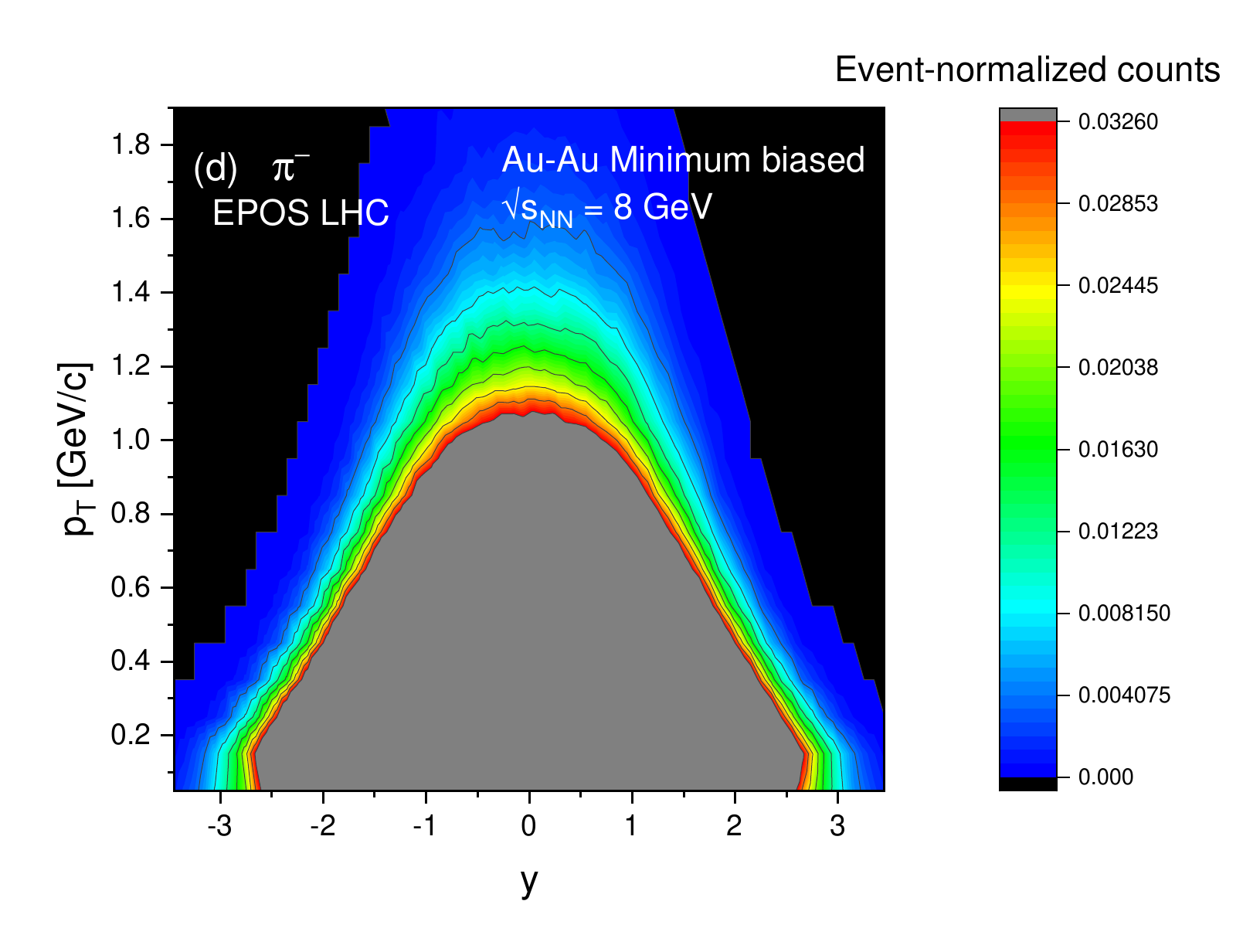}
\includegraphics[width=0.32\textwidth]{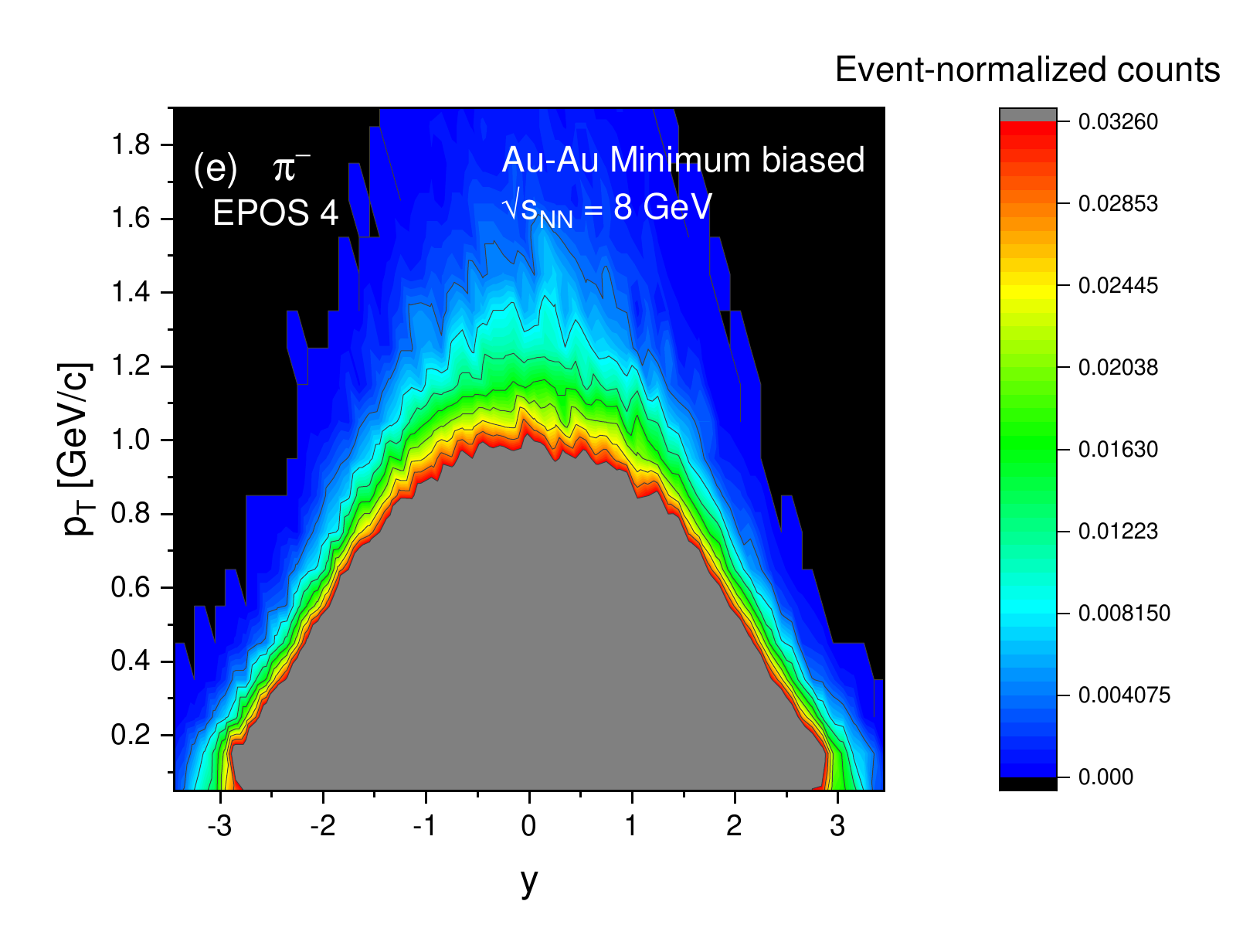} 
\includegraphics[width=0.32\textwidth]{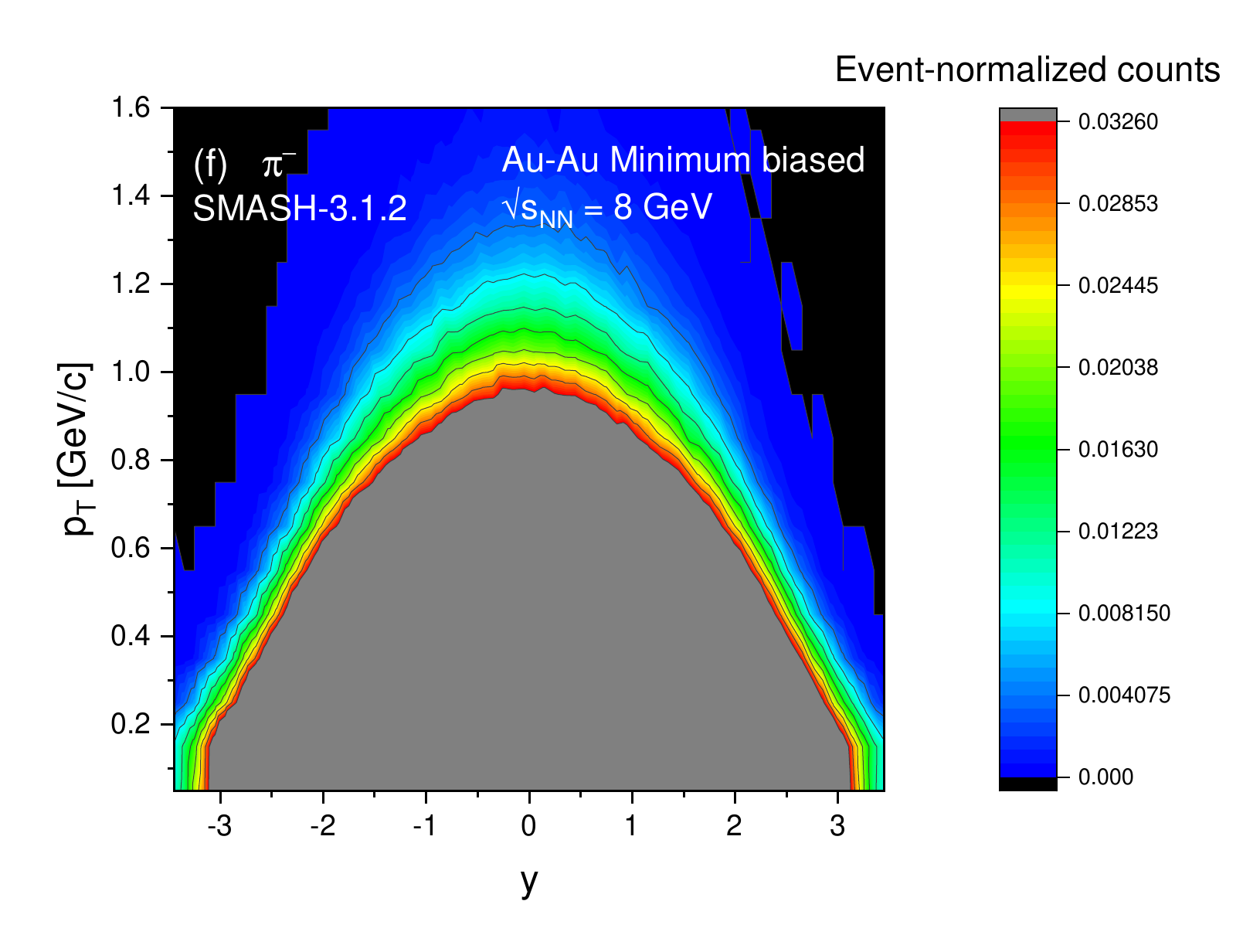}\vspace{-0.35cm}
\includegraphics[width=0.32\textwidth]{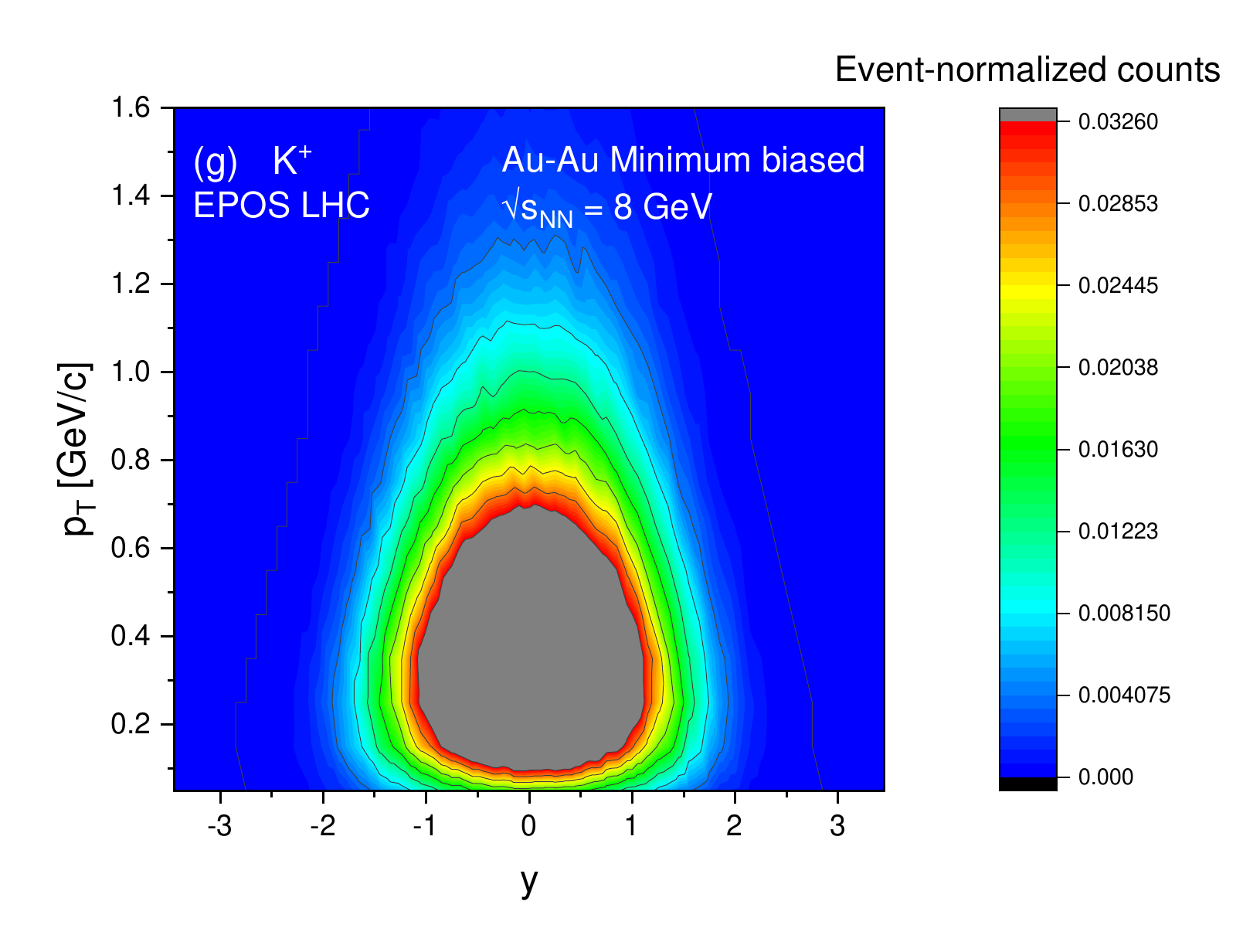}
\includegraphics[width=0.32\textwidth]{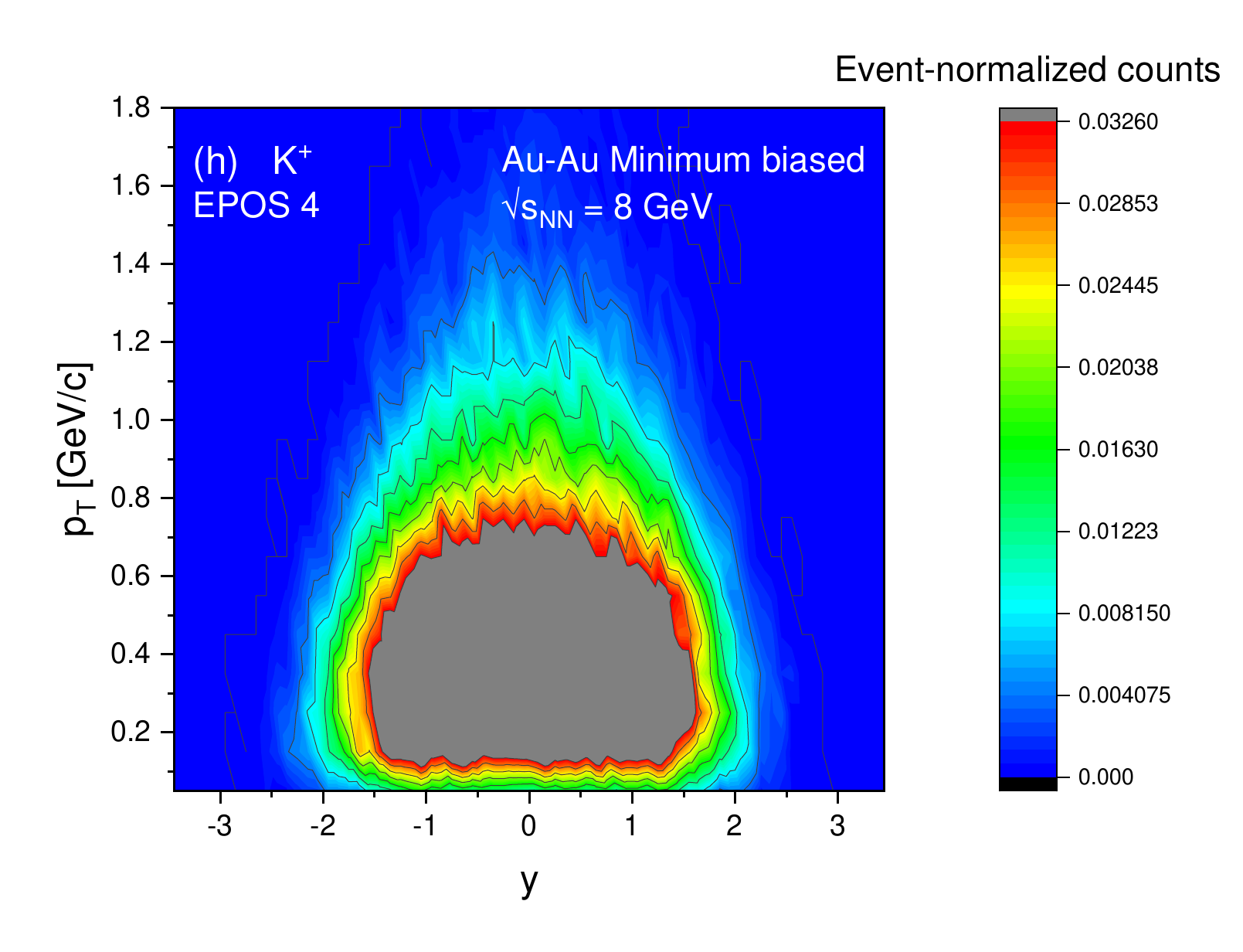} 
\includegraphics[width=0.32\textwidth]{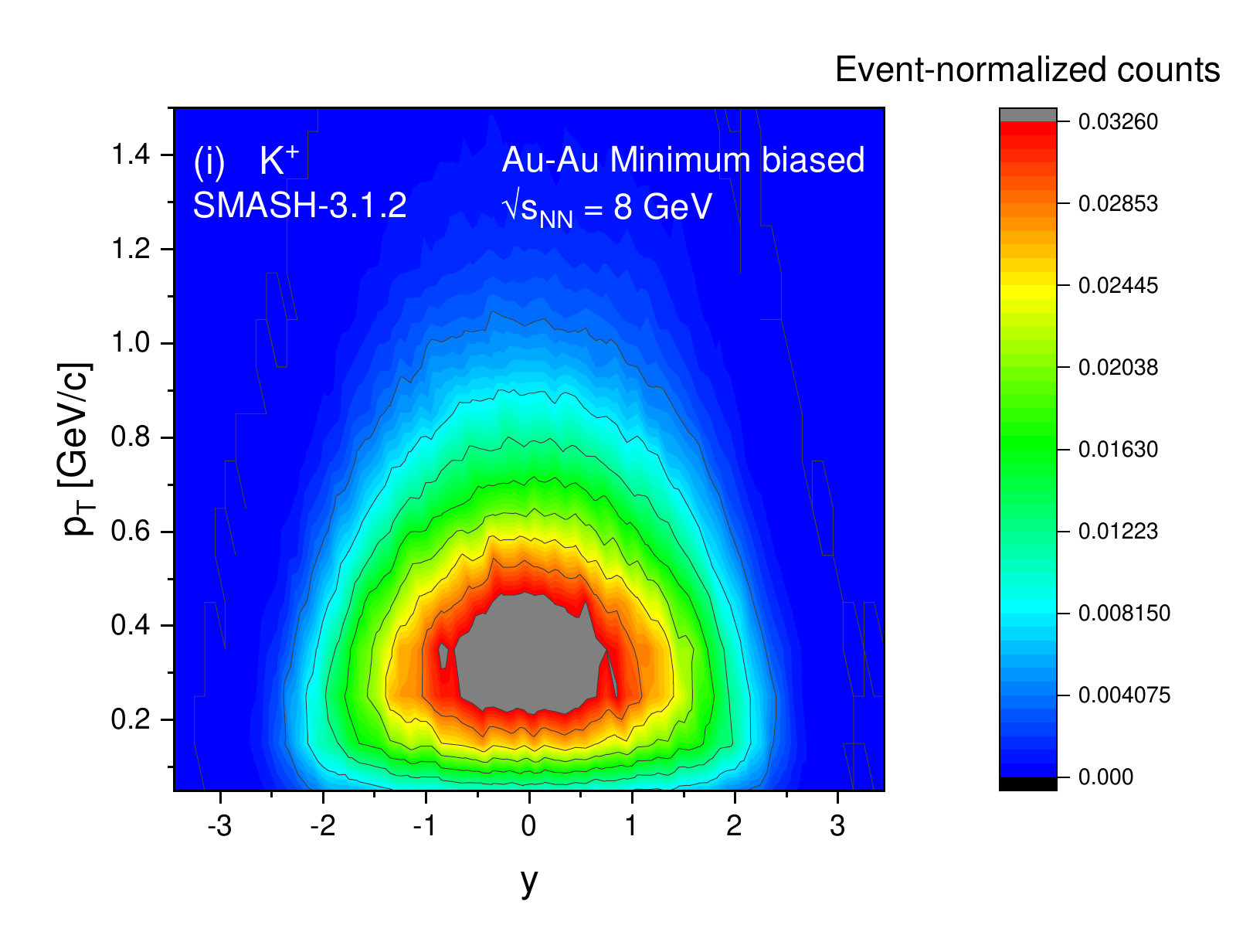}\vspace{-0.35cm}
\includegraphics[width=0.32\textwidth]{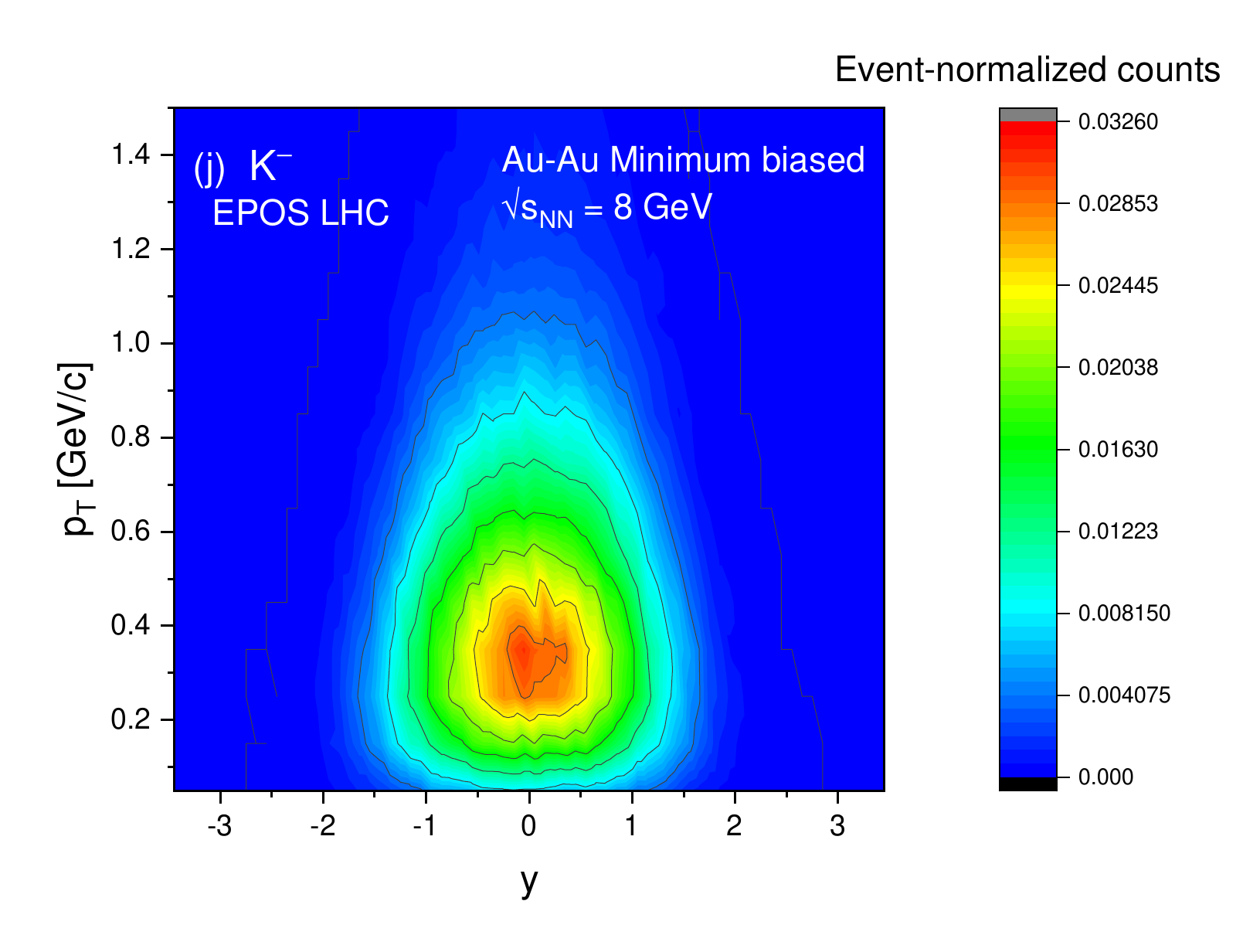}
\includegraphics[width=0.32\textwidth]{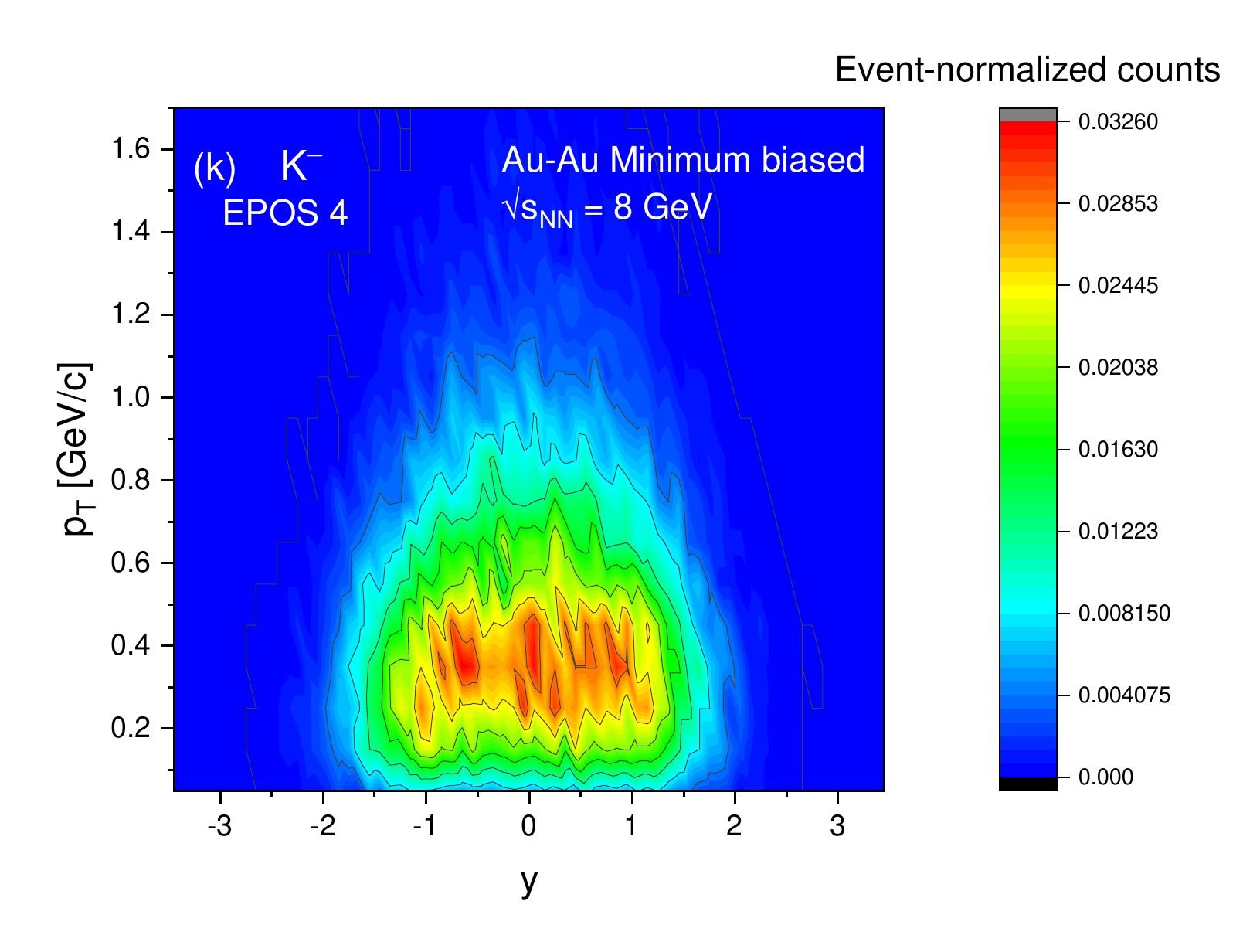} 
\includegraphics[width=0.32\textwidth]{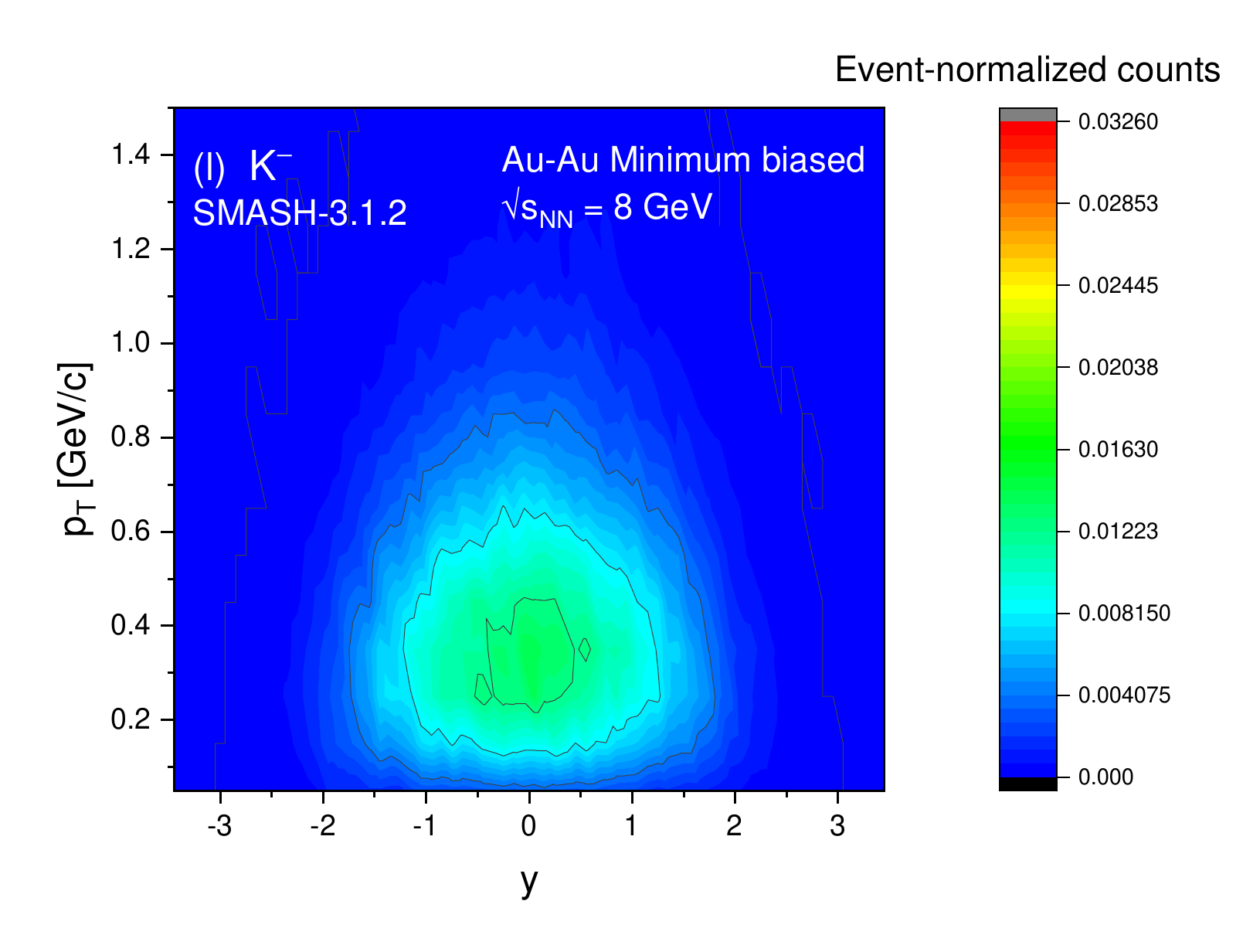}\vspace{-0.35cm}
\includegraphics[width=0.32\textwidth]{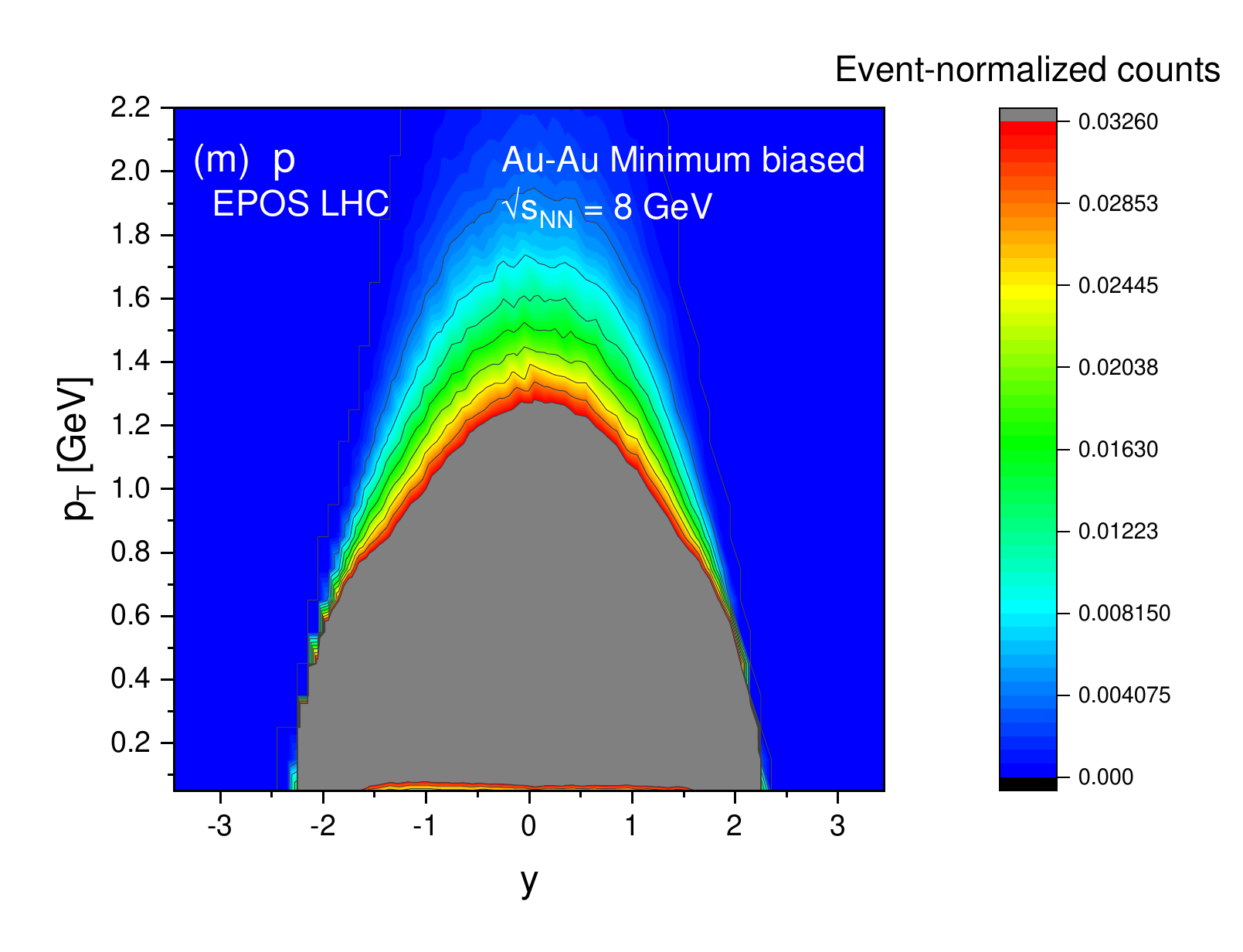}
\includegraphics[width=0.32\textwidth]{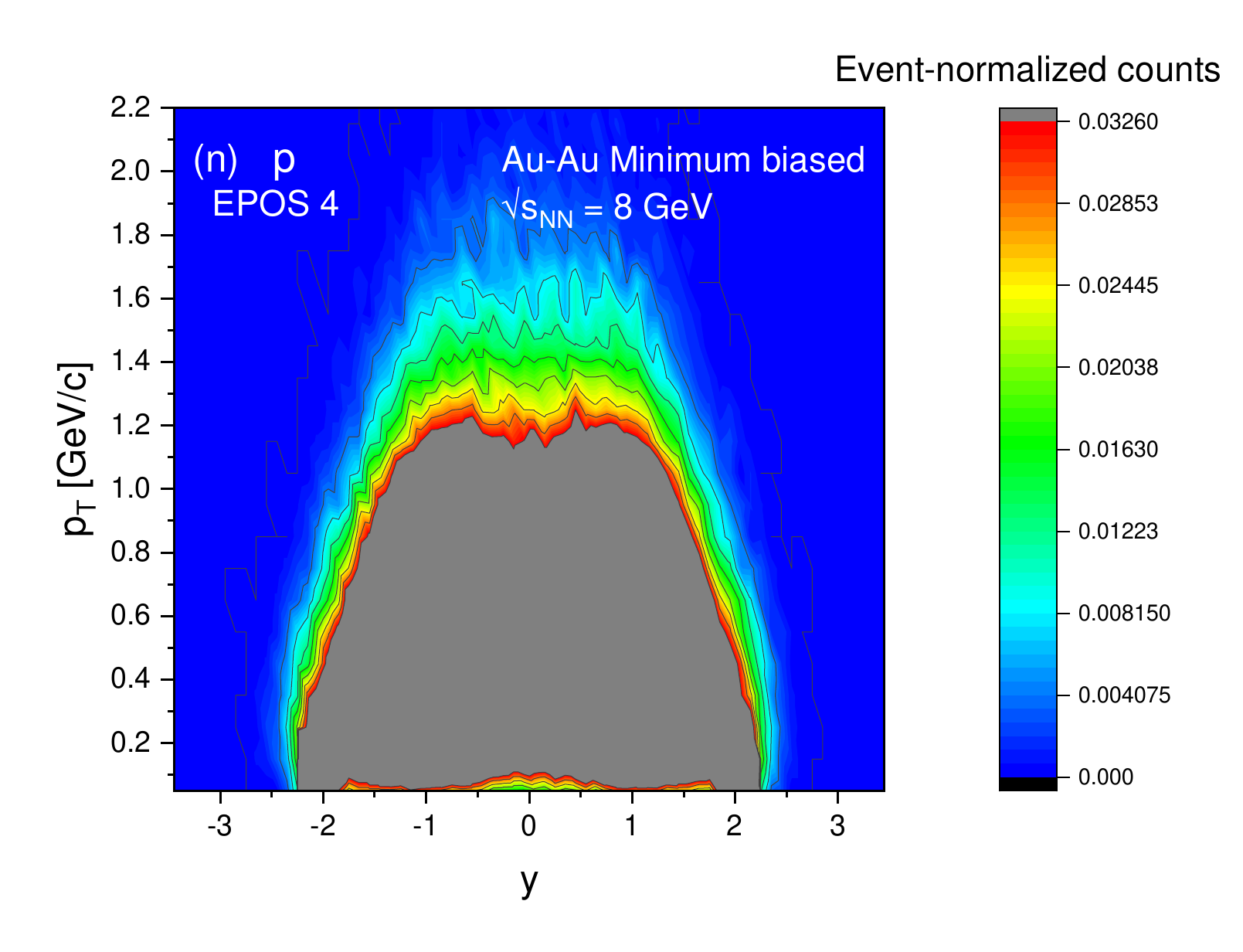} 
\includegraphics[width=0.32\textwidth]{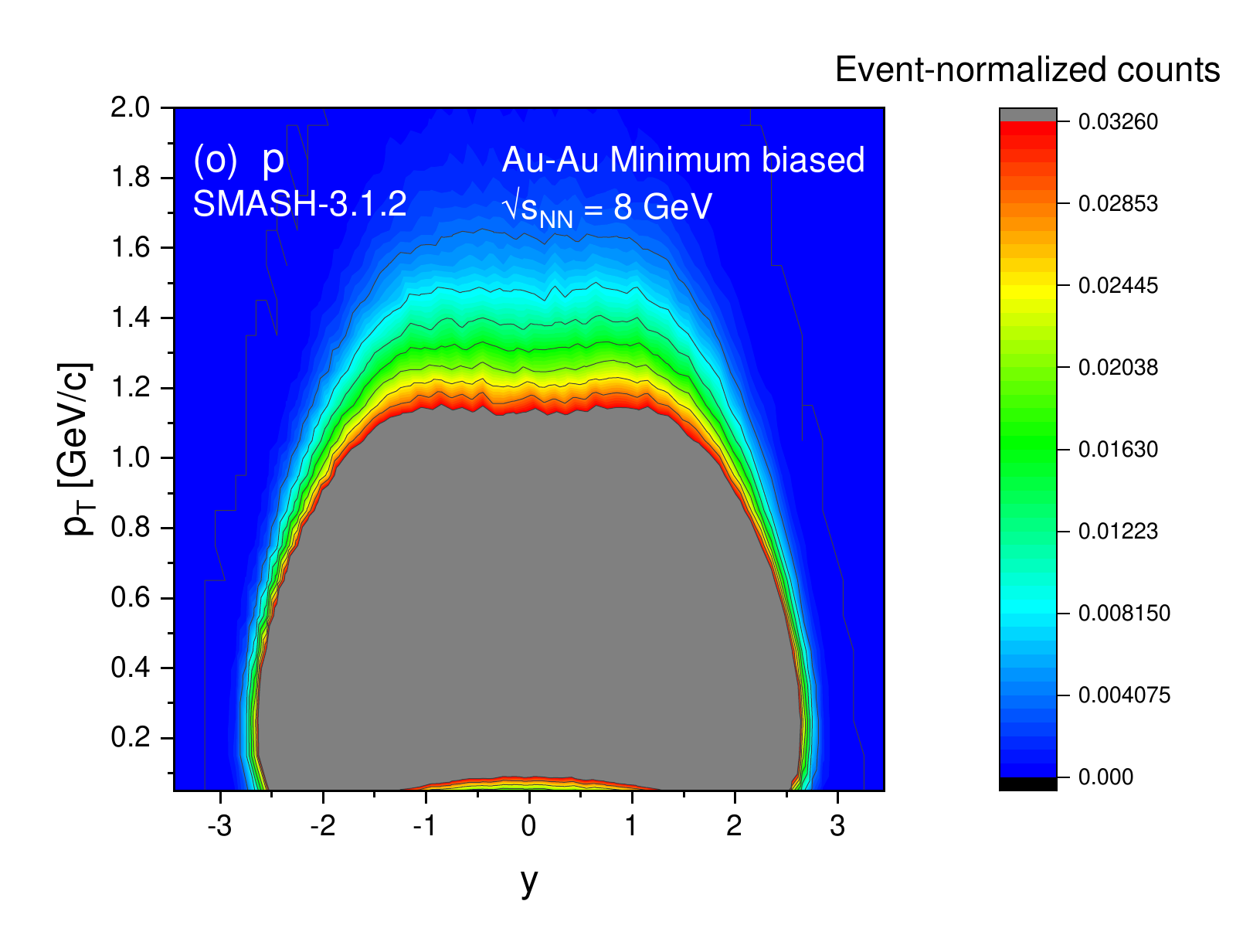}\vspace{-0.35cm}
\includegraphics[width=0.32\textwidth]{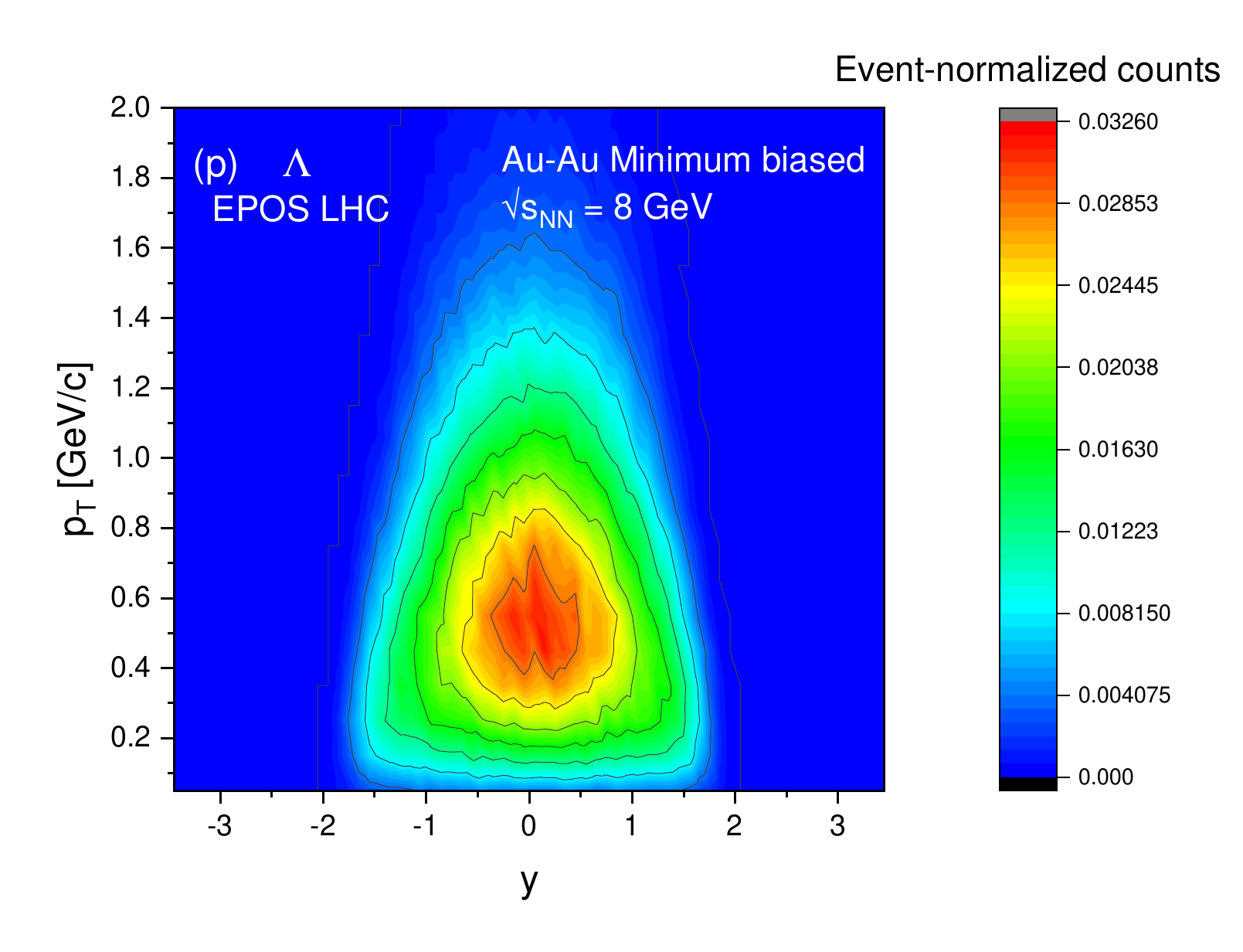}
\includegraphics[width=0.32\textwidth]{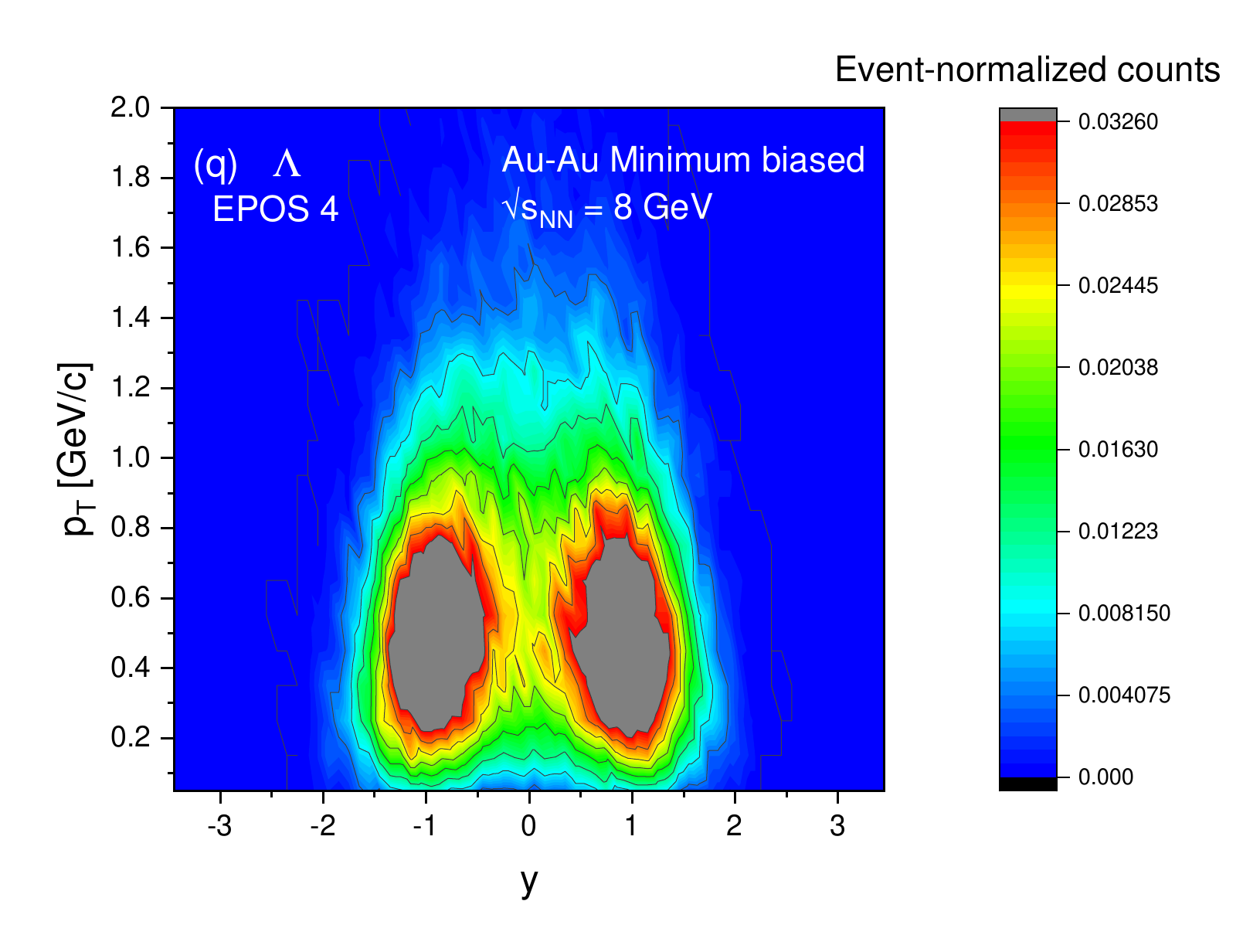} 
\includegraphics[width=0.32\textwidth]{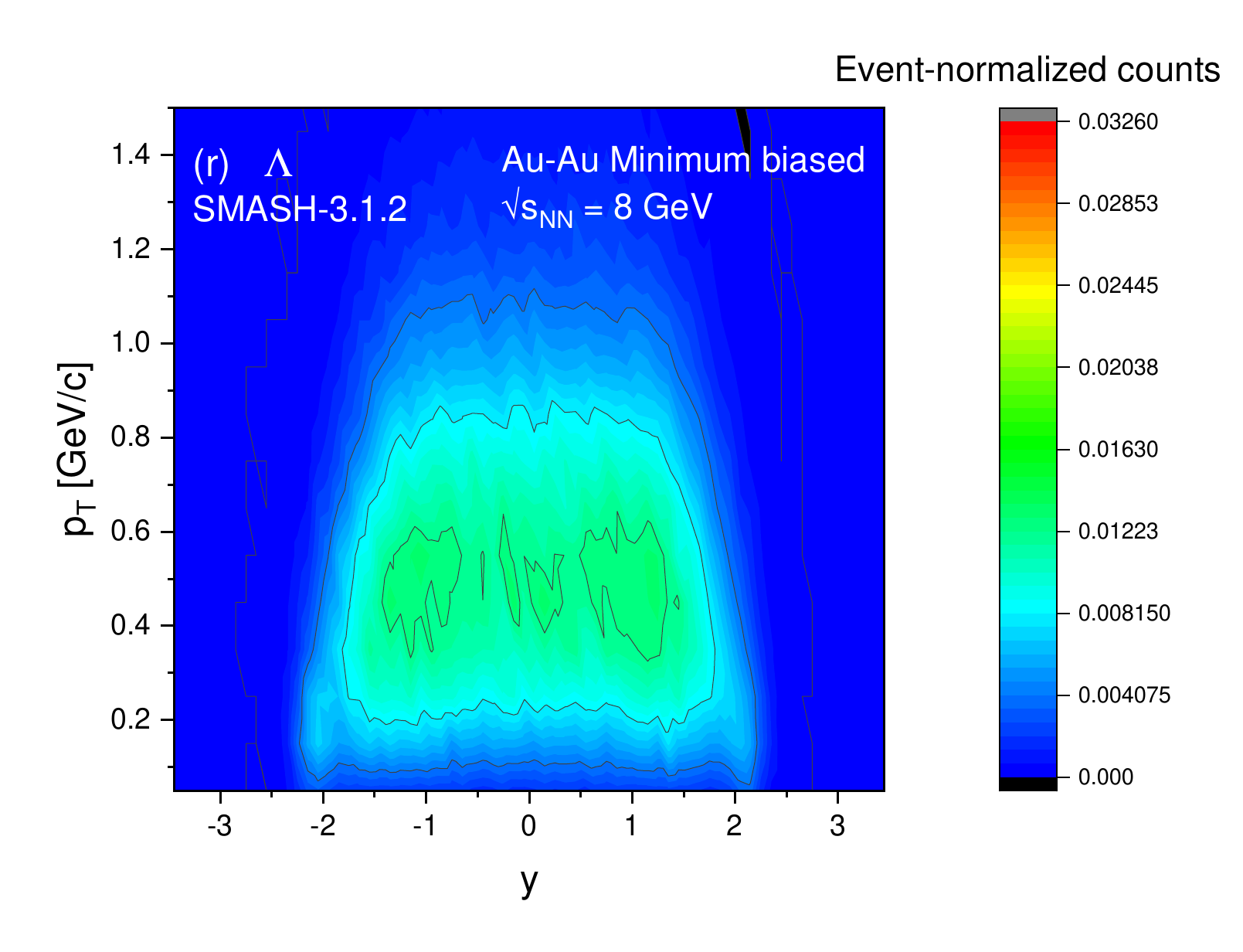}\vspace{-0.35cm}
\caption {Two-dimensional $p_T$ vs rapidity density maps (color scale = event-normalized counts) of various identified hadrons in minimum-bias collision of $Au+Au$ in $\sqrt{s_{NN}}=8~GeV$. The arrangement of the panels is similar to Figs. 7 and 8.}
\end{figure*}  
\subsection{$p_T-y$ maping}
Figs. 7(a-c) give the $\pi^+$ 2D $p_T-y$ distributions at $6~GeV$ in EPOS-LHC, EPOS-4 and SMASH respectively, and Figs. 7(d-f) give the $\pi^-$ distributions. The yield, in any case, is highly concentrated at low transverse momenta ($p_T < 0.8~GeV/c$), and is concentrated in a bright band at midrapidity. The magnitude also falls gradually with the rise of $y$ and $p_T$. EPOS-LHC and EPOS-4 show somewhat more extended distributions of $p_T$ at midrapidity ($y\approx0$), and still have yield extending to approximately $1.5~GeV/c$, whereas SMASH is slightly confined to the soft sector. The profiles of the rapidities all have the same symmetry and are concentrated at $y\approx0$. The same overall features are found at $\sqrt{s_{NN}} = 7~GeV$ (Figs. 8(a-f)) with an intermediate-$p_T$ enhancement in EPOS models being stronger. EPOS-LHC can still maintain a visible yield to a maximum of $1.8~GeV/c$, and SMASH still focuses on $p_T$ below $1.3~GeV/c$. With $\sqrt{s_{NN}}=8~GeV$ (Figs. 9(a-f)), the contrast is slightly increased: the EPOS models exhibit non-negligible population at up to a few GeV/c, and SMASH is steep and concentrated to lower transverse momenta. Distributions of rapidity are quite universal between models at any energy. These maps prove the fact that pion production at this energy scale is dominated by resonance decay (e.g., $\Delta\xrightarrow{}N\pi,~\rho\xrightarrow{}\pi \pi$) and soft hadronic reactions. These small differences between models are because the string and core dynamics present in EPOS provide a small, but systematic improvement in the intermediate $p_T$. Charged pions are thus a strong baseline observable that is sensitive to multiplicity normalization, but exhibits limited discriminating power between a hadronic and a hybrid description.

The $K^+$ distributions are presented in Figs. 7(g-i) and the $K^-$ distributions in Figs. 7(j-l) at $\sqrt{s_{NN}}=6~GeV$. In $K^+$, the maps are visible further in $p_T$ than in pions, extending up to some $1.5~GeV/c$, and the $p_T$ spectra and rapidity coverage of EPOS-LHC and EPOS-4 are farther and broader than the kaons in SMASH, which are steeply confined below $1.1~GeV/c$ and narrower around $y\approx0$. There is a significant reduction in pair-production channels of $K^-$, which are suppressed in baryon-rich systems, and the same hierarchy applies: EPOS models have broader $p_T$ tails and slightly broader rapidity distributions. At $7~GeV$ (Figs. 8(g-l)), the yields of kaons grow, and EPOS models have stronger populations in intermediate-$p_T$, and SMASH is steep and localized. The rapidity range extends slightly in EPOS structures, especially with $K^+$. The separation is most apparent at $8~GeV$ (Figs. 9(g-l)): EPOS-LHC has a lot of yield at scale up to a matter of approximately $2~GeV/c$, and has visually more extensive rapidity coverage, whereas SMASH distributions are thin and weak. This is a physically manifested behavior, which reflects the interaction of mechanisms of strangeness production. The production of $K^+$ is the most dominant due to the related processes ($NN\xrightarrow{}N\Lambda K^+$) that are boosted in baryon-enriched collisions at the expense of $K^-$ production. EPOS models also provide additional transverse momentum through string excitation and early collective effects, while SMASH provides momentum through late hadronic rescattering. The kaon $p_T-y$ maps are thus sensitive probes to strangeness dynamics and early phase collectivity, and model variations become apparent in the intermediate-$p_T$ region, and in the $K^+ /K^-$ asymmetry.

Figs. 7(m-o) present the proton $p_T-y$ distributions at $\sqrt{s_{NN}}=6~GeV$, respectively for EPOS-LHC, EPOS-4, and SMASH. There is a distinct difference between the models. SMASH gives the steepest falloff in transverse momentum: the majority of the protons are concentrated below $p_T \approx 1~GeV/c$. Nonetheless, SMASH has the widest coverage in rapidity space, and the yields have a significant extension forward and backward rapidities. In comparison, EPOS-LHC and EPOS-4 produce slightly flatter $p_T$ spectra, with visible populations at higher energies (up to $2~GeV/c$). Still, the rapidity distributions are more concentrated at midrapidity ($y \approx 0$), which shows stronger stopping. These differences are sharpened at $\sqrt{s_{NN}} = 7~GeV$ (Figs. 8(m-o)). SMASH is still characterized by large rapidity distributions, with protons further displaced in the $y$ direction with increasing energy, although the spectra are steep in $p_T$. The resulting EPOS models give smaller rapidity peaks that are concentrated more near $y \approx 0$, but with more transverse momentum tails that are harder and yield into the intermediate-$p_T$ regime ($1.5-3~GeV/c$). The separation is greatest by $\sqrt{s_{NN}} = 8~GeV$ (Figs. 9(m-o)). SMASH models the broadest distribution of rapidity of protons among all models, yet it confines them to soft $p_T$ ($<1~GeV/c$). EPOS-LHC and EPOS-4 are narrower in rapidity localization, but have much harder spectra in transverse momentum, and large populations spilling up to the $2-2.5~GeV/c$. These systematic discrepancies highlight the complementary nature of the models' physics. In SMASH, the baryon transport is purely hadronically rescattered, which results in weaker stopping and a wider rapidity distribution, but weaker transverse acceleration. Due to the incorporation of core-corona dynamics and string excitation, the EPOS models have more powerful stopping (smaller rapidity peaks) and more powerful transverse push (flatter $p_T$ spectra) \cite{werner2007core, aichelin2009centrality}. In short, the hierarchy is: Rapidity coverage - wider in SMASH, and transverse momentum hardness - more powerful in EPOS. This two-fold pattern is a good discriminator. Stopping and collective expansion are encoded in protons, which are the main bearers of baryon number. The $p_T-y$ distributions thus indicate the nature of the system that is either evolving as solely hadronic, strongly transported medium (SMASH) or more strongly stopping and transverse dynamics (EPOS).

Figs. 7(p-r) display $\Lambda$ $p_T-y$ distributions at $\sqrt{s_{NN}}=6~GeV$. Every model exhibits yields concentrated at low $p_T$ around the middle of rapidity; however, SMASH achieves the sharpest and narrowest distributions. In contrast, EPOS-LHC and EPOS-4 have non-negligible populations extending up to about $1.5-2~GeV/c$ and cover a more extended rapidity range. The separation is more pronounced at $7~GeV$ (Figs. 8(p-r), EPOS models form visible shoulders in the intermediate-$p_T$ region at midrapidity, which are not present in SMASH \cite{werner2007core, aichelin2009centrality, weil2016particle}. The difference is greatest at $8~GeV$ (Figs. 9(p-r)). The distributions of SMASH are restricted to low $p_T$ and $y\approx 0$, whereas EPOS-LHC and EPOS-4 have broad and flat distributions with $p_T$ up to about $2~GeV/c$ as well as larger rapidity distributions. This hierarchy arises because the treatment of strange baryon production and freeze-out differs. In SMASH, $\Lambda$ particles are produced through hadronic associated mechanisms ($NN\xrightarrow{}N\Lambda K$) and subsequently rescatter extensively, resulting in a soft and centrally peaked distribution. EPOS models can include early-string excitations and core-corona dynamics, giving rise to harder parent states ($\Sigma^*$, $\Xi$) whose decays cascade to harder $\Lambda$ states, and which freeze out earlier with a stronger transverse push. Consequently, EPOS models make predictions in $\Lambda$ distributions which are harder in $p_T$ and broader in rapidity. $\Lambda$ baryons are the most sensitive probes of the underlying dynamics of all species investigated. They are susceptible to collective expansion due to their greater mass and to their high level of coupling to the medium. In case the flatter and broader $\Lambda$ distributions obtained by future NICA measurements are in line with the prediction of EPOS, this would be direct evidence of the appearance of collective behavior in baryon-rich collisions. On the other hand, the finding of SMASH-like patterns would prefer an all-hadronic evolution picture.
\subsection{NCQ-scaled elliptic flow}
\begin{figure*}
\centering
\includegraphics[width=0.32\textwidth]{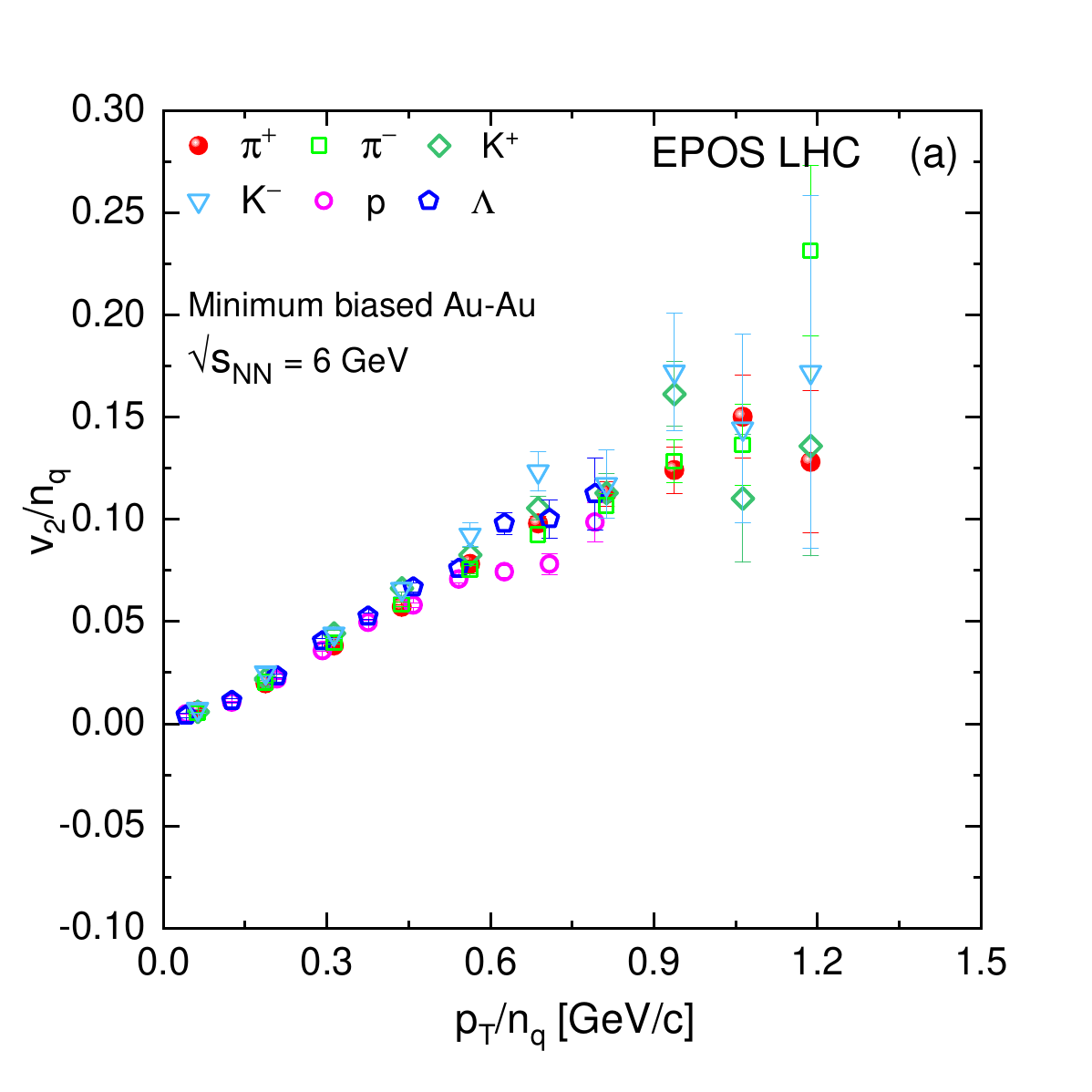}
\includegraphics[width=0.32\textwidth]{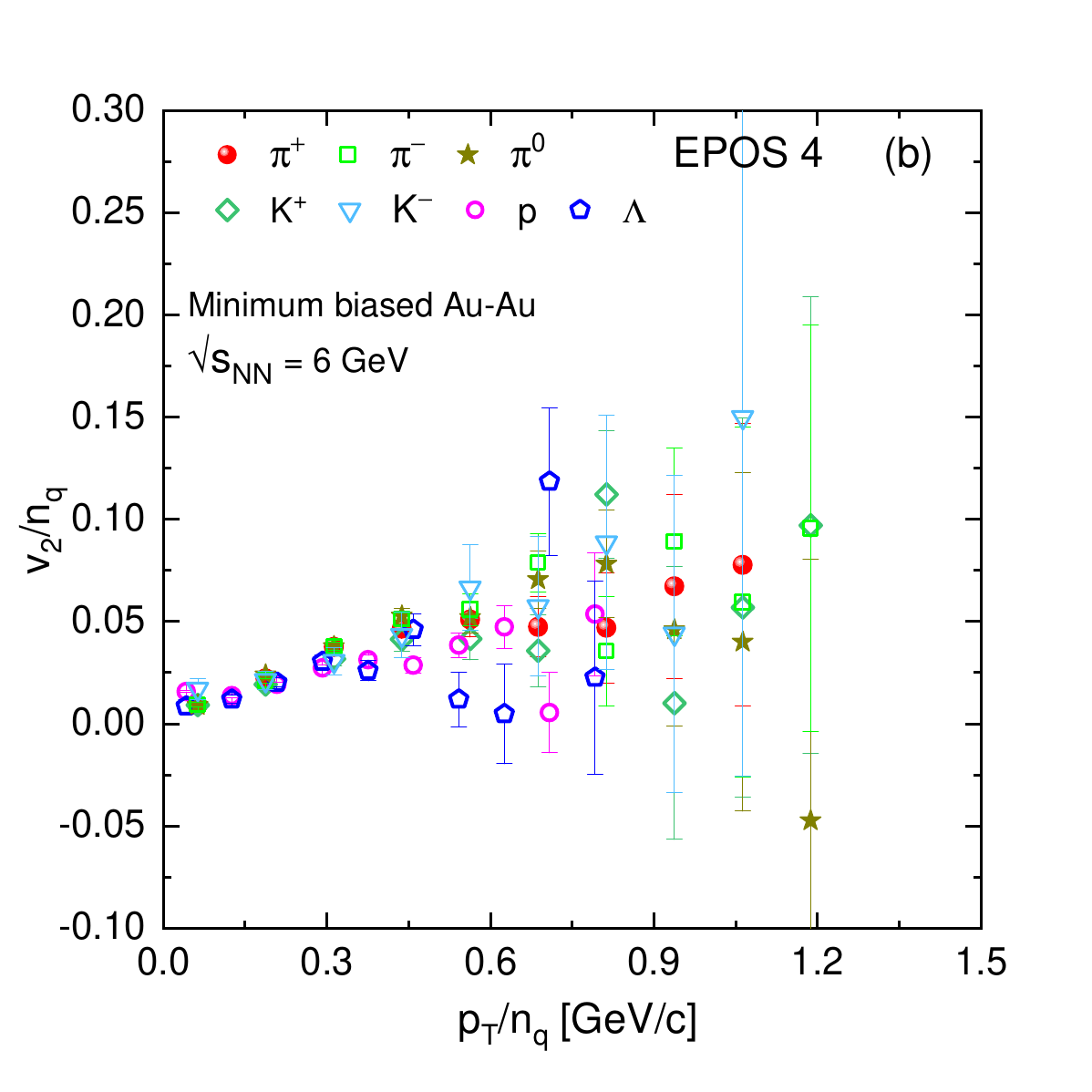} 
\includegraphics[width=0.32\textwidth]{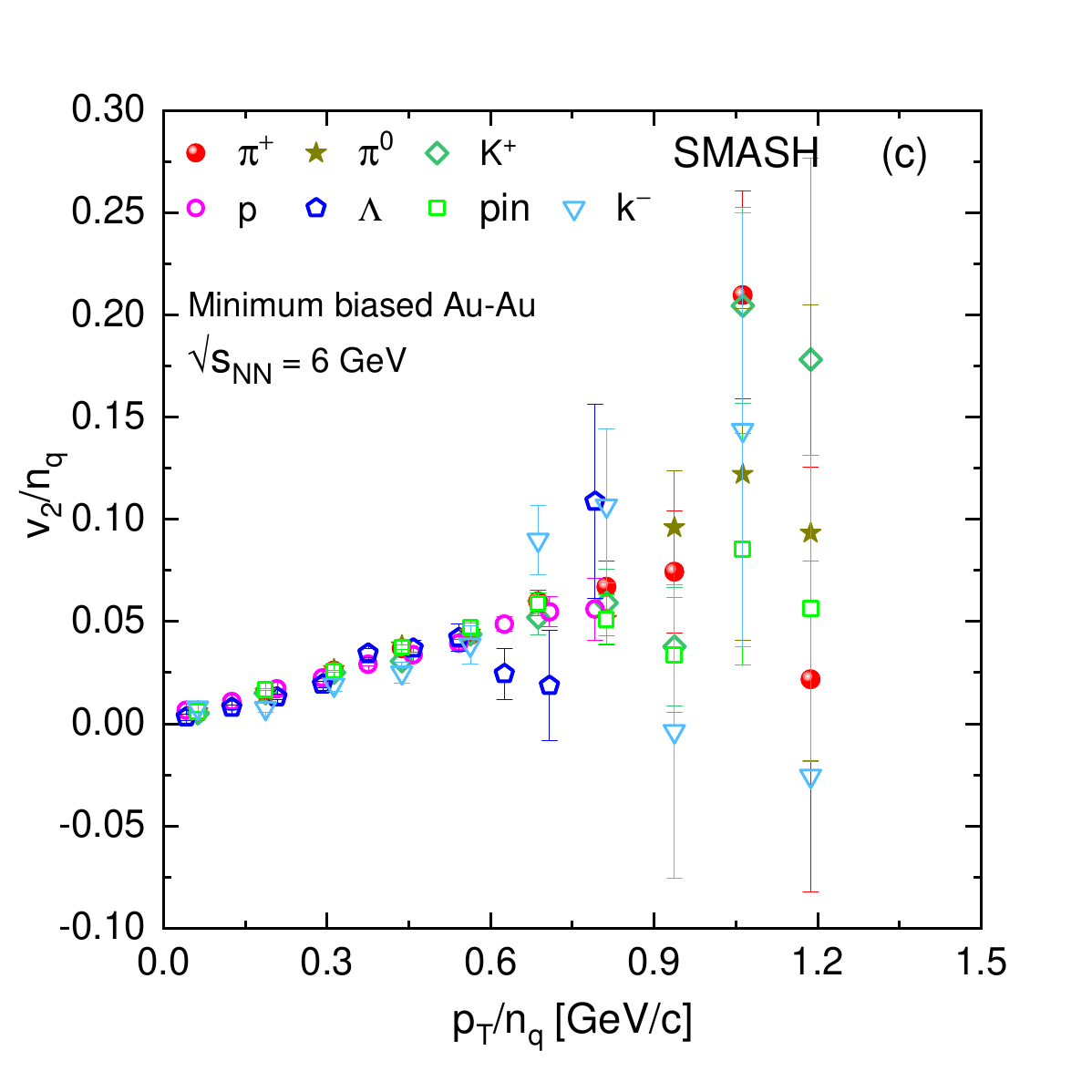}\vspace{-0.35cm}
\includegraphics[width=0.32\textwidth]{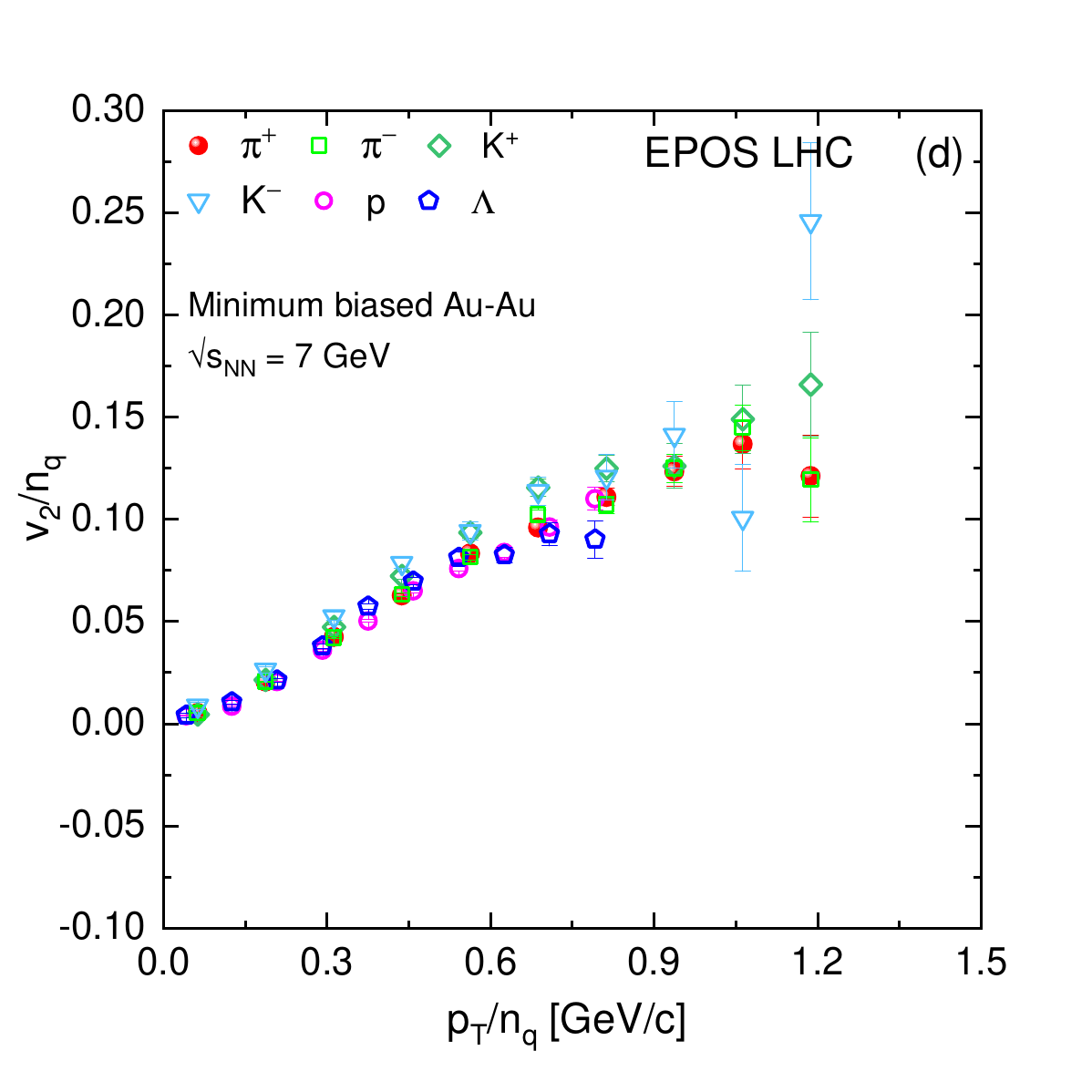}
\includegraphics[width=0.32\textwidth]{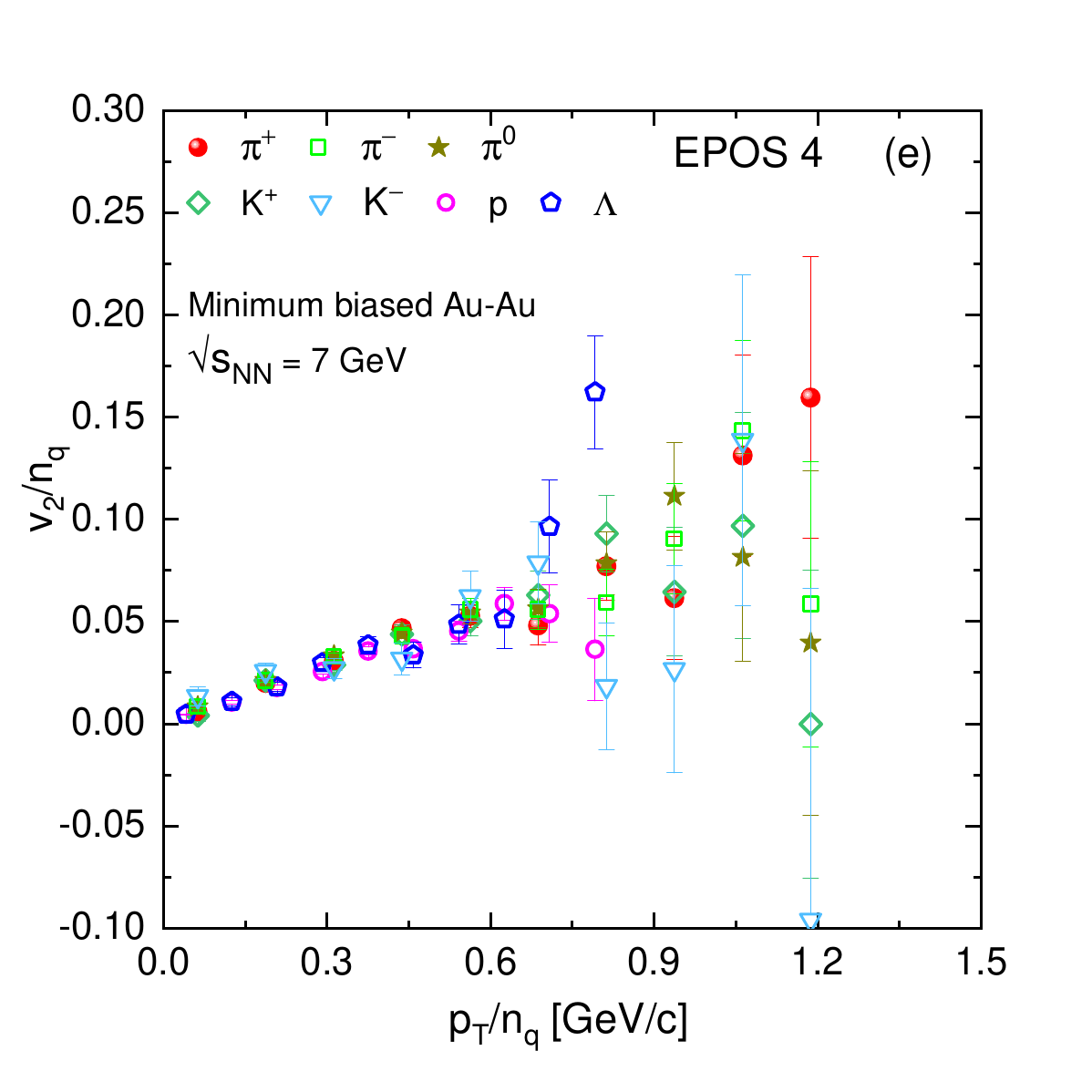} 
\includegraphics[width=0.32\textwidth]{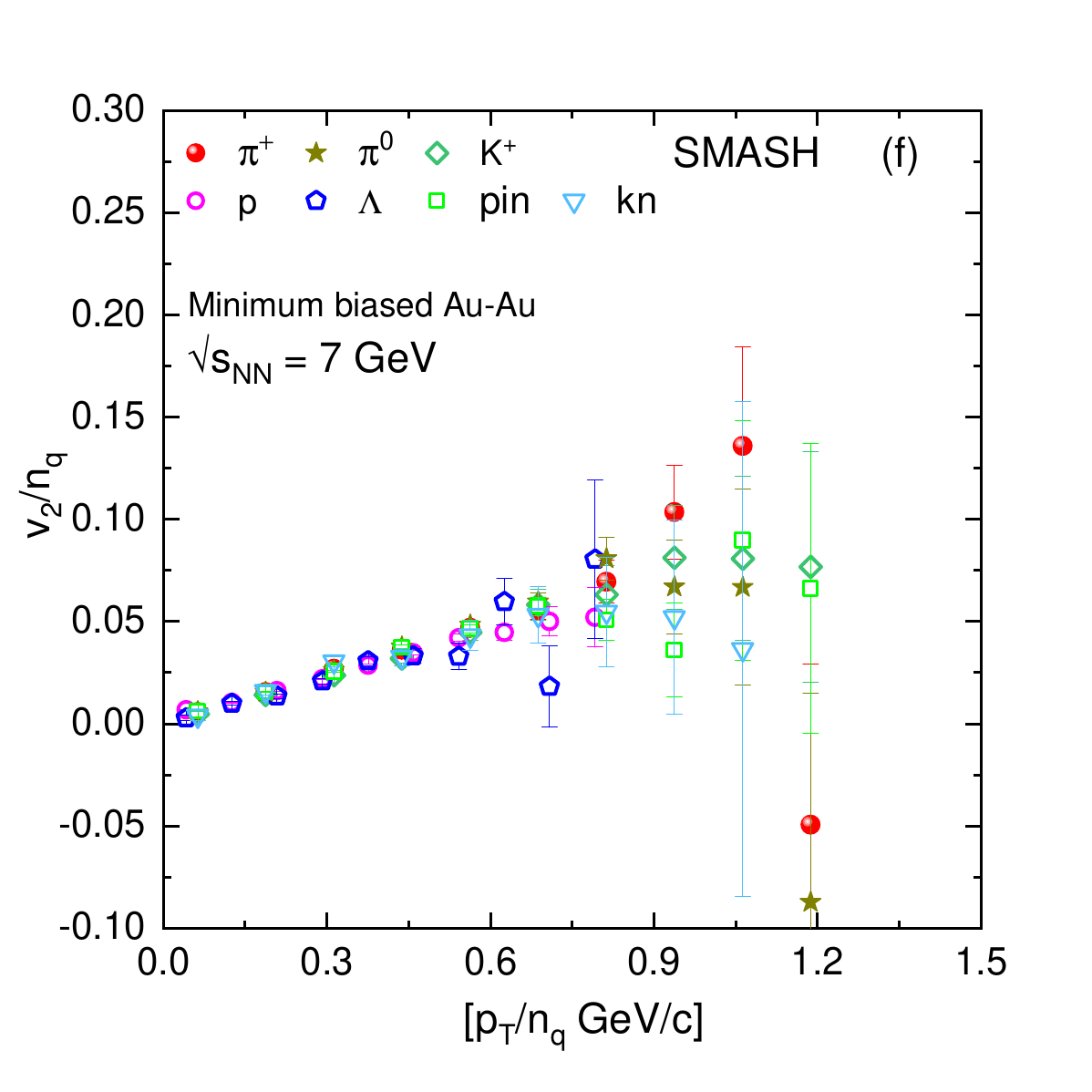}\vspace{-0.35cm}
\includegraphics[width=0.32\textwidth]{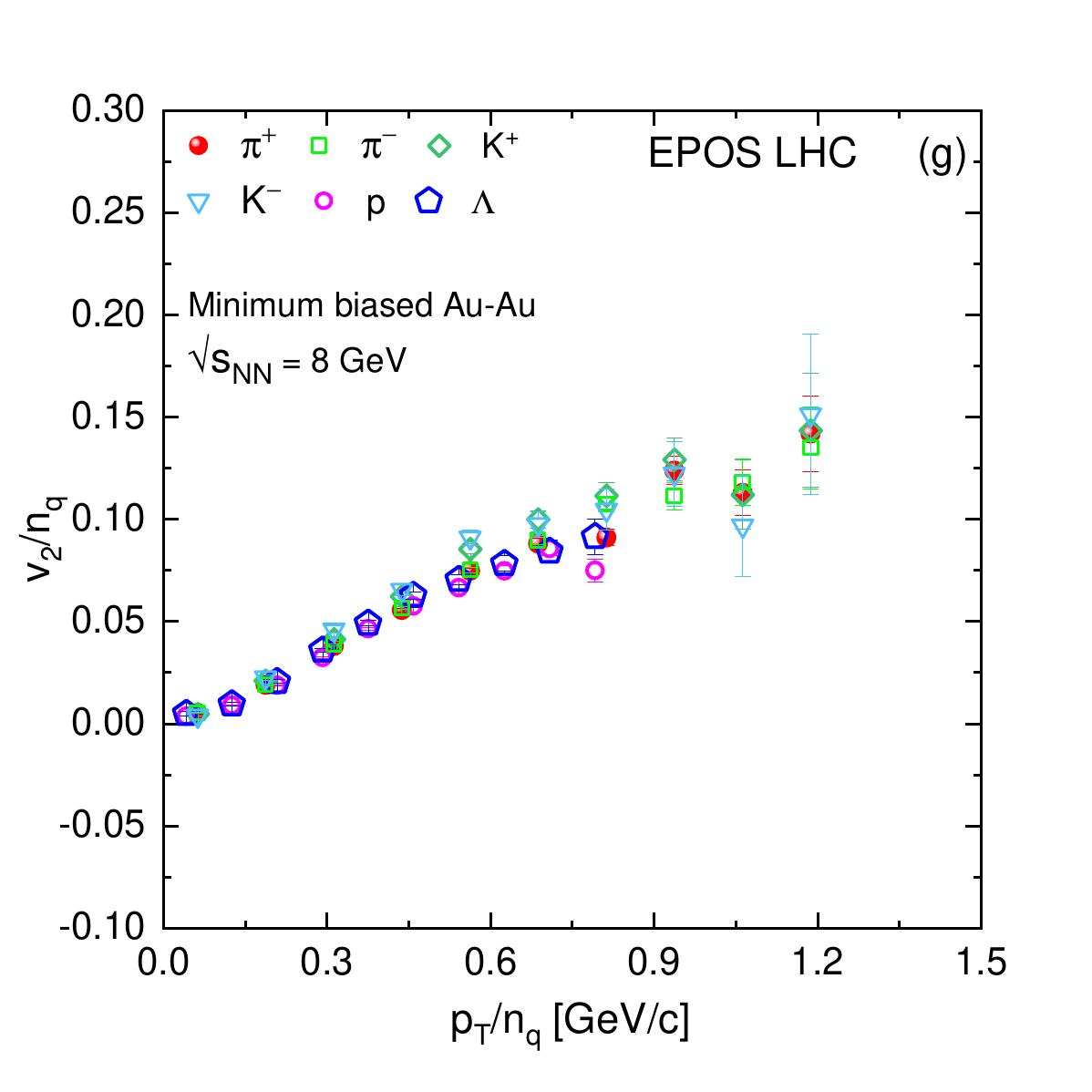}
\includegraphics[width=0.32\textwidth]{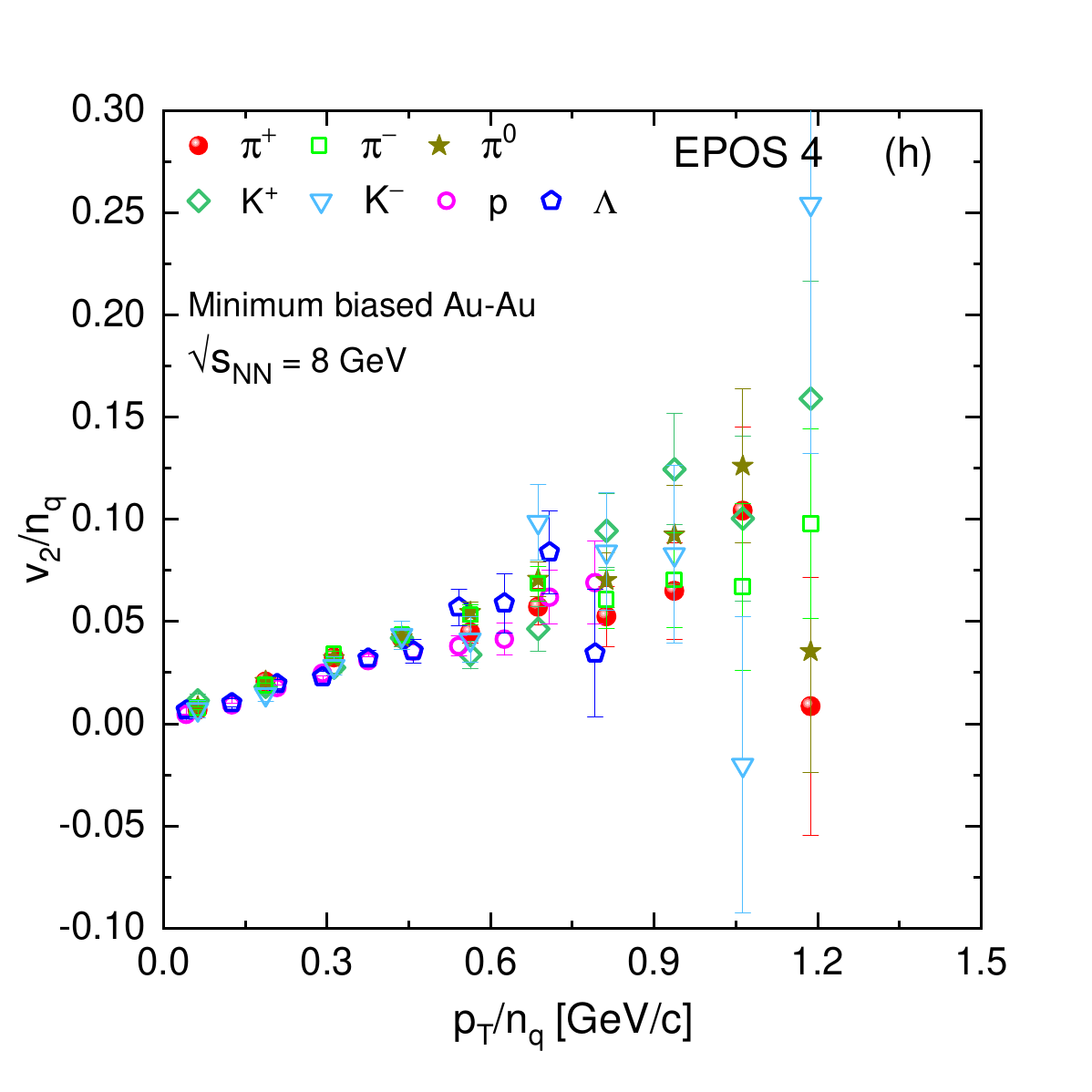} 
\includegraphics[width=0.32\textwidth]{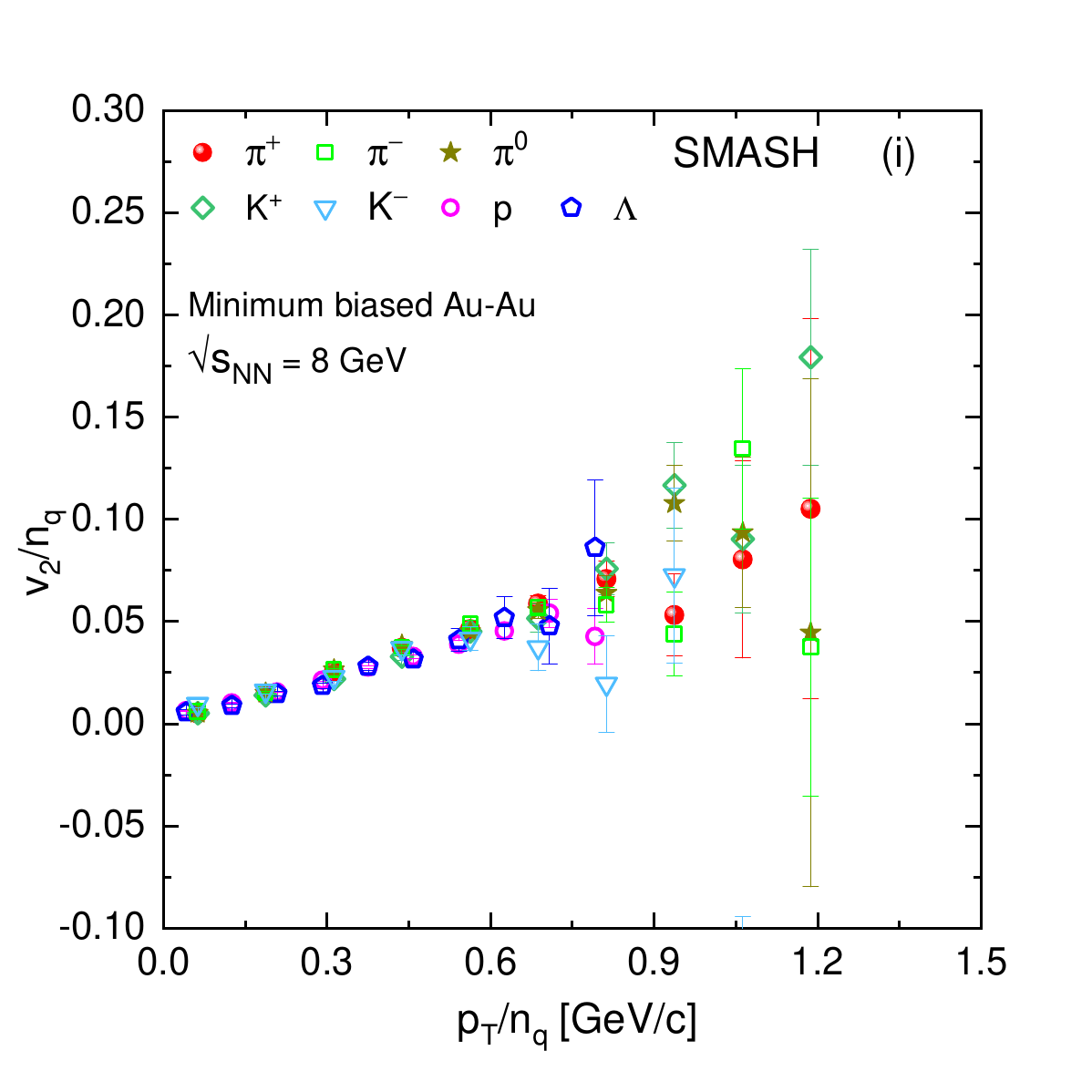}\vspace{-0.35cm}
\caption {The NCQ-scaled elliptic flow ($v_2/n_q$) as a function of NCQ-scaled $p_T$ ($p_T/n_q$). There are panels displaying model predictions arranged by energy and model. These NCQ-scale plots will investigate the presence of a quark number collapse in various model frameworks within the NICA energy range.}
\end{figure*}

Anisotropic flow coefficients were computed with an event-plane–type estimator using Q-vectors constructed from all final-state particles with $p_T$ weights; for each harmonic $n$, though we used only $n=2$ in the current manuscript, the event-plane angle was defined from the event flow vector (Q-vector) \cite{poskanzer1998methods}:
\begin{align}
Q_n = \sum_{j} w_j \, e^{i n \phi_j},
\end{align}
The sum goes over the $j$ particles used in the event plane determination, $w_j$ is the weight, defined as $w_j = p_{T,j}$, and $\phi$ is the azimuthal angle around the beam direction of the particle’s transverse momentum. Eq. (2) implies \\

~~~~~~~~~~~~~~~$Q_{n,x} = \sum_{j} w_j \cos(n\phi_j)$ and
\begin{align}
Q_{n,y} = \sum_{j} w_j \sin(n\phi_j),
\end{align}
The event-plane angle is then defined as
\begin{align}
\Psi_n = \frac{1}{n}\,\operatorname{atan2}\!\left(Q_{n,y},\,Q_{n,x}\right),
\end{align}
Differential flow is accumulated in $p_T$ bins as 
\begin{align}
v_n(p_T) = \left\langle \cos\!\left[n\left(\phi-\Psi_n\right)\right] \right\rangle_{p_T\ \mathrm{bin}},
\end{align}
averaged over particles in a single event/collision. No eta-gap, subevent method, or event-plane resolution correction was applied in the present implementation, and the same estimator was used for all models to ensure a consistent baseline comparison.

In Fig. 10, the elliptic flow divided by the number of constituent quarks (NCQ), $v_2/n_q$, versus the scaled transverse 
momentum, $p_T/n_q$, of $\pi^+, \pi^-, \pi^0, K^+, K^-, p$ and $\Lambda$ at $\sqrt{s_{NN}}=6, 7$ and $8~GeV$ has been displayed. These plots directly challenge the possibility that the flow patterns formed in the models can be scale-invariant with 
so-called quark-number scaling in RHIC and LHC, which has been regarded as scaled-up evidence 
of a partonic scale \cite{star2003particle, adler2003elliptic}. The most remarkable aspect of our findings is that, despite an obvious intuition, it is EPOS-LHC that shows the best scaling behavior to NCQ, and EPOS-4 and SMASH show partial scaling. This hierarchy offers a crucial understanding of the mechanism of the 
baryon-rich regime relevant to NICA through the dynamics of different models carried out in the 
early time and hadronization process. EPOS-LHC provides a smooth and 
increasing curve of $v_2/n_q$ vs $p_T/n_q$ on a panel of mesons, with $\pi$ and $K$ data points falling reasonably on the same trend. In this model, it means that the quark-level anisotropy is carried to mesons in a scaling of quark-number manner. Comparing EPOS-4 with 
EPOS-LHC, meson curves increase more gradually and do not coincide with each other, which 
indicates a disparity in the distribution of resonance feed-down and hydrodynamic-like expansion 
of the momentum. SMASH also represents the partial scale behavior of mesons because of the absence of partonic stage. The importance is that meson NCQ 
scaling has the first level of discrimination: only the models that have some effective partonic-like 
dynamics (EPOS-LHC) can produce this scaling property. Disparities are even 
greater in the baryon sector. The position of protons and $\Lambda$s in EPOS-LHC when scaled is far closer 
to the meson band, implying that the anisotropy per quark is the same whether the quarks become 
mesons or baryons. It is the hallmark of NCQ scaling that is characterized by this approximate 
collapse. Baryons in EPOS-4 are systematically lower than the meson curves, and the scaled values 
are smaller at intermediate $p_T/n_q$. This implies that the quark anisotropy in EPOS-4 is not strongly homogeneous, maybe because the viscous damping is stronger in the hydrodynamic stage or because hadronizing is more dissimilar and thus weaker than it would be 
with homogeneous distributions of anisotropy. In SMASH, the 
values of $v_2/n_q$ for baryons at intermediate $p_T/n_q$ are slightly suppressed compared to meson curve. This is in agreement with the fact that in an entirely hadronic transport image, both baryon halting and bradyonic rescattering are predominant, 
isotropizing baryon momentum distributions and disrupting quark-number scaling. The physical 
implication is obvious, baryons are the most stringent scaling test, and their behavior indicates that 
only EPOS-LHC scales NCQ collapse between mesons and baryons. The general trend is clear: EPOS-LHC produces the most promising NCQ scale, EPOS-4  and SMASH demonstrate the partial scaling with their evident deviations. This hierarchy is simply a direct reflection of the underlying physics; SMASH does not 
have a partonic phase or a coalescence mechanism; EPOS-4 does have a hydrodynamic-like core, 
but its physics results in a partial suppression of scaling; EPOS-LHC does have an effective 
implementation, which preserves quark-level anisotropy between hadron species. The importance 
of such an outcome is two-fold. To begin with, by showing that the NCQ scaling, which, since time 
immemorial, was considered the hallmark of quark-gluon plasma formation in RHIC, can be 
preserved even in the baryon-abounding part of $\sqrt{s_{NN}}=6-8~GeV$, but only in some 
dynamical images. Second, it gives a very discriminating observable to NICA: in case experiments 
show NCQ scaling at these energies, it would show frameworks such as EPOS-LHC and not favour 
purely hadronic transport. The novelty is that it is the first systematic prediction of NCQ scaling 
to these previously unexplored energies and so places the question of partonic collectivity in a new 
place where it has never been studied before. 
\subsection{Hadron ratios}
\begin{figure*}
\centering
\includegraphics[width=0.32\textwidth]{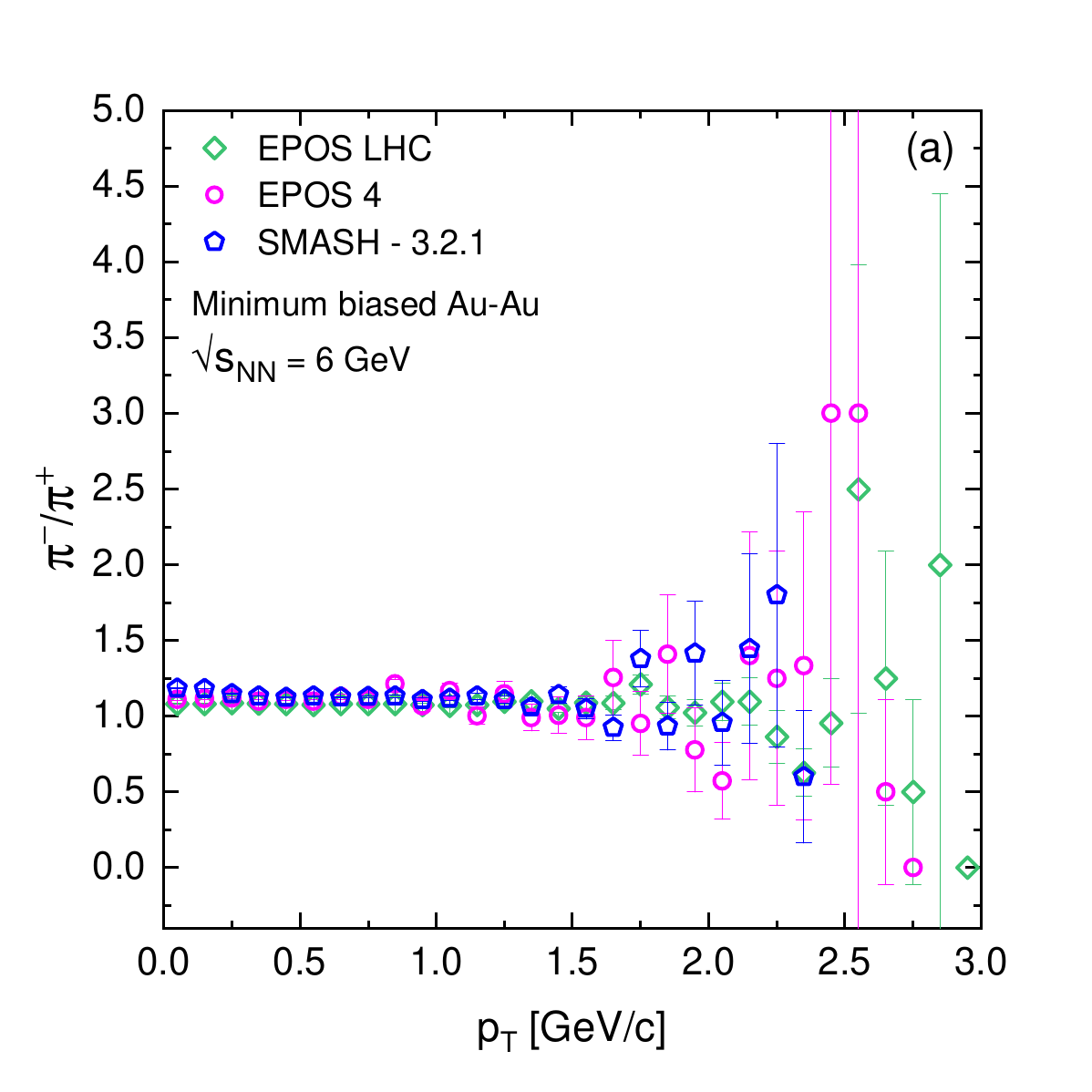}
\includegraphics[width=0.32\textwidth]{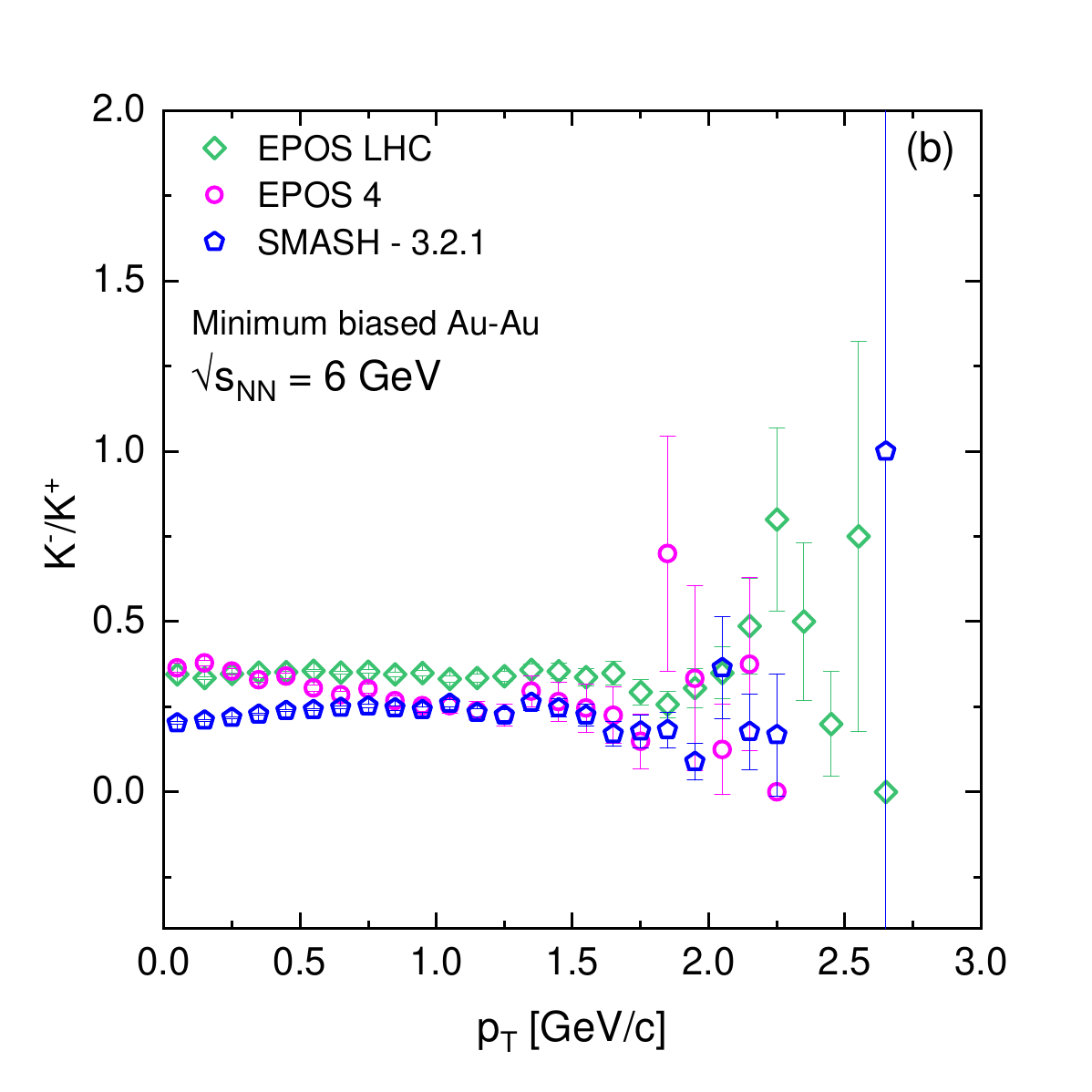}\vspace{-0.35cm} 
\includegraphics[width=0.32\textwidth]{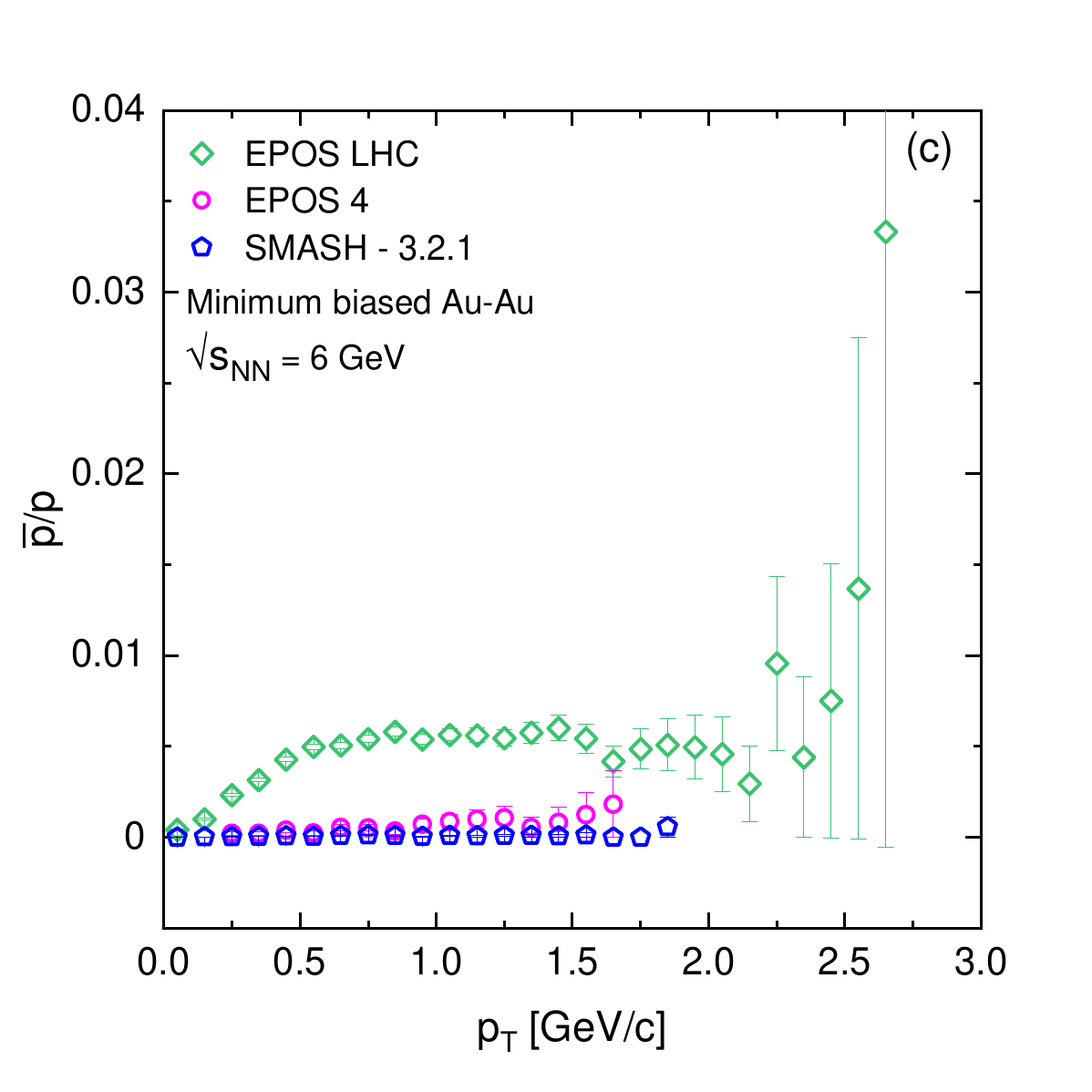}
\includegraphics[width=0.32\textwidth]{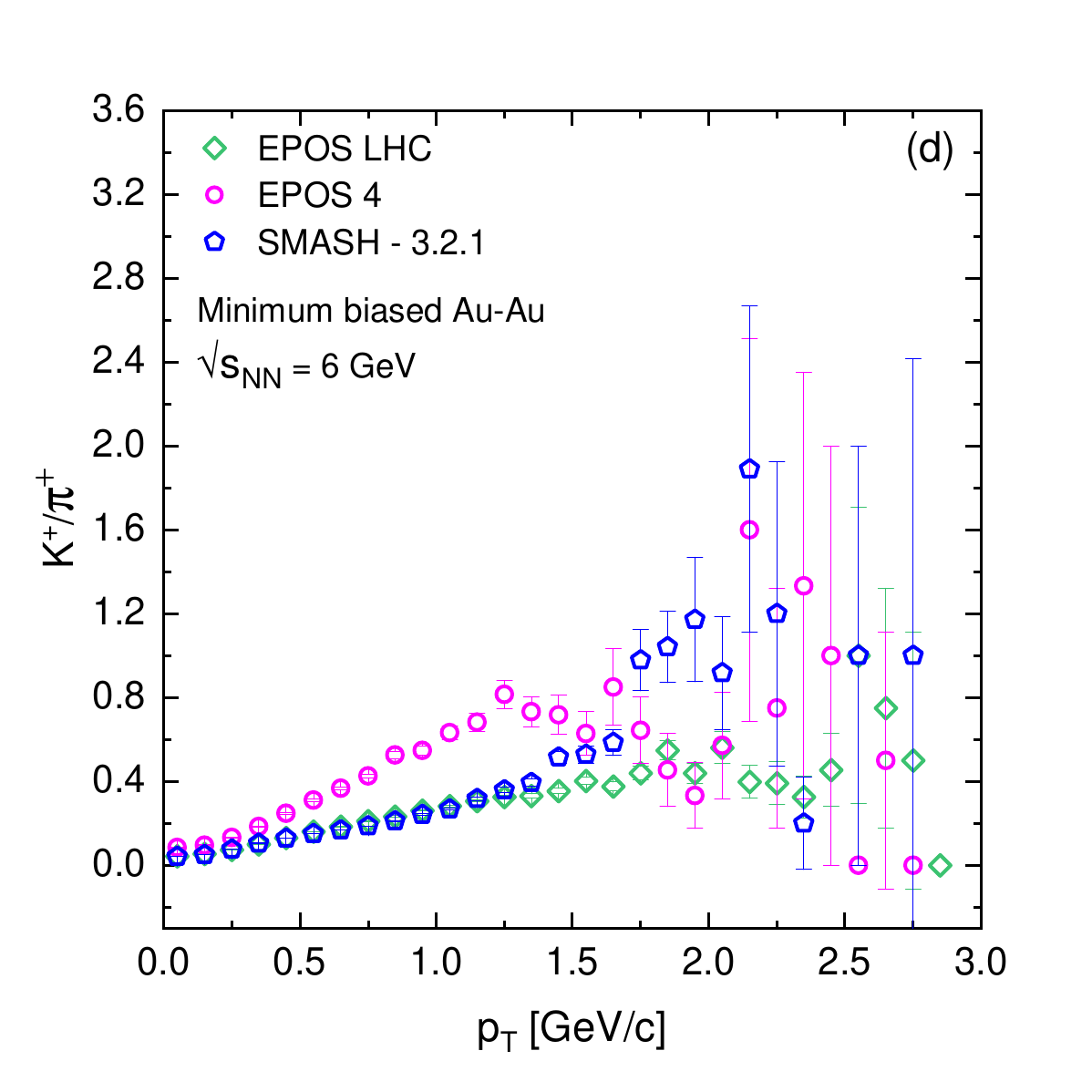}\vspace{-0.35cm}
\includegraphics[width=0.32\textwidth]{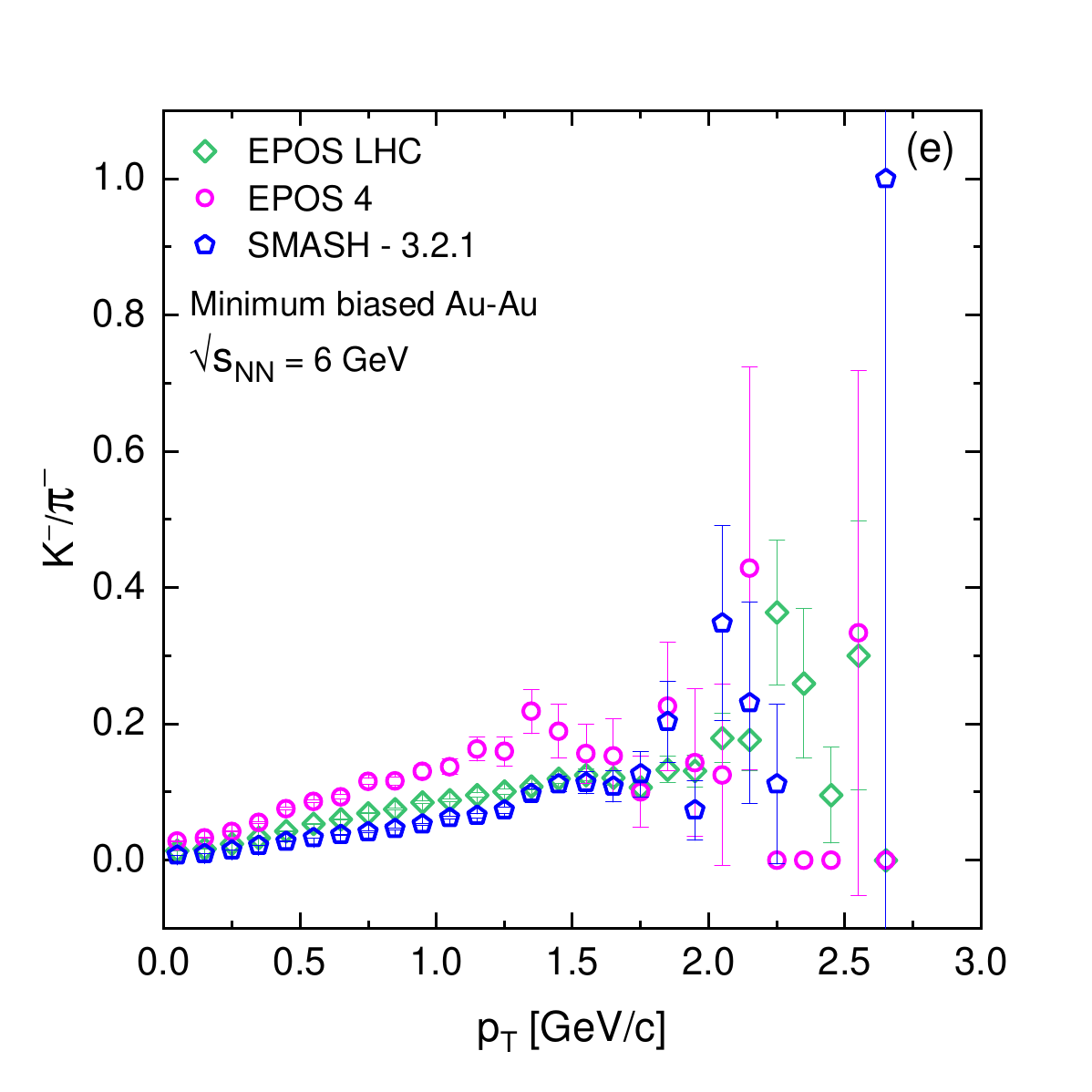}
\includegraphics[width=0.32\textwidth]{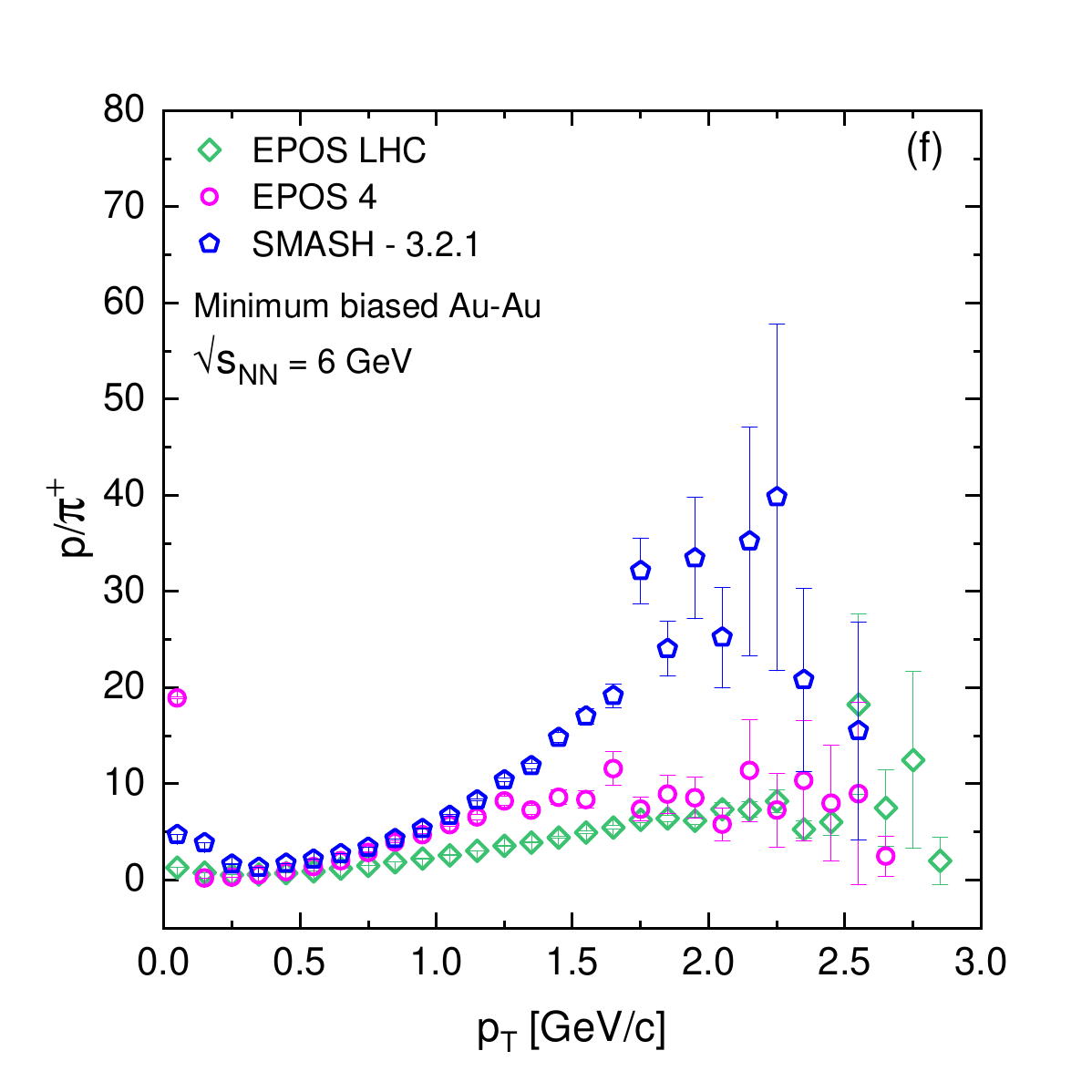}\vspace{-0.35cm}
\includegraphics[width=0.32\textwidth]{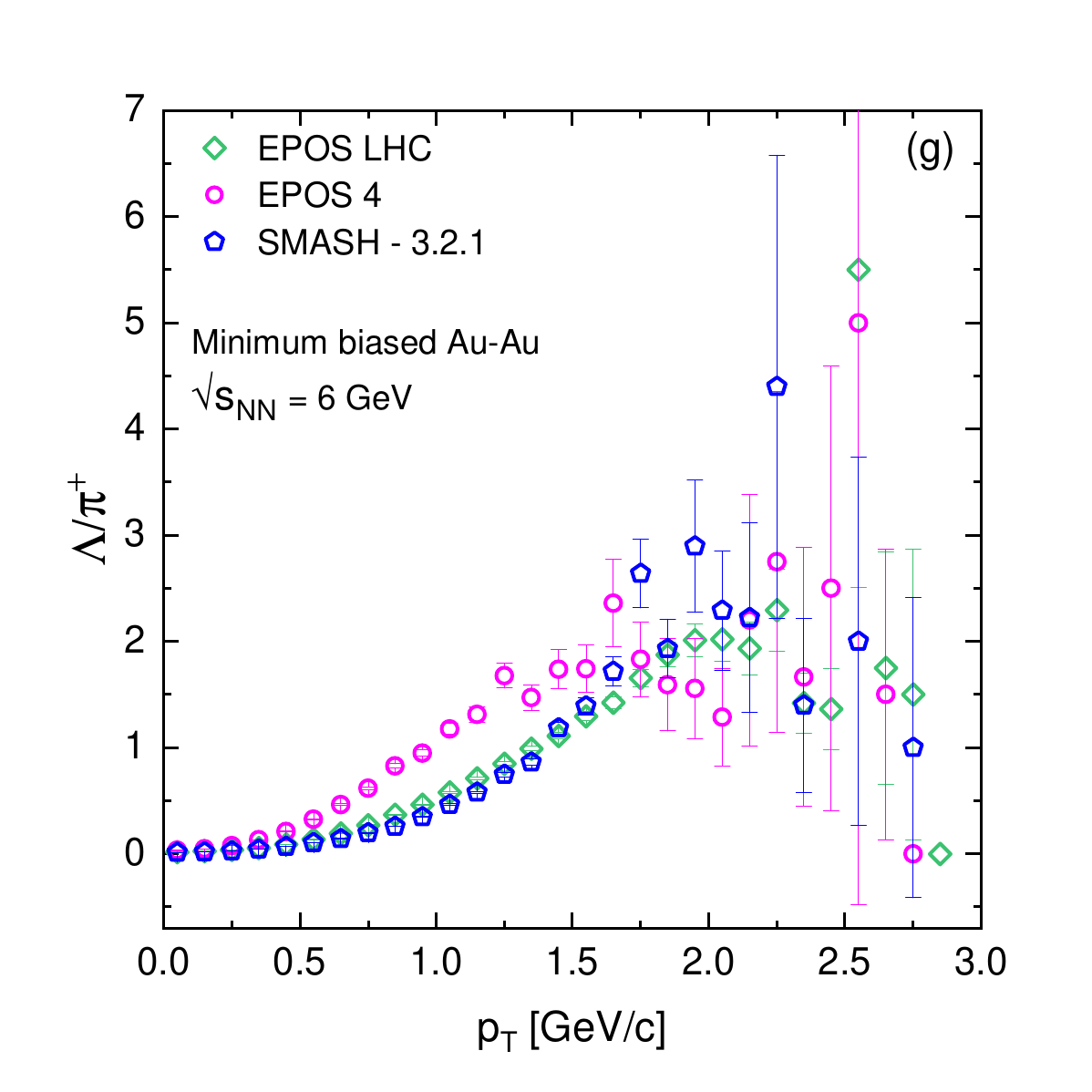}\vspace{-0.35cm}
\caption {Ratios of particle yields versus transverse momentum of minimum-bias $Au+Au$ collisions at $\sqrt{s_{NN}}=6~GeV$.}
\end{figure*}
\begin{figure*}
\centering
\includegraphics[width=0.32\textwidth]{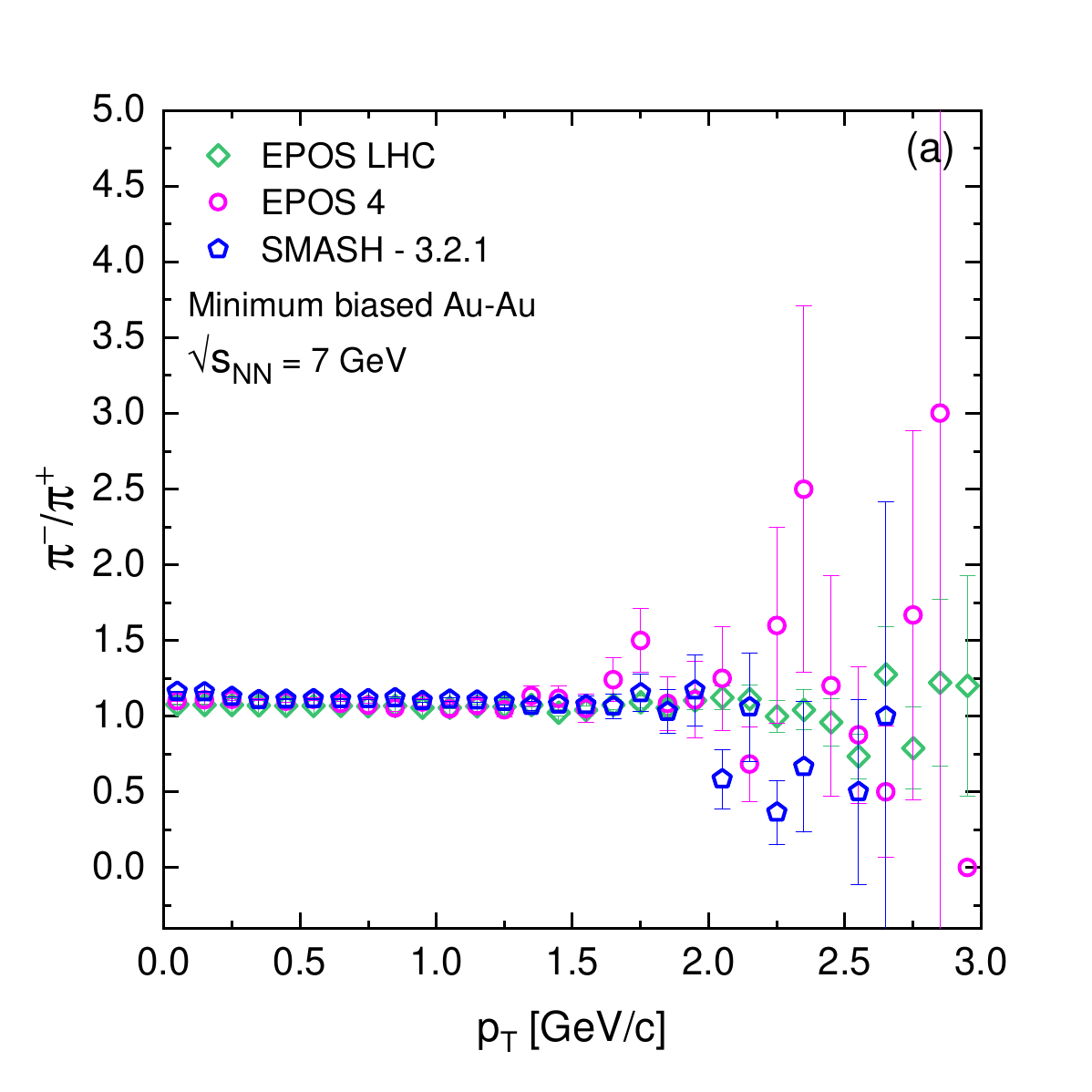}
\includegraphics[width=0.32\textwidth]{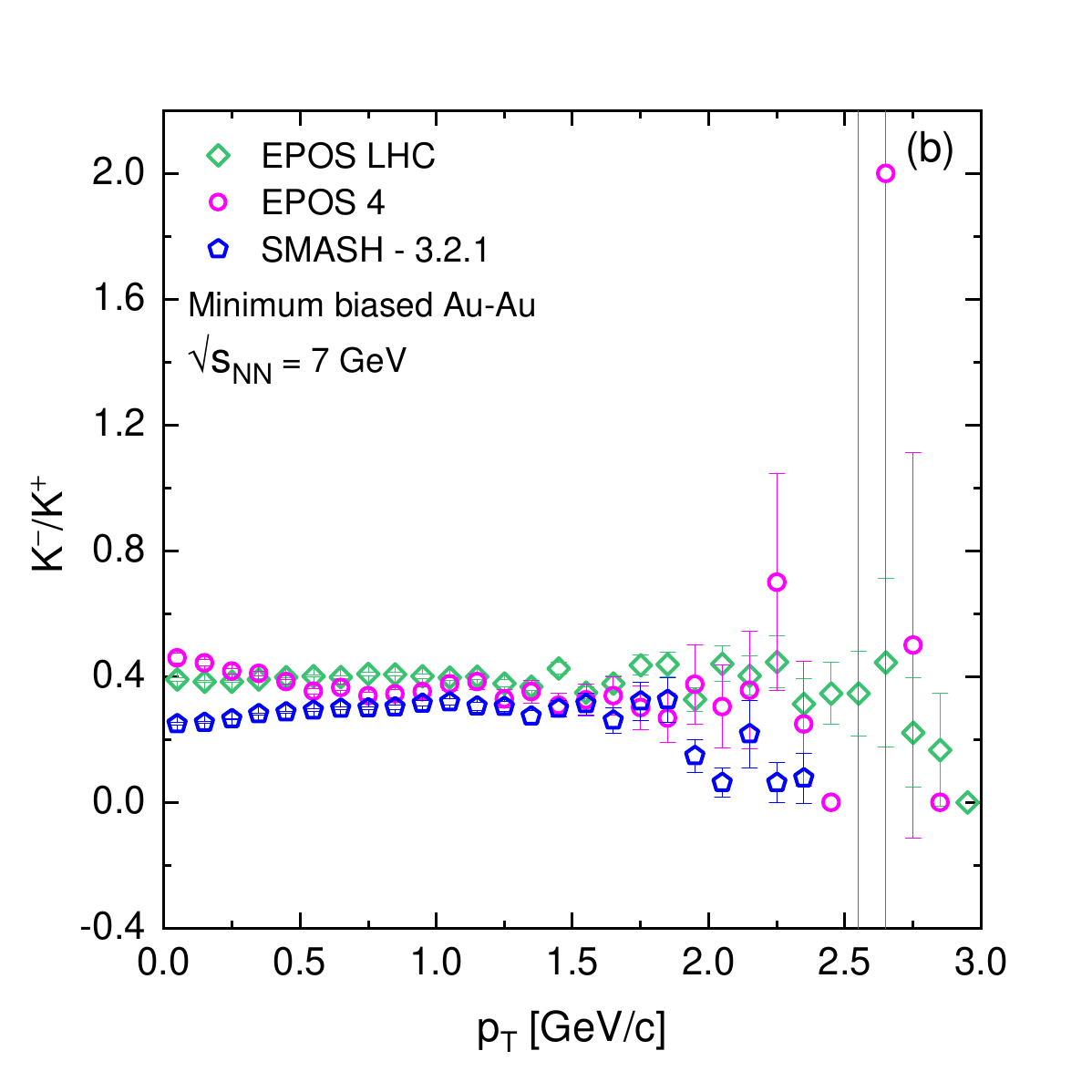}\vspace{-0.35cm} 
\includegraphics[width=0.32\textwidth]{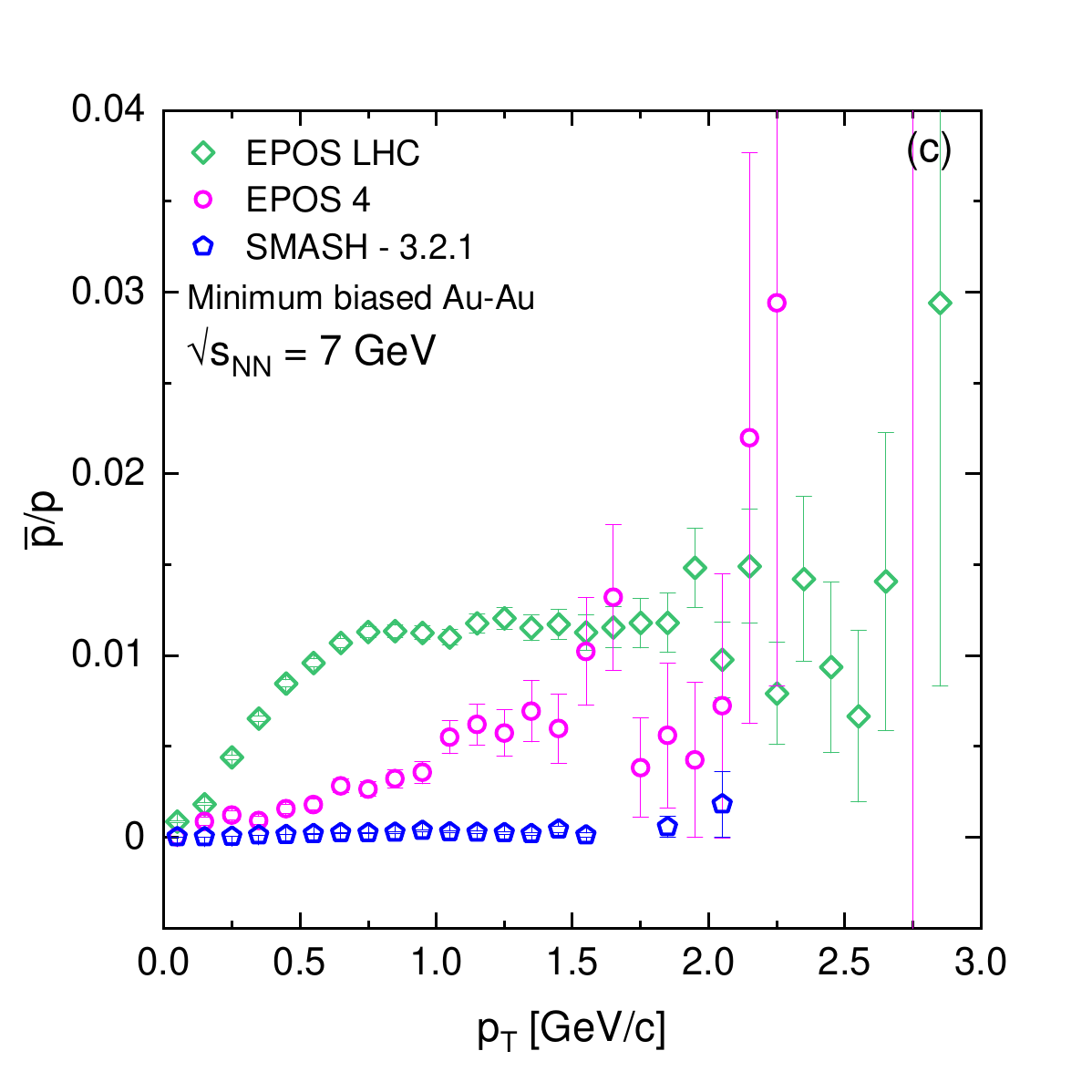}
\includegraphics[width=0.32\textwidth]{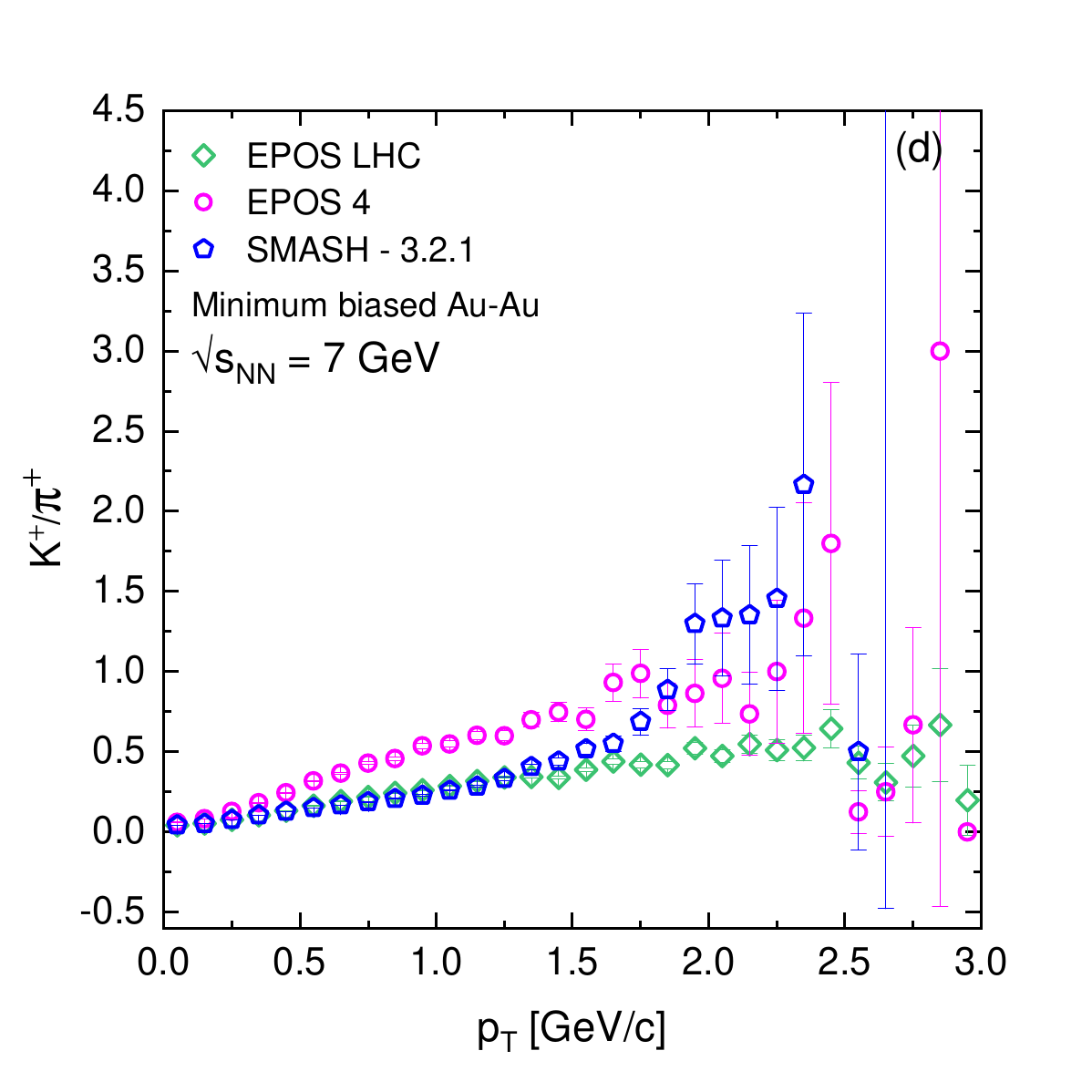}\vspace{-0.35cm}
\includegraphics[width=0.32\textwidth]{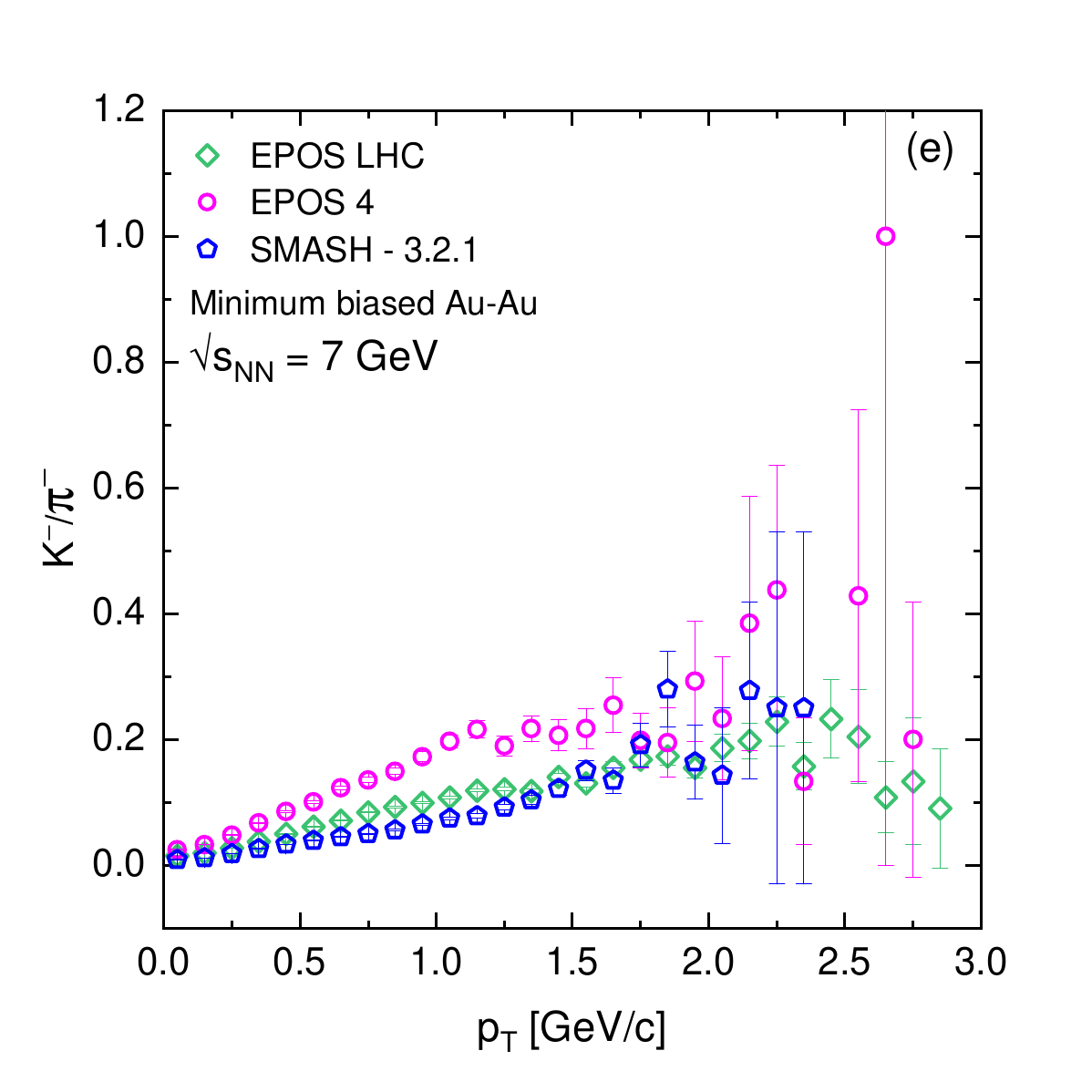}
\includegraphics[width=0.32\textwidth]{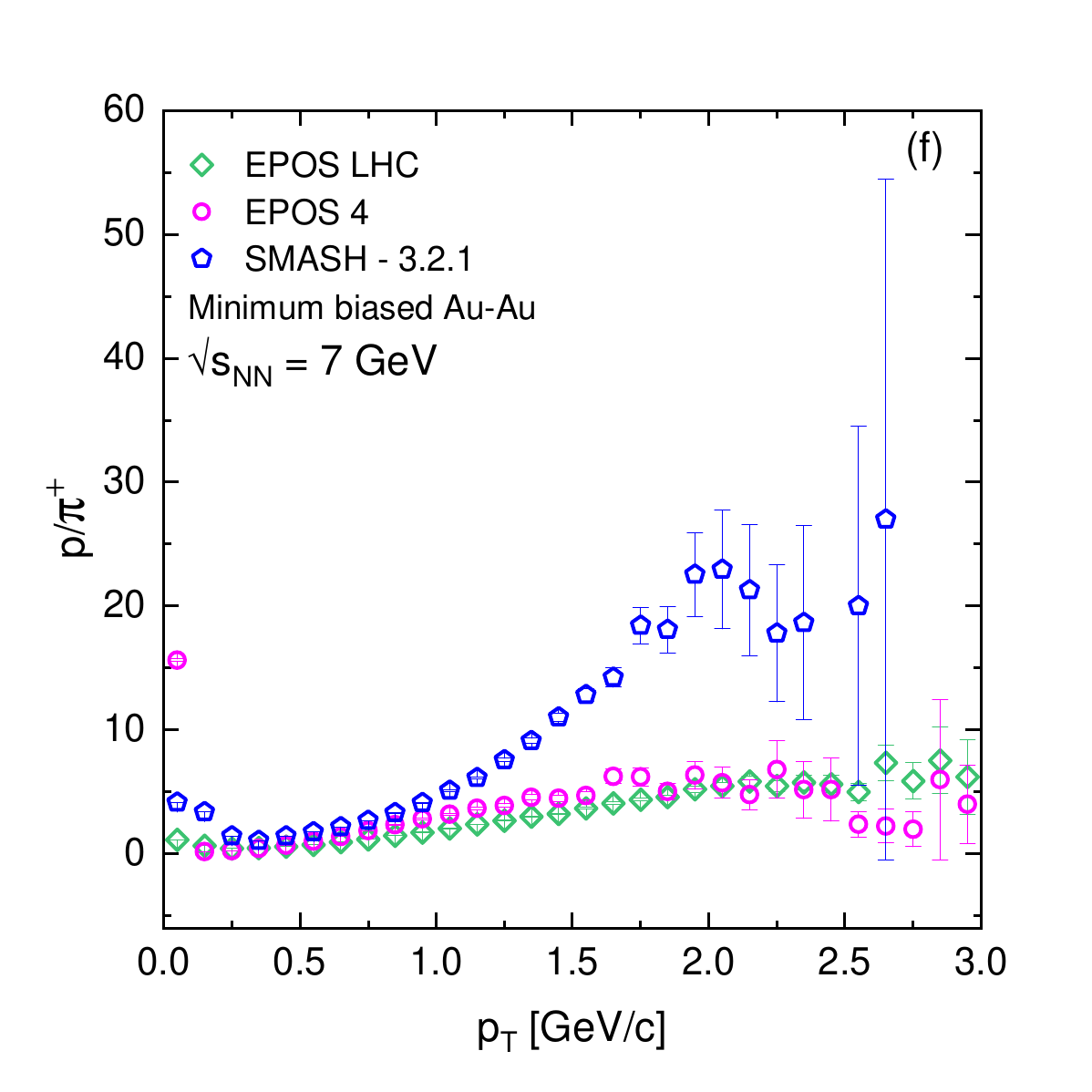}\vspace{-0.35cm}
\includegraphics[width=0.32\textwidth]{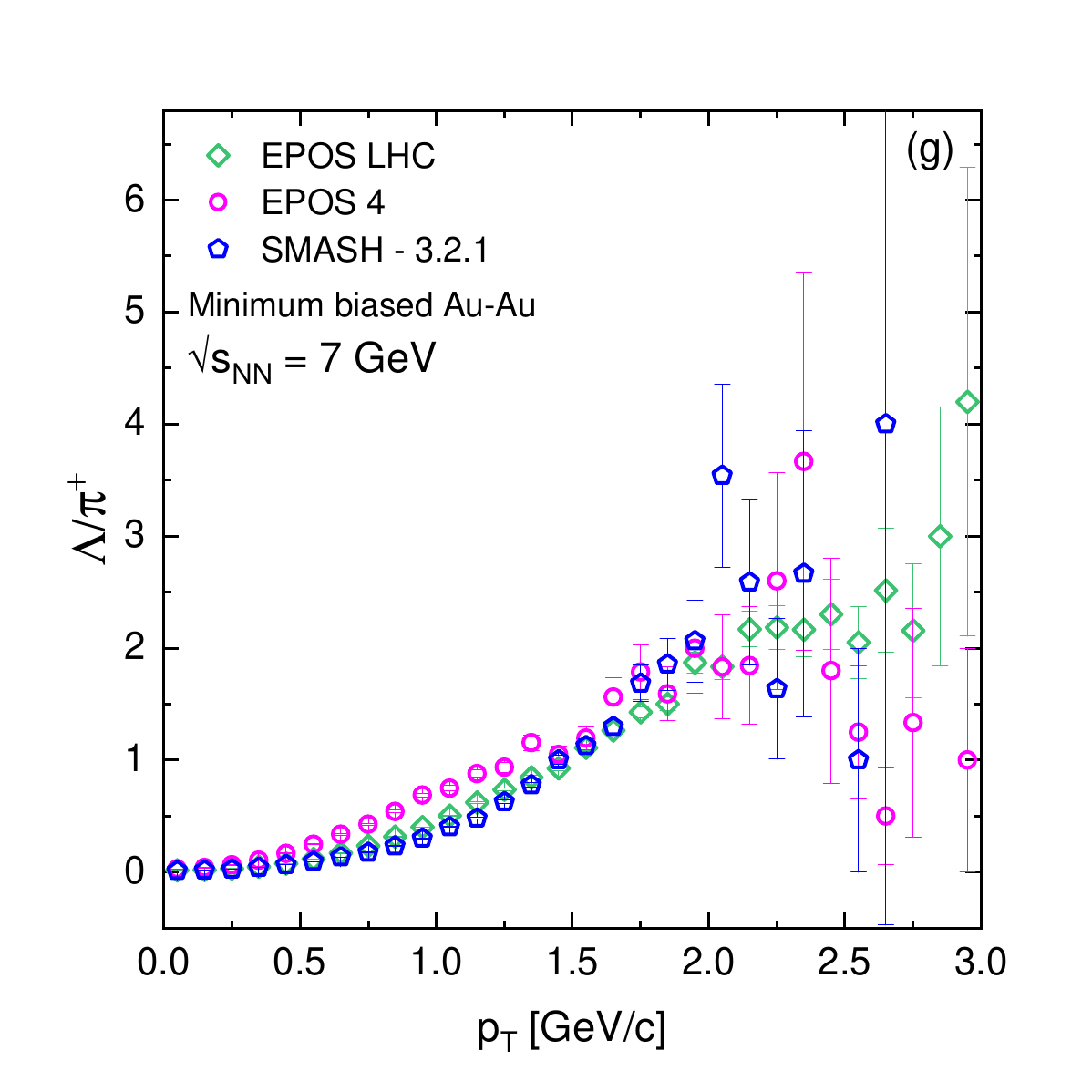}\vspace{-0.35cm}
\caption {Ratios of particle yields versus transverse momentum of minimum-bias $Au+Au$ collisions at $\sqrt{s_{NN}}=7~GeV$.}
\end{figure*}
\begin{figure*}
\centering
\includegraphics[width=0.32\textwidth]{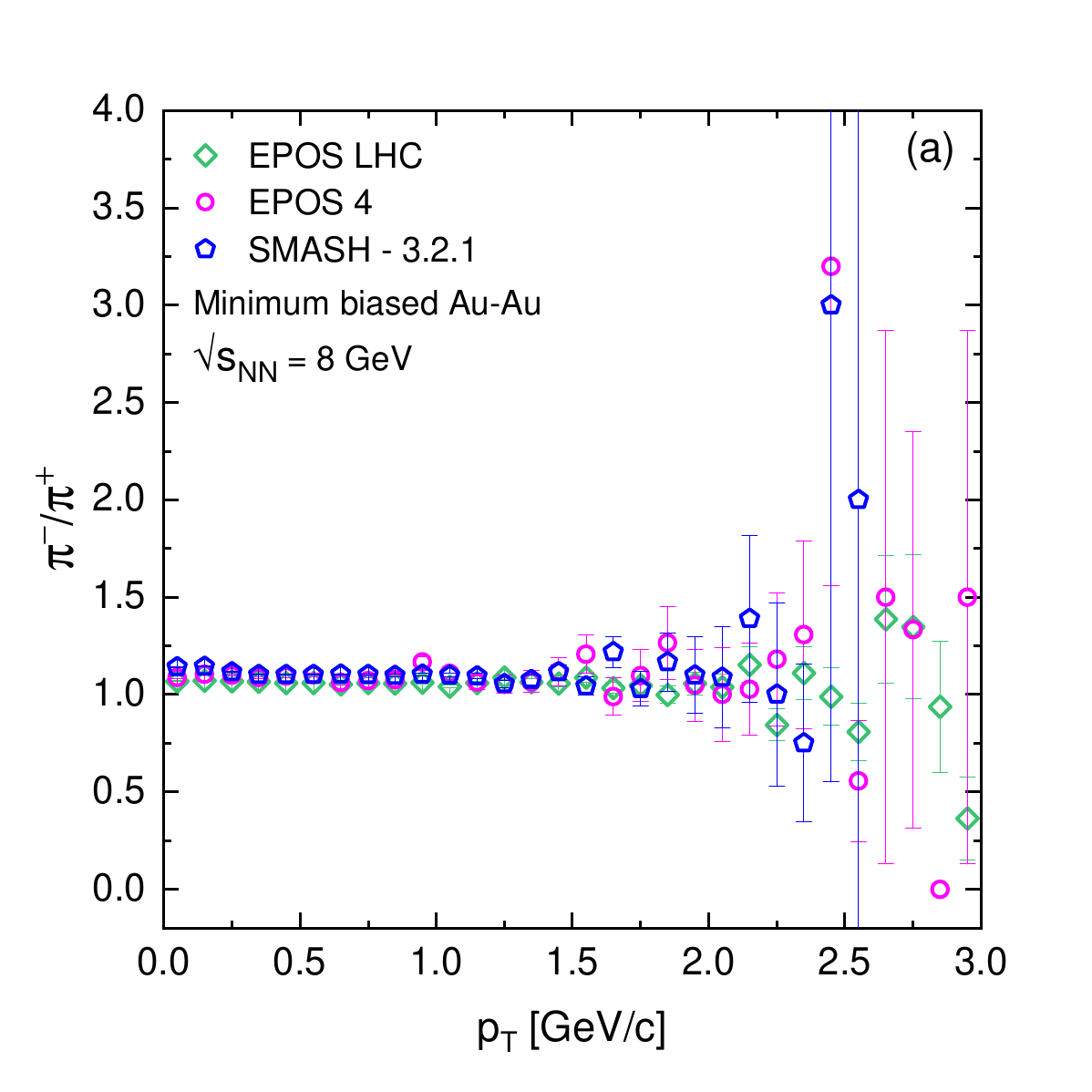}
\includegraphics[width=0.32\textwidth]{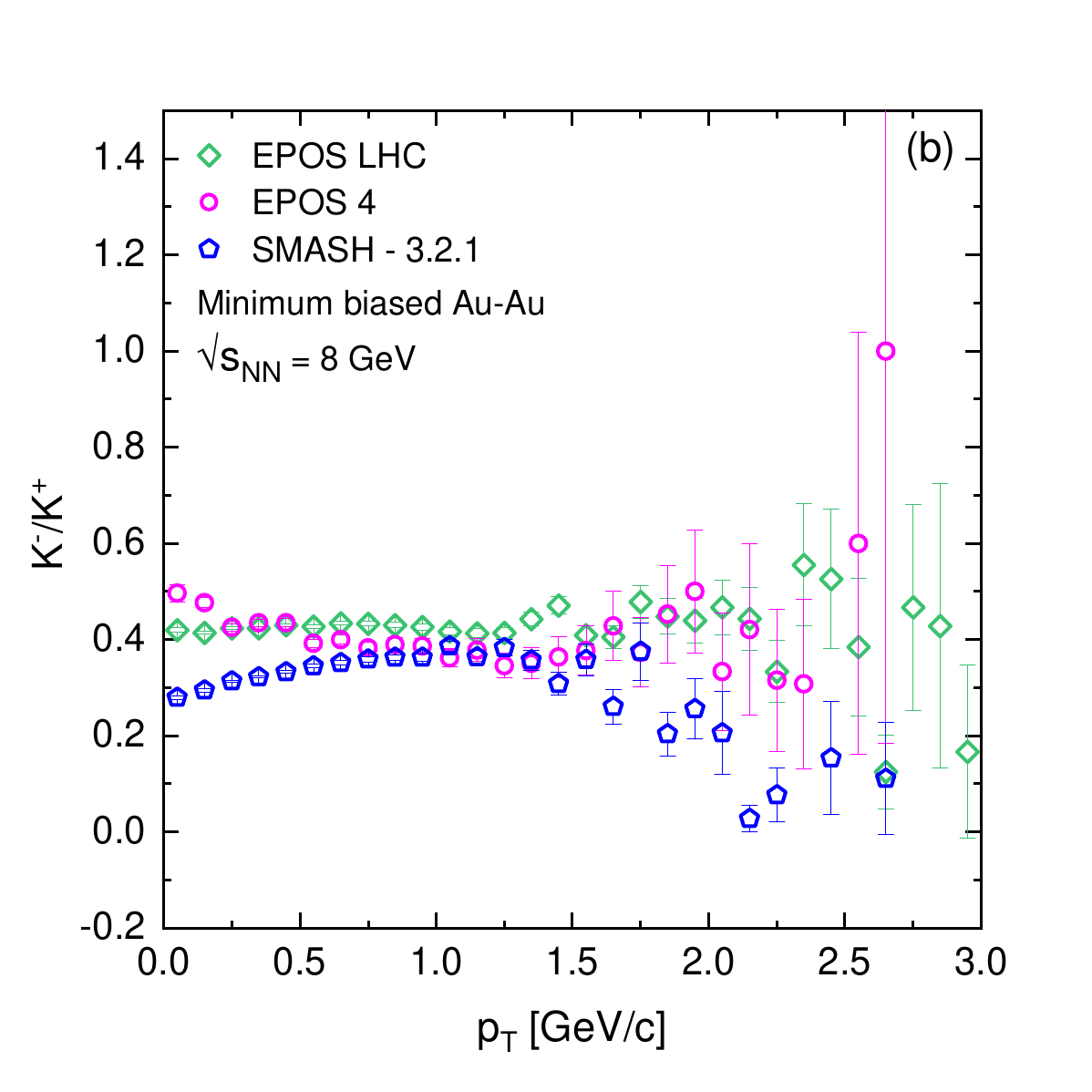}\vspace{-0.35cm} 
\includegraphics[width=0.32\textwidth]{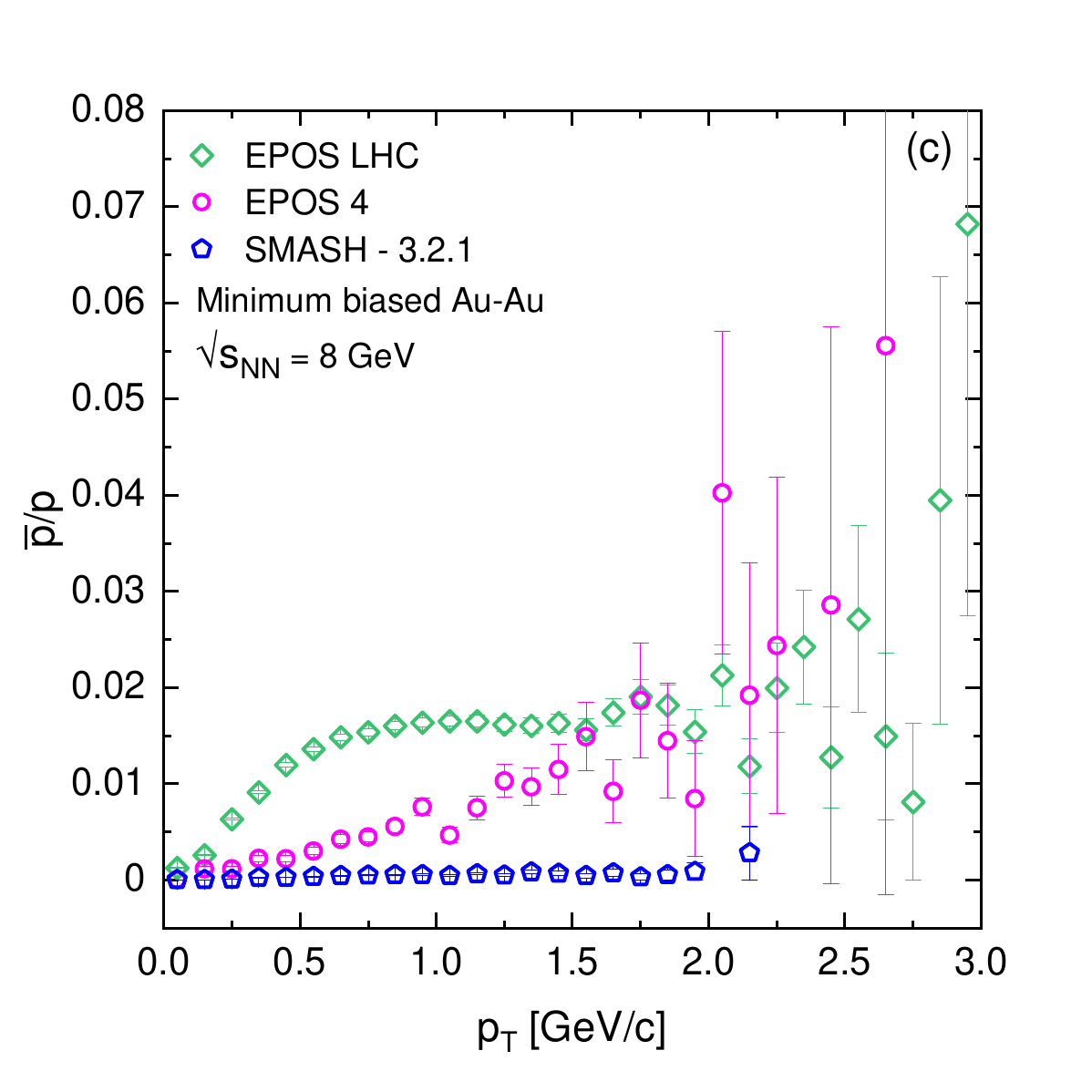}
\includegraphics[width=0.32\textwidth]{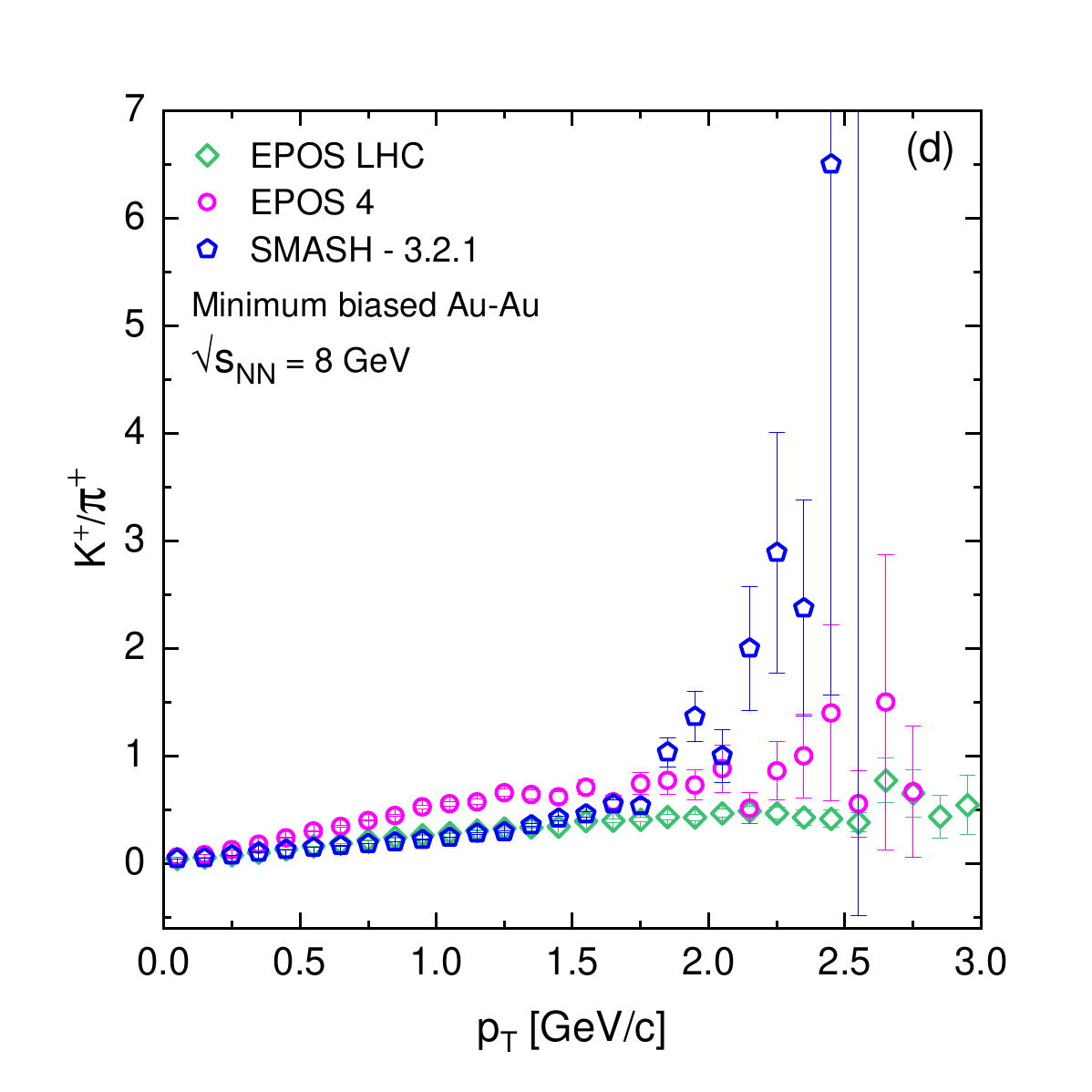}\vspace{-0.35cm}
\includegraphics[width=0.32\textwidth]{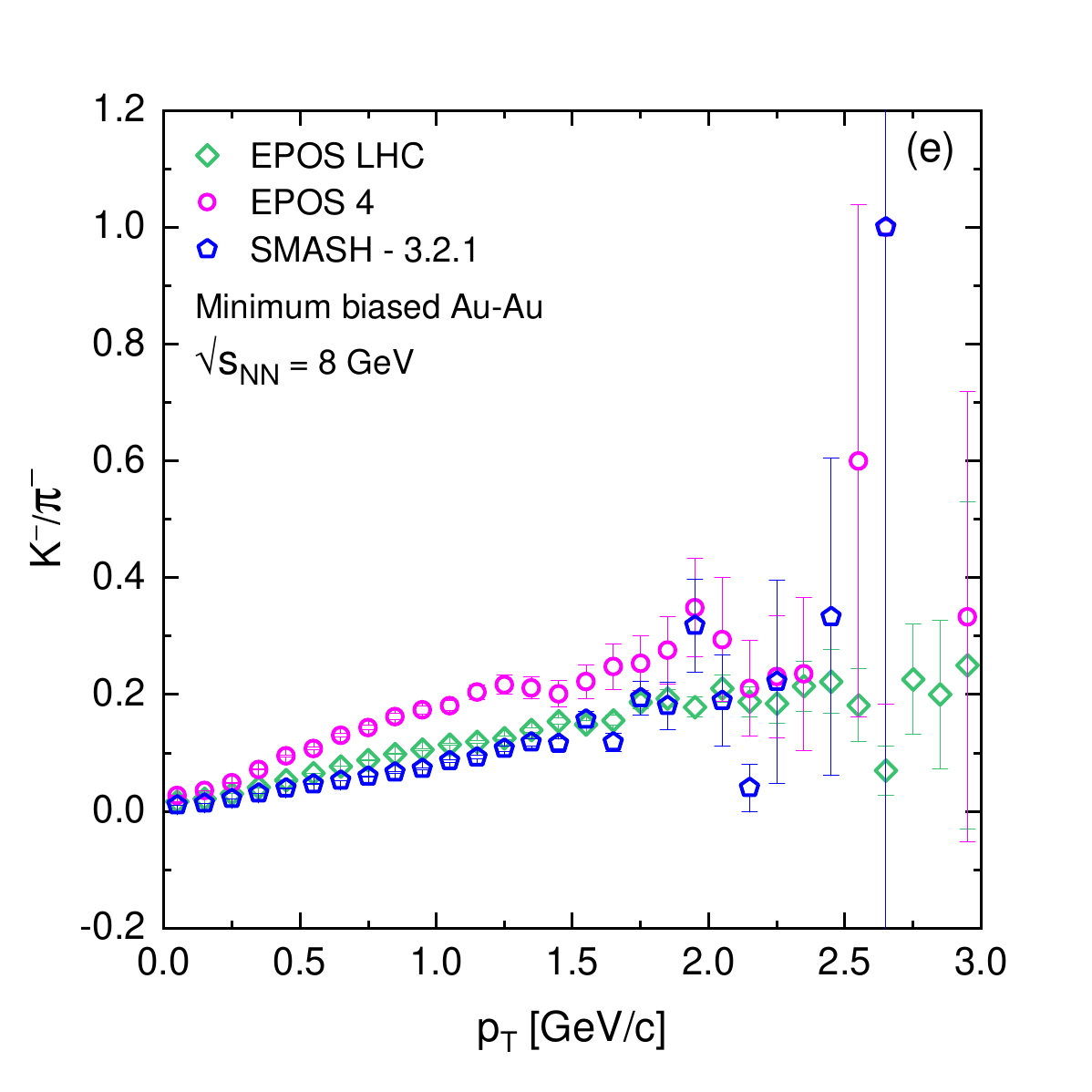}
\includegraphics[width=0.32\textwidth]{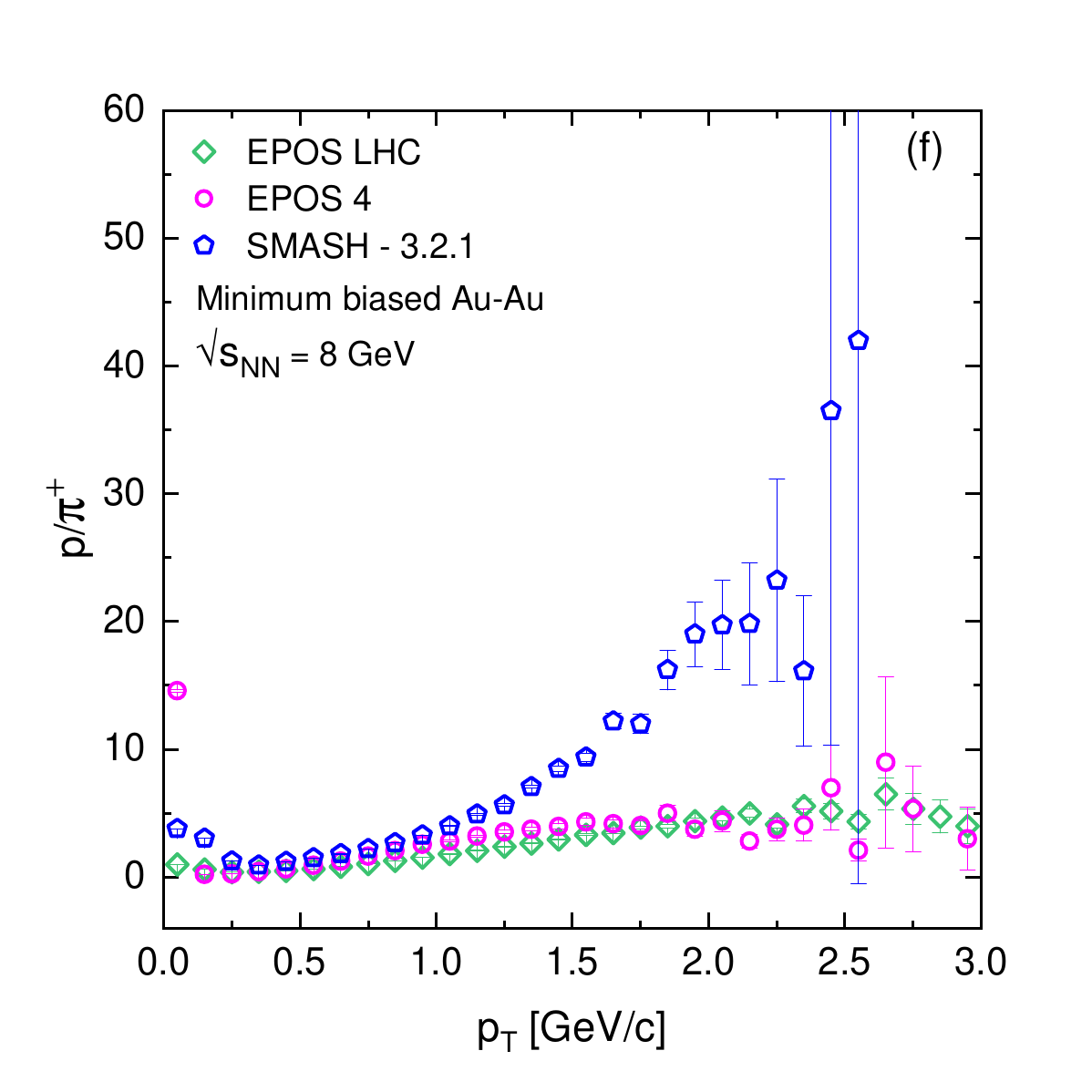}\vspace{-0.35cm}
\includegraphics[width=0.32\textwidth]{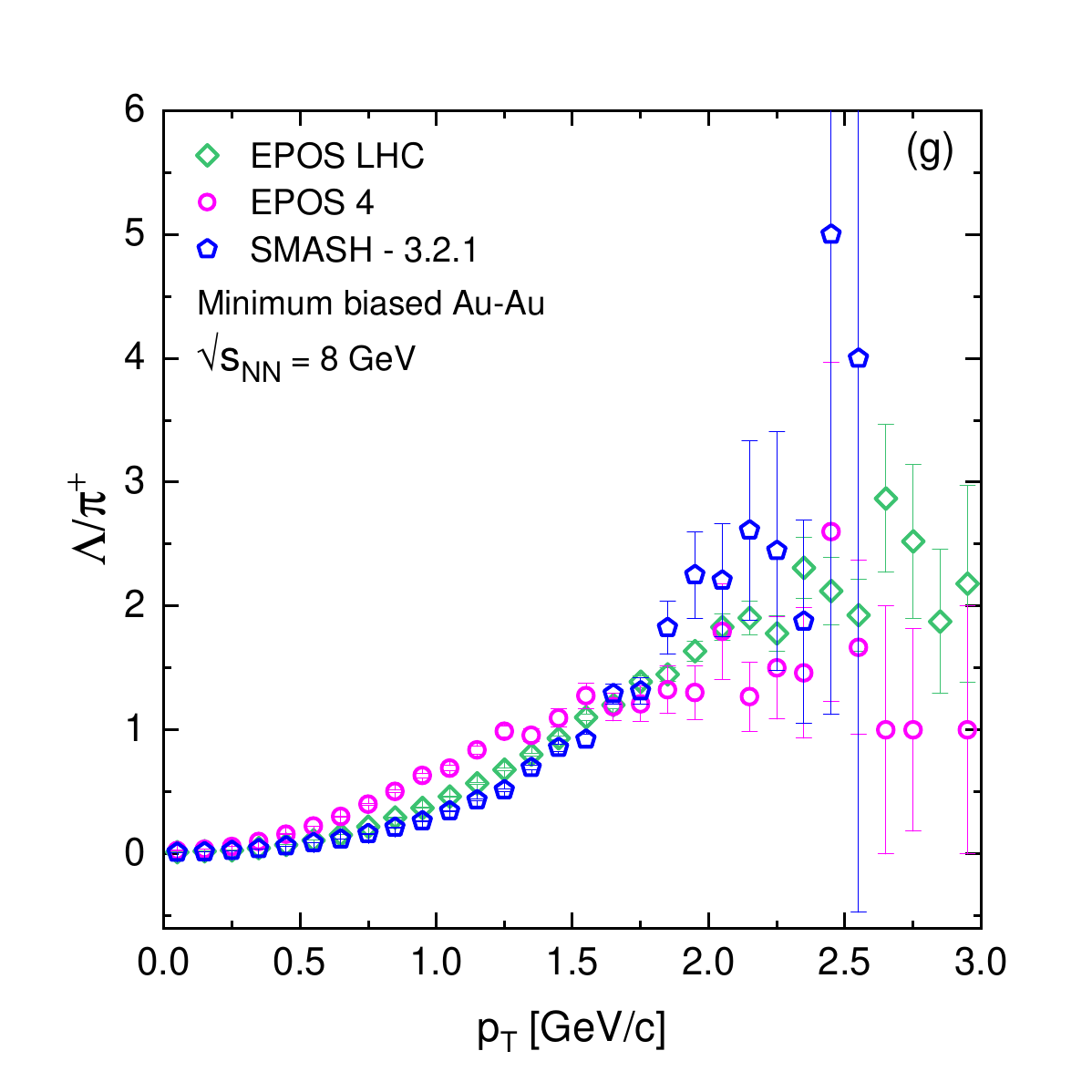}\vspace{-0.35cm}
\caption {Ratios of particle yields versus transverse momentum of minimum-bias $Au+Au$ collisions at $\sqrt{s_{NN}}=8~GeV$.}
\end{figure*}

The $\pi^-/\pi^+$ ratio as a function of $p_T$, shown in panels (a) of Figs. 11-13, is near unity at all energies 
investigated and in all three models. This is due to the stability present in the production of 
pions since the resonance decays are the ones that maintain the isospin balance. The neutron 
overpopulation in Au +Au collisions would theoretically cause a slight increase in $\pi^-$ over $\pi^+$, 
but this is insignificant and diluted in the central rapidity region. This approximate symmetry is 
replicated in both SMASH and EPOS frameworks, which is evidence that isospin conservation is 
carefully observed in both. The physical explanation of the absence of model dependence is that 
the resonance spectrum coincides in all frameworks, and its effect on pion yields is overwhelming 
compared to the other microscopic variations. Physically, such a ratio is a cross-check, but not a 
diagnostic, as it checks the simplest aspects of treating isospin, but not the inner workings. 
Nevertheless, its importance should not be under-emphasized: a constant $\pi^-/\pi^+$ ratio will give 
one confidence that any deviation that is observed in other more complex ratios (involving 
strangeness or baryons) is actually due to dynamical physics, and not merely due to trivial isospin 
effects. The novelty in this case is that this stability in the hitherto-unexplored regime of $6-8~GeV$ 
is explicitly confirmed, and thus can be used as a firm foundation in the interpretation of the 
more sensitive ratios.

The $K^-/K^+$ ratio (panels (b) of Figs. 11-13) is heavily suppressed (by far less than unity) throughout $p_T$ at $6~GeV$ in all models, due to the relative dominance of associated $K^+$ production ($NN\xrightarrow{}N\Lambda K^+$) and the energetic price of producing $K^-$ through $s\bar{s}$ pair production. At $7$ and $8~GeV$, this ratio generally rises, and more particularly at intermediate $p_T$, but it stays typically below $1$ in all models. EPOS-LHC and EPOS-4 forecast a more systematic increase in $K^-/K^+$ at intermediate $p_T$ than SMASH - the EPOS codes put more yield in the hard kaon tail since string fragmentation and any core component inject more transverse momentum to strange mesons. As a hadronic transport, SMASH exhibits greater low-$p_T$ suppression of $K^-$ and a steeper increase with $p_T$ where it is created. Physically, variations are monitored to assess the pair-production efficiency and in-medium $K^-$ absorption. EPOS is effective in enhancing early pair production (and decreasing absorption) compared to SMASH. In the case of NICA, the direct measurement of $ K^-$/$K^+$ against $p_T$ will directly limit strangeness pair production and absorption models in a baryon-rich medium.

The ratio $\bar{p}/p$, as presented in panels (c) of Figs. 11-13 for $\sqrt{s_{NN}} = 6, 7$ and $8~GeV$, is very small in all cases and exhibits an obvious model and energy dependence. At $6~GeV$ the ratio is almost zero in SMASH and EPOS-4 in the entire $p_T$ range, and EPOS-LHC only gives a very small value, slowly increasing with $p_T$ to at most a few percent. This corresponds to the strong suppression of the anti-proton production in the Baryon-Rich environment at the lowest energy. At $7~GeV$, both the EPOS variants now produce a visible, but still small, component of anti-protons. The $\bar{p}/p$ ratio from EPOS-LHC and EPOS-4 reaches a few percent and shows a slight increase with $p_T$, while SMASH is compatible with almost zero in the entire $p_T$ range. At $8~GeV$, the trend is fortunately continued: EPOS-LHC and especially EPOS-4 develop somewhat larger ratios of $\bar{p}/p$, again slowly growing with $p_T$, but with the ratios well below ten percent also in the highest $p_T$ bins. SMASH continues to predict almost no anti-protons for any $p_T$.

Overall, these plots show that all three models are consistent in showing a very strong suppression of anti-protons at NICA energies, consistent with the expectations of a high net-baryon density and large annihilation probability. EPOS-LHC and EPOS-4 are the ones that systematically yield slightly more anti-protons than SMASH, especially at higher $p_T$ and higher $\sqrt{s_{NN}}$, while SMASH seems to yield only protons. Any non-negligible $\bar{p}/p$ signal in data in the $6-8~GeV$ region would therefore provide a strong constraint on the way these models treat baryon-antibaryon creation and annihilation.

$K^+/\pi^+$ ratio (panels (d) of Figs. 11-13) increases with $p_T$ at all energies; the increase in ratio is more substantial with an increase in collision energy between $6$ and $8~GeV$, consistent with the Refs. \cite{alt2008pion, alice2013centrality}. EPOS-4 systematically yields higher values of $K^+/\pi^+$ at intermediate $ p_T$ ($0.8-3~GeV/c$) than SMASH. On the physical side, the $p_T$ increase is a reflection of the fact that heavier particles are kinematically preferred at higher transverse momenta, and that the production of kaons at the expense of pions in the intermediate band should be enhanced by collective radial push (or harder string fragments). The energy evolution (increase between $6 - 8~GeV$) is as expected since the greater the energy available, the higher the yield of strangeness and the more semi-hard processes are allowed. Model differences therefore encode differences in the efficiency in producing and accelerating strange mesons: EPOS models convert more longitudinal energy to transverse momentum and/or make more abundant strangeness at earlier times than SMASH. Practically, the $K^+/\pi^+$ vs $p_T$ is a sensitive experimental tool to trace the origin of the transverse collective behavior and to restrict the mechanism of strangeness production in the baryon-rich regime.

The qualitative variations of $K^-/\pi^-$ with $p_T$ (panels (e) of Figs. 11-13) are the same as those of $K^+/\pi^+$ but with smaller absolute values since the $K^-$ are less common. The ratio is small at $6~GeV$ and rises gradually with $p_T$; at $7$ and $8~ GeV$ the enhancement of the intermediate-$p_T$ is more apparent, particularly with EPOS-LHC and EPOS-4. The models are quantitatively different: both EPOS frameworks predict significantly larger $K^-/\pi^-$ at intermediate $p_T$, due to more pair production and less in-medium absorption of $K^-$ than SMASH. SMASH forecasts weaker $K^-$ spectra and hence weaker $K^-/\pi^-$ with respect to $p_T$. This ratio isolates the physics of pair production and absorption, and is thus complementary to the $K^-/K^+$ ratio. It is specifically valuable for testing the physics of $K^-$ production and survival in baryon-rich matter.

The behaviour of $p/\pi^+$ ratio as a function of $p_T$, as given in panels (f) of Figs. 11-13 for $\sqrt{s_{NN}} = 6-8~GeV$, is strongly model dependent. At all three collision energies, the ratio increases with $p_T$ in all three models, but the increase is much stronger in SMASH than in the EPOS versions. In the very first $p_T$ bin (below about $0.1~GeV/c$), EPOS-4 shows a large enhancement of $p/\pi^+$, which is indicative of a large number of very soft protons, while SMASH and EPOS-LHC give smaller values in this lowest bin. For $p_T$ at and above $0.15~GeV/c$, however, the values of SMASH are systematically larger than those of EPOS-LHC and EPOS-4 and increase up to $p_T$ of around $2.5-3.0~GeV/c$. The EPOS curves increase only moderately over the same range and do not exceed the SMASH. This pattern is observed consistently at $6$, $7$, and $8~GeV$. Overall, the $p/\pi^+$ ratio is thus a very sensitive observable to differentiate hadronic transport dynamics of SMASH, producing relatively more protons in the high $p_T$ tail, from hybrid/string dynamics of EPOS-LHC and EPOS-4, producing enhanced very soft protons with a reduced baryon fraction at higher transverse momentum.

This trend is also consistent in all three energies, which confirms that the $p/\pi^+$ ratio is an especially sensitive observable to discriminate between hadronic and hybrid/string-based dynamics in the NICA energy domain. In all three models, the ratio increases with $p_T$, but the strength of this increase is very different. SMASH develops a much steeper enhancement of $p/\pi^+$ towards intermediate and high $p_T$ than EPOS-LHC and EPOS-4, while EPOS-4 shows a distinctive excess of very soft protons which is confined to the lowest $p_T$ bin. As a result, the features of the detailed $p_T$ dependence of $p/\pi^+$ in NICA measurements are expected to differentiate between two scenarios: a very steep rise of $p/\pi^+$ already from $p_T$ of about $0.15~GeV/c$ upwards would be compatible with SMASH-like hadronic transport dynamics, while a more moderate rise and an enhanced very low $p_T$ baryon component would be more compatible with the EPOS one. The $p/\pi^+$ ratio at NICA energies can therefore give important information on whether the effective dynamics in the region of low-energy and high-baryon-density is closer to a purely hadronic transport picture or closer to a hybrid/string scenario with an additional early stage component.

In Figs.~4-6(f), the proton transverse-momentum spectra are presented for minimum bias events without detector-motivated acceptance cuts in rapidity or pseudorapidity. At these low collision energies, fully inclusive spectra of this measure can find a huge contribution in the lowest $p_T$ bins from baryon transport (and, depending on the generator record, also from projectile/target remnant-like nucleons) that naturally contributes to the enhancement of the near-zero $p_T$ proton population relative to produced mesons. This is the same low-$p_T$ mechanism that leads to the enhancement of the $p/\pi^+$ ratio in the very low-$p_T$ region in the EPOS-4 calculations.

The ratio $\Lambda/\pi^+$ as a function of $p_T$ is displayed in panels (g) of Figs. 11-13, which show that in the intermediate window of transverse momentum, the EPOS-4 calculations give a systematically bigger Lambda-to-pion ratio than EPOS-LHC and SMASH. At the same time, EPOS-LHC and SMASH are close to each other (EPOS-LHC slightly bigger than SMASH in most bins). This behaviour can be observed for all three collision energies investigated ($\sqrt{s_{NN}} = 6, 7$ and $8~GeV$). The $\Lambda/\pi^+$ ratio is sensitive to competition between mechanisms of baryon production and meson production: At low $p_T$ the ratio is dominated by thermal production and resonance feed-down, while at intermediate $p_T$ the ratio is sensitive in particular to the dynamics of hadronization (recombination/coalescence vs. string fragmentation) and the size of the radial collective push experienced by heavier hadrons. The larger intermediate-$p_T$ $\Lambda/\pi^+$ predicted by EPOS-4 therefore indicates that this model produces either relatively stronger radial push for baryons, a hadronization pattern that favours strange baryon formation at moderate $p_T$, or enhanced contributions from sources that feed Lambdas in this $p_T$ region. By contrast, SMASH, a hadronic transport model with particle production dominated by resonance dynamics and hadronic rescattering, does not have strong partonic/coalescence contributions and thus yields a lower $\Lambda/\pi^+$ at intermediate $p_T$. EPOS-LHC, which is tuned at much higher energies and with a reduced effective core fraction at these low collision energies, ends up close to SMASH. This clear separation of models in the intermediate-$p_T$ $\Lambda/\pi^+$ provides a good observable for near-term experiments at NICA (and similar facilities): a measurement of $\Lambda/\pi^+$ versus $p_T$ and centrality should give a definitive test of the competing model assumptions for hadronization and baryon transport.

\section{Conclusion}
This work presents a focused set of predictions for minimum-bias $Au+Au$ collisions at $\sqrt{s_{NN}} = 6, 7$, and $8~GeV$ using three different modelling philosophies (EPOS-LHC, EPOS-4, SMASH). Our conclusions, which are based on the simulation output itself, are:

\noindent \textbf{Pions as reference benchmark:} Charged-pion rapidity and $p_T$ spectra are almost identical in the case of EPOS-LHC, EPOS-4, and SMASH, for all three energies. In the presence of resonance feed-down (and the same treatment of resonance kinematics in the codes), pion observables are robust but poorly discriminating: pions are great for normalizing yields and probing global conservation. Still, they will not alone show the presence of partonic collectivity at early times.

\noindent \textbf{Strangeness ($K^{\pm},~\Lambda$):} Strange hadrons are strongly model dependent. EPOS-4 systematically predicts increased midrapidity $K^+$ and $\Lambda$ yields and harder $K^+$ and $\Lambda$ $p_T$ spectra, indicating increased pair/string production and early transverse push from the core. On the other hand, with the hadronic resonance-dominated production and in the presence of strong rescattering, SMASH gives rise to steeper (softer) spectra with more concentrated rapidity distributions in central regions. $K^-/K^+$ is still $<1$ at these energies but increases with energy and gets a relatively larger hard-$p_T$ contribution in EPOS models. These trends make kaon and $\Lambda$ measurements ($p_T$ spectra and rapidity shapes) among the very sensitive probes of non-hadronic dynamics in the NICA window.

\noindent \textbf{Baryons and baryon/meson ratios:}
The baryon-to-meson ratios have obvious and complementary model dependencies. For the $p/\pi^+$ ratio, a strong enhancement in the very first $p_T$ bin (very low $p_T$) (and thus a large number of very soft protons) is found in EPOS-4, while EPOS-LHC and SMASH start from smaller values. For $p_T$ at and above about $0.15~GeV/c$, the $p/\pi^+$ ratio increases with $p_T$ in all three models, but the increase is much more pronounced in SMASH, so that for intermediate and high $p_T$, SMASH predicts the largest $p/\pi^+$ values, with both the EPOS variants remaining at lower values. For the $\Lambda/\pi^+$ ratio, all three models provide small values at low $p_T$, the ratio increases with $p_T$ and reaches the largest values at $p_T$ of around $1.5 - 2~GeV/c$ in EPOS-4, which is systematically above EPOS-LHC, while SMASH provides the smallest values in this intermediate region. At the highest $p_T$ bins, SMASH is close to or slightly above the EPOS predictions. Taken together, these patterns show that EPOS-4 is relatively more favorable to the baryon production in the very soft and intermediate-$p_T$ region, whereas SMASH tends to enhance baryons with respect to mesons at higher values of the transverse momentum.

\noindent \textbf{Rapidity and $p_T-y$ mapping:} Two-dimensional $p_T$ versus rapidity maps exaggerate model differences: EPOS models distribute a significant yield to higher $p_T$ over a wider rapidity interval (notably, shoulders and plateaus near midrapidity), whereas SMASH yields are narrowly peaked at low $p_T$ and central rapidity \cite{werner2007core, weil2016particle}. Physically, the string fusion/core formation of EPOS and earlier freeze-out of harder parents result in wider longitudinal and transverse distributions; the hadronic rescattering of SMASH tends to trap yields around midrapidity and low $p_T$. These maps, therefore, are powerful diagnostics of when and where the transverse momentum is generated in the system.

\noindent \textbf{NCQ scaling ($v_2/n_q$ vs $p_T/n_q$):} Surprisingly, EPOS-LHC has the best approximate NCQ scaling of the three models, SMASH and EPOS-4 have partial scaling. This hierarchy leads to the conclusion that even in the baryon-rich $6-8~GeV$ range, a model that conveys quark-level anisotropy coherently into hadron formation (EPOS-LHC here) will reproduce the scaling signature often identified with the notion of partonic collectivity, while transport models like SMASH do partially. Thus, NCQ scaling (or lack thereof) at NICA energies is an essential discriminator for the experiment.

\noindent \textbf{Experimental recommendations:} Based on the model separations, we recommend that an early NICA physics program focus on: (a) accurate $\Lambda$ and $K^{\pm}$ spectra vs $p_T$ and rapidity (but with emphasis on the intermediate $p_T$ range $0.8-3~GeV/c$); (b) baryon/meson ratios $p/\pi^+$ and $\Lambda/\pi^+$ as function of $p_T$ and $y$; (c) $K^+/K^-$ as function of $p_T$ to constrain pair production and absorption; and (d) measurement of $v_2$ for several species and construction of NCQ-scaled plots. These observables will be most efficient for testing for the presence of the early collective effects and partonic-like hadronization mechanisms.

\noindent \textbf{Limitations and Outlook:} The present study compares three commonly used codes with their standard physics choices, without any data-driven retuning, at NICA energies. Differences between models can be attributed to conceptual physics choices (hadronic transport vs hybrid/string+hydro), and to implementation details (resonance lists, hadronic afterburners, hadronization recipes). Future data will need to be analysed systematically using studies on centrality, acceptance, and variations in model parameters. Future work will also include (i) a detailed look at centrality dependence, and (ii) studying variations of hadronization and afterburner setups, to try and determine which ingredients drive which observable.

In summary, the $ 6-8~GeV$ window has several observables that exhibit high discriminatory power between purely hadronic and hybrid/string/hydrodynamic pictures. The predictions presented here provide a practical and testable baseline for the NICA program and similar initiatives. Specific measurements of strange hadrons, baryon-to-meson ratios in the intermediate $p_T$ region, and NCQ scaling will be particularly conclusive.
\\

\noindent {\bf Acknowledgment:}
This research work was supported by Princess Nourah bint Abdulrahman University Researchers Supporting Project number (PNURSP2026R106), Princess Nourah bint Abdulrahman University, Riyadh, Saudi Arabia.
\\

\noindent {\bf Data availability:}
The data used to support the findings of this study are included within the article and are cited at relevant places within the text as references.
\\

\noindent {\bf Compliance with Ethical Standards:}
The authors declare that they have complied with the ethical standards regarding the content of this paper.
\\

\noindent {\bf Declaration of Interest Statement:}
We declare no conflicts of interest regarding the publication of this manuscript.
\\
\\

\noindent \textbf{Appendix A}

Fig.~A shows an illustrative benchmark of our results for minimum bias elliptic flow, comparing our results with published results from RHIC data in this same energy regime. We compare our model $v_2$ as a function of $p_T$ at $8~GeV$ with the minimum bias (0-80\%) Au+Au measurement made by the STAR experiment at $7.7~GeV$ \cite{adamczyk2013elliptic}. The energies are slightly different, and more importantly, the experimental result is obtained by implementing detector acceptance and analysis selection (including track quality selections, a given reconstruction event-plane procedure, and a definition of centrality based on reconstructed charged multiplicity), which is not reproduced in our generator-level workflow. Therefore, this comparison attempts to determine if the magnitude of the predicted $v_2$ and its dependence on the transverse momentum fall within a sufficiently wide range of reality for this low energy range, and is not intended as an analysis-in-matching tuned validation. Fig. A is included in the appendix so that we can cross-check the requested data-facing with the main focus of the paper being on the comparative model systematics at $6-8~GeV$ and across several identified hadron observables.
\begin{figure*}
\centering
\includegraphics[width=0.32\textwidth]{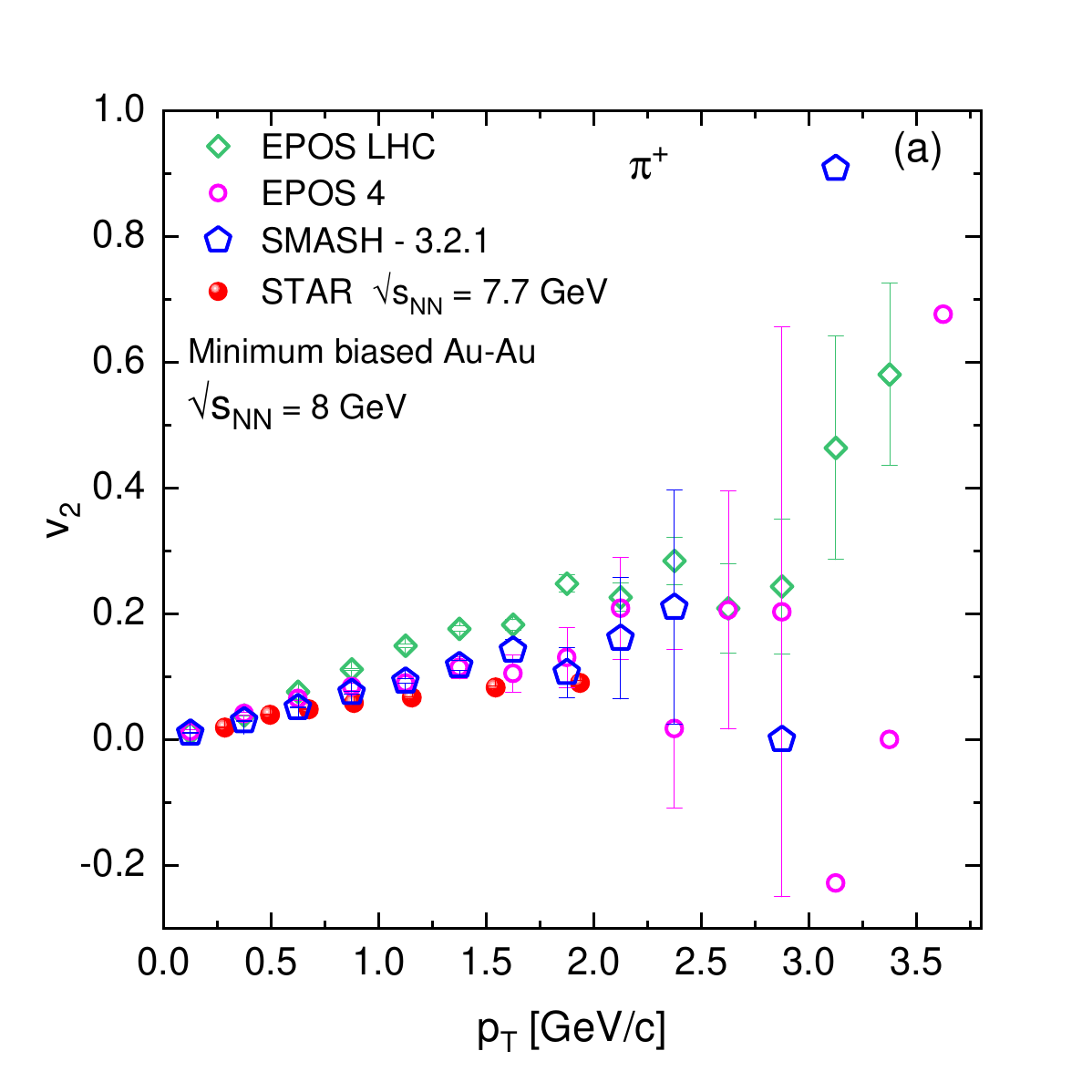}
\includegraphics[width=0.32\textwidth]{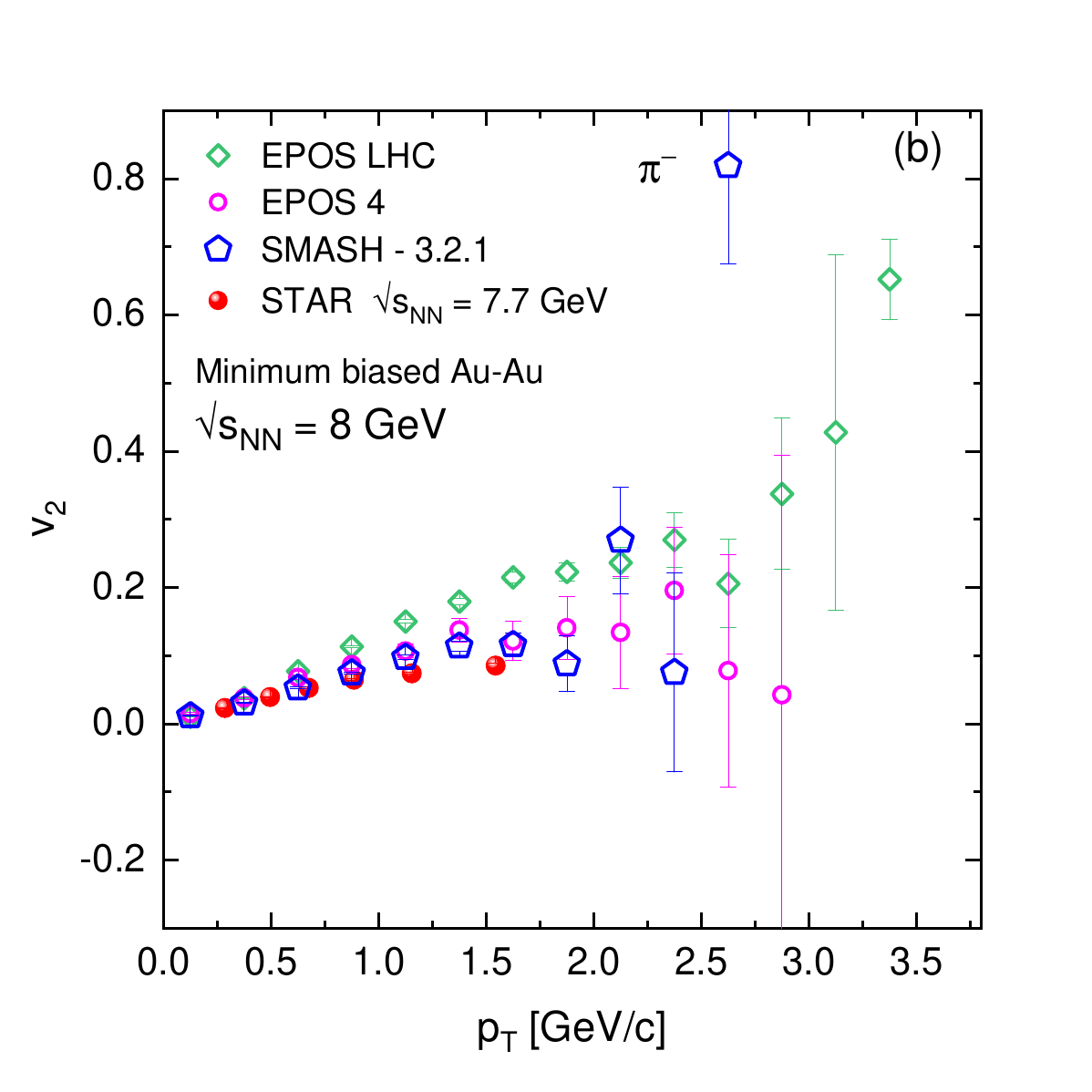}\vspace{-0.35cm} 
\includegraphics[width=0.32\textwidth]{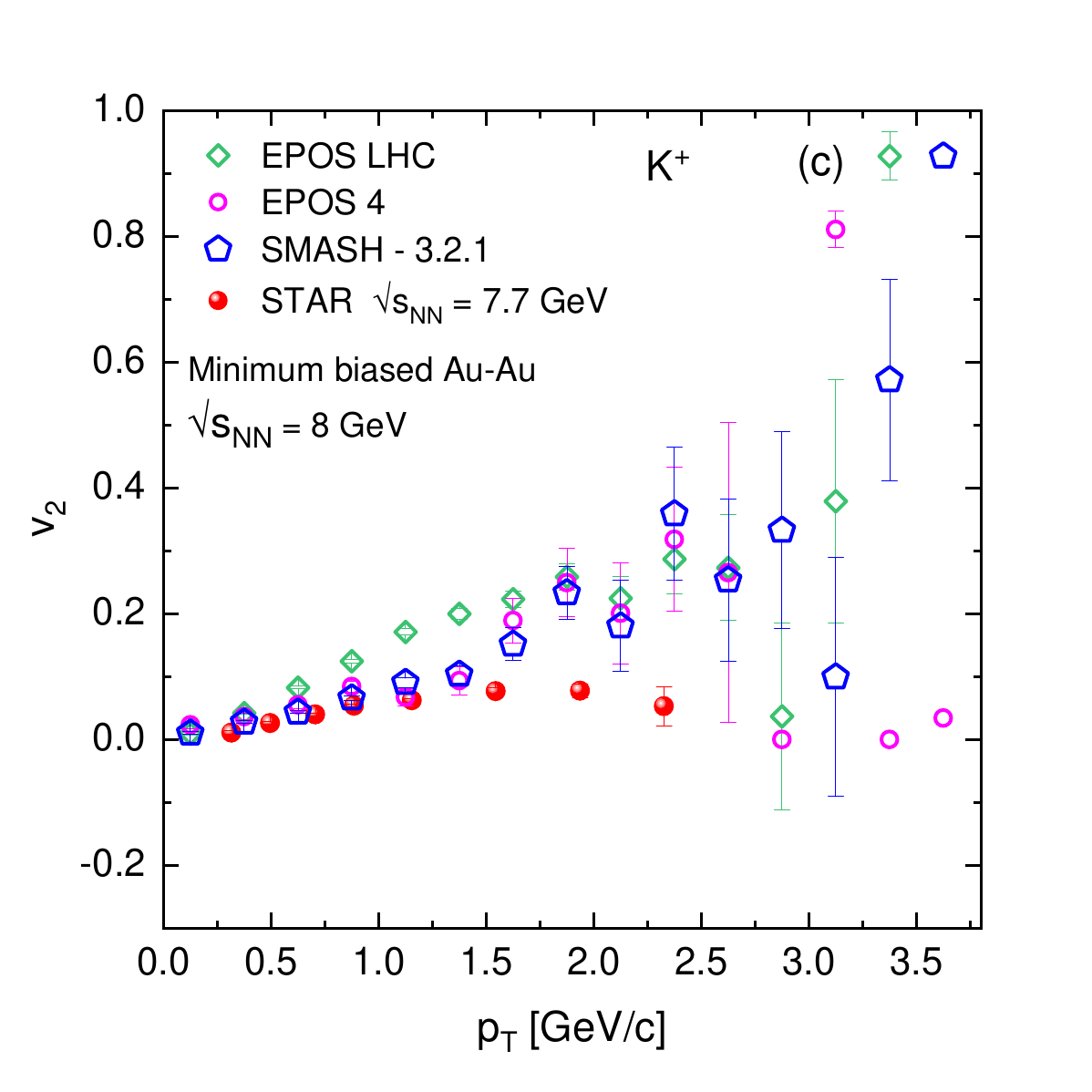}
\includegraphics[width=0.32\textwidth]{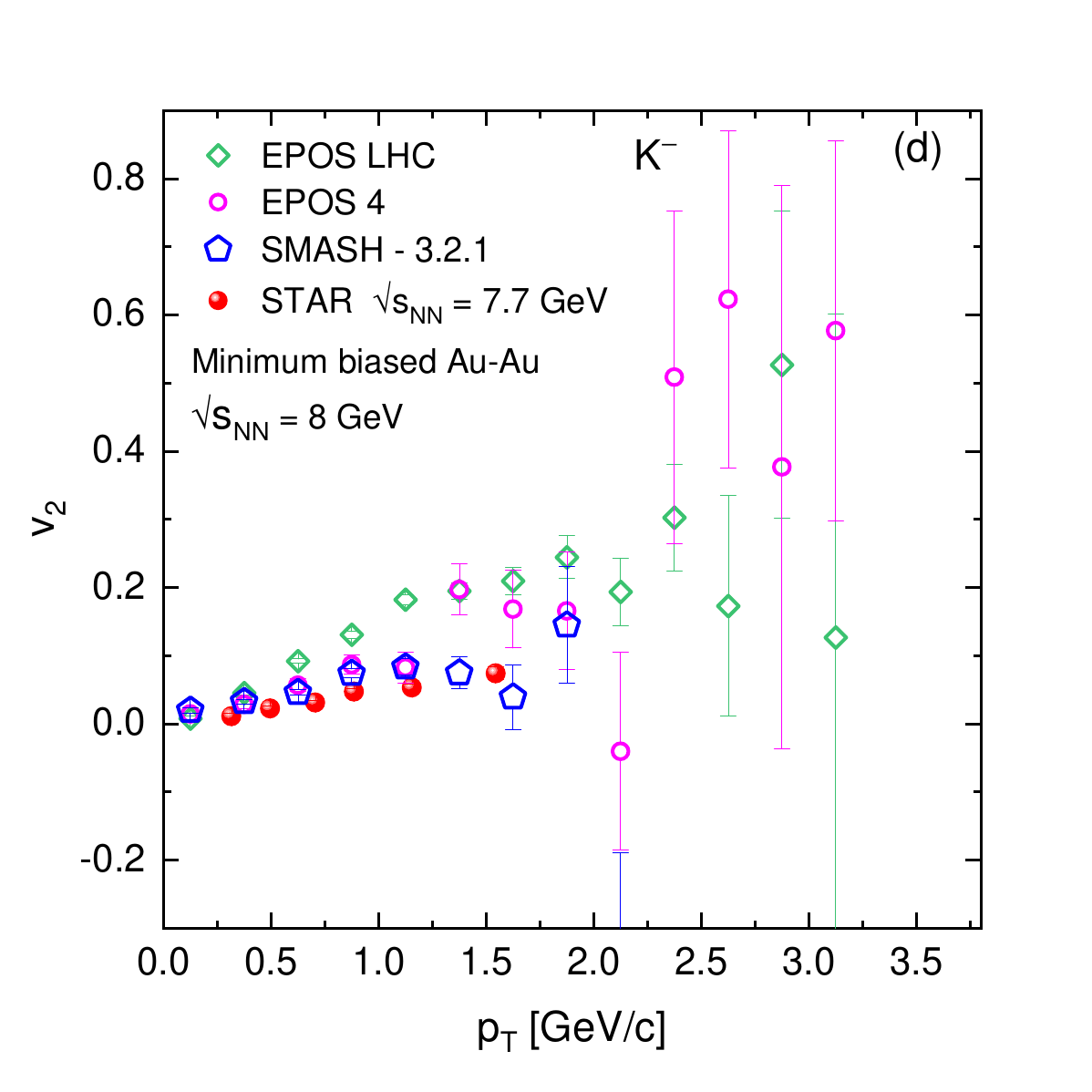}\vspace{-0.35cm}
\includegraphics[width=0.32\textwidth]{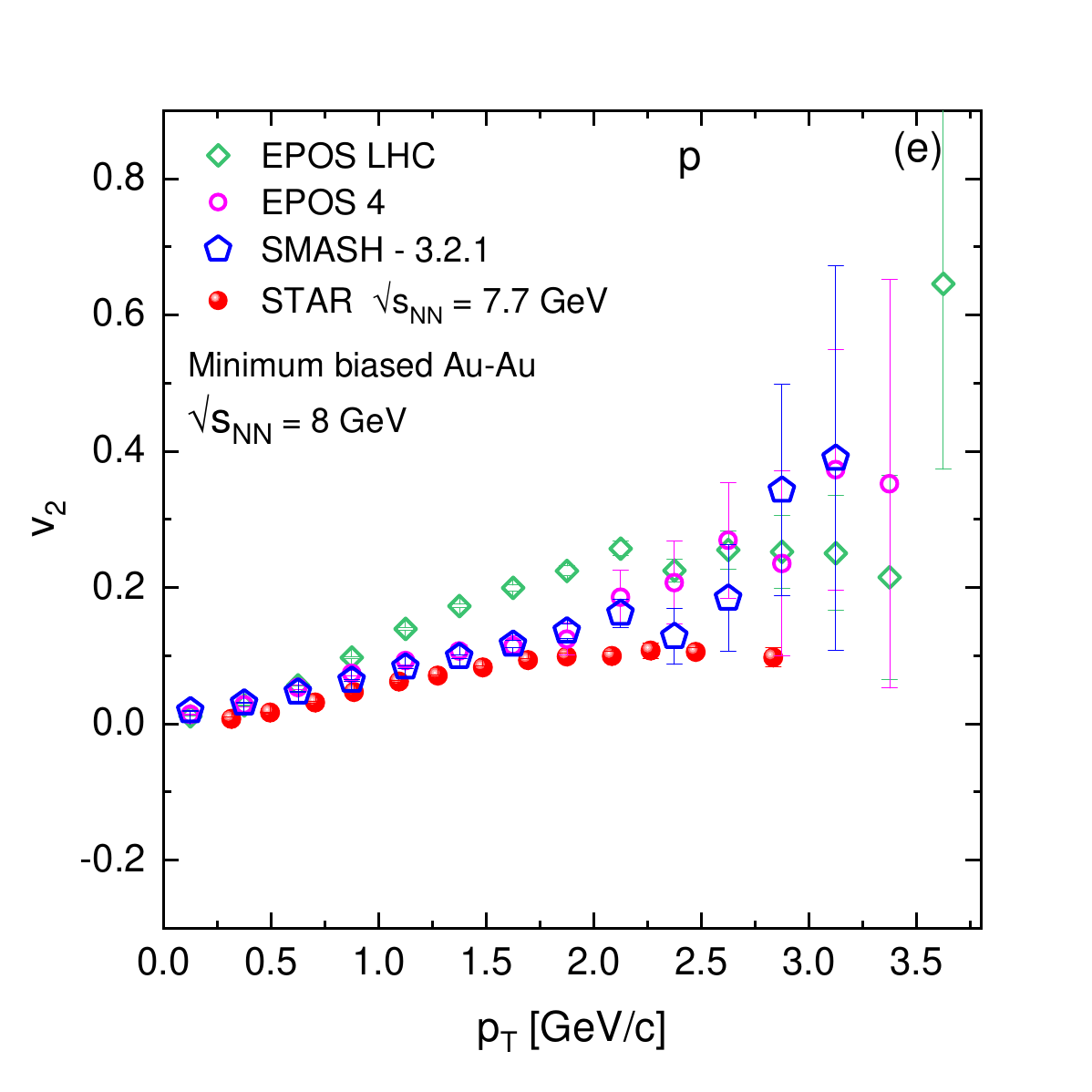}
\includegraphics[width=0.32\textwidth]{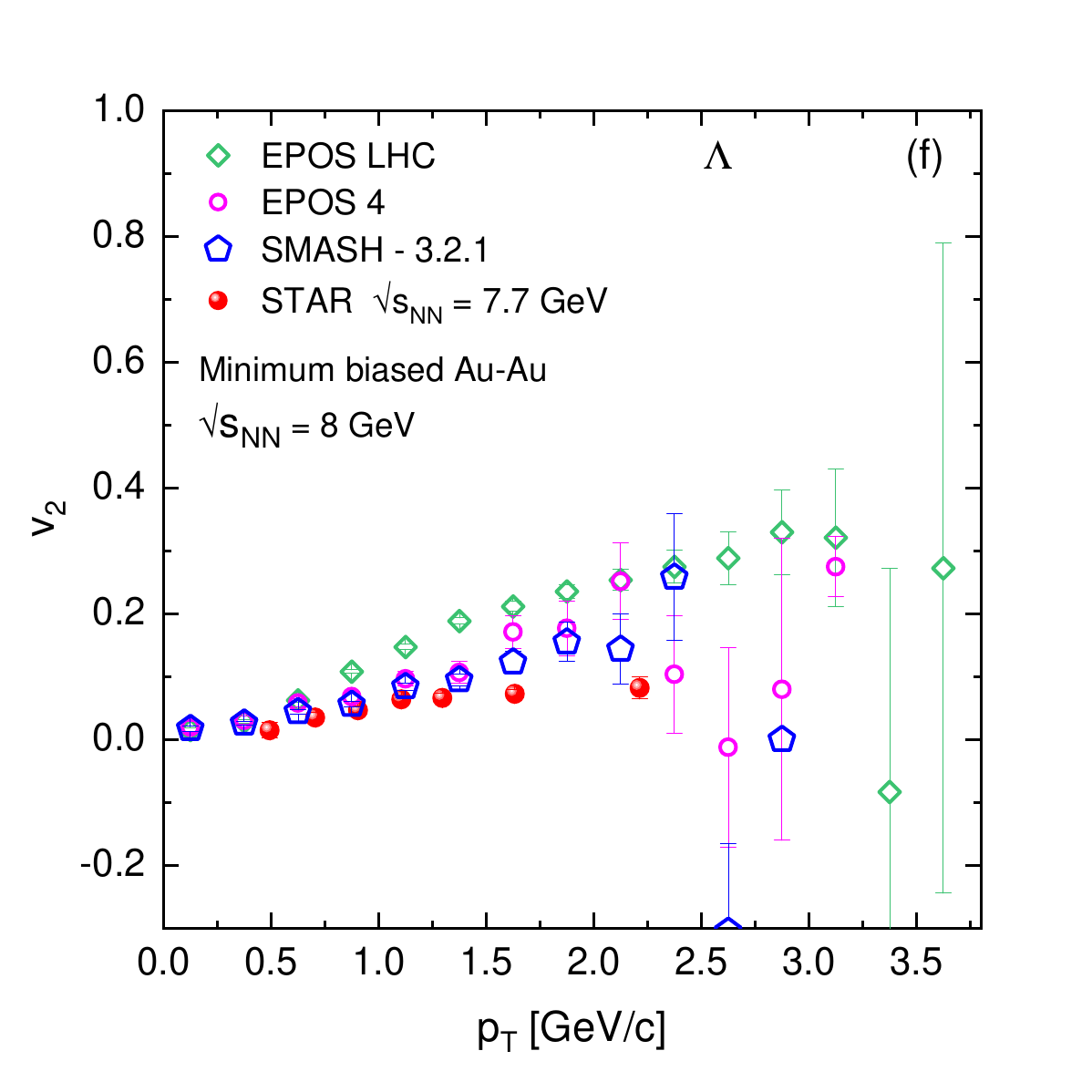}
\caption*{\textbf{Figure A:} Elliptic flow $v_2$ vs. transverse momentum for minimum bias $Au+Au$ collisions: model results at $8~GeV$ compared with the minimum bias (0-80\%) $Au+Au$ data at $7.7~GeV$ from the STAR \cite{adamczyk2013elliptic}. For the STAR points, the statistical and systematic uncertainties reported by the STAR were combined in quadrature for visualization purposes in this figure. The comparison here is qualitative as the STAR measurement incorporates detector acceptance, track quality selections, event plane reconstruction, and a multiplicity-based centrality definition, which are not used in our generator-level workflow.}
\end{figure*}
\bibliographystyle{aip}
\bibliography{references}
\end{multicols}
\end{document}